\RequirePackage[T1]{fontenc}
\documentclass[12pt]{article}

\usepackage[height=8.85in,width=6.45in]{geometry}

\usepackage[utf8]{inputenc}
\usepackage{amsmath}
\usepackage{amssymb}
\usepackage{mathtools}
\numberwithin{equation}{section}
\usepackage{slashed}
\usepackage{braket}
\usepackage[svgnames]{xcolor}
\usepackage[colorlinks,linktocpage=true,citecolor=DarkGreen,linkcolor=FireBrick,breaklinks=true]{hyperref}
\usepackage{cite}
\usepackage{graphicx}
\usepackage{tikz}
\usepackage{tikz-cd}
\usepackage{times}
\usepackage[scaled]{couriers}
\usepackage{bm}
\usepackage{subfig}
\usepackage{mathrsfs}

\usepackage{xcolor}
\usepackage{mdframed}

\renewenvironment{figure}[1][]{
  \begin{originalfigure}[#1]
    \begin{mdframed}[linecolor=black!0,backgroundcolor=black!1]
}{
    \end{mdframed}
  \end{originalfigure}
}

\def\arg{\mathop{\mathrm{arg}}}
\def\index{\mathop{\mathrm{index}}}

\def\spin{\text{spin}}

\def\CS{\mathrm{CS}}

\def\bZ{\mathbb{Z}}
\def\bR{\mathbb{R}}

\def\cD{\mathcal{D}}

\def\Arf{\mathrm{Arf}}

\def\vev#1{\langle #1 \rangle}
\def\braket#1#2{\langle#1|#2\rangle}

\DeclareMathOperator{\tr}{tr}
\def\Re{\mathop{\mathrm{Re}}}
\def\Im{\mathop{\mathrm{Im}}}

\def\Hom{\mathop{\mathrm{Hom}}}
\def\index{\mathop{\mathrm{index}}}
\def\signature{\mathop{\mathrm{\sigma}}}
\def\sign{\mathop{\mathrm{sign}}}
\usepackage{slashed}
\DeclareMathOperator{\pf}{pf}

\def\CA{{\cal A}}
\def\CB{{\cal B}}
\def\CC{{\cal C}}
\def\CD{{\cal D}}

\def\CG{{\cal G}}
\def\CH{{\cal H}}
\def\CI{{\cal I}}

\def\CK{{\cal K}}
\def\CL{{\cal L}}
\def\CM{{\cal M}}
\def\CN{{\cal N}}
\def\CO{{\cal O}}

\def\CQ{{\cal Q}}

\def\CS{{\cal S}}
\def\CT{{\cal T}}

\def\CZ{{\cal Z}}

\def\BA{{\mathbb A}}
\def\BB{{\mathbb B}}
\def\BC{{\mathbb C}}

\def\BR{{\mathbb R}}

\def\BZ{{\mathbb Z}}

\def\U{\mathrm{U}}
\def\SU{\mathrm{SU}}

\def\SO{\mathrm{SO}}

\def\Spin{\mathrm{Spin}}

\def\SL{\mathrm{SL}}
\def\Mp{\mathrm{Mp}}
\def\GL{\mathrm{GL}}

\def\spin{\text{spin}}

\def\beq#1\eeq{\begin{align}#1\end{align}}

\def\Ch{\overline{\Gamma}}
\def\d{{\rm d}}
\def\i{{\mathsf i}}
\def\DA{{\mathsf A}}
\def\DF{{\mathsf F}}
\def\DN{{\mathsf N}}
\def\Da{{\mathsf a}}
\def\Df{{\mathsf f}}
\def\Dn{{\mathsf n}}
\def\DV{{\mathsf V}}
\def\DB{{\mathsf B}}
\def\DC{{\mathsf C}}

\def\DG{{\mathsf G}}
\def\DU{{\mathsf U}}
\def\Dp{\star}
\def\hodge{*}
\def\ch{{\rm ch}}
\def\L{{\mathsf L}}
\def\Sw{{\mathsf w}}
\def\Sq{{\mathsf q}}
\def\SC{{\mathsf C}}
\def\SR{{\mathsf R}}
\def\Arf{ \mathrm{Arf}}
\def\M{\mathrm{M}}

\begin{document}

\begin{titlepage}

\begin{flushright}
IPMU-20-0028\\
TU-1098
\end{flushright}

\vskip 3cm

\begin{center}

{\Large \bfseries  Anomaly inflow and $p$-form gauge theories}

\vskip 1cm
Chang-Tse Hsieh$^{1,2}$, Yuji Tachikawa$^1$ and Kazuya Yonekura$^3$
\vskip 1cm

\begin{tabular}{ll}
$^1$&Kavli Institute for the Physics and Mathematics of the Universe, \\
& University of Tokyo,  Kashiwa, Chiba 277-8583, Japan\\
$^2$&Institute for Solid State Physics, University of Tokyo, Kashiwa, Chiba 277-8581, Japan \\
$^3$&Department of Physics, Tohoku University, Sendai 980-8578, Japan
\end{tabular}

\vskip 1cm

\end{center}

\noindent
Chiral and non-chiral $p$-form gauge fields have gravitational anomalies and anomalies of Green-Schwarz type.
This means that they are  most naturally realized as the boundary modes of bulk topological phases in one higher dimensions. 
We give a systematic description of the total bulk-boundary system which is analogous to the realization of a chiral fermion on the boundary of a massive fermion.
The anomaly of the boundary theory is given by the partition function of the bulk theory, which we explicitly compute in terms of the Atiyah-Patodi-Singer $\eta$-invariant.

We use our formalism to determine the $\SL(2,\bZ)$ anomaly of the 4d Maxwell theory.
We also apply it to study the worldvolume theories of a single D-brane and an M5-brane in the presence of
orientifolds, orbifolds, and S-folds in string, M, and F theories.

In an appendix we also describe a simple class of non-unitary invertible topological theories whose partition function is not a bordism invariant, illustrating the necessity of the unitarity condition in the cobordism classification of the invertible phases.

\end{titlepage}

\setcounter{tocdepth}{2}

\newpage

\tableofcontents

\section{Introduction and summary}

Non-chiral and chiral $p$-form gauge fields in $d$ dimensions are known to have various anomalies.\footnote{
In this paper, $d$ is the spacetime dimension on which an anomalous theory resides.
The corresponding bulk topological phase correspondingly has the spacetime dimension $d+1$.} 
Let us recall some well-known examples:
\begin{enumerate}
\item A compact scalar can be regarded as a 0-form field. 
In two dimensions, it has two $\U(1)$ symmetries corresponding to the momentum and the winding number in the $S^1$ target space.
There is a mixed anomaly between them.
The compact scalar is dual to a free fermion when the radius of the compact scalar is appropriately chosen.
Then the mixed anomaly of the compact boson can be identified with the mixed anomaly of the vector and axial $\U(1)$ symmetries of a free fermion, see e.g.~\cite{AlvarezGaume:1987vm}.\footnote{%
This classic question was recently revisited in \cite{Thorngren:2018bhj,Yao:2019bub}.
}
\item The 2-form field in ten-dimensional $\CN=1$ supergravity theories with $E_8 \times E_8$ or $\SO(32)$ gauge groups contributes to the gauge and gravitational anomalies via the Green-Schwarz mechanism~\cite{Green:1984sg}.
\item A four-dimensional free Maxwell field is a 1-form gauge field. It has $\U(1)$ 1-form electric and magnetic symmetries~\cite{Gaiotto:2014kfa}. There is a mixed anomaly between these 1-form symmetries.
\item A chiral compact scalar is dual to a chiral fermion in two dimensions. It has gravitational anomaly as well as $\U(1)$ anomaly.
\item A 2-form chiral field in six dimensions gives a generalized Green-Schwarz contribution to gauge and gravitational anomalies~\cite{Green:1984bx,Sagnotti:1992qw}.
\item In general, $p$-form chiral fields in $d=2p+2$ dimensions for even $p$ have gravitational anomalies and anomalies of Green-Schwarz type. 
The perturbative gravitational part was determined in~\cite{AlvarezGaume:1983ig},
and the global part was investigated in a series of papers \cite{Monnier:2010ww,Monnier:2011mv,Monnier:2011rk,Monnier:2012xd,Monnier:2013kna,Monnier:2013rpa,Monnier:2014txa,Monnier:2016jlo,Monnier:2017klz,Monnier:2017oqd,Monnier:2018nfs,Monnier:2018cfa} by S.~Monnier and his collaborators.
\item When $p$ is odd, a single $p$-form gauge field does not allow chirality projection,
but once we consider the duality action it effectively becomes chiral and has anomalies.
For example, the electromagnetic duality, or more generally the $\SL(2,\BZ)$ duality group, of Maxwell theory in 4 dimensions has anomalies~\cite{Seiberg:2018ntt,Hsieh:2019iba}.
\end{enumerate}
The purpose of this paper is to give a systematic treatment of  these theories and their anomalies, including global anomalies as well as perturbative anomalies, when they can be formulated on spin manifolds.\footnote{
Among the theories listed above, the Green-Schwarz mechanism in 10-dimensions is very subtle at the nonperturbative level. See \cite{Freed:2000ta} for the case of Type I superstring theory.
We do not discuss this case in detail in this paper, but we will discuss an analog of it in Type IIB superstring theory.
}
In particular, we give a careful definition of a chiral $p$-form field in $d=2p+2$ dimensions for $d=2$, $6$ and $10$ on spin manifolds, 
which leads to a precise computation of the anomaly.
The essential idea is to use a Chern-Simons-type bulk theory in $d+1$ dimensions on a space with a boundary, where the chiral $p$-form field resides as a boundary mode.
The bulk $d+1$-dimensional theory is essential to make the theory well-defined via anomaly inflow mechanism~\cite{Callan:1984sa,Faddeev:1985iz}.
We also discuss some applications of the formalism to string theories, M-theory, and F-theory.

\subsection{Outline}
We would like to give an outline of the discussions in this paper. 
We neglect many subtle but important details here, and only try to give an overall picture of the paper.

\paragraph{$p$-form gauge fields as boundary modes at the level of differential forms:}

Examples listed above are either non-chiral or chiral fields. 
However, a non-chiral $p$-form field $B$ can be described as a chiral theory
if we include both $p$-form field $B_1$ and $(d-p-2)$-form field $B_2$ in the theory, and impose the self-duality condition of the schematic form $\d B_1 \sim \hodge \d B_2$,
where $\hodge$ is the Hodge dual. Therefore, the general case is a chiral theory.
In the following discussions, we only talk about $p$-form fields, but there is also another $d-p-2$-form field in the case of a non-chiral theory.

If the theory has a $p$-form field $B$, it has a $\U(1)$ $(d-p-2)$-form symmetry whose conserved current is given by $j \sim \hodge \d B$.
There is another $\U(1)$ $p$-form symmetry whose conserved current is $\d B$, but for a chiral field $\d B \sim \hodge \d B$ (or $\d B_1 \sim \hodge \d B_2$),
it is equivalent to a current of the form $\hodge \d B$. Let $C$ be the background $(d-p-1)$-form field of the symmetry.
The coupling between $B$ and $C$ is schematically given by $\int C \wedge \d B$.
In the Green-Schwarz mechanism and its generalization, the anomalies are produced by taking the background $C$ to be a Chern-Simons form
of gauge and gravitational fields. Thus, once we obtain a complete description of the anomalies of the 
$(d-p-2)$-form symmetry, we immediately get the complete understanding of the Green-Schwarz mechanism. 

Therefore, what we need to understand is a description of a chiral field $B$ which is coupled to a general higher-form background field $C$.
Such a theory can have an anomaly of the higher-form symmetry as well as gravity. 
In the modern understanding (see e.g.~\cite{Freed:2014iua,Monnier:2019ytc}), an anomalous theory is most naturally realized as a boundary mode of a $(d+1)$-dimensional
symmetry protected topological (SPT) phase or invertible field theory in $(d+1)$ spacetime dimensions. 
Here, an invertible field theory, originally introduced in \cite{Freed:2004yc}, is characterized by the condition that its Hilbert space on any closed manifold is one-dimensional.
Therefore, to define an anomalous theory in $d$ dimensions, we seek a bulk theory in $(d+1)$ dimensions and the behavior of the theory on a manifold $Y$ with boundary $X = \partial Y$.
However, we remark that many of the following discussions are also applicable to the cases of non-invertible theories, or in other words topologically ordered phases. 
These cases are also interesting in the context of fractional quantum Hall effects and six-dimensional superconformal field theories.

The bulk $(d+1)$-dimensional theory is given by a $(p+1)$-form field $A$ with the following schematic action in Euclidean signature:
\beq
- S  \sim  2\pi  \int_Y \left(  -\frac{1}{2e^2} \d A \wedge \hodge \d A+     \i \frac{\kappa}{2} A \wedge \d A +\i  C \wedge \d A \right).\label{eq:schematicS}
\eeq
Here $A$ is a dynamical field, and $C$ is the background field of the higher-form symmetry.
$e^2$ is a positive parameter which we will take to be very large, $e^2 \to \infty$. 
$\kappa \in \BZ$ is an integer parameter.
The bulk theory is the generalization of the topologically massive gauge theory of Jackiw-Deser-Templeton \cite{Deser:1981wh,Deser:1982vy} to higher dimensions.
The equation of motion is given by
\beq
(-1)^{p+1 } \d \hodge \d A +  \i \kappa e^2 \d A  =0.
\eeq
 
On this theory, we impose the boundary condition, which we denote as $\L$, that the restriction of $A$ to the boundary should vanish:
\beq
\L:\, A|_{\partial Y} =0.\label{eq:LBintro}
\eeq
Under this boundary condition, we can find a localized chiral field. 
Let $\tau \leq 0$ be the coordinate which is orthogonal to the boundary. The boundary is at $\tau=0$ and the bulk is $\tau < 0$. 
Under the above boundary condition $\L$, there is a localized solution of the form
\beq
A =  \d ( e^{|\kappa e^2| \tau}  )   \wedge B,
\eeq
where $B$ is a $p$-form which depends only on the coordinates of the boundary. For this to be a solution of the equations of motion, $B$ is required to satisfy
\beq
\hodge \d B +  \i (-1)^{p+1}  {\rm sign}(\kappa) \d B = 0, \qquad
\d ( \hodge \d B ) = 0. \label{eq:Emode}
\eeq
These are precisely the equations of a chiral $p$-form field in $d$ dimensions. (The imaginary unit $\i$ is just an artifact of the Euclidean signature metric.)
In this way, we can realize the chiral $p$-form field as a boundary mode of a bulk $(d+1)$-dimensional theory \eqref{eq:schematicS}.

This realization is mostly in parallel to that of a chiral fermion as a boundary mode of a bulk massive fermion under a local boundary condition.
See \cite{Witten:2019bou} for a systematic treatment of such a fermion system and the anomaly.
The above discussion is valid for any value of nonzero $\kappa$.
The choice $\kappa = \pm 1$ gives an invertible field theory in the bulk and an anomalous theory on the boundary.
For more general $\kappa$, we get what is called relative field theory on the boundary~\cite{Freed:2012bs},
generalizing the Chern-Simons/Wess-Zumino-Witten correspondence, which was described e.g.~in \cite{Dunne:1990jh,Dunne:1990hh,Gukov:2004id} in $d=2$.

We note that in a holographic context, this program of realizing a $p$-form field on a boundary of a bulk theory which is massive was studied in previous works.
In \cite{Aharony:1998qu,Witten:1998wy,Belov:2006jd}, the bulk was taken to be topological field theory with boundary conditions consistent with the topological theory.
The importance of keeping the bulk mass very large but finite with different boundary conditions to realize an additional $\U(1)$ field was emphasized in \cite{Gukov:2004id,Belov:2004ht}.
In this paper we simply consider the case that the boundary $X = \partial Y$ of the bulk $Y$ is at finite distance,
and the boundary condition \eqref{eq:LBintro} produces a massless $p$-form gauge field propagating on the boundary.
This is sufficient for our purpose of the study of $p$-form gauge fields.

We also note that there have been many attempts to define the chiral $p$-form theory only by using the manifold $X$ of dimension $d$ without taking the manifold $Y$ of dimension $d+1$, 
see e.g.~\cite{Pasti:1996vs,Sen:2019qit} and many others.
But each attempt has its own merits and demerits. 
When we would like to analyze the anomaly of a chiral $p$-form theory, in particular at the non-perturbative level,
we need to introduce the manifold $Y$ whose boundary is $X$.
Our point is that we can then use a massive theory with the action  \eqref{eq:schematicS}  on $Y$ to give a satisfactory definition of the chiral $p$-form theory on $X$.
From this point of view, it is not surprising that it is extremely difficult to define the theory only by using $X$ without taking $Y$;
basically the obstruction to defining the theory by using only $X$ is the anomaly of the theory.

\paragraph{Nontrivial topology and the necessity of quadratic refinements:}

The above analysis was done at the level of differential forms. 
More precisely, the gauge field $A$ can have nontrivial topology and it is not just a differential form.
A 0-form field is a compact scalar with periodicity $A \sim  A+1$. (In this paper we normalize fields to have integer values for flux integrals.)
A 1-form field is a $\U(1)$ connection (multiplied by $\frac{\i}{2\pi}$). 
The generalization to general $p$-forms is known under the name of differential cohomology \cite{CS,Hopkins:2002rd}.
By using this formalism, we can make sense of the pairing between two fields $A$ and $B$ which is schematically given by
\beq
(A, B) = \int A \wedge \d B \in \BR/\BZ.
\eeq
We call it the differential cohomology pairing. It takes values in $\BR/\BZ$ as in the usual Chern-Simons invariant.
We cannot define it as a real number in $\BR$;
in other words, the integer part is ambiguous by gauge transformations. 

The coupling between the dynamical field $A$ and the background field $C$ in \eqref{eq:schematicS} is well-defined.
The problem is the second term in \eqref{eq:schematicS}, which is $2\pi \i $ times $ \frac{\kappa}{2} A \wedge \d A$.
If $\kappa$ is even, this term also makes sense by using the pairing $(A,A)$. 
However, the dimension of the Hilbert space of the theory \eqref{eq:schematicS} on a closed $d$ dimensional manifold $X$
behaves roughly like $|\kappa|^{\frac{1}{2}\dim H^{p+1}(X)}$, or $|\kappa|^{\dim H^{p+1}(X)}$ for a non-chiral theory. Therefore, to realize an invertible field theory,
we need to take $|\kappa|=1$. Then we have to make sense of one half of $(A,A) $ as an element in $ \BR/\BZ$. 
Let us denote it as $\CQ(A)$,
\beq
\CQ(A) \sim  \frac{1}{2} \int A \wedge \d A  .
\eeq
It is characterized by the property that 
\beq
\CQ(A+B) - \CQ(A) - \CQ(B) + \CQ(0) = (A,B).
\eeq
We call such a $\CQ$ as a quadratic refinement of the differential cohomology pairing. 
It is required to make the action \eqref{eq:schematicS} well-defined.

The importance of using quadratic refinements to analyze chiral $p$-form field theories was recognized in \cite{Witten:1996hc} and further studied in the literature. 
See e.g.~\cite{Witten:1999vg,Moore:1999gb,Freed:2000ta,Hopkins:2002rd,Gukov:2004id,Belov:2006jd,Monnier:2016jlo,Monnier:2017klz,Monnier:2017oqd,Monnier:2018nfs,Monnier:2018cfa}
for a partial list. 
We would like to emphasize that the quadratic refinement is simply necessary
for us to write down the  action of the bulk theory \eqref{eq:schematicS}.

\paragraph{Quadratic refinements from spin structures in $d+1=3,7,11$:}

To formulate a quadratic refinement for general dimensions of the form $d+1=4\ell+3$ with arbitrary $\ell$,
one needs what is called a Wu structure~\cite{Hopkins:2002rd,Monnier:2016jlo}.
However, for dimensions relevant to applications to string theories and condensed matter physics, 
the spacetime dimensions $d$ can be restricted to $d+1\leq 11$.
In this range of $d$, a spin structure of manifolds  gives a canonical quadratic refinement
whose expression is also naturally motivated by string theory considerations.
These are the quadratic refinements we use in the present paper.
 
Let us briefly comment on them. 
They all involve the Atiyah-Patodi-Singer (APS) $\eta$-invariant \cite{Atiyah:1975jf,APS2,APS3} in one way or another.

For $d+1=3$, the bulk field $A$ is just a 1-form $\U(1)$ field.
Then we can use the APS $\eta$-invariant of the Dirac operator coupled to $\U(1)$ for the definition of $\CQ(A)$. 

For $d+1=7$, we can motivate the fact that the quadratic refinement follows from the spin structure by the following M-theory consideration. 
A chiral 2-form field is realized on an M5-brane, and if we put the M5-brane on top of the Ho{\v r}ava-Witten wall~\cite{Horava:1996ma,Horava:1995qa},
we get the E-string theory~\cite{Ganor:1996mu,Seiberg:1996vs} which is a strongly coupled superconformal theory.
It has two different vacuum moduli spaces, one of which describes the M5-brane away from the wall, and the other describes the moduli space of an $E_8$-instanton.
The instanton breaks $E_8$ to $E_7$, and there are chiral fermions in the 56-dimensional representation of $E_7$.
Then the anomaly of the chiral 2-form is matched with the anomaly of the chiral fermions.
Moreover, the 3-form potential in M-theory, which plays the role of the background field $C$ for the 2-form chiral field $B$,
can be related to the Chern-Simons 3-form of the $E_8$  gauge group \cite{Witten:1996md,Diaconescu:2003bm}, or its $E_7$ subgroup.
By using these facts, it is possible to define the quadratic refinement by using the $\eta$-invariant of the Dirac operator coupled to the 56-dimensional representation of $E_7$.

For $d+1=11$, our consideration is relevant for RR-fields in string theory, including the 4-form chiral field in Type IIB string theory.
RR-fields are described by K-theory~\cite{Minasian:1997mm,Witten:1998cd,Moore:1999gb}.  
Elements of K-theory groups are basically vector bundles, and we can define the quadratic refinement by using the $\eta$-invariant of a Dirac operator
coupled to an appropriate K-theory element.

For $d+1=5$, we can consider a Maxwell theory as a chiral theory if there is nontrivial $\SL(2,\BZ)$ duality background. 
The anomaly of this theory is most naturally described by the $T^2$ compactification of the 2-form chiral field in six dimensions. 

\paragraph{Computation of the bulk partition function:}
Let us restrict our attention to the case $\kappa = \pm 1$.
Having defined the quadratic refinement, the theory \eqref{eq:schematicS} now has an explicit action.
By the modern general understanding, the anomaly of the chiral theory $B$ which appears on the boundary is characterized by
the partition functions of the bulk theory on closed manifolds. In the low energy limit, we neglect the first term in \eqref{eq:schematicS}.
At the one-loop level, the partition function of the theory \eqref{eq:schematicS} on a $(d+1)$-dimensional closed manifold $Y$ turns out to be given by
\beq
\CZ_{\text{1-loop} } \sim \exp 2\pi \i \left( -\frac{ \kappa}{2} \int_Y C \wedge \d C   -\frac{ \kappa}{8} \cdot   2\eta(\widetilde{\CD}^{\rm sig}_Y)  \right),
\eeq
where the operator $\widetilde{\CD}^{\rm sig}_Y$ is related to the second term of \eqref{eq:schematicS} as $A \wedge \d A \sim A \wedge \hodge (\widetilde{\CD}^{\rm sig}_Y A)$,
and $\eta(\widetilde{\CD}^{\rm sig}_Y)$ is the associated $\eta$-invariant. 
The term proportional to $\widetilde{\CD}^{\rm sig}_Y$ comes from the one-loop determinant
of the kinetic term $A \wedge \d A  $, after taking the limit $e^2 \to \infty$. 
This is analogous to the case that the partition function of a massive fermion is given by the corresponding $\eta$-invariant~\cite{AlvarezGaume:1984nf,Witten:2015aba}.

The APS index theorem states the following. Suppose that the manifold $Y$ is a boundary
of one-higher dimensional manifold $Z$, $\partial Z =Y$. Then the signature $\sigma(Z)$ of $Z$ (which is a topological invariant of $Z$) is given by
\beq
\signature(Z)  = \int_Z L +  2 \eta( \widetilde{\CD}^{\rm sig}_Y),
\eeq
where $L$ is the Hirzebruch $L$-polynomial, which is a polynomial of the Riemann curvature such that $\int_Z L$ gives the signature of $Z$ if $Z$ is closed. 
Therefore, by neglecting the topological invariant $\sigma(Z)$ for the moment, the one-loop partition function may be written as
\beq
\CZ_{\text{1-loop} } \sim \exp\left(   2\pi \i  \kappa \int_Z \left( - \frac{1}{2} \d C \wedge \d C   + \frac{ 1}{8} L \right) \right).\label{eq:1loopresult}
\eeq
The second term $\frac{ 1}{8} L$ is precisely the perturbative gravitational anomaly of the chiral $p$-form field obtained by \'Alvarez-Gaum\'e and Witten \cite{AlvarezGaume:1983ig}.
The first term is the perturbative anomaly of the higher-form symmetry of the Green-Schwarz type. 
Notice that whether the boundary mode $B$ is self-dual or anti-self-dual
depends on the sign of $\kappa$ as can be seen in \eqref{eq:Emode}. Corresponding to this fact,
the above anomaly also depends on the sign of $\kappa$.

At the non-perturbative level, the above analysis must be done more carefully, the detail of which  will be presented  in this paper. 
We find\footnote{%
Basically the same formula was found previously by S. Monnier and his collaborators in a series of papers where the spacetime was equipped with Wu structure. 
We will only use the spin structure in the following. 
More comments on this point will be given in the paragraph preceding Sec.~\ref{sec:PT1}.
} that the complete expression of the anomaly is given by \begin{equation}
\CZ(Y) = \exp\left(2\pi \i \kappa\left( -   \widetilde{ \CQ}(\check C)     - \frac{1}{8} \cdot 2\eta(\widetilde{\CD}^{\rm sig}_Y) +  \Arf_\Sw(Y)  
\right)\right),
\label{mainFormula}
\end{equation}
where $\widetilde{\CQ}(\check C):=\CQ(\check C)-\CQ(0)$,
$\eta(\widetilde{\CD}^{\rm sig}_Y)$ is the $\eta$-invariant as above,
and $\Arf_\Sw$ is a correction term which arises from a sum over the torsion fluxes.
We remark that $\check C$ here is not exactly the same as the $C$ appearing in \eqref{eq:1loopresult}.
In the case $d=6$, they are different by a shift by a quantity constructed from the metric tensor.
Consequently, the anomaly at $\check C =0$ is not the same as $  \frac{1}{8}L$ even at the perturbative level.
The additional metric dependence is incorporated in $ \Arf_\Sw(Y) $. 
At the nonperturbative level, there seems to be no canonical zero point in the space of $C$ (i.e.~the point which can be regarded as $C=0$).
In applications to M-theory, it is related to the phenomenon of shifted quantization found in \cite{Witten:1996md}.

We will show that the last two terms combine to simplify and gives:
\beq
- \frac{1}{8} \cdot 2\eta(\widetilde{\CD}^{\rm sig}_Y) +  \Arf_\Sw(Y) = \left\{
  \begin{array}{ll}
 \eta(\CD_Y^{\rm Dirac} ),& (d+1=3), \\
 28\eta(\CD_Y^{\rm Dirac}), & (d+1=7) , \\
 -\eta(\CD_Y^{{\rm Dirac}\otimes TY})+3\eta(\CD_Y^{\rm Dirac} ), & (d+1=11). \\
 \end{array} \right.
 \label{eq:simp-intro}
\eeq
where $\CD_Y^{\rm Dirac}$ is the usual Dirac operator without coupling to additional bundles
and $\CD_Y^{{\rm Dirac}\otimes TY}$ is the Dirac operator on the spinor bundle tensored with the tangent bundle.
The simplification happens due to interesting physics in each dimension: 
in $d=2$, a chiral boson is equivalent to a chiral fermion;
in $d=6$, a chiral $2$-form can be continuously deformed to 28 fermions using the E-string;
and in $d=10$, the anomaly cancellation of the Type IIB string says that the anomaly of a chiral $4$-form cancels against the contribution of a spin-$3/2$ fermion and a spin-$1/2$ fermion.

More abstractly, our computation can be formulated as follows.
The low-energy limit of the bulk action is  \begin{equation}
-S = 2\pi \i ( \kappa \widetilde\CQ(\check A)  + (\check A, \check C)).
\end{equation} 
We can then complete the square by shifting the dynamical field by defining $\check A':=\check A+\kappa \check C$:
\begin{equation}
-S = 2\pi \i \kappa (\widetilde\CQ(\check A') - \widetilde \CQ(\check C)).
\end{equation}
Therefore, we have the factorization \begin{equation}
\CZ(Y,\check C) = \exp(-2\pi \i \kappa \widetilde \CQ(\check C) ) \CZ(Y,\check C=0).
\end{equation}
Now, $\CZ(Y,\check C=0)$ is the partition function of a $(d+1)$-dimensional theory 
which is invertible and depends only on the spin structure.
From the general results of \cite{Freed:2016rqq,Yonekura:2018ufj},
such a theory is uniquely determined by its anomaly polynomial,
since the spin bordism group $\Omega_{d+1}^\text{spin}=0$ in dimensions $d+1=3,7,11$ are zero.
Then a careful examination of the metric dependence gives the result
\eqref{eq:simp-intro}.

\paragraph{Applications:}
The formulas presented above can be used to study the consistency of various string/M/F theory  backgrounds.
In particular, we will discuss the following applications. Consider a brane with dimension $d$, such as D$(d-1)$-branes and M5-branes for which $d=6$.
It is coupled to some $d$-form field under which the brane is electrically charged. We denote the curvature of that  $d$-form field as $F_{d+1}$. This is a $(d+1)$-form field.
A naive application of the Dirac quantization condition 
implies that $F_{d+1}$ must obey the quantization condition $\int_Y  F_{d+1} \in \BZ$, where $Y$ is a $(d+1)$-dimensional closed submanifold of the spacetime.
However, it has been observed for decades that the fluxes $\int_Y F_{d+1}$ in various string, M and/or F-theory backgrounds obtained from stringy considerations are not integers, against the requirement from the naive Dirac quantization condition.

The point is that, in the presence of the anomaly of the worldvolume theory, the Dirac quantization condition is modified. 
We denote the partition function of the bulk invertible field theory on $Y$ as 
\beq
\CZ(Y)= \exp(2\pi \i \CA (Y) ).
\eeq
We call $\CA$ as the anomaly of the boundary theory.
Then the consistency of the brane requires
the modified Dirac quantization condition given by\footnote{
Throughout the paper we assume that the normal bundle to the brane is trivial. 
The following discussion may require modifications or refinements in the presence of a nontrivial normal bundle, as already known in the case of M5-branes~\cite{Freed:1998tg}.
}
\beq
\int_Y F_{d+1} + \CA(Y) \in \BZ. \label{eq:AFcondition}
\eeq
This implies that the integral $\int_Y F_{d+1}$ is not necessarily an integer, but its fractional part is controlled by the anomaly of the worldvolume theory.

One of the consequences of the above formula is as follows. Suppose that $Y$ is a boundary of a $d+2$ dimensional manifold $Z$.
The anomaly polynomial $\CI$ is defined as $ \CA(\partial Z) = \int_Z \CI$.
Then the above equation implies the modified Bianchi identity 
\beq
\d F_{d+1} + \CI =0.
\eeq
This equation reproduces some supergravity equations such as $\d F_5 = F_3 \wedge H_3 $ in Type IIB supergravity
and $\d F_7 = \frac{1}{2} F_4 \wedge F_4 - I_8$ in 11-dimensional supergravity, where $I_8$ is a polynomial of the Riemann curvature.\footnote{%
 $I_8$ of the 11-dimensional supergravity was determined in this way in an early paper on M-theory \cite{Duff:1995wd}.
}
Notice that worldvolume fermions do not contribute to terms like $F_3 \wedge H_3 $ and $\frac{1}{2} F_4 \wedge F_4$,
so it is essential to incorporate the anomalies of $p$-form fields under higher form symmetries,
or equivalently the Green-Schwarz contributions.

The modified Bianchi identity depends only on the perturbative anomaly.
As explicit examples which require nonperturbative treatment, we can consider O-planes in string theory,  orbifolds in M-theory, and S-folds in F-theory.
If we take a submanifold which surrounds such singularities, we get real projective spaces $\mathbb{RP}^{d+1}$ and lens spaces $S^{d+1}/\BZ_k$.
For simplicity we consider cases in which $d$ is even.
As $H^{i}(S^{d+1}/\BZ_k,\BZ)$ for $0<i<d+1$ is torsion, 
the topology of the background $(d-p-2)$-form field $C$ 
is controlled by a torsion group $H^{d-p-1}(S^{d+1}/\BZ_k,\BZ)$.
This means that the computation of the anomaly requires a
careful analysis than the perturbative one discussed above. 
More concretely, \begin{itemize}
\item We perform this analysis and and confirm that the condition \eqref{eq:AFcondition} is satisfied for O$q$-planes with $q \geq 2$,
as an extension of the work \cite{Tachikawa:2018njr};
\item we  compute the fractional charge of the M-theory orbifold $\BR^3 \times \BR^8/\BZ_k$ in this manner and reproduce the result of \cite{Bergman:2009zh,Aharony:2009fc};
\item and we confirm the condition \eqref{eq:AFcondition} for the case of S-folds in F-theory~\cite{Garcia-Etxebarria:2015wns,Aharony:2016kai} by computing the anomaly of the $\SL(2,\BZ)$ duality group of Maxwell theory. 
\end{itemize}
All the results of our detailed computations show perfect consistency with known results in the literature, computed using various string dualities.

\subsection{Organization of the paper}
The rest of the paper is organized as follows. In section~\ref{sec:DiffCoh}, we review the basics of differential cohomology.
We emphasize that it is not more than a precise description of what physicists think how $p$-form gauge fields should behave.
Unfortunately, however, it is not widely used in physics community,
and we explain the basic properties which are necessary for this paper. 
In section~\ref{sec:nonchiral}, we study non-chiral $p$-form fields
by using the formalism of differential cohomology. In the case of non-chiral theory, there is an alternative description 
of the total bulk-boundary system which does not introduce a bulk dynamical field $A$ as in \eqref{eq:schematicS}. 
This alternative description is simpler when available and sheds additional insight into the theory.
We also discuss its applications  to O3-planes. 
In section~\ref{eq:QR}, we give a precise definition of the quadratic refinement in each dimension $d+1=3,7$ and $11$.
Then in section~\ref{sec:theory}, we use the quadratic refinement to define the precise version of the system \eqref{eq:schematicS} on a manifold $Y$ with boundary $X = \partial Y$
and the local boundary condition $\L$ given by \eqref{eq:LBintro}. We also give a general discussion about how to think about anomalies of the theory
which is realized by a local boundary condition imposed on a bulk theory. 
In section~\ref{sec:anomaly}, we compute the partition function of the bulk
theory on closed manifolds and derive the formulas for the anomalies. We obtain two expressions for the anomaly. One expression involves 
the signature $\eta$-invariant $\eta(\widetilde{\CD}^{\rm sig}_Y)$ and the Arf invariant which comes from the sum over topologically nontrivial sectors.
Another expression involves only the $\eta$-invariant of a Dirac operator acting on fermions.
We apply these formulas for the anomalies in section~\ref{sec:Mthy} to M5-branes in M-theory,
and study the orbifold singularity $\BR^3 \times \BR^8/\BZ_k$ in detail. 
We also draw some lessons about O2-planes and the Maxwell field on D4-branes.
Finally, in section~\ref{sec:EMdual}, we study the anomaly of the $\SL(2,\BZ)$ duality group of the $d=4$ Maxwell theory,
and discuss its applications to S-folds in F-theory. 

We also have several Appendices.
Appendix~\ref{app:NC} summarizes our notations and conventions.
In Appendix~\ref{sec:Mthconv}, we study some sign factors in M-theory which are necessary for the precise computations.
In two Appendices~\ref{sec:lens} and \ref{sec:etacomp}, we present computations of cohomology pairings and $\eta$-invariants on lens spaces $S^{2m-1}/\BZ_k$ in different methods.
In the first of the two, we use an appropriate disk bundle over the complex projective space, whose boundary is the lens space in question.
We obtain analytic expressions for cohomology pairings and $\eta$-invariants mod 1 which are useful for the applications to M-theory orbifold.
In the second of the two, we use the orbifolds of the torus and the equivariant index theorem to compute the $\eta$-invariants as real numbers.
Finally, in the last Appendix~\ref{sec:nonunitary}, we provide a simple set of non-unitary topological invertible phases whose partition function is not a bordism invariant. 
This illustrates the necessity of the unitarity condition in the cobordism classification of the invertible phases.

\section{$p$-form gauge fields and differential cohomology}\label{sec:DiffCoh}

This paper is concerned with  somewhat subtle topological properties of higher-form gauge fields, such as the Maxwell field, the RR $p$-form fields $C$ and the NSNS 2-form fields $B$.
Most physicists think that they know how they should behave in topologically nontrivial situations.
Its precise mathematical formulation is the differential cohomology~\cite{CS,Hopkins:2002rd}.
We briefly review this concept to the extent needed in this paper, following the original paper \cite{CS}. See also~\cite{Freed:2006yc,Cordova:2019jnf} for a review for physicists.
We do not try to make the discussion completely rigorous, and also we neglect many important details,
for which we refer the reader to \cite{Hopkins:2002rd}.

\subsection{Differential cohomology}\label{sec:DC1}
We denote the space of differential $p$-forms on a manifold $X$ as $\Omega^{p}(X)$, and the space of closed differential forms (i.e.~$ \omega \in \Omega^p(X)$ with $\d \omega =0$) as $\Omega_{\rm closed}^{p}(X)$.
When we talk about a $p$-form gauge field $ A$, the physically important information is the following.
\begin{itemize}
\item The field strength $\DF \in \Omega_{\rm closed}^{p+1}(X)$. It is roughly $\d A$.
\item Holonomy function $M \mapsto \chi(M) \in \U(1)$ for $p$-dimensional subspaces $M$ which are closed ($\partial M=0$). It is denoted as $\chi(M) = \exp( 2\pi \i \int_{M} \DA)$ where $\DA$ is roughly equivalent to $A$.
\item The relation between the field strength and holonomy, $\chi(\partial N)= \exp(2\pi \i \int_N \DF)$, for $(p+1)$-dimensional subspaces $N$ with boundary $\partial N$.
\end{itemize}
A pair $(\DF, \chi)$ of field strength $\DF$ and holonomy function $\chi$ is a called a differential character.
This is the precise meaning of a $p$-form gauge field.

For some purposes, $\chi$ is not convenient to deal with. It is more convenient to consider the corresponding ``gauge field'' $\DA$.
To obtain it, we may first extend $\chi$ to subspaces $M$ which are not necessarily closed, $\partial M \neq 0$.
This extension can be done in an arbitrary manner. We denote the space of $p$-dimensional subspaces of $X$ which are not necessarily closed as $C_p(X)$ (i.e.~the space of chains),
and the space of closed subspaces as $Z_p(X) \subset C_p(X)$.
So $\chi$ is now extended from a function $\chi: Z_p(X) \to \U(1) $ to a function $\chi: C_p(X) \to \U(1)$ in an arbitrary way.
Next, we also take the logarithm of $\chi$ as $\DA = \frac{1}{2\pi \i } \log (\chi)$ in an arbitrary way. This is a function from $C_p(X)$ to $\BR$, which we denote by using the notation of integral as $M \mapsto \int_M \DA \in \BR$.

From a function $\DA: C_p(X) \to \BR$, we can define $\delta \DA : C_{p+1}(X) \to \BR$ (i.e.~coboundary) as $\int_N \delta \DA = \int_{\delta N} \DA$.
Then from the equation 
\beq
 \exp(2\pi \i \int_N \DF) = \chi(\partial N) = \exp(2\pi \i \int_{\partial N} \DA)
\eeq
we get
$
\int_N (\DF - \delta \DA) \in \BZ.
$
Therefore, there is a function $\DN: C_{p+1}(X) \to \BZ$ such that
\beq
\delta \DA = \DF - \DN. \label{eq:DC1}
\eeq
We denote the triplet $(\DN, \DA, \DF)$ as 
\beq
\check A = (\DN, \DA, \DF).
\eeq
This has some gauge redundancies. First, we extended $\chi$ from the set of closed subspaces to arbitrary subspaces.
Corresponding to this, $\DA$ has an ambiguity given as $\DA \to \DA+\delta \Da$, where $\Da$ is a function $\Da : C_{p-1}(X) \to \BR$.
Also, when taking the log of $\chi$, there is an ambiguity of shifting $\DA$ by integers as $\DA \to \DA + \Dn$, where $\Dn$ is a function $\Dn : C_{p}(X) \to \BZ$.
Therefore, the triplet $ (\DN, \DA, \DF)$ has an ambiguity given by
\beq
 (\DN, \DA, \DF) \to  (\DN -\delta \Dn, \DA + \delta \Da + \Dn, \DF). \label{eq:DC2}
\eeq
This can be regarded as a gauge transformation of $\check A = (\DN, \DA, \DF)$. 
The $\check A$ up to this kind of gauge transformations contain the same information as the original pair $(\DF, \chi)$ of the field strength and the holonomy. 

The set of triplets $ (\DN, \DA, \DF)$ up to gauge transformations
is called differential cohomology, and  is denoted by $\check H^{p+1}(X)$:
\beq
\check H^{p+1}(X) = \{ \check A = (\DN, \DA, \DF) :~ (\DN, \DA, \DF) \sim  (\DN - \delta \Dn, \DA + \delta \Da + \Dn, \DF) \}.
\eeq
For example, $\check H^2(X)$ is the space of ordinary $\U(1)$ gauge fields on $X$. One can check that the set $\check H^1(X)$
is the space of compact scalar fields which take values in $\U(1)$.

Let us introduce a few standard notations. Let $\BA$ be an abelian group (such as $\BA = \BZ, \BR,\BR/\BZ$, etc.).
Then, the space of linear functions $C_p(X) \to \BA$ is denoted as $C^p(X, \BA)$.
We can define the coboundary $\delta$: $C^p(X, \BA) \ni x  \mapsto \delta x \in C^{p+1}(X, \BA)$ as before.
We denote the space of closed elements $x \in C^p(X, \BA)$, $\delta x =0$ as $Z^{p}(X,\BA)$.
A generic element $x \in C^p(X, \BA)$ is called a cochain, and if $x$ satisfies $\delta x =0$ (i.e.~$x \in  Z^{p}(X,\BA)$), it is called a cocycle.
If $x = \delta y$ for some $y\in C^{p-1}(X,\BA)$, it is called a coboundary. Because $\delta \delta=0$,
we have $\delta C^{p-1}(X,\BA) \subset Z^p(X,\BA)$ and we define the cohomology with the coefficients in $\BA$ as
$
H^p(X,\BA) = Z^p(X,\BA)/\delta C^{p-1}(X,\BA)
$.

Now let us study a few properties of $\check H^{p+1}(X)$. The Stokes theorem $\int_{\partial L} \omega = \int_L \d \omega$ for a differential form $\omega \in \Omega^p(X)$
means that $\delta \omega = \d \omega$. By using the fact that $\d \DF =0$, we get
\beq
\delta \DN = \delta \DF - \delta^2 \DA =0.
\eeq
Thus $\DN \in Z^{p+1}(X,\BZ)$, and hence it gives an element 
\beq
[\DN]_\BZ \in H^{p+1}(X,\BZ).
\eeq
From the gauge transformation rule above, we see that $[\DN]_\BZ \in H^{p+1}(X,\BZ)$ is invariant under gauge transformations. 
We call this the (integer) flux of the gauge field $\check A$.

By the embedding of $\BZ$ into $\BR$, we can obtain an element $\DN_{\BR} \in Z^{p+1}(X,\BR)$ from $\DN \in Z^{p+1}(X,\BZ)$,
and correspondingly we get $[\DN]_\BR \in H^{p+1}(X,\BR)$. Then, \eqref{eq:DC1} implies that the de~Rham cohomology $[\DF] \in H^{p+1}(X,\BR)$ of $\DF \in \Omega_{\rm closed}^{p+1}(X)$
is the same as $[\DN]_\BR $,
\beq
[\DF] = [\DN]_\BR.
\eeq
Therefore, $ [\DN]_\BZ$ contains more refined information than the flux $[\DF]$ at the differential form level,
because $ [\DN]_\BR$ can be obtained from $ [\DN]_\BZ$ but the converse is not true.
For example, $[\DN]_\BZ$ can be a non-zero torsion element, for which $[\DN]_\BR=0$.
One of the reasons that we introduce differential cohomology in this paper is that we want to study
the cases in which $[\DN]_\BZ \neq 0$ with $[\DF]=0$.

Let us consider two special cases to get more insight. One is a topologically trivial gauge field, and the other is a flat gauge field. 
\paragraph{Topologically trivial field.}
Suppose that $[\DN]_\BZ =0 \in H^{p+1}(X,\BZ)$. This means that $\DN = \delta \Dn$ for some $\Dn$, and hence we can set $\DN =0$ by a gauge transformation.
Moreover, the de~Rham cohomology of $\DF$ is zero, $[\DF] =[\DN]_{\BR}=0$.
Thus we can take $\DA$ to be a differential form such that $\DF = \d \DA$.\footnote{
In more detail, a proof is as follows. Since $[\DF] =0$, there exists a differential form $\DA_0 \in \Omega^p(X)$ such that $\DF = \d \DA_0$.
By using \eqref{eq:DC1} with $\DN=0$, we get $\delta (\DA - \DA_0) =0$ and hence $\DA-\DA_0 \in Z^p(X,\BR)$. By de~Rham theorem,
there exists a closed differential form $\DA_1 \in \Omega_{\rm closed}^p(X)$ such that $\DA-\DA_0 = \DA_1 + \delta \Da$ for some $\Da \in C^{p-1}(X,\BR)$.
Thus, up to gauge transformations, we get $\DA = \DA_0 + \DA_1  \in \Omega^p(X)$.
}

 In any topologically trivial open set $U \subset X$ in $X$,
we have $H^{p+1}(U,\BZ)=0$ and hence we can always locally represent the gauge field $\check A$ by a differential form.
This is what is usually done in physics. 

It might also be interesting to study gauge transformations \eqref{eq:DC2} which keep $\DN=0$ and $\DA$ being a differential form.
To keep $\DN=0$, we must have $\delta \Dn =0$. To keep $\DA$ to be a differential form, we need $\Df : = \delta \Da + \Dn$ to be a differential form, $\Df \in \Omega^p(X)$.
Because $\delta \Df = \delta^2 \Da + \delta \Dn =0$, this $\Df$ is a closed form, $\Df \in \Omega_{\rm closed}^p(X) $.
Then the triplet $\check a := (\Dn, \Da, \Df)$ with $\Df = \delta \Da + \Dn$ defines an element of the differential cohomology group $\check H^p(X)$ with one lower degree than $\check A$.

\paragraph{Flat gauge field.} A gauge field $\check A$ with $\DF =0$ is called a flat gauge field.
In this case, we have $\delta \DA = \DN$. By using $\BR \to \BR/\BZ \cong \U(1)$, we get $\DA_{\BR/\BZ} \in C^p(X, \BR/\BZ)$
which satisfies $\delta \DA_{\BR/\BZ} =0 $. Therefore, it defines an element $[\DA]_{\BR/\BZ} \in H^p(X, \BR/\BZ)$.

Conversely, if we are given an element $[\DA]_{\BR/\BZ} \in H^p(X, \BR/\BZ)$, we can recover the triplet $\check A = (\DN, \DA, 0)$.
The reason is as follows. From $[\DA]_{\BR/\BZ} \in H^p(X, \BR/\BZ)$ we get $\DA_{\BR/\BZ} \in Z^p(X,\BR/\BZ)$ which is defined up to 
gauge transformations $\DA_{\BR/\BZ} \to \DA_{\BR/\BZ} + \delta \Da_{\BR/\BZ}$ for $\Da_{\BR/\BZ} \in C^{p-1}(X,\BR/\BZ)$.
Then we uplift $\DA_{\BR/\BZ}$ to $\DA \in C^p(X,\BR)$ in an arbitrary way. The ambiguity of doing so is given by a gauge transformation
$\DA \to \DA + \Dn$ for some $\Dn \in C^p(X,\BZ)$. Therefore, the total ambiguity is just given by $\DA \to \DA + \delta \Da + \Dn$.
By using $\DA$, we define $\DN : = -\delta \DA$ which can be regarded as an element of $C^{p+1}(X,\BZ)$
because $\delta \DA_{\BR/\BZ} =0$. In this way we uniquely get an element of the differential cohomology $\check A = (\DN, \DA , 0)  \in \check H^{p+1}(X)$ from $[\DA]_{\BR/\BZ} \in H^p(X, \BR/\BZ)$.

As a byproduct, we get a map $\beta$ defined by
\beq
\beta: H^p(X, \BR/\BZ) \ni [\DA]_{\BR/\BZ} \mapsto [\DN]_\BZ \in H^{p+1}(X,\BZ).
\eeq
This map is called the Bockstein homomorphism associated to the exact sequence $0 \to \BZ \to \BR \to \BR/\BZ \to 0$. 
Namely, the Bockstein homomorphism $\beta$ is a map which gives
the flux $ [\DN]$ from a flat gauge field $[\DA]_{\BR/\BZ} $.
Notice also that for any cyclic group $\BZ_k$, we can embed $\BZ_k \to \U(1) \cong \BR/\BZ$ which gives a map $H^p(X, \BZ_k) \to H^p(X, \BR/\BZ)$. 
By using it,
we can also define the Bockstein homomorphism
\beq
\beta: H^p(X, \BZ_k) \to H^{p+1}(X,\BZ).
\eeq
This special case is also sometimes important, especially when we consider a $p$-form discrete $\BZ_k$ gauge field.
However, we remark that for continuous gauge fields, the more general homomorphism is the one from the differential cohomology to integer cohomology which
takes the integer flux of the gauge field,
\beq
\beta: \check H^{p+1}(X) \ni \check A \mapsto [\DN]_\BZ \in H^{p+1}(X,\BZ).
\eeq

\subsection{Product in differential cohomology}\label{sec:DC2}
From two differential forms $\omega_1 \in \Omega^{p_1}(X)$ and $\omega_2 \in \Omega^{p_2}(X)$,
we can get their wedge product $\omega_1 \wedge \omega_2 \in \Omega^{p_1+p_2}(X) $.
The wedge product has the properties that $\d (\omega_1 \wedge \omega_2) = \d \omega_1 \wedge \omega_2 + (-1)^{p_1}\omega_1 \wedge \d \omega_2$ and $\omega_1 \wedge \omega_2 = (-1)^{p_1 p_2} \omega_2 \wedge \omega_1$.

We want to define a product between differential cohomology elements $\check A_1 \in \check H^{p_1+1}(X) $ and $\check A_2 \in \check H^{p_2+1}(X) $, denoted as $\check A_1 \Dp \check A_2 \in \check H^{(p_1+1)+(p_2+1)}(X) $,
which has the property that its field strength is given by $\DF_{A_1 \Dp A_2} = \DF_{A_1}  \wedge \DF_{A_2}$. Here we use the notation that
for a differential cohomology element $\check B $, we distinguish the corresponding triplet by using a subscript as
\beq
\check B = (\DN_B, \DA_B, \DF_B).
\eeq
To define the product $\check A_1 \Dp \check A_2 $, we need some preliminary technical discussions.

It is known in cohomology theory that from two cochains $x_1 \in C^{p_1}(X, \BA_1)$, $x_2 \in C^{p_2}(X, \BA_2)$
and a homomorphism $\BA_1 \times \BA_2 \to \BA_3$ (such as $\BZ \times \BR \ni (n,r) \mapsto nr \in \BR$), there is a product, called the cup product, $x_1 \cup x_2 \in C^{p_1 +p_2}(X, \BA_3)$.
It satisfies 
\beq
\delta( x_1 \cup x_2) = (\delta x_1) \cup x_2 + (-1)^{p_1} x_1 \cup (\delta x_2).\label{eq:DC4}
\eeq
Also, it is known that there exists a homomorphism $(x_1,x_2) \mapsto P(x_1,x_2) \in C^{p_1 +p_2-1}(X, \BA_3)$ (defined for any $p_1,p_2$) such that the equation
\beq
(-1)^{p_1p_2} x_2 \cup x_1 -  x_1 \cup x_2  = P(\delta x_1, x_2) + (-1)^{p_1} P(x_1,\delta x_2) + \delta P(x_1,x_2). \label{eq:DC3}
\eeq
holds.\footnote{$P(x_1,x_2)$ in \eqref{eq:DC3} is often denoted by $x_1\cup_1 x_2$ and is called the cup-1 product, in the case of cellular cochains.}
By using these properties, we can see that for cohomology elements $y_1 \in H^{p_1}(X,\BA_1)$ and $y_2 \in H^{p_2}(X,\BA_2)$,
we can obtain a well-defined product $y_1 \cup y_2 \in H^{p_1+p_2}(X,\BA_3)$ with $y_2 \cup y_1 = (-1)^{p_1p_2} y_1 \cup y_2$.

The cup product $\cup$ is not unique on cochains. For example, we could define another $\cup'$ as $x_1 \cup' x_2 : = (-1)^{p_1 p_2} x_2 \cup x_1$.
Any two cup products $\cup$ and $\cup'$ on cochains which are related as in 
\beq
x_1 \cup' x_2 -  x_1 \cup x_2  = Q(\delta x_1, x_2) + (-1)^{p_1} Q(x_1,\delta x_2) + \delta Q(x_1,x_2),
\eeq
for some $Q$ give the same cup product for cohomology.\footnote{A cup product satisfying \eqref{eq:DC4} is a chain map $\cup : \bigoplus_{q} C^q \times C^{p-q} \to C^p$,
where $\delta$ on $\bigoplus_{q} C^q \times C^{p-q}$  is defined by $ \delta ( x_1 , x_2)=( \delta x_1 , x_2) \oplus (-1)^{p_1}( x_1 , \delta x_2)$.
We have $\delta \cdot \cup = \cup \cdot \delta$.
Then $Q$ is a chain homotopy between two chain maps, $  \cup' - \cup = Q \cdot \delta + \delta \cdot Q$.
}

In particular, differential forms $\omega$ can also be regarded as cochains. 
By abuse notations, we denote a differential form $\omega$ and its associated cochain $\iota(\omega)$ by the same symbol $\omega$. 
We identify $\delta  $ with $ \d $.
The $\wedge$ product on differential forms is related to a $\cup$ product on cochains as
\beq
\omega_1 \wedge \omega_2 - \omega_1 \cup \omega_2 = Q(\delta \omega_1, \omega_2) + (-1)^{p_1} Q(\omega_1,\delta \omega_2) + \delta Q(\omega_1,\omega_2). \label{eq:DC5}
\eeq
for some $Q$.
We note that  $Q(\omega_1,\omega_2)$ is a cochain but is not a differential form.

By using the above facts, we can define the product $\check A_1 \Dp \check A_2 \in \check H^{p_1+1}(X) $ of two differential cohomology elements $\check A_1 \in \check H^{p_1+1}(X) $ and $\check A_2 \in \check H^{p_2+1}(X) $.
First, $\DF$ and $\DN$ are simple to define:
\beq
\DF_{A_1 \Dp A_2} &= \DF_{A_1} \wedge \DF_{A_2}, \\
\DN_{A_1 \Dp A_2} &= \DN_{A_1} \cup \DN_{A_2},
\eeq
which also implies $[\DN_{A_1 \Dp A_2}]  = [\DN_{A_1}] \cup [\DN_{A_2}]$.

The definition of $\DA_{A_1 \Dp A_2}$ is slightly more complicated. We first present the definition and check that it defines a differential cohomology element, 
and then discuss two special cases (topologically trivial fields and flat fields)
in which the expression becomes much simpler. 

The definition is given as follows.
\beq
\DA_{A_1 \Dp A_2} = \DA_{A_1} \cup \DN_{A_2} + (-1)^{p_1+1} \DF_{A_1} \cup \DA_{A_2} + Q(\DF_{A_1} , \DF_{A_2}). \label{eq:PD}
\eeq
One can check that the defining equation of differential cohomology,
\beq
\delta \DA_{A_1 \Dp A_2} = \DF_{A_1 \Dp A_2} - \DN_{A_1 \Dp A_2},
\eeq
is satisfied by a straightforward computation using \eqref{eq:DC4} and \eqref{eq:DC5}:
\beq
\delta \DA_{A_1 \Dp A_2} &= (\delta \DA_{A_1}) \cup \DN_{A_2}  +  \DF_{A_1} \cup (\delta \DA_{A_2}) +  \delta Q(\DF_{A_1} , \DF_{A_2}) \nonumber \\
& =  ( \DF_{A_1} - \DN_{A_1} ) \cup \DN_{A_2}  + \DF_{A_1} \cup ( \DF_{A_2} - \DN_{A_2} ) +  ( \DF_{A_1}  \wedge \DF_{A_2} - \DF_{A_1}  \cup \DF_{A_2}) \nonumber \\
&= \DF_{A_1}  \wedge \DF_{A_2} -  \DN_{A_1} \cup \DN_{A_2}.
\eeq
One can also check that gauge transformations of $\check A_1$ and $\check A_2$ affect $\check A_1 \Dp \check A_2$ only by gauge transformations,
so that $\DA_{A_1 \Dp A_2}$ is well-defined. It is also known that 
\beq
\check A_2 \Dp \check A_1= (-1)^{(p_1+1)(p_2+1) }\check A_1 \Dp \check A_2 + \text{ (gauge transformation)},
\eeq
although its derivation is more complicated. 
This is consistent with the corresponding property of wedge and cup products, $\DF_{A_2} \wedge \DF_{A_1} =(-1)^{(p_1+1)(p_2+1) } \DF_{A_1} \wedge \DF_{A_2}$ and $[\DN_{A_2}] \cup [\DN_{A_1}] =(-1)^{(p_1+1)(p_2+1) } [\DN_{A_1}] \cup [\DN_{A_2}] $.

From $ \check A_1  \Dp \check A_2 $, we can define the holonomy function as $\chi_{A_1 \Dp A_2} (M) = \exp ( 2\pi \i \int_M \DA_{A_1 \Dp A_2} ) $ for $(p_1 + p_2+1)$-dimensional subspaces $M \in Z^{p_1 + p_2+1}(X)$.
This gives a differential character.

Now let us discuss two special cases which may give further insight into the product defined above.

\paragraph{Topologically trivial field.}
Suppose that $\check A_1$ (but not necessarily $\check A_2$) is a topologically trivial gauge field. This means that we can take $\DN_{A_1} =0$ and we can assume $\DA_{A_1}$ to be a differential form
up to gauge transformations. Then we also have $\DF_{A_1} = \d \DA_{A_1}$. In this case, \eqref{eq:DC5} gives 
\beq
 Q( \d \DA_{A_1}, \DF_{A_2} )  =    \DA_{A_1}  \wedge  \DF_{A_2} -   \DA_{A_1} \cup  \DF_{A_2}  - \delta Q(\DA_{A_1} , \DF_{A_2} ).
\eeq
Therefore, we can simplify $\DA_{A_1 \Dp A_2}$ defined in \eqref{eq:PD} as 
\beq
\DA_{A_1 \Dp A_2} &= \DA_{A_1} \cup \DN_{A_2} + (-1)^{p_1+1} (\delta \DA_{A_1}) \cup \DA_{A_2} +   \DA_{A_1}  \wedge  \DF_{A_2} -   \DA_{A_1}  \cup  \DF_{A_2}  - \delta Q(\DA_{A_1} , \DF_{A_2} ) \nonumber \\
& =   \DA_{A_1}  \wedge  \DF_{A_2}  +  \delta \left( (-1)^{p_1+1}  \DA_{A_1} \cup \DA_{A_2} +Q(\DA_{A_1} , \DF_{A_2} )  \right).
\eeq
This means that, up to gauge transformations, $\DA_{A_1 \Dp A_2} $ can be regarded as a differential form:
\beq
\DA_{A_1 \Dp A_2} =  \DA_{A_1}  \wedge  \DF_{A_2}  + \text{ (gauge transformation)}. \label{eq:DC6}
\eeq
This is the usual Chern-Simons type product of two higher-form gauge fields. 

Notice that even if $\check A_1$ is not topologically trivial, its fluctuation in the same topological sector can always be represented by a topologically trivial field $\check B_1$
as $\check A_1 + \check B_1$, $\DN_{B_1}=0$. Then we can use the above formula for $\check B_1$. This observation is sometimes useful.

\paragraph{Flat gauge field.}
Next let us consider the case that $\check A_1$ (but not necessarily $\check A_2$) is a flat gauge field, $\DF_{A_1} =0$. In this case, \eqref{eq:PD} is simplified as
\beq
\DA_{A_1 \Dp A_2} = \DA_{A_1} \cup \DN_{A_2} .
\eeq
Notice that $\check A_1 \Dp \check A_2$ is also flat, $\DF_{A_1 \Dp A_2} =0$. Then $\DA_{A_1}$ and $\DA_{A_1 \Dp A_2}$ give elements in the cohomology 
$[\DA_{A_1}]_{ \BR/\BZ}  \in H^{p_1}(X, \BR/\BZ)$ and $[\DA_{A_1 \Dp A_2}]_{ \BR/\BZ}  \in H^{p_1+p_2+1}(X, \BR/\BZ)$,
with the relation
\beq
[\DA_{A_1 \Dp A_2}]_{ \BR/\BZ} = [\DA_{A_1}]_{ \BR/\BZ} \cup [\DN_{A_2}]_\BZ. 
\eeq
Therefore, it is determined by the ordinary cohomology theory.

For computations, the Poincar\'e-Pontryagin duality is convenient.
It states that the pairing between $H^p(X, \BR/\BZ)$ and $H^{d-p}(X,\BZ)$ on a closed oriented $d$-dimensional manifold is a perfect pairing.
This means the following. Consider the pairing 
\beq
H^p(X, \BR/\BZ) \times H^{d-p}(X,\BZ) \ni (x,y ) \mapsto \int_X x \cup y \in \BR/\BZ.
\eeq
If $\int_X x \cup y$ is zero for all $y$ for a given $x$, then that $x$ is zero. Also if $\int_X x \cup y$ is zero for all $x$ for a given $y$, then that $y$ is zero.
By using this fact, it is possible to show that the differential cohomology pairing $(\check A_1, \check A_2) \mapsto \int_X \check \DA_{A_1 \Dp A_2}$
is also a perfect pairing for not necessarily flat $\check A_{1,2}$.

Now suppose that both $\check A_1$ and $\check A_2$ are flat, $\DF_{A_1} = \DF_{A_2}=0$. In this case,
the cohomology element $[\DA_{A_1 \Dp A_2}]_{ \BR/\BZ}$ actually depends only on $[\DN_{A_1}]_\BZ $ and $[\DN_{A_2}]_\BZ $.
This can be shown as follows. Suppose that $\DA_{A_1}$ and $\DA'_{A_1}$ give the same $[\DN_{A_1}]_\BZ$, that is, $[\DN_{A_1}]_\BZ = -[\delta \DA_{A_1}]_\BZ = -[\delta \DA'_{A_1}]_\BZ$.
This implies that there exists a cochain with integer coefficients $x \in C^{p_1}(X,\BZ)$ such that we have $\delta (\DA'_{A_1} -  \DA_{A_1} ) = \delta x$. This in turn implies that $y:=\DA'_{A_1} -  \DA_{A_1} -x$ is closed, $y \in Z^{p_1}(X,\BR)  $.
Let us consider the cup product
\beq
 \DA'_{A_1}  \cup \DN_{A_2} =  \DA_{A_1}  \cup \DN_{A_2} + x  \cup \DN_{A_2} + y  \cup \DN_{A_2}.
\eeq
The second term of the right hand side, $x  \cup \DN_{A_2} $, is a cochain with integer coefficients $\BZ$ and hence it does not contribute when the coefficients are reduced to $\BR/\BZ$.
The third term $y  \cup \DN_{A_2}$ is actually exact, $y  \cup \DN_{A_2} = -y \cup \delta \DA_{A_2} = - (-1)^{p_1} \delta ( y \cup \DA_{A_2})$ where we have used the fact that $y$ is closed, $\delta y =0$,
and also the fact that $\delta \DA_{A_2} = \DF_{A_2} - \DN_{A_2} = - \DN_{A_2}$.
Therefore, we get
\beq
[\DA'_{A_1}  \cup \DN_{A_2}]_{\BR/\BZ} =  [\DA_{A_1}  \cup \DN_{A_2} ]_{\BR/\BZ}.
\eeq
This shows that $ [\DA_{A_1} ]_{\BR/\BZ} \cup  [\DN_{A_2} ]_{\BZ}$ depends only on $[\DN_{A_1}]_\BZ $ and $[\DN_{A_2}]_\BZ $ and hence we get a map
\begin{multline}
H^{p_1+1}(X,\BZ) \times H^{p_2+1}(X,\BZ) \ni ([\DN_{A_1}]_\BZ ,[\DN_{A_2}]_\BZ ) 
\\
 \mapsto    [\DA_{A_1} ]_{\BR/\BZ} \cup  [\DN_{A_2} ]_{\BZ} \in H^{p_1+p_2+1}(X,\BR/\BZ).
\end{multline}
This is called the torsion pairing between $[\DN_{A_1}]_\BZ $ and $[\DN_{A_2}]_\BZ $. Let us denote it as $T([\DN_{A_1}]_\BZ ,[\DN_{A_2}]_\BZ ) $.
Since it is derived from the differential cohomology pairing, it satisfies $T([\DN_{A_1}]_\BZ ,[\DN_{A_2}]_\BZ ) =(-1)^{(p_1+1)(p_2+1)} T([\DN_{A_2}]_\BZ ,[\DN_{A_1}]_\BZ )$.
By Poincar\'e-Pontryagin duality, it is also a perfect pairing between torsion subgroups $H^{p+1}_{\rm tor}(X,\BZ) $ and $H^{d-p}_{\rm tor}(X,\BZ) $.

\subsection{Chern-Simons as differential cohomology}\label{sec:DC3}
For gauge fields of a compact Lie group $G$, we can define Chern-Simons invariants associated to characteristic classes.
$G$ need not be connected, and it may even be a discrete group like $\BZ_k$.
These generalizations of the Chern-Simons action to non-connected groups, including finite groups were first discussed in physics literature
by Dijkgraaf and Witten in   \cite{Dijkgraaf:1989pz}, and it is now customary to refer to finite group gauge theories with this type of actions as Dijkgraaf-Witten theories.
Differential cohomology gives a unified description of these cases.

We use the following fundamental fact. There exists a space $BG$, called the classifying space of $G$, with the following property.
There is a $G$-bundle $P_G$ on $BG$ such that 
any $G$-bundle on an arbitrary space $X$ can be realized as a pull back of $P_G$ under some map $f: X \to BG$.\footnote{
Such a space can be obtained as follows. First let us discuss the case $G = \U(n)$. In this case, let us show that we can take $B\U(n) = G_{n}(\BC^N)$ with sufficiently large $N$, where $ G_{n}(\BC^N)$ is the Grassmannian manifold
which is the set of complex $n$-dimensional planes inside $\BC^N$. The reason is as follows. Suppose we have a $\U(n)$ bundle on $X$. A $\U(n)$ bundle is equivalent to an $n$-dimensional complex vector bundle $E$.
It is not too hard to show (e.g.~by using partition of unity argument associated to local patches of $X$) that $E$ can be embedded into a trivial $N$-dimensional bundle $\underline{\BC}^N (= \BC^N \times X)$ for some sufficiently large $N$, i.e.~$E \subset \underline{\BC}^N$.
Then, for each $p \in X$, we get an $n$-dimensional subspace $E_x \subset \BC^N$. This defines a map $f : X \to B\U(n) = G_{n}(\BC^N)$. From this construction,
it is clear that $E$ is a pullback of the tautological $n$-dimensional bundle of $G_n(\BC^N)$ whose fiber is just the $n$-dimensional plane.

For an arbitrary compact Lie group $G$, we take a faithful unitary $n$-dimensional representation of $G$. This gives an embedding $G \to \U(n)$.
Let $P_{\U(n)}$ be the universal $\U(n)$ bundle on $G_{n}(\BC^N)$. Then, we can consider the $G$-bundle $P_{\U(n)} \times_{G} G$ whose base is $BG = P_{\U(n)}/G$.
This gives an example of a classifying space of $G$. 
}

Just for simplicity of presentation, let us also assume that the $G$ gauge field (connection) on $X$ is also obtained by the pull back of some universal connection on $P_G$ if we choose $f$ appropriately.
This assumption is not essential. (The reason is that the difference of Chern-Simons invariants between two connections in the same topological class
can be expressed by an integral of a gauge invariant polynomial of the curvature tensor without any topological subtleties. Thus it is enough to define the Chern-Simons invariant
for a single connection in each topological sector.)

Now let us discuss how a general Chern-Simons invariant is constructed as differential cohomology.
Because any bundle on any $X$ can be obtained as a pullback of the bundle on $BG$, it is enough to construct a differential cohomology element on $BG$.
Then we can simply pull it back to $X$ by the map $f: X \to BG$.

Thus we try to construct a differential cohomology element on $BG$. 
First, we take an element $c_\BZ \in H^{p+1}(BG,\BZ)$, called a characteristic class of $G$-bundles.
Next, we consider the reduction of $c$ to real coefficient $c_\BR \in H^{p+1}(BG,\BR)$.
It is known that the Chern-Weil theory gives a   representative of $c_\BR$  in terms of a polynomial of the curvature of the $G$-connection.
Namely, there  is a gauge invariant polynomial $c(F)$ of the curvature $F$ of the $G$-connection such that its de~Rham cohomology $[c(F)]$
gives $c_\BR = [c(F)]$. We denote it as $\DF_c= c(F)$. (If $c_\BR=0$, this polynomial is simply zero.)
Let us also take an arbitrary $\DN_c \in Z^{p+1}(BG,\BZ)$ such that its cohomology $[\DN_c]_\BZ \in H^{p+1}(BG,\BZ)$ gives $c_\BZ = [\DN_c]_\BZ$.
In the real coefficients we have $[\DF_c ] - [\DN_c]_\BR =0$, so there exists an $\DA_c \in C^p(BG,\BR)$ such that $\delta \DA_c = \DF_c - \DN_c$.
The triplet $\check A_c := (\DN_c, \DA_c, \DF_c)$ gives a differential cohomology element.

The choice of $\DN_c$ does not matter because different choices of $\DN_c$ are related by gauge transformations.
However, the choice of $\DA_c$ may matter, depending on the degree $p$ and the group $G$. 
Let $\DA'_c$ be another one with $\delta \DA'_c = \DF_c - \DN_c$.
Then we have $\delta (\DA'_c - \DA_c) =0$ and hence $x:= \DA'_c - \DA_c$ is closed, $x \in Z^p(BG, \BR)$.
Now, suppose that $H^p(BG,\BR) =0$. Then, $x$ is exact and there exists $y \in C^{p-1}(BG,\BR)$ such that $x = \delta y$.
In this case, $\DA'_c$ and $\DA_c$ differ by just gauge transformations. Therefore, from $c \in H^{p+1}(BG,\BZ)$,
we uniquely get $\DA_c$ up to gauge transformations. If $H^p(BG,\BR) \neq 0$,
then $\check A_c$ is not unique and we have to specify additional data.

The elements of the real cohomology $H^p(BG,\BR)$ are represented by gauge invariant polynomials of the curvature.
The curvature $F$ is a 2-form and it is nonzero only if $G$ is a continuous group. Therefore, $H^p(BG,\BR)=0$
if $p$ is odd or $G$ is discrete. These are the cases which usually appear in applications. 
The case of odd $p$ is the usual Chern-Simons, and the case of discrete $G$ was first discussed by Dijkgraaf and Witten as an example in  \cite{Dijkgraaf:1989pz}.

In summary, there is a way to construct a differential cohomology element $\check A_c \in \check H^{p+1}(BG)$
from a given characteristic class $c_\BZ \in H^{p+1}(BG)$. This construction is unique (up to gauge transformations) if $H^p(BG,\BR)=0$.
After constructing $\check A_c$ on $BG$, we can obtain the corresponding differential cohomology element on $X$ by
the pullback $f^* \check A_c$ by $f: X \to BG$ which we may denote just as $\check A_c$ by abusing the notation.
The holonomy $\exp(2\pi \i \int \DA_c)$ is the Chern-Simons invariant.

\subsection{Twisted coefficients}\label{sec:DC4}
Let us make a brief comment on a generalization concerning the coefficients.
Up to now, we have discussed coefficients $\BZ, \BR$ and $\BR/\BZ$ which do not depend on the position of the manifold $X$.
However, it is also possible to consider the following generalization. Let us consider a $\GL(k,\BZ)$ bundle on $X$, 
and let us take a representation $\rho$ of $\GL(k,\BZ)$ acting on $\BZ^n$.
By using $\rho$, we can define an associated bundle $\widetilde{\BZ}^n_\rho$ on $X$ whose fiber is $\BZ^n$ and whose transition functions are described by
the transition function of the $\GL(k,\BZ)$ bundle represented by $\rho$.
By tensoring  $\widetilde{\BZ}^n_\rho$ with $\BR$ and $\BR/\BZ$,
we also get $\widetilde{\BR}^n_\rho$ and $\widetilde{\BR}^n_\rho / \widetilde{\BZ}^n_\rho$.
 
 A chain, which is an element of $C_p(X)$ can always be decomposed into a sum of small pieces. Each piece $\Delta^p$
 is then topologically trivial, and hence the bundle $\rho$ can be trivialized on $\Delta^p$.
 Then, for a twisted coefficient $\widetilde{\BA}_\rho$ (where $\BA = \BZ^n, \BR^n, \BR^n/\BZ^n$), we can define $C^p(X, \widetilde{\BA}) $.
 An element $x \in C^p(X,\widetilde{\BA}_\rho)$ maps $\Delta^p$ to the bundle $\widetilde{\BA}_\rho$, where the bundle is trivialized on $\Delta^p$.
 The relation such as $\int_{\Delta^p} \delta x = \int_{\partial \Delta^p} x$ is also well defined by using that local trivialization. 
 $Z^p(X, \widetilde{\BA}_\rho)$ and $H^p(X,\widetilde{\BA}_\rho)$ are defined in the same way.
 
 We can also consider differential forms twisted by $\rho$. They are simply sections of $(\wedge^p T^*X ) \otimes \widetilde{\BR}^n_\rho$.
 We denote them as $\Omega^p(X, \widetilde{\BR}^n_\rho)$. Differential $\d : \Omega^p(X, \widetilde{\BR}^n_\rho) \to \Omega^{p+1}(X, \widetilde{\BR}^n_\rho)$
 is also defined without any change because transition functions are constant and hence $\d$ does not act on transition functions. 
 
 From the above facts, we can define differential cohomology with twisted coefficients, which may be denoted by $\check H^{p+1}(X,\widetilde{\BZ}^n_\rho)$.
 Most of the discussions are unchanged. One of the points which may need clarification is as follows.
 Suppose we have $\check A_1 \in H^{p_1+1}(X, \widetilde{\BZ}^n_{\rho_1} )$ and $\check A_2 \in H^{p_2+1}(X, \widetilde{\BZ}^n_{\rho_2})$.
 If the tensor product $\rho_1 \otimes \rho_2$ contains $\rho_3$, we can define a product $ \vev{\check A_1 \Dp \check A_2}_{\rho_3}$.
 
 Let us give an example. We consider a bundle with structure group $\SL(2,\BZ)$.
 Let $\rho$ be its defining 2-dimensional representation. Then $\rho \otimes \rho$ contains a trivial representation 
which is described as follows. If $(m_1, n_1) \in \BZ^2$ and $(m_2,n_2) \in \BZ^2$ are acted by $\SL(2,\BZ)$,
we can take $m_1n_2 - n_1m_2$ which is invariant under $\rho$. This is the trivial representation.
Then we can define the product of $\check A_1\in H^{p_1+1}(X, \widetilde{\BZ}^2_{\rho} )$ and $\check A_2 \in H^{p_2+1}(X, \widetilde{\BZ}^2_{\rho} )$
as an element of (untwisted) differential cohomology $\vev{\check A_1 \Dp \check A_2} \in \check H^{p_1+p_2+2}(X) $.
In this case, exchanging $\check A_1$ and $\check A_2$ gives an additional sign
\beq
\vev{\check A_2 \Dp \check A_1} = - (-1)^{(p_1+1)(p_2+1)} \vev{\check A_1 \Dp \check A_2} 
\eeq
because $m_1n_2 - n_1m_2$ is antisymmetric between $(m_1, n_1)$ and $(m_2, n_2)$.
The $\SL(2,\BZ)$ twisted case will be important when we discuss the Maxwell theory with $\SL(2,\BZ)$ duality group in Sec.~\ref{sec:EMdual}.
 
Another point which may require clarification is how to define the holonomy $\chi$ from a differential cohomology with twisted coefficients.
By tensoring $\Delta^p \in C_p(X)$ with $\widetilde{\BZ}^n_\rho$ (using local trivialization as before), we define 
a chain with twisted coefficients $C_p(X, \widetilde{\BZ}^n_\rho)$.  Now, given a representation $\rho$, we take the dual representation $\rho'=(\rho^{-1})^T$.
Then, as a domain of the holonomy function, we take $Z^p(X, \widetilde{\BZ}^n_{\rho'})$. Then we can define $\exp(2\pi \i \int_M \DA_{A} )$
for $M \in Z_p(X, \widetilde{\BZ}_{\rho'}) $ and $\check A \in \check H^{p+1}(X, \widetilde{\BZ}^n_\rho)$ by using the invariant inner product between $\rho$ and $\rho'$. 

For example, suppose that we want to integrate a differential cohomology element on an unoriented submanifold $M$. 
Such a manifold is not an element of $Z_p(X)$. However, it can be regarded as an element of $Z_p(X, \widetilde{\BZ}_o)$, where $\widetilde{\BZ}_o$ is a local coefficient system on $X$ which agrees with the orientation bundle of $M$ when restricted to $M$. 
Then, if we consider the differential cohomology $\check H^{p+1}(X, \widetilde{\BZ}_o)$, we can integrate its elements  on $M$.
In this way we can obtain the holonomy $\chi(M) =\exp(2\pi \i \int_M \DA_{A})$. This case is relevant to string theory,
 since some of the $p$-form fields in string theory are represented by elements of such twisted differential cohomology $\check H^{p+1}(X, \widetilde{\BZ}_o)$.
 We discuss examples in Sec.~\ref{sec:O3}.

\section{Non-chiral $p$-form gauge fields and their anomaly}\label{sec:nonchiral}
In this section we consider a dynamical $p$-form gauge field $\check A \in \check H^{p+1}(X)$ which is coupled to background $(p+1)$-form and $(d-p-1)$-form
gauge fields $\check B \in \check H^{p+2}(X)$ and $\check C \in \check H^{d-p}(X)$ in a $d$-dimensional manifold $X$.
$\check B$ is interpreted as a background field for the $p$-form electric symmetry, and $\check C$ is interpreted as a background field for the $(d-p-2)$-form 
magnetic symmetry~\cite{Gaiotto:2014kfa}.

By using the formalism of differential cohomology, we describe a precise coupling between the dynamical field $\check A$ and the background fields $\check B$ and $\check C$.
In particular, we will derive a mixed anomaly between electric and magnetic higher-form symmetries.
This is essentially the Green-Schwarz mechanism, and it was also discussed in the context of higher-form symmetries in \cite{Gaiotto:2014kfa}.
In this section we take into account subtle topological effects which are not captured at the level of differential forms.

One typical question for which a precise formulation might be useful is as follows. 
A $\U(1)$ Maxwell field can be described by a differential cohomology element in $\check H^2(X)$.
Let $\check A \in \check H^2(X)$ be the $\U(1)$ Maxwell gauge field on a D-brane with worldvolume $X$. 
Let $\check B \in \check H^3(X)$ be the NSNS 2-form $B$ field.
The NSNS $B$ field can be regarded as a background field for the electric symmetry of the Maxwell field $A$ on the worldvolume.
At the level of differential forms, the combination $\d A+B$ is gauge invariant, and therefore 
the field strength $H = \d B = \d (B+\d A)$ must be exact when pulled back to the D-brane worldvolume.
This gives a constraint on which cycle a D-brane can wrap. 
However, at the level of integer cohomology $H^3(M,\BZ)$ rather than differential forms, it was argued~\cite{Freed:1999vc,Witten:1998xy} that 
the pullback of $H$ to the worldvolume is $[H]_\BZ=W_3$, where $W_3$ is a torsion element of $H^3(M,\BZ)$, or its twisted version $H^3(M,\widetilde{\BZ})$.
We want to have a correct understanding of this kind of topological phenomena which is finer than the discussion at the level of differential forms.

We will still neglect any K-theoretic nature of higher-form fields which may be necessary for some string theory applications;
we will see that ordinary differential cohomology can already describe many properties relevant for D-branes.

\subsection{Differential form analysis}
Let us first neglect topological subtleties and describe the system at the level of differential forms. 
The $p$-form dynamical field is denoted as $ A$. 
The theory has higher-form symmetries,
the electric $p$-form symmetry and the magnetic $(d-p-2)$-form symmetry.
We denote their corresponding background fields  as $B$ and $C$, respectively.
The fields $A, B$ and $C$ are here regarded as differential forms.
The action of the theory, including the background fields, is given by
\beq
- S  = - \frac{2\pi }{2g^2} \int_X (\d A +B) \wedge \hodge (\d A+B) +  2\pi \i \int_X C \wedge \d A. \label{eq:naiveaction}
\eeq
where $g$ is a free positive parameter, corresponding to the coupling constant in the case of Maxwell theory,
and $\hodge$ is the Hodge star.  

The gauge transformation of the fields $B$ and $C$ are given by
\beq
B \to B+ \d b, \qquad C \to C + \d c.
\eeq
For the first term of \eqref{eq:naiveaction} to be gauge invariant, we require $A$ to transform as $A \to A -b$.
However, the second term is not invariant under this transformation of $A$.

One might try to make the second term invariant under $B \to B + \d b$ and $A \to A - b$ by modifying the second term as $2\pi \i \int C \wedge (\d A+B)$.
However, now this term is not invariant under the transformation $C \to C + \d c$. Therefore, one of the gauge symmetries of $B$ and $C$
is always violated. This is the anomaly at the perturbative level.

For example, if $d=2$ and $p=0$, the above anomaly is a well-known anomaly of a compact scalar (which is denoted by $A$ here) whose target space is $S^1$.
A compact scalar has two $\U(1)$ symmetries. One is associated to the momentum in the target space $S^1$, and the other is associated to the winding on $S^1$.
There is a mixed anomaly between these two $\U(1)$ symmetries. 
At a special radius of $S^1$, the compact scalar is dual to a free Dirac fermion,
and the above anomaly is the mixed anomaly between the vector $\U(1)$ symmetry and the axial $\U(1)$ symmetry of the fermion.\footnote{%
As we discuss later, the anomaly of the bosonic theory is $\d C \wedge  B$.
On the other hand, the fermion side is as follows. 
Let us say that the left-movers couple to $A_L$ and the right-movers couple to $A_R$.
Then their anomalies is given by $\frac12  \int \d A_L \wedge A_L-\frac12  \int \d A_R \wedge A_R $.
We note that the factors of $\frac12$ needs to be taken care of using quadratic refinements utilizing spin structures as we review in Sec.~\ref{eq:QR}.
$B$ and $C$ on the compact boson side is known to be given by $B=A_L + A_R$
and $2C=A_L - A_R$. 
The mechanism realizing it is a bit subtle. Let us start from the free fermion theory. We can obtain the boson theory by
summing over spin structures, or in other words the $(-1)^F$ gauge field, of the fermion theory. Each of left and right $\U(1)$ symmetries $\U(1)_{L,R}$ has a mixed anomaly between
the $(-1)^F$ symmetry, which can be seen by putting the fermion on a Riemann surface with unit flux of the $\U(1)_L$ (or $\U(1)_R$)
gauge field, and see that the path integral measure contains a single zero mode which is odd under $(-1)^F$. 
Since we are gauging $(-1)^F$ as a dynamical field, we want to avoid this anomaly. This can be done by taking
the symmetry groups as vector and axial $\U(1)$ symmetries, $\U(1)_V$ and $\U(1)_A$, whose gauge fields $A_V$ and $A_A$ are related to $A_{L,R}$ as
$A_L=A_V+A_A$ and $A_R = A_V- A_A$. Notice that $\U(1)_L \times \U(1)_R = [ \U(1)_V \times \U(1)_A]/\BZ_2$, so $ \U(1)_V \times \U(1)_A$ is a $\BZ_2$ extension
of $\U(1)_L \times \U(1)_R$.
Now there is no mixed anomaly between $(-1)^F$ and $\U(1)_V \times \U(1)_A$,
and we can sum over $(-1)^F$. The sum over $(-1)^F$ is equivalent to a sum over (say) $\BZ_2 \subset \U(1)_V$, because
$(-1)^F$ and $(-1) \in \U(1)_V$ has the same effect on the fermion. Therefore, the symmetry group after summing over $(-1)^F$ is $\U(1)_V/\BZ_2$
instead of $\U(1)_V$, and the corresponding gauge field is $B= 2A_V$. The $\U(1)_A$ is unchanged and we rename the gauge field as $C = A_A$.
Thus we finally get $B=A_L+A_R$ and $2C = A_L - A_R$.
For more discussions, see \cite{Thorngren:2018bhj}.
}

Another example is $d=10$, $p=2$ and we take $A$ to be the NSNS $B$-field of the heterotic string theories (although it is denoted as $A$ here).
There, we take the fields $B$ and $C$ to be certain Chern-Simons forms of the heterotic gauge and gravitational fields. Then we get Green-Schwarz mechanism which produces
a gauge and gravitational anomaly from the NSNS field.\footnote{Beyond the perturbative level, the Green-Schwarz 
anomaly cancellation is very nontrivial and the formalism of quadratic refinement using the spin structure
may be necessary. See \cite{Freed:2000ta} for the case of Type I superstring theory. To the best of the authors' knowledge, the case of $E_8 \times E_8$ has not been studied.}

In the recent understanding of anomalies, we extend the spacetime manifold $X$ to one higher dimensional manifold $Y$ with $\partial Y =X$,
and also extend all background fields to $Y$.
Then we construct a gauge invariant action on $Y$. The anomaly is understood not as a violation of gauge transformations,
but as the dependence of the action on how to take the extension to $Y$. 

Let us follow this understanding. 
We  replace the second term of \eqref{eq:naiveaction} by a new term
\beq
2\pi \i \int_Y \d C \wedge (\d A +B),
\eeq
Notice that this is completely gauge invariant. However, it depends on the choice of the extension to $Y$.
To quantify this dependence, let us take another $Y'$ to which the background fields are extended in a certain way. 
The difference of the term defined by using the extensions $Y$ and $Y'$ is given by
\beq
2\pi \i \int_Y \d C \wedge (\d A +B) - 2\pi \i \int_{Y'} \d C \wedge (\d A +B) = 2\pi \i \int_{Y_{\rm closed}} \d C \wedge (\d A +B),
\eeq
where $Y_{\rm closed} = Y \cup \overline{Y}'$ is a closed manifold obtained by gluing $Y$ and the orientation reversal $\overline{Y}'$ of $Y'$ along the common boundary $X$.
By Stokes' theorem we have $\int_{Y_{\rm closed}} \d C \wedge \d A =0$, so this difference depends only on the background fields $B$ and $C$ and not on the dynamical field $A$.

We define the anomaly $\CA$ as $\CA(Y_{\rm closed}) = \frac{1}{2\pi} \arg \CZ(Y_{\rm closed}) \in \BR/\BZ$, where $\CZ(Y_{\rm closed})$ is the bulk partition function
on a closed manifold $Y_{\rm closed}$.
By the above discussion, it is given by
\beq
\CA(Y_{\rm closed}) =  \int_{Y_{\rm closed}} \d C \wedge B =  \int_{Y_{\rm closed}} (-1)^{d-p} C \wedge \d B.
\eeq
This is the anomaly at the differential form level.

From the above result, it is easy to guess the anomaly for topologically nontrivial fields beyond the differential form level. 
First of all, the fields $A$, $B$ and $C$ should be regarded as elements of differential cohomology, $\check A$, $\check B$ and $\check C$.
Then we can guess that the anomaly should be given as
\beq
\int_{Y_{\rm closed}} (-1)^{d-p}  \DA_{C \Dp  B}.\label{eq:mixedEM1}
\eeq
See \eqref{eq:DC6} for the expression of the product in differential cohomology for topologically trivial fields.

The anomaly will be indeed given by \eqref{eq:mixedEM1}.
However, precise descriptions and interpretations of the couplings to $\check B$ and $\check C$
are subtle (even without considering the anomaly) and we discuss them in the following subsections.

\subsection{Background field couplings and the anomaly}\label{sec:nonchiral2}
Now we try to make the action \eqref{eq:naiveaction} precise. 
We first consider the coupling to the background fields by
setting one of $\check B$ or $\check C$ to be zero.
Then we will turn on both $\check B$ and $\check C$.

\paragraph{Electric symmetry background.} 
We first focus on nontrivial $\check B$, which means that we set $\check C=0$ for the time being.
The kinetic term contains the combination $\d A+B$ at the differential form level. How can we make sense of this term in differential cohomology?
When $\check B=0$, we naturally expect that the kinetic term is written by $\DF_A$.
 
We may interpret $\d A+B$ to correspond to  
\beq
 \DF_A + \DA_B  . 
\eeq
For the kinetic term to be well-defined, this combination must make sense as an element of $\Omega^{p+1}(X)$.

The requirement that $ \DF_A +  \DA_B$ is an element of differential forms means that 
$\DA_B$ is required to be a differential form.
Thus, we have $\DN_B=0$, $\DF_B = \d \DA_B$ and hence $\check B$ must be topologically trivial. 

A natural question is how to interpret the case of nontrivial $\check B$. 
When $[\DN_B]_\BZ \neq 0$, it is not possible to set $\DN_B=0$ even if we use gauge transformations. 
One might conclude that the theory cannot be coupled to topologically nontrivial $\check B$.
However, we interpret it in a different way: we simply declare that the partition function in such situations is zero.
In more detail, recall that the partition function $\CZ$ of the theory
is a function of $\check B$ as well as other fields such as the background metric.
Then we interpret the above observation as follows. 
The theory can be coupled to topologically nontrivial $\check B$,
but we define the partition function to be zero, $\CZ=0$, unless $[\DN_B]_\BZ = 0$.
This definition might look ad hoc, but we will explain in Sec.~\ref{sec:remarks}  from various points of view why such a definition should be chosen.

If $[\DN_B]_\BZ =0$, $\DA_B$ can always be represented by differential forms up to gauge transformations
as explained in Sec.~\ref{sec:DC1}.
After setting $\DN_B=0$ and $\DA_B$ to a differential form, the remaining gauge transformation is
$ \DA_B \to  \DA_B + \delta \Da + \Dn$ where $ \Da \in C^{p}(X,\BR)$ and $\Dn \in Z^{p+1}(X,\BZ)$. They must be constrained in such a way that $\Df:= \delta \Da + \Dn$ is a differential form.
Then the triplet $(\Dn, \Da, \Df)$ gives a differential cohomology element.
Let us denote this element as $\check a = (\DN_a, \DA_a, \DF_a) = (\Dn, \Da, \Df)$.
The gauge transformation of $\DA_B$ is given by $\DA_B \to \DA_B + \DF_a$ and $\DN_B \to \DN_B - \delta \DN_a$.

Now the coupling of the above $\check B$ to the theory can be done simply by taking the kinetic term as
\beq
- S = - \frac{2\pi}{2g^2} \int (\DF_A + \DA_B) \wedge \hodge (\DF_A + \DA_B),
\eeq
where $\DA_B$ is understood to be a differential form.
The remaining gauge symmetry of $\check B$ is preserved by
\beq
\DA_B \to  \DA_B + \DF_a, \qquad \check A  \to \check A - \check a \label{eq:remaingauge}
\eeq
for differential cohomology elements $\check a \in \check H^{p+1}(X)$. 
Notice that $\check A$ itself may be viewed as the gauge degrees of freedom of $\check B$. This viewpoint will be important later.

\paragraph{Magnetic symmetry background.}
Next let us take $\check C \neq 0$ but $\check B=0$. 
In this case, the second term of \eqref{eq:naiveaction} can be made precise by differential cohomology product,
\beq
2\pi \i \int_X \DA_{C \Dp A}.\label{eq:productagain}
\eeq
Thus the coupling is straightforward in the framework of differential cohomology.

\paragraph{Both backgrounds.}
Finally we introduce both $\check B$ and $\check C$.
The fact that $\check B$ on $X$ must be of the form
\beq
\check B = (0, \DA_B, \d \DA_B )
\eeq
follows in the same way as before.
Let us first take the topological coupling as
\beq
 2\pi \i \int_X \DA_{C \Dp A} ; \label{eq:TC1}
\eeq
we will later modify this coupling.
This is not invariant under the gauge transformation \eqref{eq:remaingauge}.
The change is given by $ -2\pi \i \int_X \DA_{C \Dp a}$.
We might try to cancel it by changing the topological coupling to $C\wedge (\d A+B)$
by introducing $\DA_C \cup  \DA_B$. However, it cannot completely cancel the above change $ -2\pi \i \int \DA_{C \Dp a}$,
and moreover it is not at all invariant under gauge transformations of $C$.

\paragraph{Bulk-boundary couplings and the anomaly.}
Thus, there is an anomaly which cannot be cancelled in $d$ dimensions.
We try to cancel it by introducing a $(d+1)$-dimensional bulk $Y$ such that $\partial Y = X$.
Such a bulk with one higher dimension is called a symmetry protected topological (SPT) phase.
Motivated by the differential form analysis around \eqref{eq:mixedEM1}, let us consider the bulk theory whose action is given by\footnote{
In this section, we are actually working not with differential cohomology elements, but with differential cocycles, meaning that we consider $\check A, \check B, \check C$  before dividing by gauge transformations.}
\beq
& 2\pi \i (-1)^{d-p} \int_Y \DA_{C \Dp B} \nonumber \\
&= 2\pi \i (-1)^{d-p}  \int_Y \left( \DA_C \cup \DN_B - (-1)^{d-p-1} \DF_C \cup \DA_B  + Q(\DF_C, \DF_B) \right).
\eeq
In the bulk, the $\check B$ needs not be restricted to be of the form $\check B = (0, \DA_B, \d  \DA_B )$ with a differential form $\DA_B$.
Only the restriction of $B$ to the boundary $\partial Y$ must have this form.

Let us perform gauge transformations. Under 
\beq
\DA_C  \to   \DA_C + \delta \Da + \Dn,
\eeq
the above bulk action changes as
\beq
 &2\pi \i (-1)^{d-p} \int_Y \left( ( \delta \Da - \Dn)  \cup \DN_B  \right) \nonumber \\
 &=  2\pi \i (-1)^{d-p} \int_X  \Da \cup \DN_B  - 2\pi \i (-1)^{d-p}\int_Y \Dn \cup \DN_B \nonumber \\
 &= 0 \mod 2\pi \i,
\eeq
where we have used the fact that on the boundary $X =\partial Y$ the $\check B$ is restricted as $\DN_B=0$,
and also used the fact that $\int_Y \Dn \cup \DN_B$ is integer. Thus the bulk action is invariant under the gauge transformation of $\DA_C$.

Next, consider the gauge transformation
\beq
\DA_B  \to  \DA_B + \delta \Da +  \Dn, \qquad \DN_B \to \DN_B -\delta \Dn.
\eeq
We require that the pair $(\Da,  \Dn)$ becomes $ ( \DA_a, \DN_a)$ for some differential cocycle $\check a = (\DN_a, \DA_a, \DF_a)$ on the boundary $X$
so that the transformation on the boundary is $\DA_B \to \DA_B +\delta \DA_a + \DN_a = \DA_B +\DF_a$.
The change of the bulk action is given by
\beq
 &2\pi \i (-1)^{d-p}\int_Y \left( \DA_C \cup ( - \delta \Dn) - (-1)^{d-p-1} \DF_C \cup (\delta \Da + \Dn) \right) \nonumber \\
 & = 2\pi \i (-1)^{d-p} \int_X \left( (-1)^{ d-p} \DA_C \cup   \Dn + \DF_C \cup \Da  \right) 
 - 2\pi \i   \int_Y \left(  \delta\DA_C \cup   \Dn - \DF_C \cup \Dn  \right).
\eeq
The second term becomes $ 2\pi \i   \int_Y ( \DF_C- \delta\DA_C) \cup   \Dn  = 2\pi \i  \int_Y (  \DN_C \cup \Dn  ) = 0 \mod 2\pi \i $.
The first term is completely on the boundary $X$ and hence we can set $\Da = \DA_a$ and $\Dn=\DN_a$,
\beq
&2\pi \i  \int_X \left(  \DA_C \cup   \DN_a +(-1)^{d-p} \DF_C \cup \DA_a  \right) \nonumber \\
&= 2\pi \i \int_X ( \DA_{C \Dp a} - Q(\DF_C, \DF_a) ).
\eeq
The first term precisely cancels the change of the topological coupling \eqref{eq:TC1} under the gauge transformation $\check A \to \check A  - \check a$.
The second term can be cancelled by introducing the counterterm on $X$ which is given by
\beq
2\pi \i \int_X Q(\DF_C, \DA_B).
\eeq
This is well-defined because $\DA_B$ is a differential form on $X$.
The gauge variation of this term under $\DA_B \to \DA_B + \DF_a$ is $2\pi \i \int_M Q(\DF_C, \DF_a)$.

\paragraph{Summary.}
We can summarize the above results as follows. The precise definition of the coupling $C \wedge \d A$ on $d$-manifold $X$ requires
extending it to $(d+1)$-manifold $Y$ with boundary $\partial Y = X$. The background fields $\check B, \check C$ (but not the dynamical field $\check A$) are extended to the bulk $Y$.
Then the coupling can be defined as
\beq
 2\pi \i \int_X \left( \DA_{C \Dp A}  +  Q(\DF_C, \DA_B ) \right) + 2\pi \i (-1)^{d-p} \int_Y \DA_{C \Dp B},
\eeq
which expands to
\begin{multline}
= 2\pi \i \int_X \left( \DA_{C \Dp A}  +  Q(\DF_C, \DA_B ) \right) \\
 + 2\pi \i (-1)^{d-p}  \int_Y \left( \DA_C \cup \DN_B - (-1)^{d-p-1} \DF_C \cup \DA_B  + Q(\DF_C, \DF_B) \right).  \label{eq:TC0}
\end{multline}
This is the gauge invariant action combining the bulk SPT phase and the boundary topological term.
The bulk SPT phase itself is described by $2\pi \i (-1)^{d-p}\int_Y \DA_{C \Dp B}$ if there is no boundary $\partial Y=0$.
The anomaly of the boundary dynamical theory $\check A$ is $\CA = (-1)^{d-p}\int_Y \DA_{C \Dp B}$.

We can look at the above result in another way. Let us start from the bulk action given by $2\pi \i (-1)^{d-p}\int_Y \DA_{C \Dp B}$.
Suppose that we want to make this action gauge invariant on a manifold $Y$ with boundary $X$.
It may not be possible unless the field $\check B$ is topologically trivial at the boundary $[\DN_B]_\BZ=0$ so that $B = (0, \DA_B, \d \DA_B)$.
Even then, there still remains the gauge transformation which reduces to $\DA_B \to \DA_B + \DF_a$ on the boundary.
This gauge degrees of freedom is described by differential cohomology element $\check a$. 
We take advantage of this fact and, roughly speaking, promote the gauge degrees of freedom $\check a$ to a physical field $\check A$ on the boundary so that the gauge
invariance is recovered. This is a kind of Stueckelberg field, but only lives on the boundary.
It also has similarity with the symmetry extension method of \cite{Witten:2016cio,Wang:2017loc,Tachikawa:2017gyf,Kobayashi:2019lep} which was discussed for discrete symmetries,
and it would be interesting to study its generalization to the continuous case.

\subsection{Some remarks}\label{sec:remarks}

\paragraph{Why is it OK to set $\CZ=0$ when $[\DN_B]_\BZ \neq 0$?} 
We saw above that for the electric background $\check B$, we had to define the partition function to be zero if it is topologically nontrivial, $[\DN_B]_\BZ \neq 0$. 
This definition might have looked rather ad-hoc.
Here we provide various arguments why this procedure is consistent.

Let us first consider the simplest case $p=0$ and $d=2$. 
In this case, the pair $(A,B)$ is usually denoted instead by $(\phi,A)$.
We note that $\phi$ is a section of an $S^1$ bundle coupled to the $\U(1)$ gauge background $A$.
Then what we found above  simply means that $A$ needs to be topologically trivial when there is a section $\phi$.
We said above that we set $\CZ=0$ when $A$ is topologically nontrivial. 
Why is this a consistent operation?
\begin{itemize}
\item One argument is to realize the compact scalar $\phi$ as the phase degree of freedom of a complex scalar field $\Phi$ with a potential term $V(|\Phi|)\le 0$ such that $V=0$ iff $\Phi=R e^{2\pi \i \phi}$.
Any topologically nontrivial $A$ forces $\Phi$ to take the value $\Phi=0$ at a number of points.
This costs a lot of energy, and by scaling $V(\Phi)$ by a large positive factor, the contribution of such configurations to the partition function $\CZ$ becomes zero.

\item Another argument is as follows. Decompose $X$ into two pieces, $X=X_1\cup X_2$,
such that $X_1$ and $X_2$ are glued along their common boundary $S^1$.
Even when $A$ is topologically nontrivial on $X$, $A$ is topologically trivial on $X_1$ and $X_2$.
As such, the path integral over $X_1$ and $X_2$ produces  nonzero states $\ket{X_1}$ and $\ket{X_2}$ on $\CH(S^1)$.
Recall that an element $\CH(S^1)$ is a functional $\Psi[\phi(x)]$ where $\phi(x)$ is a function $\phi:S^1\to S^1$.
The space of functions splits into disconnected components labeled by the winding number.
Then $\ket{X_1}$ as a wave functional is concentrated on a single winding number, say $n_1$,
and similarly, $\ket{X_2}$ is concentrated on another winding number $n_2$.
The topology enforces that $n_1-n_2$ is the Chern number of the $\U(1)$ bundle $A$ on $X$.
Therefore, when $A$ is topologically nontrivial, we have $n_1\neq n_2$.
This means that $\ket{X_1}$ and $\ket{X_2}$ are supported on different winding numbers, and 
$\CZ(X)=\braket{X_1}{X_2}=0$.
Therefore setting $\CZ(X)=0$ is not an ad hoc procedure; it follows from the basic gluing axiom of quantum field theory.

\item Yet another argument uses  fermionization. 
Recall that a compact scalar in $d=2$ is equivalent to a non-chiral Dirac fermion when the radius of the scalar is appropriately chosen.
In this description, what happens when $A$ is topologically non-trivial is that there are fermionic zero modes.
Therefore, the partition function vanishes, $\CZ=0$ unless we insert fermion operators to absorb the zero modes.

\item The preceding argument can be modified so that it can be applied at an arbitrary radius of the compact scalar, not just on the free fermion radius.
Recall that there is a mixed anomaly between the momentum $\U(1)$ symmetry and the winding $\U(1)$ symmetry.
This means that when the  background for the momentum $\U(1)$ is topologically nontrivial and has the Chern number $n$,
the spacetime has the anomalously-induced winding number $n$, 
so that the correlation function vanishes
unless we insert a set of operators with total winding number $-n$.
We note that in the path integral language, the argument given here simply means that to have a nonzero partition function,
we need to insert the operators of nonzero winding number to cancel the winding number introduced by the momentum $\U(1)$ background gauge field.
Therefore, the partition function vanishes $\CZ=0$ if there are no insertions.
This argument is admittedly somewhat circular, since a precise formulation of the mixed anomaly requires the fact the electric background is topologically trivial
on the space $X$ in which $\phi$ lives.
\end{itemize}

Let us come back to the case of general $d$ and $p$.
One way to convince ourselves that setting $\CZ=0$ when $[\DN_B]_\BZ\neq 0$
 is the correct definition is by the following argument.
The dynamical $p$-form theory $\check A$ is expected to have the electromagnetic dual description in terms of a $(d-p-2)$-form field. 
For example, when $d=2$ and $p=0$, it is the T-duality of a compact scalar field. For $d=4$ and $p=1$,
it is the usual electromagnetic duality of Maxwell theory. 
The duality for general $d$ and $p$ may be shown along the lines of arguments in \cite{Witten:1995gf}.
Under the duality, the roles of $\check B$ and $\check C$ are exchanged.
Below, we are going to show that the partition function is zero if the topological class of the magnetic background $\check C$ is nonzero, $[\DN_C]_\BZ \neq 0$.
Then by electromagnetic duality, it is reasonable to require that the partition function is zero if the electric background is topologically nontrivial, $[\DN_B]_\BZ \neq 0$.

Let us show that the partition function is zero if $[\DN_C]_\BZ \neq 0$. 
For this purpose, we set the dynamical gauge field as $\check A = \check A_0 + \check A'$, where $\check A'$ is a flat field, i.e.~$\DF_{A'}=0$, such that 
it gives an element of cohomology $[\DA_{A'}] \in  H^{p}(M, \BR/\BZ)$.
Then we perform path integral over $\check A'$. Because $\check A'$ is flat, it does not affect the kinetic term $\DF_A \wedge \hodge \DF_A$,
so the flat $\check A'$ only affect the topological coupling. For flat $\check A'$, it was shown in Sec.~\ref{sec:DC2} that the product gives
\beq
2\pi \i \int_X \DA_{C \Dp A'} =   2\pi \i (-1)^{d-p} \int_X  [\DN_C] \cup [\DA_{A'}].
\eeq
By Poincar\'e-Pontryagin duality, the product between $ [\DN_C]  \in H^{d-p}(X, \BZ)$ and $[\DA_{A'}]  \in H^{p}(X, \BR/\BZ)$ is
non-degenerate. Therefore, by integrating over all $[\DA_{A'}]  \in H^{p}(X, \BR/\BZ)$,
the partition function becomes zero unless $ [\DN_C] =0$.

The electromagnetic duality was derived in \cite{Witten:1995gf} in the following manner.
Here we present it with a coupling to a single background field. 
We start with a dynamical $p$-form field $\check A$,
a dynamical $(p+1)$-form field $\check G$,
a dynamical $(d-p-2)$-form field $\underline{\check A}$,
and a background $(p+1)$-form field $\check B$.
The action is \begin{equation}
-S=-\frac{2\pi}{2g^2}\int_X (\DF_A+\DA_G) \wedge * (\DF_A+\DA_G) 
+ 2\pi \i \int_X \DA_{\underline{A} \Dp G}
- 2\pi \i \int_X \DA_{\underline{A} \Dp B}.
\end{equation}
This action can be studied in two different ways.
The first approach goes as follows.
We gauge-fix $\check A=0$ by using the gauge transformation $\DA_G \to \DA_G + \DF_a$ and $\check A \to \check A - \check a$.
Then $\DA_G$ is a differential form, and can be integrated out. We then get \begin{equation}
-S=-\frac{2\pi g^2}2 \int_X \DF_{\underline{A}}\wedge *\DF_{\underline{A}}
-  2\pi \i \int_X \DA_{\underline{A} \Dp B}.
\end{equation}
The second approach is to integrate out $\underline{A}$.
From the perfectness of the differential cohomology pairing, 
we see that $\check G$ is gauge-equivalent to $\check B$,
and in particular there is a $p$-form field $\check a$ such that $\DA_G=\DA_B+\DF_a$.
We shift $\check A$ by $\check a$, and find that the action is now \begin{equation}
- S = - \frac{2\pi}{2g^2} \int_X (\DF_A + \DA_B) \wedge \hodge (\DF_A + \DA_B).
\end{equation}
As one sees, the argument is somewhat circular, since when we made both $\check A$ and $\check G$ dynamical, 
we needed to use the fact that the path integral concentrates to the configurations where $\check G$ is topologically trivial.
Still we believe our argument helps in demonstrating the overall consistency.

We also note here that a similar analysis can be carried out in the case of $\bZ_n$ $p$-form gauge theories.
There, too, the partition function becomes zero when either the electric or the magnetic background fields are topologically nontrivial \cite{Kapustin:2014lwa,Kapustin:2014zva,Witten:2016cio,Wang:2017loc,Tachikawa:2017gyf,Kobayashi:2019lep}.

\paragraph{Why is it not OK to require $\check B=0$ on the boundary?} 
So far, we argued that it is natural to set the partition function of the boundary theory to be zero when the electric background field is topologically nontrivial, $[\DN_B]_\BZ \neq 0$.
This may raise the following question. 
If we restrict the field $\check B$ to be completely zero at the boundary,
the action is invariant under gauge transformations which preserves $\check B=0$ at the boundary. Then it seems that we do not need any degrees of freedom at the boundary.
Is it possible to put an SPT phase on a manifold with boundary, but no boundary degrees of freedom? What forbids us from doing so? 
We do not try to give a complete answer to this question, but let us sketch an idea why it is forbidden. 

We are working with Euclidean signature metric, and hence there is no distinction between space and time.
 Then it is possible to see the boundary as a time slice rather than a spatial boundary.
If we look the boundary as a time slice, we get a Hilbert space of the bulk SPT phase which is described by an invertible field theory.
This point of view is considered to be essential 
 for general understanding of anomalies~\cite{Freed:2014iua}, and indeed played a crucial role~\cite{Yonekura:2016wuc,Witten:2019bou} in a general understanding of chiral fermion anomalies.
The defining property of an invertible field theory is that it has a one dimensional Hilbert space on any background on the time slice.
In particular, we do not want to restrict to the background allowed for the bulk invertible phase to $[\DN_B]_\BZ=0$ on the time slice;
we should allow completely general $\check B$ for an invertible field theory.
We note that restricting to $[\DN_B]_\BZ=0$ could have been possible if the Hilbert space had zero dimension for $[\DN_B]_\BZ \neq 0$. 
But this contradicts with the definition of invertible field theory.\footnote{
For more general theories which are not invertible, it is possible that the Hilbert space dimension becomes zero for certain backgrounds.
For example, let us consider a 3-dimensional abelian Chern-Simons theory with level $\kappa$, coupled to a background field $B$.
(Here we use a sloppy description without using differential cohomology and precise quadratic refinement.)
The action is $ \frac{2\pi \i \kappa}{2} \int A \wedge \d A + 2\pi \i \int B \wedge \d A $ where we normalized fields so that fluxes are integers.
The equation of motion of $A$ is $\kappa \d A =- \d B$. Now consider the Hilbert space $\CH(\Sigma) $ on a Riemann surface $\Sigma$.
If the flux $\int_\Sigma \d B$ of the background is not a multiple of $\kappa$, the above equation of motion implies that the Hilbert space is empty, $\dim \CH(\Sigma)=0$.
The case $\kappa=1$ is an invertible field theory, and in this case $\int_\Sigma \d B$ is always a multiple of $\kappa=1$ and $\dim \CH(\Sigma)=1$.
}
So we would like to consider general $\check B$, and try to preserve the gauge invariance by introducing some degrees of freedom on the boundary.

Then, we can justify the restriction to $[\DN_B]_\BZ=0$ on the boundary $X = \partial Y$ only if the boundary theory has the property that its partition function is zero when $[\DN_B]_\BZ \neq 0$.
In the previous subsections, we have argued that this is the case for the $p$-form gauge theory.
Strictly speaking, the fact that the partition function is zero for $[\DN_B]_\BZ \neq 0$ was imposed by hand, but we have given various physical justifications of this claim.
More generally,
whether we can take the partition function to be zero on a certain class of the background should be answered
by the principle that such a choice is consistent with the axioms of quantum field theory (i.e.~locality, unitarity, etc.).\footnote{
As an example which violates the axioms of quantum field theory,  suppose that the partition function is zero on $S^d$ with topologically trivial backgrounds.
Moreover, for simplicity, we assume that the theory is a topological quantum field theory.
In that case, it is possible to show by the axioms of topological quantum field theory that all partition functions are zero, and in particular, partition functions on $S^1 \times X$ are zero for
arbitrary $X$.
(See e.g.~Sec.~3 of \cite{Yonekura:2018ufj} for how to prove this claim.)
This is inconsistent if the theory has any Hilbert space of nonzero dimensions at all.
More nontrivial versions of this kind of argument have been used to get useful constraints on the partition function of topological field theories~\cite{Cordova:2019bsd,Cordova:2019jqi}.
}
For the $p$-form theory, we do not try to
prove the consistency of taking the partition functions to be zero for $[\DN_B]_\BZ \neq 0$,
but the consistency is suggested by e.g.~the electromagnetic duality.

\subsection{Applications to D3-brane in O3-plane background}\label{sec:O3}
Now we consider some applications of the above results to D-branes in string theory. 
On a single D-brane, there is a Maxwell field which is described by $\check A \in \check H^2(X)$ where $X$ is the worldvolume.
It is also coupled to the background NSNS $B$-field and RR $C$-field, although there is a subtle detail which we will explain later.
First we need to make a few preliminary remarks.

So far we discussed the case of coefficients which do not depend on the positions on $X$. But it is possible to consider
twisted coefficients as discussed in Sec.~\ref{sec:DC4}. There is no change in the discussions of this section
beyond what has already been mentioned in Sec.~\ref{sec:DC4}.

Next, we have to distinguish an ordinary $\U(1)$ field and a field which appears as a $\spin^c$ connection.
A $\spin^c$ connection is a $\U(1)$ part of the connection of $[\Spin(d) \times \U(1) ] /\BZ_2$, where $\Spin(d)/\BZ_2=\SO(d)$
is the spacetime Lorentz symmetry (in Euclidean signature). The gauge field which appears on a D-brane is actually a $\spin^c$ connection.
For the purpose of the present discussion, it is enough to consider a $\spin^c$ connection as a sum of an ordinary $\U(1)$ connection
and a certain background field constructed as follows. 

We have Stiefel-Whitney classes $w_q \in H^q(X, \BZ_2)$ of a manifold $X$. 
Then we can canonically define a differential cohomology element $\check w_q $ corresponding to $w_q$ as follows.
First, we can regard $\BZ_2$ as half integers $\frac{1}{2}\BZ$ mod integers $\BZ$. Next, we can uplift $w_q$ to a real cochain $\DA(w_q) \in C^q(M,\BR)$.
This uplift is unique up to gauge transformations $\DA(w_q) \to \DA(w_q) + \delta \Da + \Dn$.
Then, define $\DN(w_q) = -\delta \DA(w_q)$. 
The $\DN(w_q)$ takes values in cochains with integer coefficients, because $w_q$ is closed as a cochain with $\BZ_2$ coefficients.
By definition of Bockstein homomorphism $\beta$ defined in Sec.~\ref{sec:DC1}, the topological class of $\DN(w_q) $ is $[\DN(w_q) ] = \beta(w_q) \in H^{q+1}(M,\BZ)$.
Then we can define the differential cohomology element as $\check w_q : =  ( \DN(w_q), \DA(w_q), 0)$. This is unique up to gauge transformations.

We consider a $\spin^c$ connection as an ordinary $\U(1)$ field $\check A \in \check H^2(X)$ coupled to the electric background given by $\check w_2$,
where $w_2$ is the second Stiefel-Whitney class of $X$. This means that the field strength of the $\spin^c$ connection is $\DF_A + \DA(w_2)$.
Let $\check B_{\rm NSNS}$ be the NSNS 2-form field of string theory.
The total electric background field is given by
\beq
\check B = \check B_{\rm NSNS} + \check w_2. \label{eq:Bshift}
\eeq
The field strength appears in the kinetic term in the combination $\DF_A + \DA_B$.

By the discussion of the previous subsections, the constraint on a single D-brane 
\beq
[H_{\rm NSNS}] _\BZ= W_3
\eeq
can be easily understood, where $[H_{\rm NSNS}] _\BZ :=[\DN_{B_{\rm NSNS}}]$ is the flux of the NSNS 2-form field in integer cohomology, and $W_3:= \beta(w_2) $. 
We emphasized that $\check B$ must be topologically trivial, $[\DN_B]_\BZ=0$, for the partition function to be nonzero.
Then, we get $[\DN_{B_{\rm NSNS}} + \DN(w_2)]_\BZ=0$. Here $[H_{\rm NSNS}] _\BZ =[\DN_{B_{\rm NSNS}}]$
and $W_3= \beta(w_2) = [\DN(w_2)]_\BZ$. $W_3$ is a 2-torsion $2W_3=0$, so we get the desired result. 

We conclude that a single D-brane can be wrapped on a cycle $X$ only if we have $[H_{\rm NSNS}] _\BZ= W_3$ on that cycle $X$.
It reproduces the result in \cite{Freed:1999vc,Witten:1998xy} by a somewhat different line of reasoning. 
The above result is valid only for the abelian gauge field, and it is modified for multiple D-branes.

Another application is the anomaly of the Maxwell field on the D-brane. 
As we have shown, the anomaly is given by
\beq
\CA = (-1)^{d-1}\int_Y \DA_{C \Dp B},
\eeq
where $d$ is the dimension of the D-brane worldvolume, and we have set $p=1$.
$\check B$ is given as in \eqref{eq:Bshift}, and we expect $\check C$ to be shifted similarly as $\check C = \check C_{\rm RR} +\cdots$,
where $\check C_{\rm RR}$ is the $d-2$-form RR-field. However, the RR-fields are actually described by differential K-theory (or its generalization in the presence of other backgrounds)
rather than ordinary differential cohomology. Thus the above anomaly is not really a complete answer. Nevertheless, let us try to understand some consequences
of the above anomaly to the extent that it works.

Let us consider a D3-brane (i.e.~$d=4$) in Type IIB string theory. 
By the S-duality of Type IIB string, we expect that the relation between 
the total magnetic background field $\check C$ and the RR background field $C_{\rm RR}$ is given by
\beq
\check C = \check C_{\rm RR}  + \check w_2.
\eeq
This is the S-dual of the relation \eqref{eq:Bshift}.

In particular, let us consider a D3-brane in the presence of an O3-plane. 
See \cite{Witten:1998xy} for the background about O3-planes used in the following discussion.

The O3-plane geometry is given by
\beq
  \BR^4 \times \BR^6/\BZ_2.
\eeq
Given a worldvolume $X$ of a D3-brane, we have to take an extension to one higher dimension $Y$ with $\partial Y = X$ to make the action (and hence the partition function) well-defined.
The dependence on $Y$ is the anomaly. In particular, let us compare the difference between $Y$ and $Y'$
in the case that the closed manifold $Y_{\rm closed} = Y \cup \overline{Y}'$ obtained by gluing them is given by
\beq
Y_{\rm closed} = \{0\} \times \mathbb{RP}^5 \subset   \BR^4 \times \BR^6/\BZ_2.
\eeq
The anomaly is $\CA = -\int_{\mathbb{RP}^5} \DA_{C \Dp B}$.

The background fields $\check B_{\rm NSNS} $ and $\check C_{\rm RR}$ are elements of differential cohomology with twisted coefficients $\check H^3(\mathbb{RP}^5, \widetilde{\BZ})$,
where the twisting of $\widetilde{\BZ}$ is such that the sign is changed by going around the nontrivial loop $\pi_1 (\mathbb{PR}^5) = \BZ_2$.
For the topological classification of these fields, the relevant cohomology group is
\beq
H^3(\mathbb{RP}^5, \widetilde{\BZ}) = \BZ_2.
\eeq
In the O-plane background, the fields are flat, $\DF_{B} = \DF_C =0$. So $ \int_{\mathbb{RP}^5} \DA_{C \Dp B}$ is given by the torsion pairing
of $[\DN_B]_{\BZ}$ and $[\DN_C]_{\BZ}$ which was discussed in Sec.~\ref{sec:DC2}. The torsion pairing is known to be a perfect pairing.
This implies that if both $[\DN_B]_{\BZ}$ and $[\DN_C]_{\BZ}$ are nonzero element of $H^3(\mathbb{RP}^5, \widetilde{\BZ}) = \BZ_2$,
then we get $ \int_{\mathbb{RP}^5} \DA_{C \Dp B} = 1/2 \mod 1$. This is a consequence of the Poincar\'e-Pontryagin duality mentioned in Sec.~\ref{sec:DC2}.
If either $[\DN_B]_{\BZ}$ or $[\DN_C]_{\BZ}$ is zero, we have $ \int_{\mathbb{RP}^5} \DA_{C \Dp B} = 0 \mod 1$.

For the $\BZ_2$ coefficients, the twisting has no effect, $\widetilde{\BZ}_2 = \BZ_2$ since $1 \equiv -1 \mod 2$.
Thus $w_2$ can be regarded as an element of $H^2(\mathbb{RP}^5, \widetilde{\BZ}_2) $.
Then we have the Bockstein homomorphism $\beta:H^2(\mathbb{RP}^5, \widetilde{\BZ}_2) \to H^3(\mathbb{RP}^5, \widetilde{\BZ}) $.
It is known\footnote{It can be shown by using part of the long exact sequence 
$H^2(\mathbb{RP}^5, \widetilde{\BZ}) \longrightarrow  H^2(\mathbb{RP}^5, \widetilde{\BZ}_2)  \longrightarrow H^3(\mathbb{RP}^5, \widetilde{\BZ}) $
and the fact that $H^2(\mathbb{RP}^5, \widetilde{\BZ})  =0$. These facts imply that $H^2(\mathbb{RP}^5, \widetilde{\BZ}_2)  \longrightarrow H^3(\mathbb{RP}^5, \widetilde{\BZ}) $
is injective, and this map is the Bockstein homomorphism $\beta$. The second Stiefel-Whitney class $w_2$ is nonzero in $\mathbb{RP}^5$
and hence $W_3 = \beta(w_2)$ is nonzero.} 
that $W_3 =\beta(w_2)$ is the nonzero element of $H^3(\mathbb{RP}^5, \widetilde{\BZ}) $. 
Thus the shifts in $\check B = \check B_{\rm NSNS} + \check w_2$ and $\check C = \check C_{\rm RR} + \check w_2$
are nontrivial.

Let us represent the elements of $H^3(\mathbb{RP}^5, \widetilde{\BZ}) = \BZ_2$ by 0 and 1 mod 2.
By using the facts discussed above, we get the following values of the anomaly $-\int_{\mathbb{RP}^5} \DA_{C \Dp B}$.
We abbreviate $[{\rm NSNS}]_\BZ = [\DN_{B_{\rm NSNS} } ]_\BZ$ and $[{\rm RR}]_\BZ = [\DN_{C_{\rm RR} } ]_\BZ$.
\beq
\begin{array}{|c|c|c|c|c|}
\hline
\BZ_2 \text{ flux } ( [{\rm NSNS}]_\BZ, [{\rm RR}]_\BZ ) & (0,0)& (1,0)&(0,1) &(1,1) \\
\hline
\text{type of O-plane} &\rm{O3}^-&\rm{O3}^+&\widetilde{\rm{O3}}^-&\widetilde{\rm{O3}}^+ \\
\hline
\int_{\mathbb{RP}^5} \DA_{C \Dp B} \mod 1 & 1/2&0&0&0 \\
\hline
\end{array} \label{eq:O3list}
\eeq
The correspondence between the fluxes and the types of O-plane was discussed in \cite{Witten:1998xy}.
This is a reasonable result, because three O-planes $\rm{O3}^+, \widetilde{\rm{O3}}^-$ and $\widetilde{\rm{O3}}^+$
are related by the $\SL(2,\BZ)$ duality of Type IIB string while $\rm{O3}^-$ is a singlet under the $\SL(2,\BZ)$.

The anomaly leads to an ambiguity of the partition function.
For the consistency of string theory, the total ambiguity must be cancelled.
There are two other sources for the anomaly. One source is the anomaly of the worldvolume fermion which we denote as $\CA_{\rm fermion} \mod 1$.
Another source is the coupling to the RR 4-form field $C_4$ (which is different from the above RR 2-form $C = C_2$). 
Naively the coupling of the D3-brane to $C_4$ is given by $
\int_X C_4$. However, we need to take an extension $Y$ such that $X = \partial Y$,
and define it as $
\int_Y F_{5}$, where $F_{5}$ is the field strength of $C_4$. Then the ambiguity, or the anomaly, from this coupling is given by
\beq 
\int_{Y_{\rm closed}}  F_{5},
\eeq
where $Y_{\rm closed}$ is as before.\footnote{For ordinary differential cohomology, $\int_{Y_{\rm closed}} \DF$ is defined to be an integer.
However, in the presence of O-planes, $\int_{Y_{\rm closed}} F_5$ is not an integer. Therefore, $C_4$ is not precisely a differential cohomology element. } 
In particular, we are now concerned with the case that $Y_{\rm closed} = \mathbb{RP}^5$ as in the above discussion.
Therefore, for the total anomaly cancellation, we must have the condition that
\beq
\int_{\mathbb{RP}^5} F_{5} + (-1)\int_{\mathbb{RP}^5} \DA_{C \Dp B} + \CA_{\rm fermion}(\mathbb{RP}^5) = 0 \mod 1. \label{eq:O3cancel}
\eeq
In \cite{Tachikawa:2018njr}, it was shown that $\int_{\mathbb{RP}^5} F_{5} + \CA_{\rm fermion}(\mathbb{RP}^5) = 0$ for the $\rm{O3}^+$-plane.
The fluxes $ ( [{\rm NSNS}]_\BZ, [{\rm RR}]_\BZ ) $ do not affect the fermions, so $\CA_{\rm fermion}$ is independent of
the types of O-planes. 

The $\rm{O3}^-$ has RR-charge $-1/4$ while $\rm{O3}^+, \widetilde{\rm{O3}}^-$ and $\widetilde{\rm{O3}}^+$ have RR-charge $+1/4$.
Then the value of $\int_{\mathbb{RP}^5} F_{5}$ is $-1/4$ for $\rm{O3}^-$ and $+1/4$ for the others. 
The anomaly cancellation of $\rm{O3}^+$ requires $\CA_{\rm fermion} = - 1/4$. 
(See \cite{Tachikawa:2018njr} for more details.) From these results and \eqref{eq:O3list}, we see that the anomaly cancellation condition \eqref{eq:O3cancel}
is satisfied for all O3-planes. This result was announced in \cite{Hsieh:2019iba}.

The above analysis has been done assuming that $\check B$ and $\check C$ are flat.
It is also interesting to note the following point. Suppose now that they are not flat. 
Let us take a 6-manifold $Z$ with a boundary $\partial Z$. The anomaly cancellation must also hold in $\partial Z$ for arbitrary $Z$, and hence we get
\beq
0 =\int_{ \partial Z} F_{5} + (-1)\int_{ \partial Z} \DA_{C \Dp B} + \CA_{\rm fermion}(\partial Z)  
= \int_Z \left( \d F_{5} - \DF_C \wedge \DF_B \right)
\eeq
where we have used the fact that $ \CA_{\rm fermion}(\partial Z) =0 $ and $ \DF_{C \Dp B} = \DF_C \wedge \DF_B $.
The above equality must hold for arbitrary $Z$, and hence we get
\beq
\d F_{5} =  \DF_C \wedge \DF_B.
\eeq
This is a well-known equation of motion of $F_5$ in Type IIB supergravity.
Therefore, this supergravity equation is necessary for the well-definedness of a D3-brane.

\section{Quadratic refinement of differential cohomology pairing}\label{eq:QR}

In the previous section we have described non-chiral $p$-form gauge fields as boundary modes of bulk SPT phases. 
The relevant anomaly was the mixed anomaly between electric and magnetic higher-form symmetries. 
Such theories can be described by differential cohomology, as we have seen.

In the rest of the paper we would like to study chiral, or (anti-)self-dual $p$-form fields. 
The spacetime dimensions must be $d=2p+2$. If the coefficients are just single $\BZ $ (and the associated $\BR, \BR/\BZ$), the (anti-)self-dual condition
is only possible in dimensions $d=4n+2$ and $p=2n$. For more general coefficients such as $\BZ^2$, other dimensions such as $d=4$
are also possible. 

If we try to formulate such theories uniformly for all dimensions of the form $d=4n+2$, we need what is called the Wu structure~\cite{Hopkins:2002rd,Monnier:2016jlo}.
However, for most purposes of string theory and condensed matter physics, we only need dimensions $d=2,6,10$.
In these cases, spin structure, rather than Wu structure, can be directly used to formulate chiral $p$-form fields. 
Requiring spin structure (or its generalization such as $\spin^c$) is more natural in physics because we already have fermions in 
relevant physical systems, 
so we will use only spin structure in this paper. 
The cost is that we have to study each case $d=2,6,10$ in somewhat dimension-dependent way,
although the underlying idea, which involves Atiyah-Patodi-Singer (APS) $\eta$-invariant, is common to them.
The case $d=4$ with the coefficients $\bZ^2$ can be deduced from the $d=6$ case by compactifications on $T^2$, which we will study in detail in Sec.~\ref{sec:EMdual}.

\subsection{Basic properties of quadratic refinements}\label{sec:basicQ}
To formulate chiral $p$-form fields in $d$ dimensions, we will need a quadratic refinement of pairing in differential cohomology in $d+1$ dimensions.
First we define the pairing of $\check A_1$ and $\check A_2$ as
\beq
(\check A_1, \check A_2) = \int_Y \DA_{A_1 \Dp A_2} \in \BR/\BZ. \label{eq:pair}
\eeq
where $Y$ is a $d+1$-dimensional manifold, and $A_1, A_2 \in \check H^{p+2}(Y)$ are $p+1$-form fields.
(The reason that we need $p+1$-forms rather than $p$-forms will become clear in Sec.~\ref{sec:theory}).
Then a quadratic refinement $\CQ$ is a map
\beq
\CQ : \check H^{p+2}(X) \ni \check A \mapsto \CQ(\check A) \in \BR/\BZ \cong \U(1)
\eeq
such that
\beq
\CQ(\check A_1 + \check A_2) - \CQ(\check A_1) - \CQ(\check A_2) + \CQ(0)= (\check A_1, \check A_2)  . \label{eq:defQ}
\eeq
This is the defining property of a quadratic refinement.
Note that we do not require that $\CQ(0)=0$.
We will also use 
\beq
\widetilde \CQ(\check A) := \CQ(\check A)-\CQ(0)
\eeq
 which is also a quadratic refinement.
From the above definition, we need $d+1 = 2p+3$ and $(\check A_1, \check A_2) = (\check A_2, \check A_1)$
which (for coefficients $\BZ$) requires $p$ to be even.

Basically, we will use $\CQ$ to write down the action of the gauge field $\check A$. Thus it is convenient to know 
how $\CQ$ behaves under infinitesimal changes $\check A \to \check A + \check B$,
where $\check B$ is infinitesimal. In particular, $\check B$ is topologically trivial and hence can be represented by a differential form 
$\check B = (0, \DA_B, \d \DA_B)$. We write
\beq
\CQ(  \check A + \check B) = \CQ(  \check A) +  ( \check A, \check B) +  \widetilde \CQ(\check B) .
\eeq
The term $ \widetilde \CQ( \check B)$ is linear in $\check B$, because if $\check B_1$ and $\check B_2$ are infinitesimal, we get
$ \widetilde \CQ( \check B_1 + \check B_2 ) = \widetilde \CQ(   \check B_1) + \widetilde \CQ( \check B_2) $ where we neglected
the higher order term $ (\check B_1, \check B_2)$.
Therefore, we expect that there exists a differential form $\Sw \in \Omega^{p+2}(Y) $ (whose explicit form will become clear in later subsections) such that
\beq
\widetilde \CQ(  \check B) =  \int_Y \Sw \wedge \DA_B.
\eeq 
Also, we recall the formula \eqref{eq:DC6} which is valid for topologically trivial $\check B$. Then we get
\beq
\CQ(  \check A + \check B) = \CQ(  \check A ) +  \int_Y (\DF_{A} + \Sw) \wedge \DA_B \label{eq:infinitesimal}
\eeq
for infinitesimal $\check B$.

If $\check A$ is topologically trivial and represented by a differential form $\check A = (0, \DA_A, \d \DA_A)$, we can simplify $\CQ(\check A)$ as follows.
For a topologically trivial $\check A$, the multiplication $s  \check A$ by an arbitrary real number $s \in \BR$ makes sense.
Then we have
\beq
\CQ( (s+ \d s) \check A) = \CQ( s \check A) + \d s \int (s\DF_{A} + \Sw) \wedge \DA_A .
\eeq
where $\d s$ is infinitesimal, and $\DF_{A} = \d \DA_A$.
By integrating it, we get
\beq
\CQ(\check A) = \CQ(0) +   \int \left( \frac{1}{2}\DA_A \wedge \d \DA_A + \Sw \wedge \DA_A \right). \label{eq:QRt}
\eeq
Thus, roughly speaking, the quadratic refinement $\CQ(\check A)$ is one half of $ (\check A, \check A) $ up to a linear term.
But dividing $ (\check A, \check A) $ by 2 does not make sense because $(\check A, \check A) $ takes values in $\BR/\BZ$ rather than $\BR$.
This is the reason why we need more sophisticated constructions. In fact, it is not possible to define $\CQ$ on arbitrary manifolds.
We will define it when manifolds have spin structures. 

We can see a few properties of $\Sw$. First, it is closed. This can be seen by performing a small gauge transformation 
$\DA_A \to \DA_A + \d \Da$ for $\Da \in \Omega^{p}(Y)$ and requiring that $\CQ$ is invariant under it. 
Next, at least in $d+1$-dimensions, its de~Rham class is in the image of integral cohomology $H^{p+2}(Y,\BZ)$,
or in other words its integrals $\int_M \Sw$ on $p+2$-cycles $M$ are integers. 
This can be seen by performing a large gauge transformation $\DA_A \to \DA_A + \DF_a$ for $\check a \in \check H^{p+1}(Y)$,
and taking $\DF_a$ to be the Poincar\'e dual of $M$. However, we remark that it is not generally true that $\Sw$ is in the image of integral cohomology
in dimensions $d+2$ and larger.\footnote{More precisely, it is known that $2\Sw$ is a differential form representative of an integral lift of the $p+2$-dimensional Wu class $\nu_{p+2}$.
On a manifold with dimension less than $2(p+2)$, we automatically have $\nu_{p+2}=0$ and hence $\Sw = (2\Sw)/2$ becomes integral.}

\subsection{Atiyah-Patodi-Singer index theorem}\label{sec:APS}
For the definition of quadratic refinement in $d+1=3,7,11$, the APS $\eta$-invariant and the index theorem~\cite{Atiyah:1975jf} play important roles.
(See also \cite{Fukaya:2017tsq,Dabholkar:2019nnc,Fukaya:2019qlf} for recent discussions on the APS theorem.) 
Let us briefly review them.

The $\eta$-invariant is defined as follows. Let $\CD_Y$ be a Dirac-type operator on a $d+1$-dimensional manifold $Y$
and $\lambda $ be its eigenvalues.
Then the definition of $\eta$, in our convention, is
\beq
\eta(\CD_Y) = \frac{1}{2} \left( \sum \sign(\lambda) \right)_{\rm reg},
\eeq
where the sum runs over all eigenvalues including multiplicities, $\sign(\lambda) =+1$ for $\lambda \geq 0$ and ${\rm sign}(\lambda)  = -1$ for $\lambda <0$,
and the subscript ``reg'' means that some appropriate regularization should be done for the infinite sum.

From the definition, one can see that $\eta(\CD_Y)$ can have jumps by integer values when some of the eigenvalues cross zero.
But $\exp(-2\pi \i \eta)$ is smooth under any smooth change of eigenvalues and hence under any smooth change of the metric and gauge field.

Let us consider a $(d+2)$-dimensional manifold $Z$ with boundary $\partial Z =Y$.
On $Z$, suppose that we have a Dirac operator $\CD_Z$ and a $\BZ_2$ grading matrix $\Ch_Z$ (which may be called the {chirality} operator on $Z$)
such that $\{\CD_Z, \Ch_Z \} =0$. It is possible to define the index of $\CD_Z$ (under some appropriate boundary condition on $\partial Y$)
as $\index \CD_Z = n_+ - n_-$, where $n_\pm$ are the number of zero modes with $\Ch_Z = \pm 1$, respectively.

Near the boundary, we assume that $Z$ is a direct product $(-\epsilon, 0] \times Y \subset Z$ such that the metric is also a direct product
and the gauge fields are pulled back from $Y$. Let $s$ be the coordinate of $(-\epsilon,0]$. Then the Dirac operator $\CD_Z$ near the boundary can be written as
$
\CD_Z = \i \Gamma^s ( \partial_s + \CD'_Y),
$ 
where $\Gamma^\mu$ are the gamma matrices. The operator $\CD'_Y$ is defined on $Y$, and one can see that it commutes with $\Ch_Z$. Therefore, 
we can restrict it to the subspace with $\Ch_Z=+1$.
We define this restriction as $\CD_Y = \CD'_Y|_{\Ch_Z=+1}$.

Under the above conventions, the APS index theorem states the following; see \cite{Yonekura:2016wuc,Tachikawa:2018njr} for the precise sign factors used here.
Let $\hat{A}(R)$ be the Dirac A-roof genus given in terms of the Riemann curvature tensor, and $\ch(F) = \tr \exp ( \frac{\i}{2\pi} F) $
be the Chern character of the gauge bundle given in terms of the curvature tensor $F$. 
The index is given by
\beq
\index \CD_Z = \int_Z \hat{A}(R)\ch(F)  + \eta(\CD_Y).
\eeq
In particular, the right hand side is an integer. 

We will also use the following fact in our later discussions. The Dirac operator $\CD_Z $ acts on the space of sections $\Gamma( \CS_Z)$
of a bundle $\CS_Z$ (which is typically the tensor product of the spin bundle and a gauge bundle). 
It splits as $\CS_Z = \CS_Z^+ \oplus \CS_Z^-$ according to the $\BZ_2$ grading by $\Ch_Z$.
Suppose that $\CS_Z^\pm$ are pseudoreal. This means that there is an antilinear map $\SC : \CS_Z^\pm \to \CS_Z^\pm$
such that $\SC^2 = -1$. Because $\SC$ preserves the $\BZ_2$ grading, $\SC$ commutes with $\Ch_Z$, $\SC \Ch_Z = \Ch_Z \SC$.
By modifying $\SC$ by $\Ch_Z$ as $\SC \to \SC \cdot \Ch_Z$ if necessary, we may also assume that $\SC$ commutes with the Dirac operator, $\SC \CD_Z = \CD_Z \SC$.
More explicitly, by using properties of the representations of Clifford algebras (see e.g.~\cite{Weinberg:2000cr} for a review), 
one can construct such $\SC$ when $\dim Z = d+2 =0 \mod 8$
and the gauge representation is pseudoreal, or when $\dim Z = d+2 =4 \mod 8$ and the gauge representation is strictly real. 

If the $\SC$ with the above properties exits, each eigenvalue of the Dirac operator appears twice in the spectrum.
The reason is that if $\Psi$ is an eigenmode with $\CD_Z \Psi = \lambda \Psi$, then $\CD_Z \SC\Psi = \lambda \SC \Psi$
and hence $\SC \Psi$ is also an eigenmode, and $\SC \Psi$ is different from $\Psi$ since $\SC^2=-1$.
Because $\SC$ commutes with $\Ch_Z$ and $\CD_Y$ acts on $\CS_Y : = \CS_Z^+|_{Y}$,
the same is true for $\CD_Y$.

In particular, the index $\index \CD_Z$ is even, and $\eta(\CD_Y)$ jumps only by even integers. 
We divide the index theorem by 2 to get
\beq
\frac{1}{2} \index \CD_Z = \frac{1}{2}  \int_Z \hat{A}(R)\ch(F)  + \frac{1}{2}  \eta(\CD_Y).
\eeq
and both sides are still integers. The exponential $\exp( - \pi \i \eta)$ is a smooth function of the metric and the gauge field.

\subsection{$d+1=3$}\label{sec:3}
Here we construct the quadratic refinement on a manifold $Y$ with $\dim Y = d+1 =3$.
The importance of the quadratic refinement, or equivalently the spin structure in this situation, was emphasized by \cite{Belov:2005ze}.
The relevant differential cohomology element $\check A \in \check H^2(Y)$ is just an ordinary $\U(1)$ gauge field.
By using this fact, we define the quadratic refinement as follows. Let $\CD_Y$ be the Dirac operator
coupled to the $\U(1)$ bundle with the gauge field $\check A$. Then we define
\beq
\CQ(\check A) = - \eta(\CD_Y).
\eeq
To show that this gives a quadratic refinement, we use the following fact. 
Any 3-dimensional spin manifold $Y$ with $\U(1)$ bundle can be extended to a 4-dimensional spin manifold $Z$ with $\U(1)$ bundle such that $\partial Z =Y$.\footnote{%
This follows from $\Omega_3^{\rm spin}(B \U(1)) =0$. See e.g.~\cite{Garcia-Etxebarria:2018ajm} for a convenient collection of results for various bordism groups.}
By the APS index theorem, we get
\beq
 - \eta(\CD_Y) = \int_Z \left( \frac{1}{2} \DF_A \wedge \DF_A + \hat{A}_1(R) \right) \mod \BZ,
\eeq
where $\hat{A}_k(R)$ is a $4k$-form part of $\hat{A}(R)$, and we have used the fact that $F = -2\pi \i \DF_A $ and hence $\ch(F) = \exp ( \DF_A) $.
The index $\index \CD_Y$ drops out when we take mod $\BZ$. By using it, we get
\beq
& \CQ(\check A_1 + \check A_2) - \CQ(\check A_1) - \CQ(\check A_2) + \CQ(0)  \nonumber \\
&=  \int_Z  \DF_{A_1} \wedge \DF_{A_2} =  \int_Z  \DF_{A_1 \Dp A_2}  
= \int_Y \DA_{A_1 \Dp A_2}.
\eeq
Thus $\CQ$ satisfies the defining equation of the quadratic refinement.

As is clear from the above derivation, we could also define $\CQ$ directly as
\beq
\CQ(\check A) = \int_Z \left( \frac{1}{2} \DF_A \wedge \DF_A + \hat{A}_1(R) \right).
\eeq
The index theorem\footnote{%
We can also use basic results in algebraic topology to show this. 
What needs to be shown is that $\int_Z \DF\wedge \DF$ is even on a closed spin 4-manifold $Z$ for $\DF \in H^2(Z,\bZ)$. 
This follows if we can show $\int_Z F\cup F=0$ mod 2 for $F\in H^2(Z,\bZ_2)$.
This is indeed true since $\int _Z F\cup F=\int_Z \mathrm{Sq}^2\, F
=\int_Z \nu_2 \cup F = 0$.
Here, we used a few facts in algebraic topology, namely that $\mathrm{Sq}^n a=a\cup a$ for $a\in H^n(M,\bZ_2)$,
that $\int_M \mathrm{Sq}^m a=\int_M \nu_m \cup a$ for the Wu class $\nu_m$,
where $\nu_1=w_1$, $\nu_2=w_2+w_1^2$, $\nu_3=w_1w_2$, $\nu_4=w_4+w_1w_3 + w_2^2+w_1^4$, \ldots,
and that $w_1=w_2=0$ on spin manifolds.
These facts can be found in the standard textbooks, e.g.~\cite{MilnorStasheff}.
\label{foot:Wu}
} guarantees that this definition does not depend on how we extend $Y$ to $Z$.
For the purposes of practical computations, it is helpful to know both of the definitions, using the $\eta$-invariant and the extension to higher dimensions.
The same remark applies to the case $d+1=7$ below.

\subsection{$d+1=7$}\label{sec:7}
The quadratic refinement on a $d+1=7$ dimensional manifold is not as simple as the case $d+1=3$. 
First let us give one definition by using the extension to 8-dimensions. 
We again use the fact that a 7-dimensional spin manifold $Y$ with $\check A \in \check H^4(Y)$ can be extended to an 8-dimensional spin manifold $Z$
in which $\check A$ is also extended.\footnote{%
This follows from $\Omega_7^{\rm spin}(K(4,\BZ) ) =0$ \cite{Stong:1985vj,Monnier:2018nfs}, where $K(4,\BZ)$ is an Eilenberg-MacLane space of the appropriate type. Note also that $B\U(1)=K(2,\bZ)$.}
Then we define
\beq
\CQ(\check A) = \int_Z \left( \frac{1}{2} \DF_A \wedge \DF_A -  \frac{1}{4}p_1(R)  \wedge \DF_A +  28 \hat{A}_2(R) \right). \label{eq:QR7}
\eeq
where $p_1(R) = - \frac{1}{2} \tr (\frac{R}{2\pi})^2$ is the first Pontryagin class represented by the Riemann curvature $R$.
It is related to $\hat{A}_1(R)$ as $\hat{A}_1(R) = - \frac{1}{24}p_1(R)$.  From this definition of $\CQ$, the property \eqref{eq:defQ} can be easily checked.
However, what is nontrivial is that this definition does not depend on how we extend $Y$ to $Z$ if $Y$ and $Z$ are spin manifolds. 
We will show it by using the index theorem following \cite{Witten:1996hc,Witten:1996md}\footnote{%
Similarly to footnote \ref{foot:Wu}, it can be shown also using basic results in algebraic topology.
What needs to be shown is that $\int_Z (\DF\wedge \DF + (p_1/2) \wedge \DF)$ is even on any closed spin 8-manifold $Z$, for $\DF\in H^4(Z,\bZ)$.
First we use the result of \cite{Thomas} which says that $p_1=\mathfrak{P}(w_2)+\iota_2 w_4$ mod 4, where $\mathfrak{P}:H^2(X,\bZ_2)\to H^4(X,\bZ_4)$ is a certain cohomology operation known as the Pontrjagin square,
and $\iota_2$ is the homomorphism $\bZ_2\to \bZ_4$ sending 1 mod 2 to 2 mod 4. 
(This is a basic relation to analyze discrete theta angle in 4d SO gauge theory \cite{Aharony:2013hda}).
On a spin manifold $w_2=0$, therefore $p_1$ is divisible by 2, and $p_1/2=w_4$ mod 2.
Then all we have to show is that $\int_Z (F\cup F+ w_4 \cup F)=0$ mod 2 for $F\in H^4(Z,\bZ_2) $. 
This follows since $\int_Z F\cup F=\int_Z \mathrm{Sq}^4\, F = \int_Z \nu_4 \cup F= \int_Z w_4 \cup F$, where we used the fact $\nu_4=w_4$ on a spin manifold.
}.

We will use the Dirac operator coupled to the 56 dimensional representation of $E_7$ gauge field. 
The reason that $E_7$ is relevant to the current problem can be motivated by some M-theory consideration~\cite{Hsieh:2019iba}, but we proceed formally by a mathematical argument.

The point is that the topological classification of $E_7$ bundles on $Z$ is the same as the topological classification of $\check{A}$, which is done by $H^4(Z,\BZ)$.
$E_7$ has homotopy groups $\pi_k(E_7) = 0$ for $k \leq 10$ and $k \neq 3$. The homotopy group $\pi_3(E_7)$ is related to the instanton number of $E_7$ bundles
which is completely captured by some characteristic class $c(E_7) \in H^4(Z,\BZ)$ defined explicitly below.   
Then $E_7$ bundles are completely classified by $c(E_7)$ on a manifold with dimensions $\dim Z \leq 11$.
Moreover, any element of $H^4(Z,\BZ)$ can be realized as $c(E_7)$ of some $E_7$ bundle.
(The precise arguments for these claims require obstruction theory. See \cite{Witten:1985bt} where the case of $E_8$ is discussed.
The discussion for $E_7$ is completely the same except for the upper bound $\dim Z \leq 11$ for the dimensions.) 

Let $\check S \in \check H^4(Z) $ be the Chern-Simons 3-form of $E_7$ gauge fields defined as a differential cohomology element as in Sec.~\ref{sec:DC3} by using the characteristic class $c(E_7)$.
Then the facts mentioned above imply that for any $\check A $, we can always find an $E_7$ bundle such that $\check A - \check S$ is topologically trivial.
So we set $\check A = \check S + \check a$, where $\check a = (0, \DA_a, \d \DA_a)$ is given by a differential form~\cite{Witten:1996md,Diaconescu:2003bm}.

We can rewrite \eqref{eq:QR7} as
\beq
\CQ(\check A) &= \int_Z \left( \frac{1}{2} (\DF_S + \d \DA_a) \wedge (\DF_S + \d \DA_a)  - \frac{1}{4}p_1(R)  \wedge (\DF_S +\d \DA_a)  +  28 \hat{A}_2(R) \right) \nonumber \\
&= \CQ(\check S)  + \int_Y \left( \frac{1}{2} \DA_a \wedge \d \DA_a + \DA_a \wedge \DF_S  - \frac{1}{4}p_1(R)  \wedge \DA_a   \right).
\eeq
The second term is defined on $Y$ and hence it is manifestly independent of the extension to $Z$.
So we only need to check that $\CQ(\check S)$ is independent of $Z$.

We define the curvature representation of $c(E_7)$ (i.e.~the $\DF_S$ of the Chern-Simons $\check S$) as 
\beq
\DF_S = \frac{1}{24} \tr_{56} \left( \frac{ \i }{2\pi} F_{E_7} \right)^2, \label{eq:DFS1}
\eeq
where $F_{E_7} $ is the $E_7$ gauge field strength, and the trace is taken in the 56-dimensional representation.
The normalization is chosen as follows. $E_7$ has a subgroup $\SU(2) \times \Spin(12)$ under which the pseudoreal representation $56$ decomposes as
$56 \to 2 \otimes 12 \oplus 1 \otimes 2^5$, where $2^5$ is one of the two spinor representations of $\Spin(12)$.
A minimal instanton of $E_7$ is realized by an instanton in the $\SU(2)$ subgroup.
The above normalization is taken so that the minimal instanton gives 1 for $ \int c(E_7) $, i.e.~when the $E_7$ background is actually an $\SU(2)$ background, we have
\beq
\DF_S  = \frac{1}{2} \tr_{2} \left( \frac{ \i }{2\pi} F_{\SU(2)} \right)^2.  \label{eq:DFS2}
\eeq
Thus $\DF_S$ can be regarded as the differential form representation of an integer cohomology element. 

We will also need $\tr_{56}  ( \frac{ \i }{2\pi} F_{E_7} )^4$ in later discussions. 
By the fact that $E_7$ bundles are completely classified by $c(E_7)$, it should be possible to expand it by $\DF_S$.
By using the above decomposition $56 \to 2 \otimes 12 \oplus 1 \otimes 2^5$ again, we see that
\beq
\tr_{56} \left( \frac{ \i }{2\pi} F_{E_7} \right)^4 = 24 (\DF_S)^2.
\eeq

Let us consider the index of the Dirac operator $\CD_Z$ coupled to the 56 dimensional representation of $E_7$ on an 8-dimensional manifold $Z$ with boundary $Y$.
By APS index theorem, it is given as
\beq
\index \CD_Z -\eta(\CD_Y)&=\int_Z \hat{A}(R) \tr_{56} \exp \left( \frac{ \i }{2\pi} F_{E_7} \right) \nonumber \\
&= \int_Z \hat{A}(R) (56 +12 \DF_S + (\DF_S)^2) \nonumber \\
&= \int_Z \left( \DF_S \wedge \DF_S - \frac{1}{2}p_1(R) \DF_S +56 \hat{A}_2(R) \right) .
\eeq

Moreover, let us recall the fact that the spinor representation is strictly real in 8-dimensions and the $56$ dimensional representation of $E_7$ is pseudoreal.
Thus the bundle on which $\CD_Z$ acts is pseudoreal. 
Therefore, as remarked in Sec.~\ref{sec:APS}, $\index \CD_Z$ is even and we get
\beq
 -\frac{1}{2}\eta(\CD_Y) &=\int_Z \left( \frac{1}{2}\DF_S \wedge \DF_S - \frac{1}{4}p_1(R) \DF_S +28 \hat{A}_2(R) \right) \mod \BZ \nonumber \\
 &= \CQ(\check S).
\eeq
The $\eta$-invariant is defined on $Y$ and hence independent of $Z$. 

In summary, we have obtained the formula
\beq
\CQ(\check A) =  -\frac{1}{2}\eta(\CD_Y) +  \int_Y \left( \frac{1}{2} \DA_a \wedge \d \DA_a + \DA_a \wedge \DF_S  - \frac{1}{4}p_1(R)  \wedge \DA_a   \right),
\eeq
for $\check A = \check S + \check a$ where $\check S$ is the Chern-Simons of $E_7$ and $\check a = (0, \DA_a, \d \DA_a)$ is a differential form.
This completes the proof that $\CQ$ is independent of how we extend $Y$ to $Z$.

We have used $E_7$ because it has a natural motivation in M-theory and the above definition of $\CQ$
has the direct relevance to the anomaly of chiral 2-form fields as we will see later in this paper. However, 
there is no problem for the above argument to use $E_6$ instead of $E_7$ since the homotopy groups
are $\pi_k(E_6) = 0$ for $k \leq 8$ and $k \neq 3$. (See e.g.~Table~1 of \cite{Garcia-Etxebarria:2017crf} for convenient summary of homotopy groups.) 
There is a subgroup $E_6 \times \U(1) \subset E_7$ (or more precisely there is a homomorphism $E_6 \times \U(1) \to E_7$)
such that the $56$ dimensional representation becomes $(27_{1} \oplus  1_{3} ) \oplus \text{c.c.}$ where the subscripts are $\U(1)$ charges, and c.c. is the complex conjugate representation.
The $\U(1)$ here can be used as a $\spin^c$ connection, so that the chiral 2-form field may be formulated not only on spin manifolds,
but also on $\spin^c$ manifolds. In our applications in this paper, we will only consider spin manifolds, so we do not study the detail of the $\spin^c$ case.

\subsection{$d+1=11$}\label{sec:11}
It is not known how to construct a quadratic refinement for ordinary differential cohomology in the dimension $d+1=11$ by using only spin structure.
However, for the application to the chiral RR 4-form field in Type IIB string theory, 
what we actually need to classify the topological classes of the RR fields
is  K-theory rather than ordinary cohomology.
 Therefore, what we need in this case is not the ordinary differential cohomology, but 
differential K-theory~\cite{Freed:2000tt,Freed:2000ta,Hopkins:2002rd}. We construct a quadratic refinement in differential K-theory in $d+1=11$ dimensions,
or more generally in any dimensions of the form $d+1 = 8\ell+3$. In particular, it includes the case $d+1=3$ of Sec.~\ref{sec:3} as a special case.
The content of this subsection is out of the main line of discussions in this paper,
and the reader can safely skip it.\footnote{
It is a nontrivial question whether the 11-dimensional bulk of Type~IIB string theory is just given by the differential K-theory of this section (even if there is no twist). There is a nontrivial topological coupling of the RR fields (in particular the chiral 4-form) in Type~IIB string theory~\cite{Debray:2021vob,Yonekura:2024bvh}. It is studied mainly when there is a nontrivial $\GL(2,\bZ)$ bundle (or its pin$^+$ generalization), but it exists even for purely gravitational backgrounds without either $\GL(2,\bZ)$ or the $B$-field~\cite{Yonekura:2024bvh}. One possibility is that it is already contained in the formulation of this section, which may be revealed along the lines of computations in \cite{Diaconescu:2000wy} for the case of Type~IIA. Another possibility is that we need to add an additional term in the action, as in the case of differential KO theory in Type~I~\cite{Freed:2000ta}. It would be interesting to investigate this problem further.
}

We need to define differential K-theory. First let us outline its ingredient. 
The topology is classified by elements $\DV \in K^{p}(Y)$ for $p=0$ or $p=-1$ mod 2. 
Once the topological class is fixed, the dynamical fluctuations in that topological class are given by differential form fields whose degrees are equivalent to $p+1$ mod 2.
We denote them as $\DC=\DC_{p+1}+ \DC_{p+3} +\DC_{p+5}+ \cdots$ where $\DC_{p+1} \in \Omega^{p+1}(Y)$ and so on. 
A subtle point is that we need some quantity which represents $\DV$ by a concrete differential geometric object. 
For $p=0$, it is a connection $\DB$ on $\DV$ represented as a difference of vector bundles.
For $p=-1$, it is a map $\DU: Y \to \U(N)$ for sufficiently large $N$. However, $\DB$ or $\DU$ is not a dynamical field,
so we need to impose some equivalence relation or gauge transformations so that the degrees of freedom of $\DB$ or $\DU$ can be gauged away.
$\DB$ or $\DU$ play an analogous role as the $E_7$ connection in the construction of the case $d+1=7$ in Sec.~\ref{sec:7}.

Now we are going to define differential K-theory $\check K(Y)$ on a manifold $Y$ as follows. We define it for $K = K^0$, but a similar definition is possible for $K^{-1}$.\footnote{
The differential K-theory here is relevant to the dynamical field of the $d+1 = 11$ dimensional bulk topological phase. From it, we will later construct a
 chiral 4-form field as part of the boundary mode in $d=10$ dimensions. 
 If the relevant generalized cohomology
 for the bulk dynamical field is $K^0(Y)$, the generalized cohomology relevant for the boundary mode will be $K^{-1}(X)$ as we will discuss at the end of Sec.~\ref{sec:topedge}.
 Conversely, if the $d+1=11$ dimensional bulk dynamical field is described by $K^1(Y)$, the boundary mode is described by $K^0(X)$.}
An element of $\check K(Y)$ consists of a triplet 
\beq
\check A = (\DV, \DB, \DC). 
\eeq
$\DV \in K(Y)$ is an element of the topological K-group $K(Y)$. $\DB$ is a unitary connection on $\DV$ represented as a difference of hermitian vector bundles.
$\DC \in \Omega^{\rm odd}(Y)$ is a sum of differential forms $\DC=\DC_1+ \DC_3 + \cdots$, where $\Omega^{\rm odd}(Y) = \bigoplus_k \Omega^{2k+1}(Y)$
and $\DC_{2k+1} \in \Omega^{2k+1}(Y)$. 

We impose the following equivalence relation to gauge away the degrees of freedom of $\DB$. Let $\DB_t~( 0 \leq t \leq 1)$ be a homotopy between $\DB_0$ and $\DB_1$.
Then we can regard $\DB_t$ as a connection of $\DV$ on $[0,1] \times Y$ which we denote as $\DB'$. Let $\DG'$ be its field strength $\DG' = \d' \DB' + \DB' \wedge \DB'$
where $\d'$ is the differential on $[0,1] \times Y$.
Then, we get
\beq
\int_{[0,1]} \ch( \DG') \in \Omega^{\rm odd}(Y).
\eeq
Here the integral is defined by $\int_{[0,1]} ( \d t \wedge \omega_1(t) + \omega_2(t)) = \int_0^1\d t \, \omega_1(t)$ for $\omega_1(t) , \omega_2(t) \in \Omega^\bullet (Y)$.
The above integral is independent of how to take the homotopy between $\DB_0$ and $\DB_1$. 
We regard $(\DV,\DB_0, \DC_0)$ and $(\DV,\DB_1, \DC_1)$ to be gauge-equivalent if they satisfy the relation
\beq
\int_{[0,1]} \sqrt{\hat{A}(R)}  \ch( \DG') = \DC_0 - \DC_1, \label{eq:DK1}
\eeq
where $  \sqrt{\hat{A}(R)} $ only depends on $Y$ and can be taken outside of the integral. This factor $  \sqrt{\hat{A}(R)}  $ is put to simplify the quadratic refinement defined later,
and it is also conventional in string theory.

The equivalence relation implies that the actual connection $\DB$ in a given topological class is not an invariant information.
We can set $\DB$ to any connection we like, as long as we also change $\DC$ accordingly. In particular, in a topologically trivial case,
we can simply set $\DB=0$.

The equivalence relation in particular implies the following. Let $\DG = \d \DB + \DB \wedge \DB$ be the field strength of $\DB$ on $Y$.
From the equivalence relation, one can check that the combination
\beq
\DF : =   \sqrt{\hat{A}(R)}   \ch( \DG) + \d \DC \in \Omega^{\rm even}(Y) \label{eq:KDF}
\eeq
is invariant under the equivalence relation. It can be proved by noticing that the differential $\d$ on $Y$ and $\d'$ on $[0,1] \times Y$
are related as $\d' = \d + \d t \partial_t$, and hence
\beq
\d \int_{[0,1]}  \ch( \DG') =  - \int_{[0,1]} \d \,  \ch( \DG')  =  \int_{[0,1]} (-\d' +  \d t \partial_t)  \ch( \DG')  =   \ch( \DG_1) -   \ch( \DG_0),
\eeq
where the sign in the first equality comes from a careful examination of the order of the differential and the integration.
By applying $\d$ to \eqref{eq:DK1} and using the above result, we get the invariance of $\DF$.

The invariant differential form $\DF$ in differential K-theory roughly corresponds to the quantity with the same symbol $\DF$ in differential cohomology. 
Also, $\DV \in K(Y)$ roughly corresponds to $[\DN]_\BZ \in H^p(Y,\BZ)$, although the relevant generalized cohomology theories are different.

In addition to the above equivalence relation, we also impose the gauge equivalence 
\beq
\DC \sim \DC + \Df \label{eq:Equiv2}
\eeq
 where $\Df$ is the field strength of differential K-theory elements in $\check K^{-1}(Y)$.
This is the analog of the gauge equivalence $\DA_A \sim  \DA_A + \DF_a$ for $\check A \in \check H^{p+1}(Y)$ and $\check a \in \check H^{p}(Y)$  in ordinary differential cohomology which have
played an important role in Sec~\ref{sec:nonchiral2}.
Here $\check K^{-1}(Y)$ may be defined as follows. The group $K^{-1}(Y)$ is a subgroup of $K(S^1 \times Y)$ such that $\DV \in K^{-1}(Y) \subset K(S^1 \times Y) $ becomes trivial when restricted to 
$\{0\} \times Y \subset S^1 \times Y$. If we regard $S^1 \times Y$ as obtained from $[0,1] \times Y$ by gluing the two ends of the interval $[0,1]$,
we can construct $\DV \in K^{-1}(Y)$ by using a unitary transition function $\DU : Y \to \U(N)$ (for some sufficiently large $N$) which is used to glue the bundles at the two ends of the interval $[0,1]$.
On this bundle, we take a connection $\DB = t \DU^{-1} \d \DU$, where $t \in [0,1]$ is the coordinate of the interval.
The Chern character of $\DU$ is defined by $\ch(\DU) = \int_{[0,1]} \tr \exp( \frac{\i}{2\pi} \DG)$, where $\DG = \d \DB + \DB \wedge \DB = \d t  \DU^{-1} \d \DU + (t^2-t) ( \DU^{-1} \d \DU)^2 $.
It is a sum of odd differential forms, $\ch(\DU) \in \Omega^{\rm odd}(Y)$. By using it, we define $\check K^{-1}(Y)$ as having elements of the form $(\DV, \DU, \DC)$,
where now $\DC$ is a sum of even dimensional forms $\DC \in \Omega^{\rm even}(Y) $. We impose a similar equivalence relation on $(\DU,\DC)$ as in the case of $\check K(Y)$
which in particular implies that $\DF : =   \sqrt{\hat{A}(R)}  \ch(\DU) + \d \DC$ is invariant.

In ordinary differential cohomology $\check H^{p+1}(Y)$, we have the holonomy function $\chi(M) = \exp(2\pi \i \int_M \DA) $. 
The corresponding quantity in differential K-theory $\check K(Y)$
is given by the APS $\eta$-invariant. Let $M$ be an odd-dimensional $\spin^c$ submanifold of $Y$, and let $\CD_M(\DB)$ be the Dirac operator 
coupled to the gauge field $\DB$ and $\spin^c$ connection. Let $R_M$ be the Riemann curvature of $M$, and let $c_1(F_M) = \frac{\i}{2\pi} F_M$, where $F_M$ is the curvature
of the $\spin^c$ connection. Then we define the holonomy function $\chi$ as
\beq
\chi(M) = \exp 2\pi \i  \left(  -\eta( \CD_M(\DB) ) + \int_M \hat{A}(R_M) e^{c_1(F_M)} \, \frac{ \DC }{  \sqrt{\hat{A}(R)}  }\right).\label{eq:Khol}
\eeq
Here $R = R_Y$ is the Riemann curvature of $Y$.
One can check by the APS index theorem that this holonomy function is invariant under the equivalence relation  $(\DV,\DB_0, \DC_0) \sim (\DV,\DB_1, \DC_1)$ defined above.
The connection $\DB$ is introduced to make this definition possible. Notice that if we neglect the $\eta$-invariant, the above 
expression is the familiar one in the worldvolume action of D-branes.

The product $\check A_1 \Dp \check A_2$ is defined as follows. 
First, we simply define $\DV_{A_1 \Dp A_2} = \DV_{A_1} \otimes \DV_{A_2}$ which is the product in K-theory, i.e.~the tensor product of bundles.
Also we define $\DB_{A_1 \Dp A_2} = \DB_{A_1} \otimes 1 + 1 \otimes \DB_{A_2}$ which is a connection on $\DV_{A_1} \otimes \DV_{A_2}$.
Finally we define
\beq
\DC_{A_1 \Dp A_2} = \frac{1}{ \sqrt{\hat{A}(R)} }  \DC_{A_1} \wedge \d \DC_{A_2} +  \DC_{A_1} \wedge \ch(\DB_{A_2}) +  \ch(\DB_{A_1}) \wedge \DC_{A_2} .
\eeq
One can check that this is invariant up to gauge  equivalence when $\check A_1$ and $\check A_2$ are changed using the equivalence relations.
In particular we have
\beq
\DF_{A_1 \Dp A_2} =  \frac{1}{ \sqrt{\hat{A}(R)} } \DF_{A_1} \wedge \DF_{A_2}.
\eeq

Before defining the quadratic refinement, we need to define  the pairing between $\check A_1$ and $\check A_2$.
First, we define the involution $\check{A} \to \check{ A}^* = (\DV^*, \DB^*, \DC^*)$.
Here $\DV^*$ is the complex conjugate bundle to $\DV$. $\DB^*$ is the complex conjugate connection.
$\DC^*$ depends on the degrees of the forms and is defined as $\DC^*_{2k-1} = (-1)^k \DC_{2k-1}$. In particular, we have $\DF_{2k}^* = (-1)^k \DF_{2k}$.
Then we define the pairing $(\check A_1, \check A_2)$ as
\beq
(\check A_1, \check A_2) = -\frac{1}{2\pi \i} \log \chi_{A_1 \Dp A^*_2}(Y),
\eeq
where $\chi_{A_1 \Dp A^*_2}$ is the holonomy function associated to $\check A_1 \Dp \check A^*_2$, and it is evaluated on the entire manifold $Y$ which we assume to be spin.
One can show that the pairing is symmetric, $(\check A_2, \check A_1) =(\check A_1, \check A_2)   $ as follows.
The point is that if the dimension of $Y$ is of the form $d+1 = 4k+3$, the spin bundle is either pseudo real or strictly real, and hence
the $\eta$-invariant has the property that $\eta(\DB) = \eta(\DB^*)$. In particular we have $\eta(\DB_1+\DB_2^*) = \eta(\DB_1^*+\DB_2)$.
Also it is straightforward to show that $  \DC_{A_1} \wedge \d \DC_{A_2}^* =   \DC_{A_2} \wedge \d \DC_{A_1}^* + \text{(exact form)}$ in $d+1 =4k+3$ and so on.

We require that the quadratic refinement $\CQ$ satisfies the property that
\beq
\CQ(\check A_1 + \check A_2) - \CQ(\check A_1) - \CQ(\check A_2) + \CQ(0)= (\check A_1, \check A_2). \label{eq:DK2}
\eeq
Such a $\CQ$ can be defined as follows if the dimension is of the form $d+1= 8\ell+3$. Given $\check A = (\DV, \DB, \DC)$, we consider $\check A \Dp \check A^*$.
The bundle  $ \DV_{A \Dp A^*}=  \DV \otimes \DV^*$ and the connection $\DB_{A \Dp A^*} = \DB \otimes 1 + 1 \otimes \DB^*$ are strictly real.
On the other hand, the spin bundle in $8\ell+3$ dimensions is pseudoreal. Therefore, for the Dirac operator $\CD_Y( \DB \otimes 1 + 1 \otimes \DB^* ) $ coupled to it,
the exponentiated $\eta$-invariant $\exp( - \pi \i \eta(\CD_Y( \DB \otimes 1 + 1 \otimes \DB^*) ) )$ is a smooth function of the metric and the gauge field. 
Then we define
\begin{multline}
\CQ(\check A) =   \frac{1}{2}\eta( \CD_Y( \DB \otimes 1 + 1 \otimes \DB^* )) \\
- \frac{1}{2}\int_Y    \left(\DC \wedge \d \DC^* +  \sqrt{\hat{A}(R)} \left( \DC \wedge \ch(\DB^*) +  \ch(\DB) \wedge \DC^* \right) \right).\label{eq:KthQuad}
\end{multline}
The factor $ \sqrt{\hat{A}(R)} $ in the definition \eqref{eq:KDF} of $\DF$ of $\check K(Y)$ was introduced to simplify the term $\DC_{A} \wedge \d \DC_{A^*}$ in this expression.
The overall sign was chosen so that the middle dimensional form $\DC_{4\ell+1}$ appears as $+ \frac{1}{2} \int \DC_{4\ell+1} \wedge \d \DC_{4\ell+1}$.
(Note $\DC_{4\ell+1}^* = - \DC_{4\ell+1}$.)

Notice that
\beq
&(\DV_{A_1} \oplus  \DV_{A_2}) \otimes (\DV_{A_1}^* \oplus  \DV_{A_2}^*)  \nonumber \\
&= (\DV_{A_1}  \otimes \DV_{A_1}^* ) \oplus  (\DV_{A_2}  \otimes \DV_{A_2}^* ) \oplus  (\DV_{A_1}  \otimes \DV_{A_2}^* ) \oplus  (\DV_{A_2}  \otimes \DV_{A_1}^* ).
\eeq
Also notice that the $\eta$-invariants of the Dirac operators coupled to $(\DV_{A_1}  \otimes \DV_{A_2}^* )$ and $(\DV_{A_2}  \otimes \DV_{A_1}^* )$ are the same
by the pseudoreality of the spin bundle. Therefore, we can see that $\CQ$ satisfies the equation \eqref{eq:DK2}.

We remark that there is similarity and difference between $d+1=7$ and $d+1=11$ .
We have used $E_7$ bundle $d+1=7$, and we have used the K-theory bundle $\DV$ in $d+1=11$.
The construction of the quadratic invariant involved the $\eta$-invariant coupled to the 56-dimensional representations of the $E_7$ bundle in $d+1=7$,
and the $\eta$-invariant of the real bundle $\DV \otimes \DV^*$ in $d+1=11$.

\paragraph{The case $d+1=3$.}
The above formulation might seem complicated, and it may be helpful to see the case of $d+1=3$ in a little more detail. 
We will see that the K-theoretic formulation essentially coincides with the formulation in Sec.~\ref{sec:3}.

We restrict our attention to the case that $\ch_0(V)$ (i.e.~the virtual dimension of $V$) is zero.
The reason is that it basically represents a ``tadpole'' and the partition function of the theory developed in Sec.~\ref{sec:theory} vanishes
unless it is zero. This is due to the fact that there is a coupling $2\pi \i \cdot \ch_0(V) \DC_{d+1} $ in the action and the path integral over $\DC_{d+1}$
requires $\ch_0(V)=0$ for the partition function to be nonzero. 
Also, the field $\DC_{d+1}$ appears only in this term and its only role is to set $\ch_0(V)=0$,
so we can neglect this field $\DC_{d+1}$ for most purposes after setting $\ch_0(V)=0$.
In $d+1=3$, we only need to consider the 1-form $\DC_1$ and we denote it as $\DC$ for simplicity.

We can always take $V = E - \underline{\BC}^N$,
where $E$ is a complex vector bundle and $\underline{\BC}^N$ is the trivial bundle.
For $\ch_0(V)=0$, the rank of $E$ is $N$.

Now consider $d+1=3$ dimensional manifold $Y$. In this case, there is a simplification. 
The fibers of the bundle $E$ have complex dimension $N$ or real dimension $2N$. Then, by dimensional reason,
$E$ has a section on a 3-manifold which is everywhere nonzero if $N>1$. (Generically, $2N$ real functions of $d+1$ coordinates $(x^1, \cdots ,x^{d+1})$ do not simultaneously vanish if $2N>d+1$.)
Therefore, by repeatedly taking nonzero sections, any complex vector bundle $E$ can be reduced to a bundle of the form $ E \cong L \oplus \underline{\BC}^{N-1}$,
where $L$ is a one-dimensional (line) bundle on which $\U(1)$ acts as the structure group.\footnote{
Equivalently, this reduction of the structure group of $E$ also follows from the fact that the embedding $\U(1) \to \U(N)$ gives the isomorphisms $\pi_k(\U(1)) = \pi_k(\U(N)) $ for $k=0,1,2$.
} 
Therefore, we can take $V = L - \underline{\BC}$.
We take the connection on $ \underline{\BC}$ to be trivial, and hence the connection $\DB$ is nontrivial only on $L$. By abuse of notation,
we denote the connection on $L$ by $\DB$.

The equivalence relation \eqref{eq:DK1} is simplified for the $\U(1)$ connection $\DB$ as
$\frac{\i}{2\pi } \DB_0+\DC_0 = \frac{\i}{2\pi }\DB_1+\DC_1$. Thus, up to the equivalence relation the differential K-theory depends only on 
\beq
\DA:= \frac{\i}{2\pi }\DB+\DC.
\eeq
Notice that we have neglected the 3-form field $\DC_3$, and hence $\DA$ is just a $\U(1)$ connection (up to the normalization by $\frac{\i}{2\pi }$).

For a map $\DU : Y \to \U(N)$, we have $\ch_1(\DU) = \frac{\i}{2\pi }\tr (\DU^{-1} \d \DU) = \frac{\i}{2\pi } \d \log {\mathsf u} $, where ${\mathsf u} = \det \DU$.
Therefore, the equivalence relation \eqref{eq:Equiv2} is given by $\DA \sim \DA + \frac{\i}{2\pi } \d \log {\mathsf u} + \d {\mathsf c}_0 $, where $c_0 \in \Omega^0(Y)$
and ${\mathsf u} : Y \to \U(1)$. This is just an ordinary gauge transformation of the $\U(1)$ gauge field $\DA$.

The field strength $\DF$ is just the field strength of $\DA$. If $M$ is a 1-dimensional circle $S^1$ with the antiperiodic spin structure, the APS index theorem or
a direct computation of the $\eta$-invariant shows that $-\eta(\CD_{S^1}(\DB)) =  \int_{S^1} \frac{\i}{2\pi} \DB$ mod 1.
(A direct computation shows that the result is valid also for the periodic spin structure and $V = L - \underline{\BC}$.) Thus the holonomy function \eqref{eq:Khol} is just given by $\chi(S^1) = \exp (2\pi \i \int_{S^1} \DA)$.

Finally, let us check that the quadratic refinement \eqref{eq:KthQuad} essentially coincides with the one in Sec.~\ref{sec:3} up to a minor difference.
(The pairing is uniquely determined from the quadratic refinement.)
To see this, note that $V \otimes V^* = - (L \oplus L^*) + \underline{\BC}^2$. Therefore, we get
$
  \frac{1}{2}\eta( \CD_Y( V \otimes V^* )) =   - \eta( L) +\eta(0) 
$
where we used $\eta(L^*) = \eta(L)$. Thus we get
\beq
\CQ(\check A) = - \eta( L) + \eta(0) + \int_Y ( \frac{1}{2} \DC \wedge \d \DC +  \frac{\i}{2\pi}\d \DB \wedge \DC)=  -\eta( \DA) + \eta(0) ,
\eeq
where $ \eta( \DA)$ is the $\eta$-invariant of the $\U(1)$ connection $\DA$ and we have used the APS index theorem on the manifold $[0,1] \times Y$
with the gauge field $\DA_t = \frac{\i}{2\pi} \DB + t \DC$, $t \in [0,1]$.
This $\CQ(\check A)$ coincides with that in Sec.~\ref{sec:3} up to the term $\eta(0)$ which is independent of $\check A$.

\subsection{Other dimensions}

We can formally regard non-chiral $p$-form fields in $d$ dimensions as a self-dual theory by introducing both a $p$-form field and a $(d-p-2)$-form field and imposing the duality relation between them.
Correspondingly, we can always define the following quadratic refinement in $(d+1)$ dimensions without using any spin structure.
We assume that the field $\check A$ consists of a pair of fields $\check A = (\check A^1, \check A^2) \in \check H^{p+2}(Y) \times \check H^{d-p}(Y)$.
Then we take the quadratic refinement as 
\beq
\CQ(\check A) = (\check A^2, \check A^1) = \int_Y \DA_{A^2 \Dp A^1} .
\eeq
The corresponding pairing $(\check A, \check B)$ between two such pairs $\check A = (\check A^1, \check A^2) $ and $\check B = (\check B^1, \check B^2) $
is given by
\beq
(\check A, \check B) = \CQ(\check A + \check B) - \CQ(\check A) - \CQ(\check B) + \CQ(0)  = (\check A^2, \check B^1) +  (\check B^2, \check A^1) .
\eeq
The discussions in the later sections can be applied to this quadratic refinement and we get another realization of the $p$-form theory on the boundary
in addition to the realization discussed in Sec.~\ref{sec:nonchiral}.

More interesting case is the case in which $d=2p+2$ and $p$ is odd, such as $p=1$ and $d=4$.
By using the totally antisymmetric tensor $\epsilon_{ij} ~(i,j = 1,2)$ with $\epsilon_{21} = -\epsilon_{12}=  1$, we can write 
\beq
(\check A, \check B) = \epsilon_{ij} (\check A^i , \check B^j), \qquad \CQ(\check A) = \frac{1}{2} \epsilon_{ij}(\check A^i, \check A^j) .
\eeq
Therefore, there is an $\SL(2,\BZ)$ symmetry acting on the index $i$ at the differential form level. 
At the more precise level of differential cohomology theory, we need to make sense of the factor $\frac 12$ which appears in the definition of $\CQ(\check A) $.
But it can be done if we consider an appropriate spin structure, which gives rise to new anomalies involving the $\SL(2,\BZ)$ symmetry.
This $\SL(2,\BZ)$ duality is the generalization of the electromagnetic duality of Maxwell theory. We will discuss more details about it in Sec.~\ref{sec:EMdual}.

\section{Chiral $p$-form fields as boundary modes} \label{sec:theory}
By using the quadratic refinement $\CQ(\check A)$, we can now construct chiral or equivalently (anti-)self-dual $p$-form fields $\check B \in \check H^{p+1}(X)$ in $(d=2p+2)$-dimensions
as the boundary modes of bulk SPT phases.\footnote{In $d=10$, the cohomology should be changed to the appropriate K-theoretic cohomology group as discussed in Sec.~\ref{sec:11}. 
Then the following discussions are valid also in that case with minor modifications.}

\subsection{The theory}
As the bulk theory, we take the theory described by a dynamical field $\check A \in \check H^{p+2}(Y)$ in $d+1 = 2p+3$ dimensions.
The Euclidean action is given by
\beq
-S  = -  \int     \frac{2\pi  }{2e^2} \DF_A \wedge \hodge \DF_A  + 2\pi \i \kappa  ( \CQ(\check A) - \CQ(0)) + 2\pi \i (\check A, \check C) , \label{eq:bulkS}
\eeq
where $e^2 >0$ and $\kappa \in \BZ$ are parameters, and $\check C \in H^{p+2}(Y)$ is a background field of the $\U(1)$ $p$-form symmetry of the theory. 
The pairing $(\check A, \check C)$ between differential cohomology elements was defined in \eqref{eq:pair}. 

The term $\CQ(0)$ depends only on the background metric and not on the dynamical field $\check A$.
The reason that we take the difference $ \CQ(\check A) - \CQ(0)$
rather than $\CQ(\check A)$ itself will be explained later.

The parameter $e^2$ has a dimension of mass, and the theory has massive particles whose mass is of order $|\kappa|e^2$.
We take $e^2$ to be very large so that there is no propagating degrees of freedom in the low energy limit. 
In the strict large $e^2$ limit and the trivial background $\check C=0$, the action is given by 
$
-S = 2\pi \i \kappa (\CQ(\check A) - \CQ(0))  \label{eq:ACS}.
$
This is a generalization of the spin abelian Chern-Simons theory studied in \cite{Belov:2005ze}. 
The metric dependence will be important later when we consider the gravitational anomaly of the boundary mode.

If we quantize the theory \eqref{eq:ACS} on a spatial manifold $S^{p+1} \times S^{p+1}$, 
the dimension of the Hilbert space is given by $|\kappa|$ in the limit $e^2 \to \infty$.\footnote{
This quantization is done in the standard way, basically following \cite{Witten:1988hf}. By \eqref{eq:infinitesimal}, and also by using the fact that $\Sw$ in that equation is zero on $S^{p+1} \times S^{p+1}$,
the equation of motion requires that the field is flat, $\DF_A =0$. On $S^{p+1} \times S^{p+1}$, the flatness also implies that $\check A$ is topologically trivial,
and hence it can be written by a flat differential form $\check A = (0,  \DA_A, 0)$. The gauge invariant degrees of freedom are $\phi_i = \int_{S^{p+1}_i} \DA_A$,
where the subscript $i=1,2$ distinguishes the two spheres $S^{p+1}$. These variables take values in $\BR/\BZ$.
The Lagrangian (in Lorentz signature) is then given by $\CL = 2\pi  \kappa   \phi_1 \partial_t \phi_2$.
The canonical quantization of this theory by regarding $\phi:=\phi_2$ as the canonical position coordinate gives $\varpi:=2\pi  \kappa \phi_1$ as the canonical momentum coordinate.
The wave functions are $\Psi_m(\phi) = e^{2\pi \i m \phi}$ for $m \in \BZ$, but $\phi_1 \sim \phi_1 +1$ or in other words $\varpi \sim \varpi+ 2\varpi \kappa $ implies that 
the states $\Psi_m$ and $\Psi_{m+\kappa}$ should be identified. Thus we get $|\kappa|$ states. 
The quantization here is rather ad hoc, but a more precise treatment by regarding wave functions as holomorphic sections of a line bundle on $T^2=\{(\phi_1,\phi_2) \}$ would give the same result.
}
Thus the bulk theory would be a topologically ordered phase if $|\kappa|>1$.
Such a topologically ordered phase is very interesting in itself. However,
in this paper we mainly focus on invertible field theories, i.e.~those theories whose Hilbert spaces are always one-dimensional, for the purpose of considering anomalies.
Thus we restrict our attention to the case $\kappa = \pm 1$. This was  the essential reason that we need the quadratic refinement:
Without a quadratic refinement, we could have defined the theory by using an action $2\pi \i  (\check A, \check A) + \cdots$ which is roughly $2\CQ(\check A)$.
But such a theory has a Hilbert space on $S^{p+1} \times S^{p+1}$ whose dimension is greater than 1.
To get a one-dimensional Hilbert space to have an invertible field theory, we need a quadratic refinement. 

If $\kappa = \pm 1$, it turns out that the partition function on an arbitrary closed manifold $Y$ has unit norm $|\CZ(Y)|=1$ (up to local counterterms),
and in particular $|\CZ(S^1 \times X)|=1$ on any spatial manifold $X$ in the low energy limit. This $\CZ(S^1 \times X)$ is the dimension of the Hilbert space on $X$,
so the Hilbert space dimension is always one.

The discussion so far is valid on a closed manifold. Now we define the theory on a manifold $Y$ with boundary $X = \partial Y$.
We need to impose an appropriate boundary condition. We adopt the Dirichlet type boundary condition which requires
that the restriction $\check A|_X$ of the bulk gauge field $\check A$ to the boundary is zero.
We denote this boundary condition as $\L$,
\beq
 \L: \check A|_{\partial Y} =0. \label{eq:L}
 \eeq
Such a boundary condition is physically sensible. (For example, it is an elliptic boundary condition which guarantees well-defined perturbation theory; see \cite{Witten:2018lgb} for a review.)

We have defined the pairing $(\check A, \check C)$ and the quadratic refinement $\CQ(\check A)$ only on closed manifolds.
Thus we need to explain how to define them on a manifold with boundary. Let us take a copy of the orientation reversal of $Y$, which we denote as $\overline{Y}$,
and make a closed manifold $Y_{\rm closed} = Y \cup \overline{Y}$ which is obtained by gluing the two manifolds along the common boundary $X$.
The boundary condition $\check A|_X = 0$ implies that we can extend the gauge field $\check A$ on $Y$ to a gauge field on $Y_{\rm closed}$
in such a way that it is zero on $\overline{Y}$, $\check A|_{\overline{Y} } =0$. After extending $\check A$ to $Y_{\rm closed}$ in this way,
we define $(\check A, \check C)$ and $\CQ(\check A)$ on $Y$ as these quantities on $Y_{\rm closed}$. 

Let us check a consistency of the above definition. If $Y$ is closed, $\partial Y =0$, then $Y_{\rm closed} = Y \cup \overline{Y}$ is just the disjoint union of $Y$ and $\overline{Y}$.
Then the action $S(Y_{\rm closed}) $ defined by using $Y_{\rm closed}$ is just the sum of the action on $Y$ and $\overline{Y}$,
$
S(Y_{\rm closed})  = S(Y) + S( \overline{Y} ).
$
For a closed manifold $Y$, the consistency of the original definition of the action and the new definition requires that $S(\overline{Y}) =0$.
This is indeed the case, because the gauge field $\check A$ is trivial on $\overline{Y}$ and the action vanishes for the trivial gauge field.
This statement is true only after subtracting $\CQ(0)$ from $\CQ(\check A)$ as in \eqref{eq:bulkS}.
In fact, the term $\CQ(0)$ would give additional gravitational anomaly of the boundary theory, and hence 
its presence would change the theory in a significant way.

\subsection{The boundary mode: differential form analysis}\label{sec:edge2}
We have defined the theory whose action is \eqref{eq:bulkS} and the boundary condition is \eqref{eq:L}.
Now we would like to see that a chiral $p$-form field $\check B \in \check H^{p+1}(X)$ is realized as the boundary mode of the bulk theory $\check A \in \check H^{p+2}(Y)$.
First we show the existence of the mode by the differential form analysis. After that, we discuss the topology of $\check B$.

We consider a topologically trivial field $\check A=(0, \DA_A, \d \DA_A )$ at the differential form level.
For a topologically trivial $\check A$, the quadratic refinement $\CQ(\check A)$ is given as in \eqref{eq:QRt},
\beq
\widetilde\CQ(\check A):=\CQ(\check A) -\CQ(0)= \int_Y \left( \frac{1}{2}\DA_A \wedge \d \DA_A + \Sw \wedge \DA_A \right), 
\eeq
where $\Sw =0$ in $d+1 =3$ and $d+1=11$, and $\Sw= -\frac{1}{4}p_1(R)$ in $d+1=7$. (In $d+1=11$, $\DA_A$ here was denoted as $\DC_A$ in Sec.~\ref{sec:11}).
The action is given by
\beq
- S  =  2\pi  \int_Y \left(  -\frac{1}{2e^2} \d \DA_A \wedge \hodge \d \DA_A  +     \i \frac{\kappa}{2}\DA_A \wedge \d \DA_A + \i (\kappa \Sw + \DF_C) \wedge \DA_A \right),
\eeq
where $\DF_C$ is the field strength of the background field $\check C$.
Here $e^2$ is taken to be very large. But for the present purpose of finding the localized mode $\check B$, it is important to keep $e^2$ finite no matter how large it is.

The equation of motion is
\beq
 \frac{(-1)^{p+1 } }{e^2} \d \hodge \d \DA_A +  \i \kappa \d \DA_A  +  \i (\kappa \Sw + \DF_C)=0. 
\eeq
From the equation of motion, one can see that the mass $m$ of the field $\DA_A$ is given by
\beq
m = e^2|\kappa|.
\eeq
We will consider the homogeneous equation of motion
\beq
(-1)^{p+1 }\d \hodge \d \DA_A +  \i m \sign( \kappa) \d \DA=0.\label{eq:homogeneous}
\eeq
The reason is that any solution of the equation of motion is given by the sum of a particular solution of the inhomogeneous equation
and a general solution of the homogeneous equation. In particular, any fluctuations are given by homogeneous solutions.

Now a localized (anti-)self-dual $p$-form field $\check B$ is obtained as follows as a homogeneous solution of the above equation.
We assume that near the boundary, $Y$ is of the product form $(-\epsilon, 0] \times X \subset Y$. The metric is also the product, and the background is a pullback from $X$.
We denote the coordinate of $(-\epsilon,0]$ as $\tau$. The boundary is at $\tau=0$.
Under the boundary condition \eqref{eq:L}, we can have the following ansatz $\DA_A^{(\L)}$ of a localized solution:
\beq
\DA_A^{(\L)} =  \d \left( e^{m\tau}  \right)   \wedge \DA_B , \label{eq:ansatz1}
\eeq
where $\DA_B$ is independent of $\tau$ and only depends on the coordinates of the boundary $X$.
This ansatz is consistent with the boundary condition \eqref{eq:L} because $\d \tau |_{\partial Y} = 0$.
In our convention, the coordinate $\tau$ is negative inside $Y$ and the above solution is localized exponentially to the boundary. 
It is localized more and more as the mass $m$ is increased.
The field strength is
\beq
\DF_A^{(\L)} = \d \DA_A^{(\L)} =  - \d \left( e^{m\tau} \right)  \wedge \DF_B ,
\eeq
where $\DF_B = \d \DA_B$. The Hodge dual is
\beq
\hodge_Y \DF_A^{(\L)}  = - m e^{m\tau} (\hodge_X \DF_B) ,
\eeq
where the subscripts in $\hodge_Y$ and $\hodge_X$ indicate that the Hodge star is taken in the manifolds $Y$ and $X$, respectively.

Let us substitute the above ansatz into the homogeneous equation of motion \eqref{eq:homogeneous}.
Then we get two equations from the terms with and without $\d \tau$,
\beq
 0 & = \hodge_X \DF_B +  \i (-1)^{p+1}  {\rm sign}(\kappa) \DF_B, \\
 0 & = \d ( \hodge_X \DF_B ). \label{eq:ansatz2}
\eeq
The first equation is the self-dual or anti-self-dual equation depending on the sign of $\kappa $.
The imaginary unit $\i$ appears because we are working in Euclidean signature. In Lorentz signature, 
this imaginary unit disappears and we get the standard (anti-)self dual equation of the chiral $p$-form field.
The second equation is the equation of motion for $\DF_B = \d \DA_B$. Therefore, we conclude that there is a localized
(anti-)self-dual field living on the boundary when we impose the local boundary condition given by \eqref{eq:L}.

\subsection{Topology of the boundary mode}\label{sec:topedge}
We have seen that there is a localized mode $\check B$ on the boundary.
The discussion so far assumed that $\check A$ and $\check B$ are topologically trivial. 
Let us now consider what kind of non-trivial topology is allowed for $\check B$.
First we remark that the topology of the bulk field is classified by $H^{p+2}(Y, X, \BZ)$ i.e.~cohomology on $Y$
which vanishes on the boundary $X = \partial Y$. The reason is that the field is trivial on the boundary due to the boundary condition $ \L: \check A|_{\partial Y} =0$.

We have assumed that the manifold $Y$ looks like $(-\epsilon, 0] \times X$ near the boundary.
The fundamental property of a localized mode is that it decays quickly inside $Y$, and in particular it is very small at $\tau = -\epsilon$.
This implies that the localized mode $\check B$ is topologically trivial at $\tau = - \epsilon$.
Also, the boundary condition $\L$ says that $\check B$ is topologically trivial at $\tau =0$.

The above facts imply the following. Let $I = [-\epsilon,0]$ be the interval on which the localized solution is concentrated. 
For a solution $\check A^{(\L)}$ which gives a localized solution $\check B$,
its topological class $[\DN_{A^{(\L)}}] \in H^{p+2}(Y,X,\BZ)$ must be trivial on the boundary of $I \times X$.
Such a topological class is always of the form
\beq
[\DN_{A^{(\L)}}] = \mu \cup [\DN_B],
\eeq
where $[\DN_B] \in H^{p+1}(X, \BZ)$ is some topological class in $X$, and $\mu \in H^1(I, \partial I, \BZ)$
is the unique cohomology class in $I$ which vanishes on the boundary $\partial I$. 
This is interpreted as the fact that the topology of $\check B$ is classified by $ [\DN_B] \in H^{p+1}(X,\BZ)$.

The discussion above was formulated in the case of ordinary differential cohomology.
More abstract version of the above discussion is as follows; readers who are not familiar with algebraic topology may skip the next paragraph.
The argument there is essential in the case of $d=10$, where we need to use differential K-theory, discussed in Sec.~\ref{sec:11}.

The topology of $\check A^{(\L)}$ is such that it is concentrated on $I \times X$ and it is trivial on $\partial I \times X$.
Then we can consider that it comes from the suspension $S X$ which is obtained from $I \times X$ by collapsing each of $\{-\epsilon\} \times X$ and $\{ 0 \} \times X$ to a point.
Let $E^{p+2}(Y)$ be the generalized cohomology group which classifies the topology of $\check A$. Then the topology of $\check A^{(\L)}$ is classified by 
$E^{p+2}(SX)$. By the axioms of generalized cohomology, we have $E^{p+2}(SX) = E^{p+1}(X)$. This is the cohomology group which classifies the topology of $\check B$.
This more abstract reasoning applies not only for the case of differential cohomology for which $E^q=H^q$, but also to the case of differential K-theory for which $E^q=K^q$.
For example, for Type IIB string theory, we consider a $d+1=11$ dimensional theory with $\check A \in \check K(Y)$ whose topology is classified by $K^0(Y)$. 
Then the topology of $\check B$ is classified by $K^{-1}(X)$. 
This $\check B$ is the RR-field in $d=10$ dimensions.

In our formulation, the manifold $X$ appears as the boundary of $Y$. In this case, 
there is another point which needs to be taken into account in the path integral of the bulk theory $\check A$.
For concreteness we discuss it for ordinary cohomology, but the discussion would be valid for generalized cohomology if an appropriate Poincar\'e duality theorem would be available.

We consider the homomorphism $\delta: H^{p+1}(X,\BZ) \to H^{p+2}(Y,X,\BZ)$ which is given by embedding $I \times X$ into $Y$ in the way described above.
This is the map which appears in the long exact sequence of the pair $(Y,X)$, 
\beq
\cdots \to H^{p+1}(Y) \to H^{p+1}(X) \to H^{p+2}(Y,X)  \to \cdots. \label{eq:exXY}
\eeq
Now, even if an element $[\DN_B] \in H^{p+1}(X,\BZ)$ is nontrivial, its image $[\DN_{ A^{(\L)}} ] \in H^{p+2}(Y,X,\BZ)$ under the above map
may be zero. 
Roughly speaking, the map $\delta$ reduces half of the elements of $ H^{p+1}(X,\BZ)$. 
Let us neglect torsion elements of cohomology groups for simplicity. Let $\BA = \ker \delta $ and let $\BB$ be such that $H^{p+1}(X,\BZ) = \BA \oplus \BB$.
The Poincar\'e duality and the above exact sequence imply that the pairing $(a,b) \in \BA \times \BB \to \int_X a \cup b \in \BZ$ is 
a perfect pairing while $\int_X a_1 \cup a_2 = 0$ for $a_1, a_2 \in \BA$.\footnote{
This is shown as follows. For simplicity we consider real coefficients $\BR$ so that cohomology groups can be regarded as vector spaces. 
First, the exact sequence \eqref{eq:exXY} implies that $\ker \delta$ is the image of $H^{p+1}(Y) \to H^{p+1}(X)$.
So let us uplift $a_1,a_2 \in \BA$ to elements of $H^{p+1}(Y)$. Then $\int_X a_1 \cup a_2 = \int_Y \delta( a_1 \cup a_2) =0$.
Next, notice that $\delta: \BB \to H^{p+2}(Y,X)$ is injective since $\BA$ is the kernel. Let $\beta_i~(i=1,2,\cdots)$ be a basis of $\BB$.
The Poincar\'e duality between $H^{p+2}(Y,X)$ and $H^{p+1}(Y)$ implies that there are dual elements $\alpha^i \in H^{p+1}(Y)$ such that 
$\int_Y  \delta(\beta_j) \cup \alpha^i = \delta^i_j $. But $\int_Y  \delta(\beta_j) \cup \alpha^i  = \int_Y  \delta(\beta_j \cup \alpha^i ) = \int_X \beta_j \cup \alpha^i$,
so we get $ \int_X \beta_j \cup \alpha^i = \delta^i_j$. This in particular implies that $\alpha^i$ regarded as elements of $\BA \subset H^{p+1}(X)$ are linearly independent, and hence $\dim \BA \geq \dim \BB$.
Because $\int_X a_1 \cup a_2 =0$ for any $a_1,a_2 \in \BA$, the Poincar\'e duality in $H^{p+1}(X)$ is possible only if $\alpha^i$ span the entire $\BA$ and $\dim \BA = \dim \BB$.
We conclude that $\alpha^i$ and $\beta_j$ are bases of $\BA$ and $\BB$ with $\int_X \alpha^i \cup \alpha^j =0$ and $ \int_X \beta_j \cup \alpha^i = \delta^i_j$.
By shifting $\beta_j$ by linear combinations of $\alpha^i$ if necessary, we can also take $\beta_j$ such that $\int_X \beta_i \cup \beta_j=0$.
}
We can also choose $\BB$ in such a way that $\int_X b_1 \cup b_2 = 0$ for $b_1, b_2 \in \BB$.

The image of $\BA$ is topologically trivial and hence the corresponding localized solution is unstable and decays to the trivial solution. Thus the bulk path integral only sums over the image of $\BB$.
In this way the topology of the boundary mode $\check B$ is restricted.

Splitting as $H^{p+1}(X,\BZ) = \BA \oplus \BB$ and summing only over one of them (say $\BB$) is the prescription used in the definition of the 
partition function of chiral $p$-form fields discussed in \cite{Witten:1996hc,Witten:1999vg,Moore:1999gb} without introducing the bulk $Y$.
It would be interesting to reproduce the precise partition function of \cite{Witten:1996hc,Witten:1999vg,Moore:1999gb} by using our formalism.

\subsection{Local boundary condition and the anomaly}\label{eq:LA}

Now we describe the anomaly of the boundary mode $\check B$.
The discussion here follows the one given in \cite{Witten:2019bou}.
There, the case that chiral fermions appear as boundary modes was discussed.
The bulk theory is a massive fermion  $\CL = - \bar{\Psi} (\slashed{D} -m) \Psi$ (with $m>0)$ with the boundary condition $\L$ specified by $\L: (1-\gamma^\tau) \Psi|_{\partial Y} =0$,
where $\gamma^\tau$ is the gamma matrix in the direction $\tau$ orthogonal to the boundary. 
The localized  chiral fermion appears in
the ansatz 
\beq
\Psi = \exp(m\tau) \chi, \qquad \gamma^\tau \chi=\chi, \qquad  \CD_X \chi=0,\label{eq:chiralfermion}
\eeq
where $\CD_X$ is the Dirac operator on $X = \partial Y$. Notice the similarity between \eqref{eq:chiralfermion} and \eqref{eq:ansatz1}, \eqref{eq:ansatz2}.
Chiral fermions and chiral $p$-form fields are realized in a similar way from a bulk theory with a mass gap.\footnote{
There is one difference between the cases of fermions and $p$-form gauge fields. 
In the case of fermions, chiral fermions are often realized in the literature as domain wall fermions in which we vary the mass parameter from positive to negative values as a function of space coordinates $m(y), ~ y \in Y$.
However, such a domain wall construction by a varying parameter is not possible in the case of $p$-form gauge fields. The parameter $\kappa$ is quantized and cannot be changed as a function of the space coordinate.
Also, the parameter $e^2$ is positive and hence it does not make sense to change $e^2$ from positive to negative values. In this case, what is physically sensible is the local boundary condition $\L$.
 Thus the strategy of \cite{Witten:2019bou} becomes especially important for the purposes of the present paper. 
}

In fact, many of the discussions in \cite{Witten:2019bou} can be made abstract and general without specifying the theory. We would like to present this abstract argument.

Abstractly, suppose that we have a theory $\CT$ in $(d+1)$ dimensions. 
We assume that there is a mass gap $m$ which is very large. We also assume that the low energy limit of $\CT$ is not topologically ordered,
which means that the ground state $\ket{\Omega}$ is unique up to a phase on any $d$-dimensional (spatial) manifold $X$. 

We put the theory $\CT$ on a manifold $Y$ with boundary $X = \partial Y$. 
We impose a local boundary condition which we denote as $\L$. 
$\L$ is assumed to preserve all relevant symmetries of the bulk theory so that the background field on $Y$ is not restricted on the boundary $X$.

We want to study the partition function $\CZ(Y,\L)$ of the theory $\CT$ on $Y$ with the boundary condition $\L$.
If we perform the path integral of the theory on $Y$ (or more abstractly by axioms of quantum field theory), we get a state vector $\ket{Y} \in \CH_X$.
Here $\CH_X$ is the Hilbert space on $X$. The local boundary condition $\L$ also corresponds to some state vector $\bra{\L} \in \CH_X^*$
in the dual space $\CH_X^*$. 
(See \cite{Witten:2019bou} for the explicit construction of $\bra{\L}$ in the case of fermions.) 
In this point of view, the partition function is given by the inner product
\beq
\CZ(Y,\L) = \braket{\L}{ Y }.
\eeq
When the bulk has a large mass gap and the boundary has a boundary mode, we want to regard this partition function
as the partition function of the boundary mode. 

The path integral in the region $(-\epsilon, 0] \times X$ near the boundary gives a Euclidean time evolution $e^{-\epsilon H}$ where $H$ is the Hamiltonian on $X$.
Because of the large mass gap, this Euclidean time evolution is dominated by the ground state,
\beq
e^{-\epsilon H} \to \ket{\Omega}\bra{\Omega} \qquad ( m\epsilon \to \infty).
\eeq
This means that the state vector $\ket{Y}$ is proportional to the ground state in the large mass-gap limit
\beq
\ket{Y} \propto \ket{\Omega}. \label{eq:proptoG}
\eeq
By using this property, we can split the partition function as
\beq
\CZ(Y,\L) = \braket{\L}{ \Omega } \braket{\Omega}{ Y }.
\eeq
Roughly speaking, $ \braket{\Omega}{ Y } $ is the bulk contribution, and $ \braket{\L}{ \Omega }$ is the boundary mode partition function.
In \cite{Witten:2019bou}, it was explicitly shown that the absolute value of $ \braket{\L}{ \Omega }$ is the absolute value of the partition function of the boundary chiral fermion.
We refer the reader to \cite{Witten:2019bou} for more precise expressions including the phase factor.
Also, it is shown in \cite{Belov:2004ht} that $ \braket{\L}{ \Omega }$ gives the Maxwell partition function when $d=4$ and 
the bulk $d+1=5$-dimensional theory is taken to be the theory which is analogous to the one given by \eqref{eq:bulkS}. 
It would be very interesting to do more general analysis in the case of the theory \eqref{eq:bulkS}.

In general, the ground state vector $\ket{\Omega}$ has phase ambiguities due to Berry phases. Therefore,
in general, it is not possible to fix the phase of $\braket{\L}{ \Omega }$. However, the combination $\CZ(Y,\L) = \braket{\L}{ \Omega } \braket{\Omega}{ Y }$
does not suffer from such phase ambiguity and we want to interpret $\CZ(Y,\L) $ as the partition function of the boundary mode.
The problem is that the definition of $\CZ(Y,\L) $ requires the $(d+1)$-dimensional manifold $Y$ rather than $X$ on which the boundary mode lives.
So let us study the dependence of $\CZ(Y,\L) $ on $Y$.

For this purpose, we need some preparation. When $Y$ has a boundary, we set $\CZ(Y) = \ket{Y} \in \CH_{\partial Y}$. 
(In particular, the Hilbert space on an empty space $\CH_\varnothing$ is taken to be $\BC$ and hence $\CZ(Y)$ is a complex number for closed manifolds.)
Now we claim that $\CZ(Y) $ has the unit norm $|\CZ(Y)|=1$, if we add appropriate local counterterms to the theory $\CT$.
The reason is as follows. The theory $\CT$ has only one state $\ket{\Omega}$ in the low energy limit, so we can regard it almost as an invertible field theory in low energies.
If we have an invertible field theory $\CT$ whose Hilbert space $\CH_X$ is always one-dimensional on any $X$, then we can 
define another invertible field theory $|\CT|$ as follows. The Hilbert space of $|\CT|$ is not only one-dimensional, but is defined to be canonically isomorphic to $\BC$
on any space $X$. We also define $\CZ$ of the theory $|\CT|$ to be the absolute value of $\CZ$ of the theory $\CT$,  $|\CZ|$.
The theory $|\CT|$ defined in this way satisfies the axioms of quantum field theory due to the property that $\CH_X$ is always one-dimensional. 
We can also take the inverse $|\CT|^{-1}$ in the obvious way.
This invertible field theory $|\CT|^{-1}$ can be regarded as a local counterterm, and the modified theory $\CT \otimes |\CT|^{-1}$ has $\CZ$ which have unit norm.
In the following we assume that we have added such a local counterterm to $\CT$.

In particular, $\braket{\Omega}{ Y } $ is a pure phase, $|\langle\Omega| Y \rangle|=1$ since $\ket{Y}$ has unit norm.
Now we can study the dependence of $\CZ(Y,\L)$ on $Y$. Let us take another manifold $Y'$ with $\partial Y' = X$.
Let $Y_{\rm closed} = Y \cup \overline{Y}'$ be the closed manifold obtained by gluing $Y$ and $\overline{Y}'$. Then we have
\beq
\frac{\CZ(Y,\L) }{\CZ(Y',\L ) } = \frac{\braket{\Omega}{ Y }}{\braket{\Omega}{ Y' }} = \braket{Y'}{ \Omega }  \braket{\Omega}{ Y } =  \braket{Y'}{  Y } = \CZ(Y_{\rm closed}),
\eeq
where we have used the fact that $\ket{Y} \propto \ket{\Omega}$ as shown in \eqref{eq:proptoG}.
Notice that the boundary contribution $ \braket{\L}{ \Omega }$ cancels out in the ratio.

We conclude that if $\CZ(Y_{\rm closed}) =1$ on any closed manifold $Y_{\rm closed}$, then $\CZ(Y,\L)$ is independent of $Y$
and hence can be regarded as the partition function of the boundary mode. In other words, nontrivial value of $\CZ(Y_{\rm closed})$
is the anomaly of the boundary theory; see Sec.~3 of \cite{Witten:2019bou} for details about more precise statement.\footnote{
In the above discussion, we have considered the specific invertible field theory $|\CT|$. However, we can consider any invertible field theory $\CT_{\rm counterterm}$
whose Hilbert spaces are canonically isomorphic to $\BC$, and modify $\CT$ as $\CT \otimes \CT_{\rm counterterm}$. Anomalies are classified by invertible field theories $\CT $
up to such counterterms $\CT_{\rm counterterm}$.  }
We also denote the anomaly as 
\beq
\CA (Y_{\rm closed}) = \frac{1}{2\pi \i} \log \CZ(Y_{\rm closed}).
\eeq
This concludes the general discussion of the anomaly of the boundary theory which is constructed from the theory $\CT$ with local boundary condition $\L$.

\section{Computation of the anomaly of the chiral $p$-form field} \label{sec:anomaly}

We have seen that the anomaly of the chiral $p$-form field is given by the partition function of the theory whose action is given by \eqref{eq:bulkS}.
For the bulk computation, we can take the low energy limit $e^2 \to \infty$ to neglect the first term.
(It was necessary to keep $e^2$ to be finite but very large for the purpose of deriving the localized boundary mode as in Sec.~\ref{sec:edge2}.
Without the boundary, we can safely take the limit $e^2 \to \infty$.)  

We also assume $|\kappa| = 1$ throughout this section. However, we remark that some of the computations in this section are valid
in the cases of $|\kappa| >  1$ which are topologically ordered. 
Those cases $|\kappa| >  1$ may be relevant to fractional quantum hall effects in $d=2$ and some superconformal field theories in $d=6$.

When $\kappa = \pm 1$, we can modify the definition of $\check C$ as $\check C \to \kappa \check C$.
This modification is not essential at all but it simplifies later equations somewhat. Then we get the action
\beq
-S  = 2\pi \i \kappa  \widetilde \CQ(\check A)  + 2\pi \i \kappa (\check A, \check C) , \label{eq:bulkS2}
\eeq
where $\widetilde{\CQ}(\check A) = \CQ(\check A) - \CQ(0)$.
We take this action as the starting point of this section and compute its partition function. 

The partition function can be computed in two steps. The first step is to find classical saddle points. The second step is to take into account the one-loop determinant.

Suppose that the gauge field $\check A$ is given as 
\beq
\check A = \check A_0 + \check a,
\eeq
where $\check a = (0, \DA_a, \d \DA_a)$ is topologically trivial. 
By the general results of Sec.~\ref{sec:basicQ} and the concrete constructions of Sec.~\ref{sec:3}, \ref{sec:7} and \ref{sec:11}, the above action is expanded in $\check a$ as
\beq
-S = 2\pi \i \kappa \left(  \widetilde{\CQ}(\check A_0) +  (\check A_0, \check C) + \int \left( \frac{1 }{2}\DA_a \wedge \d \DA_a  + ( \DF_{A_0} +  \Sw + \DF_{C} ) \wedge \DA_a \right) \right) \label{eq:qS},
\eeq
where $\Sw =0$ in $d+1 =3$ and $d+1=11$, and $\Sw= -\frac{1}{4}p_1(R)$ in $d+1=7$. 
In particular, it stops at the quadratic order in $\check a$ and
the theory is free without interactions if the metric and $\check C$ are background fields.
Thus the classical action at the saddle points and the one-loop determinants around them are sufficient to obtain the complete nonperturbative result.

We point out that a careful path-integral analysis of  abelian Chern-Simons theories was presented in \cite{Manoliu:1996fx,Manoliu:1998pe} for $d+1=3$.
We also point out that the general formula of the anomaly in terms of the signature index and the Arf invariant, \eqref{eq:etaArf}, was previously found by Monnier from a different approach, see e.g.~\cite{Monnier:2011rk,Monnier:2013kna}.
There are two major differences of our approach from Monnier's: 
the first is that we started from the coupled bulk-boundary system with a known action in the bulk and a specific boundary condition $\L$ realizing a chiral $p$-form field on the boundary, 
whereas Monnier tried to characterize the holonomy of the line bundle on which the partition function of the boundary theory takes values in by an indirect argument.
The second is that Monnier worked in general spacetime dimensions equipped with Wu structures,
whereas we only consider spacetime dimensions where a spin structure induces a canonical Wu structure and a quadratic refinement.
This also allows a significant simplification of the final formula which is given simply in terms of the $\eta$-invariant of the fermion, \eqref{eq:28}, for $d+1=3$ and $7$.

\subsection{The structure of the partition function}\label{sec:PT1}
Classical saddle points are found by solving the equations of motion which can be easily obtained from \eqref{eq:qS}.
If $\check A_0$ is a classical solution, we get from  \eqref{eq:qS} that
$
  \DF_{A_0} +  \Sw + \DF_{C} =0.
$
Let us set 
\beq
\check A_1 = \check A_0 +  \check C.
\eeq
For a classical solution $\check A_0$, the action simplifies to
\beq
-S =   2\pi \i \kappa \left( -\widetilde{\CQ}( \check C) +  \widetilde{\CQ}(\check A_1)  +  \int  \frac{1 }{2}\DA_a \wedge \d \DA_a \right),    \label{eq:qS2}
\eeq
where we have used $ \widetilde{\CQ}(\check A_1) = \widetilde{\CQ}(\check A_0) + \widetilde{\CQ}( \check C) + (\check A_0,  \check C)   $.
$\check A_1$ is constrained by the condition that 
\beq
\DF_{A_1} = -  \Sw.
\eeq

The moduli space of classical solutions $\CM$ has the following structure, as was nicely illustrated in \cite{Belov:2006jd}.
If $\check A_1$ and $\check A'_1$ are two classical solutions,
then $\check A_1 - \check A'_1$ is a flat gauge field. Then the moduli space is split based on the topological class as 
\beq
\CM = \sqcup_{j \in J} \CM_j,
\eeq
where the elements of each $\CM_j$ have a fixed topology $[\CN_{A_1}]_{\BZ} \in H^{p+2}(Y,Z)$ such that $[\CN_{A_1}]_{\BR}  = - [ \Sw]$, where $[ \Sw]$
is the de~Rham cohomology class of $ \Sw$. 
The index set $J$ is thus a torsor over  $H^{p+2}(Y,\BZ)_{\rm tor}$ and is (not necessarily canonically) isomorphic to it. 
Here  $H^{p+2}(Y,\BZ)_{\rm tor}$  is the kernel of the map $H^{p+2}(Y,\BZ) \to H^{p+2}(Y,\BR)$.
Then each $\CM_j$ is a torsor over 
the space of topologically trivial flat gauge field $H^{p+1}(Y ,\BZ) \otimes \BR/\BZ$,
and in particular is (not canonically) isomorphic to it.
In particular, this is independent of $ j \in J$. The classical action is constant on each connected component $\CM_j$,
and hence the path integral over $H^{p+1}(Y ,\BZ) \otimes \BR/\BZ$ gives a fixed factor which we denote as $\CN_0 >0$.

Let us take $\DA_a$ to be topologically trivial gauge fields which are orthogonal to flat gauge fields.
The part of the action which contains $\DA_a$ can be rewritten as
\beq
-S  \supset \int  \frac{2\pi \i \kappa }{2}  \DA_a \wedge \hodge ( \hodge \d \DA_a),
\eeq
where we have inserted $\hodge^2$ which is identity in odd dimensional manifold.
The path integral over $\DA_a$ gives a one-loop factor ${\det}'( - \i \kappa \hodge \d)^{- \frac 12}$.
Here, the prime in ${\det}'$ means that we omit flat gauge fields which are zero modes of $\d$.
We will discuss more about this determinant later. 

The total partition function is given by
\beq
\CZ(Y) = \left(\CN_0 \exp( -2\pi \i \kappa  \widetilde{\CQ}( \check C) ) \right) \left( \sum_{j \in J} \exp( 2\pi \i \kappa  \widetilde{\CQ}( \check A^{(j)}_1) )   \right) {\det}'( - \i \kappa \hodge \d)^{- \frac 12}  . \label{eq:BPT}
\eeq
The first factor comes from the integral over topologically trivial flat fields. 
The second factor comes from the sum over different topological sectors which are labelled by $j \in J$,
and $\check A^{(j)}_1  \in \CM_j$ is an arbitrary point of $\CM_j$.
Finally, the third factor is the one-loop determinant.

The background field $\check C$ for the $p$-form symmetry appears in the above result only as a factor $ \exp( -2\pi \i \kappa  \widetilde{\CQ}( \check C) )  $.
Therefore, we can already see that the $p$-form symmetry has the anomaly described by this factor. 
The other factors give the gravitational anomaly of the chiral $p$-form field. Let us study these factors.

For the computation of $ {\det}' ( - \i \kappa \hodge \d)^{- \frac 12} $,
we first review the signature index theorem. Readers who are not interested in technical details may skip the next subsection and go to Sec.~\ref{sec:pertCS}.
The reason that the signature index theorem becomes important is that the phase part of $ {\det}' ( - \i \kappa \hodge \d)^{- \frac 12} $ will be given by
the exponential of the $\eta$-invariant of the Dirac-type operator associated to the signature index.
The fact that the signature index theorem is important for the anomaly of a chiral $p$-form field was first found in perturbation theory in \cite{AlvarezGaume:1983ig}.

\subsection{Signature index theorem}\label{sec:Signature}

As preliminaries to the computation of the one-loop factor, we review some technical details of the signature index theorem.
One of the purposes is to study all the sign factors carefully. It is also useful for practical computations of the signature $\eta$-invariant.

On $2m$-dimensional oriented manifolds, the signature operator can be defined as follows by using a Dirac operator.

Let $\CS = \CS^+ \oplus \CS^-$ be the spin bundle, where $\CS^\pm$ are spin bundles with positive and negative chirality, respectively. 
Then $\CS \otimes \CS^*$ is isomorphic to $\bigoplus_k \wedge^k T^* W$. More explicitly, a section $\Phi$ of the bi-spinor bundle $\CS \otimes \CS^*$ can be identified with
a sum of $k$-forms $\omega^{(k)}$ as
\beq
\Phi = \sum_{k=0}^{2m} \frac{ \i^k}{k!} \Gamma^{I_1 \cdots I_k} \omega^{(k)}_{I_1 \cdots I_k},
\eeq
where $\Gamma^{I_1 \cdots I_k} $ is defined by anti-symmetrizing the product of gamma matrices $\Gamma^{I_1} \cdots \Gamma^{ I_k} $
(e.g.~$\Gamma^{1 \cdots k} = \Gamma^1 \cdots \Gamma^k$). The factor $\i^k$ is introduced for later convenience. 

We have a formula
\beq
\Gamma^I \Gamma^{I_1 \cdots I_k} =\Gamma^{ I I_1 \cdots I_k}  +  \sum_{j=1}^k (-1)^{j-1} g^{I I_j} \Gamma^{I_1 \cdots \widehat{I_j} \cdots I_k},   
\eeq
where the hat on $\widehat{I_j}$ means that we omit that index. Using this formula, we see that
\beq
\i \Gamma^I D_I \Phi &= \sum_{k=0}^{2m} \frac{ \i^k}{k!}  \Gamma^{ I_1 \cdots I_k} ( k D_{I_1} \omega^{(k-1)}_{I_2 \cdots I_k} - D^I  \omega^{(k+1)}_{I I_1 \cdots I_k} )
\nonumber \\
&= \sum_{k=0}^{2m} \frac{ \i^k}{k!}  \Gamma^{ I_1 \cdots I_k}  (\d \omega^{(k-1)} + \d^\dagger \omega^{(k+1)})_{ I_1 \cdots I_k}.
\eeq
This implies that for the sum
\beq
\omega = \sum_{k=0}^{2m} \omega^{(k)}
\eeq
the action of the Dirac operator $\CD^{\rm sig} = \i \Gamma^I D_I$ on $\Phi$ is equivalent to the action of $(d+d^\dagger)$ on $\omega$,
\beq
\CD^{\rm sig} \Phi \quad \Longleftrightarrow \quad (\d + \d^\dagger) \omega.
\eeq

Let us define the chirality operator in $d=2m$ dimensions as
\beq
\Ch = \i^{-m} \Gamma^1 \cdots \Gamma^{2m},
\eeq
which has the standard properties $\Ch^\dagger = \Ch, \quad \Ch^2=1$. We have a formula
\beq
\Ch \Gamma^{I_1 \cdots I_k} = \frac{ \i^{-m} (-1)^{\frac{1}{2} k(k+1)}  }{ ({2m}-k)! } 
 \epsilon^{I_1 \cdots I_k I_{k+1} \cdots I_{2m} }\Gamma_{I_{k+1} \cdots I_{2m}}.
\eeq
By using this formula, we get
\beq
\Ch \Phi &=  \sum_{k=0}^{2m} \frac{\i^k }{k!}  \cdot  \frac{ \i^{-m} (-1)^{\frac{1}{2} k(k+1)}  }{ ({2m}-k)! } 
 \epsilon^{I_1 \cdots I_k I_{k+1} \cdots I_{2m} }\Gamma_{I_{k+1} \cdots I_{2m}}   \omega^{(k)}_{I_1 \cdots I_k}  \nonumber \\
 &= \sum_{k=0}^{2m}     \frac{ \i^{{2m} -k}  }{ ({2m}-k)! }  \cdot \i^{k(k-1) + m}
\Gamma_{I_{k+1} \cdots I_{2m}}   (\hodge \omega^{(k)})^{I_{k+1} \cdots I_{2m}},
\eeq
where $\hodge $ is the Hodge dual. Thus the chirality operator $\Ch$ acting on $\Phi$ is equivalent to the Hodge operator up to a phase factor.
Let $K$ be the operator which gives the degree of a form, such that $K \omega^{(k)} = k \omega^{(k)}$. Then $\Ch$ is equivalent to
$\hodge  \cdot \i^{K(K-1)+m}$,
\beq
\Ch \quad \Longleftrightarrow \quad  \hodge  \cdot \i^{K(K-1)+m}, \label{eq:chiralhodge}
\eeq
where $\hodge  \cdot \i^{K(K-1)+m}$ maps $\omega^{(k)}$ to
$ \i^{k(k-1)+m} \hodge \omega^{(k)} $.

In particular, if the dimension is $  4\ell$ (i.e.~$m = 2\ell$), the chirality operator on $2\ell$-form coincides with the Hodge dual $ \hodge$.
The signature is defined as the number of zero modes of $\d +  \d^\dagger$ with $\hodge = +1$ minus the number of zero modes with $\hodge = -1$.
Thus the index of the Dirac operator $\CD^{\rm sig}$ coincides with the signature.

The characteristic class relevant for the above index theorem is given by
\beq
L = \hat{A}(R) \tr_{S} \exp\left( \frac{\i}{2\pi} R \right) =  \prod_{i=1}^{m} x_i \coth( x_i/2) \label{eq:Hir},
\eeq
where the trace is taken in the spin representation $S$, and $\pm x_i$ are Chern roots, i.e.~formal eigenvalues of the Riemann curvature 2-form $R$.
We have, $ \hat{A}(R) =  \prod_{i=1}^{m}\frac{ x_i /2}{\sinh( x_i/2)}$

On a closed manifold $Z$, the signature is given by
\beq
\index \CD^{\rm sig} = \int_Z L.
\eeq
If the $2m$-dimensional manifold $Z$ has a boundary $\partial Z = Y$, the relevant $\eta$-invariant is defined as follows.
From now on we put subscript $Z$, $Y$ etc.~to quantities on the respective manifolds. 
See Sec.~\ref{sec:APS} for the general description of the APS index theorem.

On $Y$, we restrict attention to the modes with $P_+ \Phi = \Phi$, where $P_+ = \frac{1}{2}(1+\Ch_Z)$.
We represent the Dirac operator $\CD^{\rm sig}_Z$ on $Z$ near the boundary $( -\epsilon, 0] \times Y$ as
\beq
\CD^{\rm sig}_Z = \i \Gamma^I_Z D_I = \i \Gamma^s_Z (\partial_s + \Gamma^s_Z\Gamma^{I'}_Z D_{I'} ) =\i \Gamma^s_Z (\partial_s + \CD'^{\rm sig}_Y),
\eeq
where $s$ is the coordinate $s \in ( -\epsilon, 0] $,  the index $I'$ runs over the directions of $Y$, and $I$ runs over the directions of $Z$.
The operator $\CD^{\rm sig}_Y = P_+ \CD'^{\rm sig}_Y  $  is the relevant operator for the $\eta$-invariant. 

We want to see how the operator $\CD^{\rm sig}_Y $ looks like on differential forms $\omega^{(k)}$.
We first decompose $\omega^{(k)}$ near the boundary as
\beq
\omega^{(k)} = \omega^{(k)}_1 + \d s \wedge \omega^{(k-1)}_2,
\eeq
where $\omega^{(k)}_1$ and $\omega^{(k-1)}_2$ do not contain $\d s$.
The condition $P_+ \Phi = \Phi$ and the fact that $\Ch_Z$ corresponds to $\hodge_Z  \cdot \i^{K(K-1)+m}$ 
implies that $\omega^{(k)}_1$ and $\omega^{(2m-k-1)}_2$ are related to each other, so we can restrict attention to $\omega^{(k)}_1$.
The $\omega^{(k-1)}_2$ is just determined by $\d s \wedge \omega^{(k-1)}_2 = \hodge_Z  \cdot \i^{K(K-1)+m} \left(\omega^{(2m-k)}_1 \right)$.
Let us define
\beq
\Phi_1 =   \sum_{k=0}^{2m-1} \frac{ \i^k}{k!} \Gamma_Z^{I'_1 \cdots I'_k} (\omega^{(k)}_1)_{I'_1 \cdots I'_k}. \label{eq:tech2}
\eeq
Then $\Phi = \Phi_1 + \Ch_Z \Phi_1$.

We are going to show the following correspondence
\beq
\CD^{\rm sig}_Y \Phi \quad \Longleftrightarrow \quad \left( \hodge_Y \d_Y \i^{ K(K-1)+m } - \d_Y \hodge_Y  \i^{K(K+1) +m } \right) \omega_1. \label{eq:b-op}
\eeq
This operator was given by Atiyah-Patodi-Singer in a slightly different convention \cite{Atiyah:1975jf}.
To show this correspondence, we need some technical computation. 

We act $\CD^{\rm sig}_Y$ on $\Phi$ to get
\beq
\CD^{\rm sig}_Y( \Phi_1 + \Ch_Z \Phi_1) &= \frac{1}{2} (1+ \Ch_Z)  \Gamma^s_Z\Gamma^{I'}_Z D_{I'} ( 1+ \Ch_Z )\Phi_1   \nonumber \\
& = \Ch_Z \Gamma^s_Z\Gamma^{I'}_Z D_{I'} \Phi_1 + \Gamma^s_Z\Gamma^{I'}_Z D_{I'} \Phi_1.
\eeq
When we expand the bi-spinor in terms of the products of gamma matrices $\Gamma^I_Z$,
the first factor $\Ch_Z \Gamma^s_Z\Gamma^{I'}_Z D_{I'} \Phi_1 $ does not contain $\Gamma^s_Z$ (because $\Ch_Z$ contains one $\Gamma^s_Z$ which cancels another $\Gamma^s_Z$), while the second factor $\Gamma^s_Z\Gamma^{I'}_Z D_{I'} \Phi_1$
contains one $\Gamma^s_Z$. Therefore, to see the action on $\omega_1$, we study $\Ch_Z \Gamma^s_Z \Gamma^{I'}_Z D_{I'} \Phi_1$.

The operator $\i \Gamma^{I'}_Z D_{I'} $ on $\Phi_1$ corresponds to $\d_Y + \d_Y^\dagger$ on $\omega_1$,
\beq
\i \Gamma^{I'}_Z D_{I'}  \quad \Longleftrightarrow \quad \d_Y + \d_Y^\dagger.
\eeq
Thus we need to study $-\i \Ch_Z \Gamma^s_Z$. We define it as $\Ch_Y$.
 
 Taking $s$ as the $I=1$ direction, we have
 \beq
 \Ch_Y := \i  \Gamma^s_Z \Ch_Z =  \i^{-m+1} \Gamma^2_Z \cdots \Gamma^{2m}_Z .
 \eeq
 It has the property that
 \beq
 \Ch_Y \Gamma_Z^{I'_1 \cdots I'_k} =  \i^{- m+1} (-1)^{\frac{1}{2} k(k-1) } \frac{1}{(2m-1-k)!}\epsilon_Y^{I'_1 \cdots I'_k I'_{k+1}  \cdots I'_{2m-1}}(\Gamma_Z)_{I'_{k+1} \cdots I'_{2m-1}} .\label{eq:tech1}
\eeq
By using this equation to \eqref{eq:tech2}, we obtain
\beq
 \Ch_Y \Phi_1 =  \sum_{k=0}^{2m-1} \frac{ \i^{2m-1-k} }{(2m-1-k)!} \cdot \i^{k(k+1)+m+2} \cdot  (\Gamma_Z)_{I'_{k+1} \cdots I'_{2m-1}} (\hodge_Y \omega^{(k)}_1)^{I'_{k+1} \cdots I'_{2m-1}},
\eeq
where $\hodge_Y$ is the Hodge dual on the boundary $Y$. Thus we get the correspondence
\beq
\Ch_Y \quad \Longleftrightarrow \quad \hodge_Y \cdot \i^{K(K+1)+m+2}.
\eeq

Therefore, $\CD^{\rm sig}_Y = \Ch_Y (\i \Gamma^{I'} D_{I'}) $ corresponds to
\beq
\CD^{\rm sig}_Y \quad \Longleftrightarrow \quad \hodge_Y \cdot \i^{K(K+1)+m+2} \cdot ( \d_Y + \d_Y^\dagger ).
\eeq
In odd dimensions, we have
\beq
\hodge_Y^2 =1, \qquad \d_Y^\dagger = \hodge_Y \d_Y \hodge_Y (-1)^{K}.
\eeq
The operator $\d_Y$ raises the degree by one, and $\d_Y^\dagger$ lowers the degree by one. Thus we get
\beq
(\hodge_Y \cdot \i^{K(K+1)+m+2} ) \cdot (\d_Y + \d_Y^\dagger) &=  \hodge_Y  \d_Y \i^{(K+1)(K+2)+m+2 } + \d_Y \hodge_Y  \i^{(K-1)K+m+2 +2K}  \nonumber \\
&=  \hodge_Y \d_Y \i^{ K(K-1)+m } - \d_Y \hodge_Y  \i^{K(K+1) +m } . \label{eq:b-op2}
\eeq
This completes the proof of the correspondence \eqref{eq:b-op}.

For the signature, we can further simplify the result by noticing the following. 
We can multiply $\Phi$ by $\Ch_Y$ from the right as $\Phi \Ch_Y$, and this operation commutes with the Dirac operator on $Y$.
This right multiplication of $\Ch_Y$ changes even differential forms $\Omega^{\rm even}(Y)$ to odd differential forms $\Omega^{\rm odd}(Y)$ and vice versa. 
Therefore, $\CD^{\rm sig}_Y$ acting on $\Omega^{\rm even}(Y)$ and $\CD^{\rm sig}_Y$ acting on $\Omega^{\rm odd}(Y)$ have completely the same spectrum.
For the computation of the $\eta$-invariant, we can thus just restrict our attention to $\Omega^{\rm even}(Y)$ or $\Omega^{\rm odd}(Y)$ and then multiply $\eta$ by a factor of 2.
In particular, for the purpose of the present paper, we will consider forms $\Omega^{m-1+ 2\bullet}(Y)$ whose degree is $m-1 \mod 2$.

There is one more simplification which comes from the following point~\cite{Atiyah:1975jf}. (We again use a slightly different convention from that of \cite{Atiyah:1975jf}.)
We decompose differential forms as
\beq
\Omega^k(Y) = H^k(Y) \oplus \d_Y \Omega^{k-1}(Y) \oplus \d_Y^\dagger \Omega^{k+1}(Y),
\eeq
where $H^k(Y)$ is the space of harmonic forms (i.e.~forms which are annihilated by $\d_Y+\d_Y^\dagger$).
Notice that operators $\hodge_Y  \d_Y$ and $ \d_Y \hodge_Y$ map these forms as
\begin{align}
 \hodge_Y  \d_Y  :&\quad \d_Y^\dagger \Omega^{k+1}(Y) \to \d_Y^\dagger \Omega^{2m-1-k}(Y) \nonumber \\
 \d_Y \hodge_Y :&\quad \d_Y \Omega^{k-1}(Y) \to \d_Y \Omega^{2m-1-k}(Y).
\end{align}
They annihilate other spaces. Now, if an operator $T$ maps one space $V_1$ to another $V_2$ isomorphically,
the operator $T \oplus T^\dagger$ acting on $V_1 \oplus V_2$ always has eigenvalues which are pairs of positive and negative eigenvalues, $(+\lambda, -\lambda)$. 
Thus they cancel each other in the definition of the $\eta$-invariant.
Therefore, we can neglect them in the computation of the $\eta$. 
The operator $ \CD^{\rm sig}_Y $ acting on $\Omega^{m-1+ 2\bullet}(Y)$ always changes the degrees of the forms, except for 
\beq
\widetilde{\CD}^{\rm sig}_Y := \hodge_Y  \d_Y \i^{K(K-1) + m } : \d_Y^\dagger \Omega^{m}(Y) \to \d_Y^\dagger \Omega^{m}(Y). \label{eq:middleD0}
\eeq
Therefore, only this part contributes to the $\eta$-invariant (except for zero modes).
We will see the physical interpretation of this fact later.

Based on the simplification discussed above, the $\eta$-invariant of $\CD_Y^{\rm sig}$ is given as follows:
\beq
\eta(\CD_Y^{\rm sig} ) = 2 \eta( \widetilde{\CD}^{\rm sig}_Y) + \sum_{\bullet}\dim H^{m-1+2\bullet}(Y,\BR).
\eeq
The factor $2$ in $ 2 \eta( \widetilde{\CD}^{\rm sig}_Y) $ is due to the fact that $\Omega^{\rm even}(Y)$ and $\Omega^{\rm odd}(Y)$
contribute the same way to the $\eta$-invariant. By definition, $ \widetilde{\CD}^{\rm sig}_Y$ acts on $\d_Y^\dagger \Omega^{m}(Y) $ which does not have zero modes.
The contribution of the zero modes $ H^k(Y)$ to the $\eta$-invariant is given by the term $\sum_{\bullet}\dim H^{m-1+2\bullet}(Y,\BR)$.

As discussed in \cite{Atiyah:1975jf}, the difference $\index \CD^{\rm sig}_Z  - \sum_{\bullet}\dim H^{m-1+2\bullet}(Y,\BR)$ has a particular geometric meaning,
which is called the signature of a manifold $Z$ with boundary $\partial Z =Y$,
\beq
\signature(Z) = \index \CD^{\rm sig}_Z  -\sum_{\bullet} \dim H^{m-1+2\bullet}(Y,\BR).
\eeq
The APS index theorem is now given by
\beq
\signature(Z) = \int_Z L + 2 \eta( \widetilde{\CD}^{\rm sig}_Y) . \label{eq:Hir1}
\eeq
The topological meaning of $\sigma(Z)$ is that we consider the image of the map $H^{m}(Z,\partial Z) \to H^{m}(Z)$ on which the intersection form $(x_1, x_2) \to \int x_1 \cup x_2$
is well-defined, and then $\sigma(Z)$ is the signature of this intersection form.
See \cite{Atiyah:1975jf} for details.

\subsection{One-loop determinant}\label{sec:pertCS}
What we have found in the previous subsection is summarized as follows. 
For a manifold $Z$ of dimension $d+2 = 2p+4$  with boundary $\partial Z = Y$, the signature of $Z$,
which is denoted as $\signature(Z)$ and defined in \cite{Atiyah:1975jf}, is given by the APS index theorem as
\beq
\signature(Z)  = \int_Z L +  2 \eta( \widetilde{\CD}^{\rm sig}_Y). \label{eq:signind}
\eeq
Here, $L$ is the Hirzebruch polynomial defined in \eqref{eq:Hir}. The operator $\widetilde{\CD}^{\rm sig}_Y$ is the one defined in \eqref{eq:middleD0},
\beq
\widetilde{\CD}^{\rm sig}_Y =  \i^{p(p+2)+2 } \hodge  \d : ~   \widetilde \Omega^{p+1}(Y) \to  \widetilde \Omega^{p+1}(Y). \label{eq:middleD}
\eeq
where $\widetilde \Omega^{p+1}(Y) = \d^\dagger  \Omega^{p+2}(Y) $ is the subspace of $\Omega^{p+1}(Y) $ which is orthogonal to the space of the closed forms $\Omega^{p+1}_{\rm closed}(Y)$.
In other words, $\widetilde \Omega^{p+1}(Y) $ is the subspace which is orthogonal to the kernel of $\widetilde{\CD}^{\rm sig}_Y$.

We want to compute the phase part of the one-loop determinant $ {\det}' ( - \i \kappa \hodge \d)^{- \frac 12} $.
From the discussion of  Sec.~\ref{sec:PT1}, we see that the determinant $ {\det}'$ is taken exactly in the space $ \widetilde \Omega^{p+1}(Y)$.
The reason is as follows. The determinant is taken over the space of fields $\DA_a \in \Omega^{p+1}(Y)$.
In this space, the space of exact forms $\d \Omega^p(Y)$ is just gauge degrees of freedom and hence we neglect it.\footnote{
For the computation of the absolute value of ${\det}' ( - \i \kappa \hodge \d)^{- \frac 12} $, it is necessary to perform gauge fixing and do  the computation more carefully.
However, for the phase contribution, we can just neglect these gauge fixings. The underlying reason is as follows. 
Introducing gauge fixings and ghosts would ultimately give an elliptic operator $\CD^{\rm sig}_Y$ acting on the space of all forms $\Omega^{p+1 +2\bullet}(Y)$. See \cite{Witten:1988hf}
for the explicit construction of it in the case of $p=0$. However, only the part \eqref{eq:middleD0} contributes to the $\eta$-invariant of $\CD^{\rm sig}$. This means that all the contributions cancels out except for $\widetilde \Omega^{p+1}(Y)$.
}
Also, the space of closed forms which are not exact are flat gauge fields which are excluded in ${\det}'$;
their contribution is already taken into account as the factor $\CN_0$ in \eqref{eq:BPT}.
Therefore, we only need to consider $\widetilde \Omega^{p+1}(Y) =  \d^\dagger  \Omega^{p+2}(Y)$.

If the coefficients are not twisted, the chiral $p$-form field and its associated bulk theory is possible only if $p$ is even, $p = 2\ell$.
Then we have $\widetilde{\CD}^{\rm sig}_Y =  - \hodge  \d$ and 
\beq
{\det}' ( - \i \kappa \hodge \d)^{- \frac 12}  = \det (\i \kappa \widetilde{\CD}^{\rm sig}_Y)^{- \frac 12} .
\eeq

Now we compute the determinant. The eigenvalues $\lambda$ of the operator $\i \kappa \widetilde{\CD}^{\rm sig}_Y$ are nonzero, and we get
\beq
\det (\i \kappa \widetilde{\CD}^{\rm sig}_Y) &=  \prod_{\lambda} \i \kappa \lambda \nonumber \\
 &=  \prod_{\lambda} |\lambda|  \exp \left( \frac{\i \pi}{2} {\rm sign}(\kappa \lambda)  \right) \nonumber \\
 &= \CN_1^{-2}  \exp \left( \i \pi  \kappa \eta(\widetilde{\CD}^{\rm sig}_Y)  \right).
\eeq
where $\CN_1^{-2} :=  \prod_{\lambda} |\lambda|   >0$ is a positive factor. 
Therefore, we conclude that 
\beq
{\det}' ( - \i \kappa \hodge \d)^{- \frac 12} = \CN_1 \exp \left( -\frac{2\pi \i \kappa}{8} \cdot   2\eta(\widetilde{\CD}^{\rm sig}_Y)  \right) . \label{eq:oneloopeta}
\eeq

\subsection{Total partition function}\label{sec:totalP}
In the partition function \eqref{eq:BPT}, the remaining factor which we have not explicitly computed yet is the factor 
$ \left( \sum_{j \in J} \exp( 2\pi \i \kappa  \widetilde{\CQ}( \check A^{(j)}_1) )   \right)$. The index set is isomorphic to $H^{p+2}(Y,\BZ)_{\rm tor}$.
Let us define it as
\beq
\left( \sum_{j \in J} \exp( 2\pi \i \kappa  \widetilde{\CQ}( \check A^{(j)}_1) )   \right) = \CN_2 \exp ( 2\pi \i \kappa \Arf_\Sw(Y)) ,
\eeq
where $\CN_2>0$.
Here the subscript $\Sw$ is to remember that the sum is over $\check A_1$ with $\DF_{A_1} = -\Sw$.
The appearance of the phase factor with a factor $\kappa$ is easily seen by complex conjugation. 

The partition function is now given by
\beq
\CZ(Y) = \CN_0\CN_1\CN_2 \exp 2\pi \i \kappa\left( -  \widetilde{\CQ}(\check C) - \frac18\cdot 2\eta(\widetilde{\CD}^{\rm sig}_Y) +  \Arf_\Sw(Y)  \right).\label{eq:PT1}
\eeq
We can see that $\CN_0\CN_1\CN_2 = 1$ (possibly up to regularization-dependent local counterterms) by the following formal argument.
We have integrated over $\check A$ whose action is $2\pi \i \kappa \widetilde \CQ(\check A)$ and obtained the above partition function.
Now let us integrate over $\check C$. It appears as $-2\pi \i \kappa \widetilde \CQ(\check C)$ in \eqref{eq:PT1}. The only difference from the action of $\check A$ is $\kappa \to -\kappa$.
Therefore, the phases coming from the integration over $\check A$ and $\check C$ cancel each other and we get $( \CN_0\CN_1\CN_2 )^2$.
On the other hand, if we go back to the action \eqref{eq:bulkS} and integrate over $\check C$ before integrating over $\check A$, we get the delta function $\delta (\check A)$.
Then we integrate over $\check A$ to get $1$. Therefore we conclude that $( \CN_0\CN_1\CN_2 )^2=1$.
As $\CN_{0,1,2}$ are all positive almost by definition, the statement follows.

A consequence of the above result $\CN_0\CN_1\CN_2 = 1$ is that $|\CZ(Y)|=1$ on any manifold $Y$.
In particular, we have $|\CZ(S^1 \times X)|=1$ for any spatial manifold $X$. This quantity $\CZ(S^1 \times X)$ (with the antiperiodic spin structure on $S^1$)
counts the dimension of the Hilbert space $\CH_X$ on $X$. Thus we have established that the Hilbert space is one-dimensional and 
the theory is an invertible field theory.

The anomaly $\CA(Y) $ is the phase part of $\CZ(Y)$ and it is given by
\beq
\CA(Y) = \kappa\left( -   \widetilde{\CQ}(\check C) - \frac{1}{8} \cdot 2\eta(\widetilde{\CD}^{\rm sig}_Y) +  \Arf_\Sw(Y) \right). \label{eq:etaArf}
\eeq
The terms $- \frac{1}{8}  \cdot 2\eta(\widetilde{\CD}^{\rm sig}_Y) +  \Arf_\Sw(Y)$ give the gravitational anomaly of the theory.
We denote it as $\CA_{\rm grav}$,
\beq
\CA_{\rm grav} =    - \frac{1}{8} \cdot 2\eta(\widetilde{\CD}^{\rm sig}_Y) +  \Arf_\Sw(Y)   .\label{eq:GRAV}
\eeq

We can simplify the above gravitational anomaly. 
We will argue that it is given as follows depending on the dimension:
\beq
\CA_{\rm grav} =  \left\{
  \begin{array}{ll}
 \eta(\CD_Y^{\rm Dirac} ),& (d+1=3), \\
 28\eta(\CD_Y^{\rm Dirac}), & (d+1=7) ,\\
-\eta(\CD_Y^{{\rm Dirac}\otimes TY})+3\eta(\CD_Y^{\rm Dirac} ), & (d+1=11) ,\\
 \end{array} \right. \label{eq:28}
\eeq
where $\CD_Y^{\rm Dirac}$ is the usual Dirac operator (without coupling to additional bundles) 
and $\CD_Y^{{\rm Dirac} \otimes TY}$ is the Dirac operator acting on the spinor bundle tensored with the tangent bundle.
The physical reason that we expect this result is as follows. 

In $d=2$, a chiral $(p=0)$-form field (a chiral compact scalar) is dual to a chiral fermion, and hence we expect their anomaly is the same.

In $d=6$, a chiral $(p=2)$-form field is not equivalent to chiral fermions. However, there is an anomaly matching
between a chiral 2-form field and 28 chiral fermions which can be shown by using $E$-string theory as discussed in \cite{Hsieh:2019iba}.

In $d=10$, the Type IIB superstring theory should be anomaly free.
This means that the anomaly of the 4-form chiral fields can be canceled by the anomaly of the the gravitino and the dilatino.
The gravitino on $X$ corresponds to the Dirac operator acting on the spinor bundle tensored with $TX$ minus an ordinary fermion.
Considering also the fact that  $TY|_{\partial Y=X}=TX\oplus \underline{\bR}$, the anomaly of the gravitino is captured by $\eta(\CD_Y^{{\rm Dirac}\otimes TY})-2\eta(\CD_Y^{\rm Dirac})$.\footnote{
The chirality of the 5-form field strength given by \eqref{eq:ansatz2} is $\hodge \DF = \i \kappa \DF $.
The $\CA_{\rm grav}$ is the gravitational anomaly in the case $\kappa=+1$, so suppose that $\hodge \DF = \i  \DF $. 
In that case, it can be shown (see e.g.~\cite{Tachikawa:2018njr}) that the chirality of the gravitino is negative, $\Ch = -1$.
Therefore, the anomaly of the gravitino is $\eta(\CD_Y^{{\rm Dirac}\otimes TY})-2\eta(\CD_Y^{\rm Dirac})$ including the sign.}
Since the dilatino has the chirality opposite to the gravitino, we need to subtract another contribution of $\eta(\CD_Y^{\rm Dirac})$, 
leading to the combination $\eta(\CD_Y^{{\rm Dirac}\otimes TY})-3\eta(\CD_Y^{\rm Dirac} )$.

The formula \eqref{eq:28} can be shown in the following way. 
We study the perturbative anomaly, or in other words the continuous dependence of \eqref{eq:GRAV} to the metric.
Then we argue that the perturbative anomaly is enough to determine the complete answer.

First we study the metric dependence of $\Arf_\Sw$. 
For this purpose, we take a manifold $Z$ which is topologically $[0,1] \times Y$,
but the metric on $Y$ is a function of $s \in [0,1]$, which we denote as $g_s$. 
Let $ \widetilde{\CQ}( \check A^{(j)}_1)(g_s)$ be the value of $ \widetilde{\CQ}( \check A^{(j)}_1)$ on the metric $g_s$. 
We have
\beq
 \widetilde{\CQ}( \check A^{(j)}_1)(g_1) -  \widetilde{\CQ}( \check A^{(j)}_1)(g_0) 
 = \int_Z \left( \frac{1}{2} \DF_{A_1} \wedge \DF_{A_1} + \Sw \wedge \DF_{A_1} \right).
\eeq
Here $ \check A_1$ is extended to $Z$. This is possible because $Z$ is topologically $[0,1] \times Y$ which is contractible to $Y$. 
Its field strength $\DF_{A_1}$ is constrained as $\DF_{A_1} = -\Sw$, so we get
\beq
 \widetilde{\CQ}( \check A^{(j)}_1)(g_1) -  \widetilde{\CQ}( \check A^{(j)}_1)(g_0) 
  = -\frac{1}{2} \int_Z  \Sw \wedge \Sw. 
\eeq
In particular, it is independent of the torsion part of the topological class of $\check A_1$, so we get
\beq
\Arf_\Sw(g_1) - \Arf_\Sw(g_0) =  -\frac{1}{2} \int_Z  \Sw \wedge \Sw. \label{eq:gdepArf}
\eeq
The metric dependence of the part $2\eta(\widetilde{\CD}^{\rm sig}_Y)$ can be determined by the signature index theorem \eqref{eq:signind}.
The signature of $Z = [0,1] \times Y$ is zero, $\signature(Z) =0$. Therefore we get
\beq
 - \frac{1}{8}  \left( 2\eta(\widetilde{\CD}^{\rm sig}_Y)(g_1) -    2\eta(\widetilde{\CD}^{\rm sig}_Y)(g_0) \right) = \frac{1}{8} \int_Z L.
\eeq
Thus the dependence of the anomaly on the metric is given by
\beq
\CA_{\rm grav}(g_1) - \CA_{\rm grav}(g_0) =  \frac{1}{8} \int_Z \left( L - 4\Sw \wedge \Sw  \right) .\label{eq:Gdepend}
\eeq

Recall that $\Sw=0$ in $d+2=4$, $12$ and $\Sw = - \frac{1}{4} p_1(R)$ in $d+2=8$.
By straightforward computation, one can check the following. For $d+2=4$,
\beq
\frac{1}{8}\left( L - 4\Sw \wedge \Sw  \right)|_{4\text{-form}} = \frac{1}{24}p_1 = -\hat{A}|_{4\text{-form}} \label{eq:relAL2}
\eeq
For $d+2=8$,
\beq
\frac{1}{8}\left( L - 4\Sw \wedge \Sw  \right)|_{8\text{-form}} = \frac{-p_1^2 +7p_2}{2^3 \cdot 3^2 \cdot 5} - \frac{p_1^2}{2^5} = -28\cdot  \frac{7p_1^2 -4p_2}{2^7 \cdot 3^2 \cdot 5} 
= -28\hat{A}|_{8\text{-form}} \label{eq:relAL}
\eeq
Similarly for $d+2=12$,
\beq
\frac{1}{8}\left( L - 4\Sw \wedge \Sw  \right)|_{12\text{-form}} = \hat{A}(R) \left. \left(  \tr \exp\left( \frac{\i}{2\pi} R \right) -4  \right) \right|_{12\text{-form}}. \label{eq:relAL3}
\eeq
The last case is the celebrated result originally shown in \cite{AlvarezGaume:1983ig}.\footnote{%
It is known that $L/8$ is  given by an integer linear combination of the $\hat A$ genus of the Dirac operator coupled to a tensor power of the tangent bundle in 
arbitrary dimensions of the form $d+2=8\ell +4$, see \cite{Liu:1994wh}.
In fact, the integrality of $L/8$ in dimensions $d+2=8\ell +4$ would probably be required by the consistency of the differential K-theory in $d+1 = 8\ell+3$ dimensions
developed in Sec.~\ref{sec:11}. The reason is that the anomaly polynomial after taking the background $\check C$ to be zero is given by $L/8$
since $\Sw=0$ in differential K-theory.
}

By using the above results \eqref{eq:relAL2}, \eqref{eq:relAL}, \eqref{eq:relAL3} and also the index theorem of the Dirac operator, we get
\beq
  \frac{1}{8} \int_Z \left( L - 4\Sw \wedge \Sw  \right) =  \left\{
  \begin{array}{ll}
 \eta(\CD_Y^{\rm Dirac} )|_{g_1-g_0}, & (d+1=3), \\
 28\eta(\CD_Y^{\rm Dirac})|_{g_1-g_0}, & (d+1=7), \\
-(\eta(\CD_Y^{{\rm Dirac}\otimes TY})-3\eta(\CD_Y^{\rm Dirac} ))|_{g_1-g_0}, & (d+1=11), \\
 \end{array} \right.
\eeq
where $X|_{g_1 -g_0}$ is an abbreviation for $X(g_1)-X(g_0)$.
By this result and \eqref{eq:Gdepend},
we conclude that \eqref{eq:28} holds at the perturbative level.

Let us denote the right-hand-side of \eqref{eq:28} as $\CA'_{\rm grav}$.
Then the combination $\CA_{\rm grav} - \CA'_{\rm grav} $ (or more precisely its exponential) defines an invertible field theory which is relevant to the anomaly of the chiral $p$-form field
and some number of fermions. 
Moreover it is independent of the continuous deformation of the metric as we have shown above, so it represents a global anomaly. 
In unitary quantum field theory, such a quantity must be a spin-cobordism invariant~\cite{Kapustin:2014dxa,Freed:2016rqq,Yonekura:2018ufj}.
However, $\Omega_3^{\rm spin} =0$, 
$\Omega_7^{\rm spin} =0$
and $\Omega_{11}^{\rm spin} =0$, from the determination of the additive structure of the spin bordism group by Anderson-Brown-Peterson \cite{ABP0,ABP}.\footnote{%
The spin bordism groups are given by \[
\begin{array}{c|ccccccccccccccccccccccccccc}
d &  0& 1&2&3&4&5&6&7 \\
\hline
\Omega_d^\text{spin} & \BZ & \BZ_2 &\BZ_2 & 0& \BZ & 0  &0 & 0\\
\hline\hline
d &8&9&10&11&12&13&14  & 15 \\
 \hline
\Omega_d^\text{spin} & 2\BZ& 2\BZ_2 & 3\BZ_2 & 0 & 3\BZ & 0&0 &0  \\
\hline\hline
d & 16 & 17 & 18 & 19 & 20 & 21 & 22 & 23 \\
\hline
\Omega_d^\text{spin} & 5\BZ & 5\BZ_2 &7\BZ_2 & 0 & 7\BZ+\BZ_2 &0 &\BZ_2 & 0 
\end{array} 
\]
according to \cite{ABP0,ABP}. 
(The authors thank the authors of \cite{BoyleSmith:2025duo} for reporting the error in this table in the previous version of the paper.
The algorithm to produce this table is explained in detail in the appendix A of \cite{BoyleSmith:2025duo}.)
} 
Therefore, it must vanish, $\CA_{\rm grav} - \CA'_{\rm grav} =0$.

We remark that if we consider non-unitary theories, invertible field theories need not be cobordism invariant.
We discuss examples in Appendix~\ref{sec:nonunitary}.

We conclude that the anomaly $\CA(Y)$ of the chiral $p$-form field $\check B$ is given by
\beq
\CA(Y) = \kappa \left( - \widetilde \CQ(\check C) +  \CA_{\rm grav} \right).
\eeq
where $\CA_{\rm grav}$ is given by \eqref{eq:28} or \eqref{eq:GRAV}.
Here $\kappa = \pm 1$ is the parameter which specifies whether the field strength is self-dual or anti-self-dual.
In Euclidean signature and $p=$ even, it is given by \eqref{eq:ansatz2},
\beq
\hodge \DF_B =  \i  \kappa \DF_B. \label{eq:selfduality}
\eeq
The imaginary unit $\i$ is just an artifact of the Euclidean signature metric.
The result for $\CA(Y)$ is valid in $d=2,6$ and $10$. 
In the cases $d=2$ and $d=6$, we have chosen the definitions of $\CQ$ in Sec.~\ref{eq:QR}
in such a way that the constant term $\CQ(0)$ (which have played no role up to now) coincides with $-  \CA_{\rm grav}$.
Thus the above result is simplified as 
\beq
\CA(Y) = -\kappa (  \widetilde \CQ(\check C)  +   \CQ(0) ) = - \kappa \CQ(\check C) \label{eq:d2and6}
\eeq
for $d=2,6$.

Finally, let us make one comment on the consistency of the above results. The equivalence between \eqref{eq:28} and \eqref{eq:GRAV} requires
the following expression for the signature. Let $Z$ be a $(d+2)$-dimensional manifold with boundary $Y$.
Then, the APS index theorem and the equality of \eqref{eq:28} and \eqref{eq:GRAV} implies that the signature $\sigma(Z) \mod 8$ is given by
\beq
\sigma(Z) = 8\Arf_\Sw +   \int_Z (2\Sw)^2 \mod 8.\label{eq:signmod8}
\eeq
A formula which is very close to this equation was proved, see Theorem~4.3 of \cite{BrumfielMorgan}.
$2\Sw$ is a differential form representative of the Wu class. The right-hand-side is independent of the metric due to \eqref{eq:gdepArf}.
We do not perform detailed comparison between \eqref{eq:signmod8} and the theorem of \cite{BrumfielMorgan},
but nevertheless the above formula might be regarded as a consistency check of our results.

\subsection{Remark on the case $d=6$}\label{sec:notshifted}
\'Alvarez-Gaum\'e and Witten computed the perturbative gravitational anomaly of a chiral $p$-form field in $d=2p+2$-dimensions~\cite{AlvarezGaume:1984nf}.
They have obtained the result that the anomaly polynomial is given by $\frac{1}{8} L$.
This corresponds to the term $\CA(Y) \supset -\frac{1}{8} \cdot 2\eta(\widetilde{\CD}^{\rm sig}_Y)$ since the signature theorem \eqref{eq:signind} relates
this term to $\frac{1}{8}L$. In $d=2$ and $d=10$, the Arf invariant $\Arf_\Sw$ is independent of the metric and hence it does not contribute to the perturbative anomaly.

However, in $d=6$, $\Arf_\Sw$ depends on the metric as can be seen explicitly by the formula \eqref{eq:gdepArf} and $\Sw = - \frac{1}{4}p_1(R)$.
Thus it contributes to the perturbative gravitational anomaly. Naively, it might look that it contradicts with the result of \cite{AlvarezGaume:1984nf}.
There is no contradiction, and the situation is as follows.

Let $Y$ be a 7-manifold and $Z$ be an 8-manifold such that $\partial Z = Y$.
We can represent $\CQ(\check C)$ as
\beq
\CQ(\check C) &= \int_Z \left( \frac{1}{2}\DF_C^2  + \Sw \wedge \DF_C + 28\hat{A}_2(R) \right)  \nonumber \\
&=\int_Z \left( \frac{1}{2}(\DF_C +\Sw)^2  - \frac{1}{8} L \right) ,
\eeq
where we have used \eqref{eq:relAL}. In M-theory, the 3-form field in 11-dimensions is usually defined so that its field strength is shifted as
\beq
\DF_C^{\rm shifted} = \DF_C +\Sw = \DF_C - \frac{1}{4}p_1(R).\label{eq:shift2}
\eeq
Then one might interpret the part $\frac{1}{2}(\DF_C^{\rm shift})^2$ as the anomaly of the higher-form symmetry,
and $-\frac{1}{8}L$ as the gravitational anomaly. At the perturbative level, there is no problem in this interpretation.
Under this interpretation, the gravitational anomaly is as obtained in \cite{AlvarezGaume:1984nf}.

We have chosen not to use the shifted quantity like $\DF_C^{\rm shifted} = \DF_C +\Sw$ in this paper.
Before explaining why we did so, let us review some topological facts.

The Pontryagin class $p_1$ can be defined for any $\SO$ bundle as an element of the integr cohomology $H^4(Y,\BZ)$.
In the case of a spin bundle, we can refine it as $p_1 = 2 c(\Spin)$, where $c(\Spin)$ is an integer cohomology class 
normalized in such a way that the minimal instanton number of the $\Spin$ group corresponds to $\int c(\Spin) =1$.\footnote{%
More precisely it can be defined by using the obstruction theory argument as reviewed e.g.~in \cite{Witten:1985bt}, based on the fact that $\pi_0(\Spin)=\pi_1(\Spin) =\pi_2(\Spin)=0$ and $\pi_3(\Spin)=\BZ$.
We can also use $\pi_{k}(B\Spin) = \pi_{k-1}(\Spin) $ and the Hurewicz theorem to find $H^4(B\Spin,\BZ) = \BZ$ and get the characteristic class $c(\Spin)$.}
The reduction of $c(\Spin)$ to $\BZ_2$ coefficients coincides with the 4-th Stiefel-Whitney class $w_4$, $c(\Spin)_{\BZ_2}= w_4$.\footnote{%
The $c(\Spin)_{\BZ_2}$ is the generator of $H^4(B\Spin,\BZ_2) = \BZ_2$. So the only possibilities are $w_4=0$ identically or $w_4 = c(\Spin)_{\BZ_2}$. 
We can consider a vector bundle whose fiber is $\BC^2$ and which has a minimal instanton number of $\SU(2)$ acting on $\BC^2$.
By viewing $\BC^2 \cong \BR^4$, it gives an example for which $w_4 =(e)_{\BZ_2}=(c_2)_{\BZ_2} \neq 0$, where $c_2$ is the 2nd Chern class of the complex bundle $\BC^2$,
and $e$ is the Euler characteristic class of $\BC^2 \cong \BR^4$.
This bundle is also a $\Spin(4)$ bundle.
A minimal instanton of an $\Spin(4)$ bundle gives an example that $w_4$ is nontrivial, so $w_4 = c(\Spin)_{\BZ_2}$.}

The above facts are valid for any spin bundle. Now let us restrict our attention to spin manifolds and the spin bundle associated to the tangent bundle.
If the dimension $D$ of the manifold is $D \leq 7$, it is known that $w_4=0$.\footnote{
On manifolds we have the Wu class $\nu = 1+\nu_1+\nu_2+\cdots \nu_{[D/2]} $, 
where $[D/2] $ is the largest integer which does not exceed $D/2$.
It is known to satisfy ${\rm Sq}(\nu) = w$, 
where ${\rm Sq}=1+{\rm Sq}^1+{\rm Sq}^2+\cdots$ is the total Steenrod square and $w=w_1+w_2+\cdots$ is the total Stiefel-Whitney class of the manifold.
See e.g.~\cite{MS} for details.
On spin manifolds, we have $w_1=0, w_2=0,w_3=0$ corresponding to $\pi_0(\Spin)=0,\pi_1(\Spin) =0,\pi_2(\Spin)=0$ respectively. Then we get $\nu_1=0,\nu_2=0,\nu_3=0$.
By dimensional reason, we conclude $\nu=1$ if $D \leq 7$ and hence $w=1$.
}
Therefore, by using the long exact sequence associated to $0 \to \BZ \to \BZ \to \BZ_2 \to 0$, there exists $\Sw_{\BZ} \in H^4(Y,\BZ)$ such that $c(\Spin) = -2\Sw_{\BZ}$, and hence $p_1 = -4\Sw_{\BZ}$.
Thus the de~Rham cohomology class of $- \frac{1}{4}p_1(R)$ can be represented as an image of the integer cohomology $\Sw_{\BZ}$.
This implies that there exists a differential cohomology $\check C_\Sw \in \check H^4(Y)$ such that $\DF_{C_\Sw} = - \frac{1}{4}p_1(R) = \Sw$ and $[\DN_{C_\Sw}] = \Sw_\BZ$.
There may not be a natural choice of such an element $\check C_\Sw$, but anyway by choosing one such $\check C_\Sw$, we can define 
\beq
\check C^{\rm shifted} = \check C + \check C_\Sw \label{eq:shift}
\eeq
so that it has the field strength $\DF_C^{\rm shifted}$.

However, on a manifold $W$ with dimension $\dim W \geq 8$ such as the 11-dimensional bulk in M-theory,
$\Sw = - \frac{1}{4}p_1(R)$ is not guaranteed to be an image of integer cohomology, 
and hence $ \check C^{\rm shifted}$ is not guaranteed to be an ordinary differential cohomology element.
It requires the use of shifted differential cohomology, which we choose not to use in this paper.

The concept of shifted differential cohomology can be avoided as far as we can analyze the system (or M-theory in the current case) in a consistent manner without it.
The flux quantization of the 3-form $\check C$ is relevant to M2-branes,
so let us consider the partition function of a single M2-brane in M-theory~\cite{Witten:2016cio}.

For simplicity we neglect quantum fluctuations of  scalar fields on the M2-brane which does not contribute to the following discussion.
This means that the position of the M2-brane is considered to be fixed.
The worldvolume of the single M2-brane contains Majorana fermions in the spin representation of $[\Spin(3 ) \times \Spin(8)]/\BZ_2$,
where $\Spin(3)$ is the spin group associated to the tangent bundle, and $\Spin(8)$ is the normal bundle of the M2-brane in 11-dimensions. 
We only consider the case in which all manifolds are oriented. (More generally, M-theory has the parity symmetry and we can consider non-orientable manifolds.)

Let $M$ be the worldvolume of the M2-brane ($\dim M=3$).
The partition function of the M2-brane is expected to be of the form
\beq
\CZ_{\rm M2}(M) = \CZ_{\rm fermion}(M) \exp(2\pi \i \int_M \DA_C),
\eeq
where $\CZ_{\rm fermion} $ is the fermion partition function. 

The definition of $\CZ_{\rm fermion}$ requires care. It is given as
\beq
\CZ_{\rm fermion}(M) = \pf (\CD_M) \exp(2\pi \i \CB),
\eeq
where $\CD_M$ is the Dirac operator acting on the fermions, $\pf$ is the pfaffian, and $\CB \in \BR/\BZ$ is some phase which we will specify later. 
The pfaffian is possible because the bundle of the fermion is pseudoreal and hence each eigenvalue appears twice. 
We define $\pf(\CD_M)$ by using the Pauli-Villars (PV) regularization with PV mass $m_{\rm PV}$.
Let $\lambda$ be eigenvalues of $\CD_M$, and let $\sum'_\lambda$ and $\prod'_\lambda$ be the sum or product over pairs of the same eigenvalues $(\lambda,\lambda)$.
Then we define~\cite{AlvarezGaume:1984nf},
\beq
\pf(\CD_M) &= {\prod_\lambda}' \frac{\i \lambda}{\i \lambda + m_{\rm PV} } = |\pf(\CD_M)|  \exp\left( {\sum_\lambda}' \frac{\pi \i}{2} \sign(\lambda) \right) \nonumber \\
&=  |\pf(\CD_M)|  \exp\left( \frac{1}{2}\pi \i  \eta(\CD_M)  \right).
\eeq
This partition function has the standard parity anomaly~\cite{Redlich:1983kn,Redlich:1983dv,Niemi:1983rq,AlvarezGaume:1984nf}.
By changing the orientation of $M$, the phase of the partition function changes by $\exp\left(-  \pi \i  \eta(\CD_M)  \right) $.

Let us take a manifold $N$ such that $\partial N = M$ and all gauge fields are extended from $M$ to $N$.
Then APS index theorem and the pseudoreality of the bundle (which implies that the index is even) gives
\beq
 \frac{1}{2} \eta(\CD_M)   \equiv - \frac{1}{2} \int_N  \left(  -\frac{8}{24}p_1(R_N) + p_1(R_{\rm normal}) \right) \mod 1,
\eeq
where $R_N$ is the Riemann curvature on $N$, and $R_{\rm normal}$ is the curvature of the $\SO(8)=\Spin(8)/\BZ_2$ normal bundle.
Let $R$ be the bulk Riemann curvature. Topologically $p_1(R) = p_1(R_N) +p_1(R_{\rm normal})   $ up to total derivative.
Then we can write
\beq
 \frac{1}{2}  \eta(\CD_M)  \equiv  \int_N  \left(   \frac{2}{3}p_1(R_M) -\frac{1}{2} p_1(R) \right) \mod 1. \label{eq:parityA}
\eeq
We partially cancel the parity anomaly by introducing the signature $\eta$-invariant $\CB = \eta( \CD_M^{\rm sig})$
which satisfies
\beq
-\eta( \CD_M^{\rm sig}) \equiv \int_N \frac{1}{3}p_1(R_N) \mod 1.
\eeq
However, the second term of \eqref{eq:parityA} cannot be  cancelled and we leave it as it is. The fermion partition function is now
\beq
\CZ_{\rm fermion}(M) =   |\pf(\CD_M)| \exp( \pi \i \CC(M)),
\eeq
where
\beq
\CC(M) := \frac{1}{2}\eta(\CD_M)  + 2\eta( \CD_M^{\rm sig}).
\eeq
The function $\exp( \pi \i \CC(M))$
is not smooth in $\BR/\BZ$, and it jumps by $\frac{1}{2}$ whenever some eigenvalue of $\CD_M$ crosses zero.

The M2-brane partition function is now given by
\beq
\CZ_{\rm M2}(M) =  |\pf(\CD_M)| \exp 2\pi \i \left( \int_M \DA_C + \frac{1}{2} \CC \right).
\eeq
Physically, the important point related to the shift \eqref{eq:shift} is as follows.
The M-theory has time-reversal symmetry, so we may want to define the M-theory 3-form in such a way that
its holonomy function $\chi^{\rm shift}(M)$ would be given by
\beq
\chi^{\rm shift}(M) :=  \exp 2\pi \i \left( \int_M \DA_C + \frac{1}{2} \CC \right).
\eeq
Then the parity symmetry acts simply as $\chi^{\rm shift}(M)  \to \chi^{\rm shift}(M)^*$.
When $N = \partial M$, we also have
\beq
\chi^{\rm shift}(M) = (-1)^{\index(\CD_N)} \exp 2\pi \i \int_N  \left(\DF_C -\frac{1}{4}p_1(R) \right), \label{eq:shift3}
\eeq
where $\index(\CD_N)$ is the APS index. 
Notice that we have already encountered  the combination $\DF_C^{\rm shifted}  = \DF_C - \frac{1}{4}p_1(R) $ in \eqref{eq:shift2}.
However, $\chi^{\rm shift}(M)$ is not smooth since $\exp( \pi \i \CC(M))$ is not. 
It changes the sign when some eigenvalue of $\CD_M$ crosses zero.
A related fact is that  we have the unwanted factor $(-1)^{\index(\CD_N)}$ in \eqref{eq:shift3}.
From these reasons, the function $\chi^{\rm shift}(M)$ cannot be regarded as a holonomy function of an ordinary differential cohomology element.
This is the physical reason behind the phenomenon of the shift.

The non-smoothness of $\chi^{\rm shift}(M)$ is not a problem:
it appears in the M2-brane partition function as
$\CZ_{\rm M2}(M) = |\pf(\CD_M)|  \chi^{\rm shift}(M)$. 
The absolute value $|\pf(\CD_M)|$ is also not smooth precisely when some eigenvalue of $\CD_M$ crosses zero,
so that the two factors $|\pf(\CD_M)|$ and $\chi^{\rm shift}(M)$ combine together to make the partition function smooth. 

In this paper, we prefer to keep the smoothness of various quantities so that we can use the basic formalism of differential cohomology reviewed in Sec.~\ref{sec:DiffCoh}. 
Thus we did not consider the shifted holonomy $\chi^{\rm shift}(M)$ or shifted differential cohomology $\check C^{\rm shift}$, and instead we consider the unshifted $\check C $ associated to $\chi(M) = \exp (2\pi \i \int_N \DA_C)$.
There is no problem in this description, although the time-reversal symmetry becomes not manifest.

\section{Applications to M-theory }\label{sec:Mthy}

On a single M5-brane in M-theory, there is a chiral 2-form field $\check B$ and two chiral fermions $\chi$.
They contribute to the anomaly of the worldvolume theory. 
We use the convention that the supercharge $Q^{\rm{M}5}$ preserved by the M5-brane has negative chirality
under the worldvolume chirality operator $\Ch^{\rm{M}5}$,
\beq
\Ch^{\M5}Q^{\M5} = - Q^{\rm{M}5}.
\eeq
The worldvolume chiral fermions $\chi$ are obtained from the worldvolume scalar as $\chi \sim [Q^{\M5} , {\phi}]$ and hence they have negative chirality 
\beq
\Ch^{\M5} \chi = - \chi.
\eeq 
The chiral 2-form field strength $\DF_B$ is obtained as $\DF_B \sim \{Q^{\M5}, \bar{\chi} \}$.
Thus, as a bi-spinor, it has $\Ch^{\M5} = -1$.
By the relation between bi-spinor and $p$-forms discussed in Sec.~\ref{sec:Signature} and in particular by \eqref{eq:chiralhodge}, we see that its field strength satisfies the duality equation
\beq
\hodge \DF_B = \i \DF_B.
\eeq
Thus, by \eqref{eq:selfduality}, it corresponds to the case $\kappa = +1$.

Throughout this section, we assume that the normal bundle to the worldvolume of the M5-brane is trivial and does not contribute to the anomaly.
However, we consider nontrivial 3-form backgrounds $\check C$ from the bulk 3-form field in M-theory. (Our $\check C$ is not shifted; see Sec.~\ref{sec:notshifted} for the details.)
A single chiral fermion with negative chirality contributes to the anomaly as $+\eta(\CD)$,
where $\CD$ is the ordinary Dirac operator.
By the APS index theorem it is related to $\hat{A}$ as
\beq
+\eta(\CD) = -\int_Z \hat{A}_2(R) \mod 1.
\eeq
The anomaly of $\check B$ is given by $- \CQ(\check C)$ as discussed around \eqref{eq:d2and6}.
The total anomaly from the fields $\check B$ and $\chi$ is given by
\beq
\CA = - \CQ(\check C) + 2\eta(\CD) = -\int_Z \left( \frac{1}{2} \DF_C \wedge \DF_C - \frac{1}{4}p_1(R) \wedge \DF_C + 30\hat{A}_2(R) \right) \mod 1.
\eeq

\subsection{Cancellation of the anomaly and the flux for M5-branes}\label{sec:M}
M-theory must be consistent. This means that the anomaly of the chiral fields $\check B$ and $\chi$ on an M5-brane should be somehow cancelled.
The anomaly cancellation for the M5-brane including the contributions of the normal bundle was discussed in \cite{Freed:1998tg}.
In this section, we restrict our attention to the cases where the normal bundle is trivial,
and adopt  the argument in \cite{Witten:1999eg} originally carried out for F1 and D1 strings instead.
We postpone the extension of our argument when the normal bundle is nontrivial to future work.

We denote the 11-dimensional bulk as $W$, and the worldvolume of the M5-brane as $X$.
The fact that the worldvolume fields $\check B$ and $\chi$ have the anomaly means that their partition function $\CZ_{\rm matter}$
depends on how to extend $X$ to $Y$ such that $\partial Y = X$. In particular, we take $Y$ to be a subspace of the bulk $W$.
We denote the partition function defined by using $Y \subset W$ as $\CZ_{\rm matter}(Y)$.

There is also another contribution to the M5-brane partition function. 
Roughly, the field strength $F_4$ of the 3-form field $C_3 (\sim \check C^{\rm shift})$ can be dualized as
\beq
\hodge F_4 \sim F_7 \sim \d C_6+\cdots,
\eeq
where $C_6$ is some 6-form field. The M5-brane is coupled to this 6-form as $2\pi \i \int_X C_6$.
When we have the extension of $X$ to $Y $ such that $\partial Y = X$, 
we express this coupling as $2\pi \i \int_Y F_7$. The total M5-brane partition function is given by
\beq
\CZ_{\M5} = \CZ_{\rm matter}(Y)\exp(2\pi \i \int_Y F_7).
\eeq
This does not depend on how to take $Y$ if its value on a closed manifold $Y_{\rm closed}$ is trivial by the same argument as in Sec.~\ref{eq:LA}. This means that
$ \CZ_{\rm matter}(Y_{\rm closed})\exp(2\pi \i \int_{Y_{\rm closed} }F_7) =1$ or equivalently
\beq
\CA(Y_{\rm closed}) +  \int_{Y_{\rm closed} }F_7 =0 \mod 1.\label{eq:ACancel}
\eeq
This is the condition for the anomaly cancellation.

Let us check this discussion in the 11-dimensional supergravity.
By taking $Z$ such that $\partial Z = Y_{\rm closed}$, the anomaly cancellation condition above becomes
\beq
0 = \int_Z \left( -\frac{1}{2} F_4 \wedge F_4 + I_8 + \d F_7 \right),
\eeq
where
\beq
F_4 &:= \DF_C - \frac{1}{4} p_1(R), \\
I_8 & :=  \frac{1}{32}p_1(R)^2 - 30\hat{A}_2(R) = -\frac{p_1^2 - 4p_2}{192}.
\eeq
For the above equation to be valid for any $Z$, we must have $-\frac{1}{2} F_4 \wedge F_4 + I_8 + \d F_7 =0$.

We assume the precise duality relation between $F_7$ and $F_4$ as
\beq
\hodge F_4 = \i F_7,
\eeq
where the factor $\i$ comes from the fact that we are working in Euclidean signature metric. 
See Appendix~\ref{sec:Mthconv} for details about the precise sign.
Therefore, we get
\beq
\d \hodge F_4 = \i \d F_7 = \i\left( \frac{1}{2} F_4 \wedge F_4 - I_8 \right).
\eeq
This equation follows from the well-known supergravity action which is roughly given by
\beq
-S = -\int_W \frac{2\pi}{2} F_4 \wedge \hodge F_4 + 2\pi \i \int_W \left( \frac{1}{6}C_3 \wedge F_4 \wedge F_4 - C_3 \wedge I_8  \right), \label{eq:SUGRA}
\eeq
where roughly $F_4 = \d C_3$. 
A more precise definition of this supergravity action is given in \cite{Witten:1996md}.

We see that the anomaly cancellation condition \eqref{eq:ACancel} is consistent with the supergravity.
In particular, \eqref{eq:ACancel} implies that the flux of $F_7$ is not quantized to be integers, but is shifted by the anomaly $-\CA$ of the degrees of freedom on the M5-brane,
\beq
\int F_7 \in - \CA + \BZ. \label{eq:anomalyFfrac}
\eeq
This type of phenomenon was studied in \cite{Tachikawa:2018njr} in the case of orientifold planes in Type II string theories, based on the earlier discussions in~\cite{Witten:2016cio}.

\subsection{M5-brane in M-theory orbifold backgrounds}\label{sec:orb}

We now consider an application of the formalism to M-theory orbifold of the form
\beq
W = \BR^3 \times (\BR^8/\BZ_k).
\eeq
We denote the coordinates as $x^I$ with $I=0,1\cdots, 10$.
We are working with Euclidean signature metric, and the coordinate $x^0$ may be regarded as a Euclidean time direction.

We assume that the M5-brane is confined within $ \{0\} \times (\BR^8/\BZ_k)$.
In particular, we are interested in the case that $Y_{\rm closed}$ is the lens space $S^7/\BZ_k$ which surrounds the
orbifold point of $ (\BR^8/\BZ_k)$. For simplicity we just denote it as $Y$,
\beq
Y =  S^7/\BZ_k.
\eeq
The quantization rule for the flux is given by
\beq
\int_Y F_7 = -\CA(Y) =  \CQ(\check C) - 2\eta(\CD) \mod 1 .\label{eq:F7int}
\eeq

Before computing the values of the anomaly, 
let us discuss the implication of the above result.
Suppose that the 11-dimensional manifold is just a flat space $W = \BR^{11}$, and instead of the orbifold singularity,
we have M2-branes with M2 charge $\Sq$ extending in the direction $\BR^3 \times \{0\}$ with the orientation given by the volume form $\d x^0 \wedge \d x^1 \wedge \d x^2$.
We denote the orthogonal direction as $\vec{z} = (x^3, \cdots, x^{10})$.
The action of the 3-form field $C_3$ (at the differential form level) is given by
\beq
-S = - \int \frac{2\pi}{2} \d C_3 \wedge \hodge \d C_3 + 2\pi \i  \Sq \int C_3 \wedge \delta(\vec z),
\eeq
where $\delta(\vec z ) = \delta(x^3) \d x^3 \wedge \cdots \wedge \delta(x^{10} )\d x^{10}$.
The equation of motion is 
\beq
\d (\hodge F_4) = \i \Sq \delta(\vec z).
\eeq
By using $\hodge F_4 = \i F_7$, we conclude that the integral of $F_7$ over $Y =S^7$ is given by
\beq
\int_Y F_7 = \Sq. \label{eq:M2Charge}
\eeq
Therefore, $\int_Y F_7$ measures the M2 charge. 
Thus we can interpret \eqref{eq:F7int} as the M2-charge of the orbifold singularity.
In particular, the anomaly gives the fractional part of the orbifold M2-charge.

Now we want to compute the anomaly $\CA= - \CQ(\check C) + 2\eta(\CD)$.
We assume that the background $\check C$ is flat, $\DF_C=0$. 
Then $\check C$ is completely characterized by the torsion part of the cohomology group $H^4(Y,\BZ)$ where $Y=S^7/\BZ_k$. It is known that
\beq
H^4( S^7/\BZ_k , \BZ) = H^3( S^7/\BZ_k , \BR/\BZ) = \BZ_k. \label{eq:4coh}
\eeq
In $\check C = ( \DN_C, \DA_C,0)$, 
$[\DA_C]_{\BR/\BZ}$ is an element of $H^3( S^7/\BZ_k , \BR/\BZ) $, 
and $[\DN_C]_{\BZ} = - [\delta \DA_C]_{\BZ}$ is an element of $H^4( S^7/\BZ_k , \BZ)$ such that 
$[\DN_C]_\BZ = \beta [\DA_C]_{\BR/\BZ}$, where $\beta$ is the Bockstein homomorphism. 

Let $\check C_1$ be a differential cocycle corresponding to a generator of \eqref{eq:4coh}. 
(We will specify more details of this generator later.)
Then we consider $\check C = \ell \check C_1$ for $\ell \in \BZ$.
By using the property that $\widetilde \CQ(\check C) := \CQ(\check C)- \CQ(0)  $ is a quadratic refinement of the differential cohomology pairing with $\widetilde{\CQ}(0)=0$, we get
\beq
\widetilde{\CQ}( \ell \check C_1)   = \widetilde{\CQ}( (\ell-1) \check C_1)  + \widetilde{\CQ}( \check C_1)  + (\ell-1)   (\check C_1 , \check C_1),
\eeq
where $(\check A, \check B) = \int \DA_{A \Dp B}$ is the differential cohomology pairing.
By induction we get
\beq
\widetilde{\CQ}( \ell \check C_1)   = \ell  \widetilde{\CQ}( \check C_1)   + \frac{\ell(\ell-1)}{2} (\check C_1 , \check C_1) . \label{eq:QnC0}
\eeq
Also recall that $\CQ(0) = -28\eta(\CD)$ by the definition of $\CQ$ in Sec.~\ref{sec:7}.
Thus the anomaly is given by
\beq
\CA = 30\eta(\CD)  - \ell\widetilde{\CQ}(  \check C_1) - \frac{\ell(\ell-1)}{2} (\check C_1 , \check C_1).
\eeq
We need to compute each term of the right hand side. 

Let us study $\widetilde{\CQ}(  \check C_1)$ in more detail.
From Sec.~\ref{sec:7}, $\CQ$ can be expressed by using the $\eta$-invariant of the 56-dimensional representation of $E_7$.
Let us recall this construction.
The $E_7$ contains a subgroup $\SU(2) \times \Spin(12) \subset E_7$ under which the 56 dimensional representation is decomposed as $2 \otimes 12 \oplus 1 \otimes 2^5$.
If we restrict the gauge field to the $\SU(2)$ subgroup, we get
\beq
c(E_7)= - c_2 ( \SU(2)).
\eeq
where $c_2$ is the second Chern class which in the de~Rham cohomology is represented by the curvature of $\SU(2)$ as $ - \frac{1}{2} \tr_{ 2 } \left( \frac{\i }{2\pi} F_{\SU(2)} \right)^2 $,
see \eqref{eq:DFS1} and \eqref{eq:DFS2}.
If we further restrict the $\SU(2)$ to the $\U(1)$ subgroup,
the bundle associated to the two-dimensional representation of $\SU(2)$ becomes a sum of $\U(1)$ bundles $\CL \oplus \CL^{-1}$, and 
\beq
c(E_7) = c_1(\CL)^2,
\eeq
where $c_1(\CL)$ is the first Chern class of $\CL$.
The bundle associated to the 56 dimensional representation of $E_7$ becomes
\beq
 ( \CL \oplus \CL^{-1})^{\oplus 12} \oplus \underline{\BC^{32}}, \label{eq:56tolines}
\eeq
where $\underline{\BC}$ is the trivial bundle.

Now we define the basic line bundle $\CL_1$ as follows. Let $\vec{z} \in \BC^4$ be a complex vector of unit length $|\vec{z}|=1$ which represents points on $S^7$.
We consider a trivial $\U(1)$ bundle $S^7 \times \BC$ on $S^7$, and divide it by $\BZ_k$ which acts as
\beq
S^7 \times \BC \ni (\vec{z}, v) \mapsto ( e^{2\pi \i /k} \vec{z}, e^{-2\pi \i  /k} v),
\eeq
Then we get the line bundle over the lens space $S^7/\BZ_k$,
\beq
\CL_1 = (S^7 \times \BC)/\BZ_k. \label{eq:L0onlens}
\eeq

The cohomology $H^4(S^7/\BZ_k, \BZ) = \BZ_k$ is generated by $c_1(\CL_1)^2$.
(The entire cohomology $H^\bullet (S^7/\BZ_k, \BZ)$ is generated as a ring from $c_1(\CL_1)$.)
We take $\check C_1$ in such a way that 
\beq
[\DN_{C_1}]=c(E_7) = c_1(\CL_1)^2.\label{eq:NC1onlens}
\eeq
Notice that $\CL_1$ is a flat bundle and hence its curvature is zero, which is consistent with the fact that the curvature of $\check C_1$ is zero.
Then by the construction of Sec.~\ref{sec:7}, the quadratic refinement $\CQ(\check C_1)$ is given by the $-1/2$ of the $\eta$-invariant coupled to \eqref{eq:56tolines}.
It is given by
\beq
\CQ(\check C_1) = - \frac{1}{2} \left(  12 \eta( \CD_1) +12\eta(\CD_{-1}) +32\eta(\CD_0) \right) = - 12\eta(\CD_1) - 16\eta(\CD_0),\label{eq:orbA}
\eeq
where $\CD_s$ is the Dirac operator coupled to $\CL_1^{\otimes s}$, and in particular $\CD_0=\CD$.
In the above equality, we have used the fact that $\eta(\CD_{-1}) = \eta(\CD_1)$ which follows from the fact that the 7-dimensional spin bundle is pseudoreal.

By using the above facts, we can represent the anomaly as
\beq
\CA = 30\eta(\CD_0)  + 12 \ell (\eta(\CD_1) - \eta(\CD_0)) - \frac{\ell(\ell-1)}{2} (\check C_1 , \check C_1).
\eeq
We want to know $(\check C_1, \check C_1)$ and $\eta(\CD_s) $.

Let us first present the results.
It turns out that 
\beq
& (\check C_1 , \check C_1) \equiv - \frac{1}{k} \mod 1 , \nonumber \\
& 12\left( \eta(\CD_1) - \eta(\CD_0) \right) \equiv \frac{k^2-1}{2k} \mod 1, \nonumber \\
& 30 \eta(\CD_0) \equiv  \frac{K_k(K_k-k)}{2k} + \frac{k^2-1}{24k}  \mod 1 , \label{eq:summaryvalue}
\eeq
where
\beq
K_k = \frac{1}{2}k(k+1)-1 .
\eeq
The total anomaly is simplified if we use
\beq
n  = \ell + K_k. \label{eq:orbflux}
\eeq
Then we get from \eqref{eq:orbA} that 
\beq
\CA \equiv \frac{k^2-1}{24k} - \frac{n(k-n)}{2k} \mod 1.
\eeq
This is the result for the anomaly.

From what we have explained around \eqref{eq:M2Charge} and \eqref{eq:anomalyFfrac}, we conclude that the fractional part of the M2-charge of the orbifold singularity $\Sq$ is given by
\beq
\Sq \equiv  - \frac{k^2-1}{24k} + \frac{n(k-n)}{2k} \mod 1.
\eeq
This is exactly as known in the literature \cite{Sethi:1998zk,Bergman:2009zh,Aharony:2009fc}.
We cannot determine the integer part because it does not contribute to the anomaly. More physically, we can put ordinary M2-branes on top of the orbifold singularity
to change the integer part of $\Sq$ without affecting the consistency of the theory.

Our remaining task is to show \eqref{eq:summaryvalue} for the values of $(\check C_1 , \check C_1),\,  12( \eta(\CD_1) - \eta(\CD_0) )$ and $30 \eta(\CD_0)$
modulo integers.
This is done in Appendix~\ref{sec:lens}.
We also have another Appendix~\ref{sec:etacomp}
where the $\eta$-invariants of the lens spaces (but not the pairing $(\check C_1 , \check C_1)$) 
are computed as real numbers $\BR$ by using different techniques than Appendix~\ref{sec:lens}.

\subsection{D4-brane in O2-plane backgrounds}\label{sec:O2}

By the result of the M5-brane anomaly in the orbifold background, we can also determine the anomaly of a D4-brane 
in an O2-plane background. In M-theory, we consider the geometry
\beq
\BR^3 \times (\BR^7 \times  S^1)/\BZ_2.
\eeq
This geometry has two orbifold singularities. In Type IIA description, it corresponds to an O2-plane,
\beq
\BR^3 \times (\BR^7 /\BZ_2 ).
\eeq
Each of the two orbifold singularities in M-theory can have two discrete fluxes.
Correspondingly, we have four types of O2-planes. By using the label $n $ ($=\ell+2 \equiv \ell$ mod 2)  defined in \eqref{eq:orbflux} for the flux of a single M-theory orbifold,
we denote the corresponding O2-plane as $\text{O}2(n_1,n_2)$ where $n_1=0,1$ and $n_2=0,1$ corresponds to the flux at each singularity.
The D2-charge of $\text{O}2(n_1,n_2)$ modulo integers is given by
\beq
\Sq(n_1,n_2) = -\frac{1}{8} + \frac{n_1(2-n_1)}{4} + \frac{n_2(2-n_2)}{4}.
\eeq
From this, we see that $\text{O}2(0,0)$ is the O2$^-$-plane, $\text{O}2(0,1)$ and $\text{O}2(1,0)$ are the O2$^+$-plane and $\widetilde{\text{O}2}^+$-plane, and $\text{O}2(1,1)$
is the $\widetilde{\text{O}2}^-$-plane.
The anomaly of a D4-brane around the O2-plane is given by the fractional part of this charge up to a sign. 

We can decompose the anomaly into the contribution of the fermions and the $\U(1)$ Maxwell field on the D4-brane.
First we determine the anomaly of the chiral 2-form field on the M5-brane in the single orbifold background.
It is given by
\beq
\CA_\text{2-form}(S^7/ \BZ_2) = 28\eta(\CD_0)  + 12 (n-2) (\eta(\CD_1) - \eta(\CD_0)) - \frac{(n-2)(n-3)}{2} (\check C_1 , \check C_1).
\eeq
The value of $\eta(\CD_0) $ mod 1, rather than $30\eta(\CD_0) $ mod 1, can also be computed, see Appendix~\ref{sec:lens} or Appendix~\ref{sec:etacomp}.
It is given for $k=2$ by
\beq
\eta(\CD_0) = - \frac{1}{32}.
\eeq
Therefore we get
\beq
\CA_\text{2-form}(S^7/ \BZ_2) =  \frac{1}{8} - \frac{n(2-n)}{4}.
\eeq
The Maxwell theory on the D4-brane is obtained by the dimensional reduction of the chiral 2-form field on the M5-brane.
Therefore, its anomaly on the $\text{O}2(n_1,n_2)$-plane background $S^6/\BZ_2 = \mathbb{RP}^6$ is given by
\beq
\CA_\text{Maxwell}( \mathbb{RP}^6) = \frac{1}{4} - \frac{n_1(2-n_1)}{4}  - \frac{n_2(2-n_2)}{4}.
\eeq
This takes the following values:
\beq
\CA_\text{Maxwell}( \mathbb{RP}^6) =
\left\{ \begin{array}{cl} 
+1/4 & \text{for }\text{O2}^-, \\
0 & \text{for }\text{O2}^+ \text{ and }   \widetilde{\text{O}2}^+,  \\
-1/4 & \text{for } \widetilde{\text{O}2}^-.
\end{array}
\right.
\eeq
Let us make a few comments on this result.

The cancellation of the anomaly against the fractional part of the flux was shown in the case of the O2$^+$-plane in \cite{Tachikawa:2018njr}. There, only the contribution
of the fermions was taken into account. This is consistent, since the anomaly from the Maxwell theory happens to vanish for the O2$^+$ case, as can be seen from the above result.

In the cases of O2$^-$ and $\widetilde{\text{O}2}^-$, there is a nonzero contribution to the anomaly from the Maxwell field.
Notice that the value of the anomaly is $\pm 1/4$. 
Naively, the Maxwell theory in five dimensions is a non-chiral theory
which seems to be  describable using the framework of Sec.~\ref{sec:nonchiral}.
But the value $\pm1/4$ means that the framework of Sec.~\ref{sec:nonchiral} is not general enough.
This is because the mixed anomaly between the electric and the magnetic higher-form symmetries discussed in Sec.~\ref{sec:nonchiral}
can produce only values which are integer multiples of $1/k$ if the background is flat and if the relevant cohomology group is $\BZ_k$.

The reason is as follows. The mixed anomaly there is given by a differential cohomology pairing of the form $(\check B, \check C)$,
where $\check B$ and $\check C$ are the background fields for the electric and magnetic higher-form symmetries. 
When these backgrounds are flat and the relevant cohomology is $\BZ_k$, we have $k \check B = k \check C = 0$ up to gauge transformation. Then 
we get $k (\check B, \check C) = ( k\check B, \check C) =0$. 
More explicitly, the relevant cohomology in the case of the O2-plane background $\mathbb{RP}^6$ is
\beq
H^4(\mathbb{RP}^6,\BZ)  = H^3(\mathbb{RP}^6, \widetilde \BZ) = \BZ_2,
\eeq
where $\widetilde \BZ$ is the coefficient system twisted by the orientation bundle on $\mathbb{RP}^6$.
Therefore, the mixed anomaly of a non-chiral theory which is formulated by ordinary differential cohomology can only produce anomalies which are $1/2$ or zero.

The discussions above imply that the Maxwell theory on the D4-brane in the presence of the O2$^-$ and $\widetilde{\text{O}2}^-$ planes 
must have  subtler topological couplings than those discussed in Sec.~\ref{sec:nonchiral}.
We note that the expected anomaly of the Maxwell theory on the D$p$-brane in the presence of the O$(6-p)$-plane background
is given by $ 2^{p-4}$ up to sign.
This follows from the fact that the difference of the charges of O$(6-p)^+$ and O$(6-p)^-$ is given by $2^{p-4}$,
and the fractional part of the flux of O$(6-p)^+$ is cancelled by the fermion anomaly alone~\cite{Tachikawa:2018njr}.
This means that for O1 and O0, the situation becomes worse.

In the case of the D4-brane, we fortunately had the lift to the M5-brane as above, which allowed us to circumvent this question.
In principle, a dimensional reduction should allow us to find the action of the Maxwell theory on the D4-brane, but this is not immediate.
Type IIA string theory is basically an $S^1$ reduction of M-theory, but the detail is very subtle at the topological level. 
For example, Type IIA is formulated by K-theory, but M-theory is not.
Their equivalence is not at all obvious, and requires careful analyses. See \cite{Diaconescu:2000wy} for details.
Similarly, the chiral 2-form field on the M5-brane 
may be formulated by ordinary differential cohomology,
and the quadratic refinement of ordinary differential cohomology pairing can give values $\pm 1/4$ and $\pm 1/8$
even if the relevant cohomology group is $\BZ_2$.
This was the technical reason why we could reproduce the anomaly $\pm 1/4$ in our method.
However, for D5 and D6 branes in the backgrounds of O1 and O0 planes,
there is no such M-theory lift, and we need to produce $\pm1/8$ and $\pm1/16$.

Here it seems important to recall the fact that the NSNS and RR fields in Type II string theories are not ordinary differential cohomology elements. 
We need to use some twisted K-theoretic formulation of these fields, as was studied in~\cite{Minasian:1997mm,Witten:1998cd,Moore:1999gb,Bouwknegt:2000qt}. 
We also note that the K-theoretic RR-fluxes produced by O-planes were studied in \cite{Bergman:2001rp,GarciaCompean:2008kf}.
It would be very interesting to find the precise formulation of the Maxwell theory, and compute the correct anomaly in that framework.

\section{Electromagnetic duality of four dimensional Maxwell theory}\label{sec:EMdual}
In $d=4$ dimensions, the Maxwell theory has the electromagnetic duality.
Let $\check b$ be the Maxwell field whose action in Euclidean signature is given by.
\beq
-S = \frac{1}{2} \int \left(  -2\pi \i \tau\, \DF^-_b \wedge \DF^-_b -  2\pi \i  \bar{\tau}\,  \DF^+_b \wedge \DF^+_b\right),
\eeq
where $\DF^\pm_b = \frac{1}{2} (\DF_b \pm \hodge \DF_b)  $, and $\tau$ is given in terms of the electric coupling $g$ and the $\theta$ angle as
$
\tau = \frac{\theta}{2\pi} + \frac{2\pi \i}{g^2}
$. In the presence of the electric and magnetic currents $j_e$ and $j_m$, it satisfies the equations of motion
\beq
j_e  =  \d \left(  \tau\,   \DF^-_b   +   \bar{\tau}\, \DF^+_b \right) , \qquad
j_m = \d \DF_a.
\eeq
These equations are formally invariant if we introduce a dual field $\check  b'$ whose field strength is given by
\beq
\DF^-_{b'} = - \tau\,   \DF^-_b   ,\qquad \DF^+_{b'}  = -  \bar{\tau}\, \DF^+_b. \label{eq:MSD}
\eeq
In terms of the dual coupling $\tau' = -1/\tau$, we get 
\beq
j_e  =  -\d  \DF_{b'}  , \qquad
j_m =  \d    \left(   \tau'\,   \DF^-_{b'}   +   \bar{ \tau}' \, \DF^+_{b'} \right).
\eeq
This duality is also justified at the quantum level \cite{Witten:1995gf}.\footnote{%
The study of the electromagnetic duality and its anomaly has a long history.
The duality does not seem to be known to Maxwell himself, since he used the electric potential $\phi$ and the vector potential $\mathbf{A}$ in his original paper \cite{Maxwell} from 1865;
his notation was cumbersome to the extent that he used different alphabets for each component of $\mathbf{A}$.
It was Heaviside \cite{Heaviside1,Heaviside2} in 1885 who eliminated $\phi$ and $\mathbf{A}$ in favor of $\mathbf{D}$, $\mathbf{E}$, $\mathbf{H}$ and $\mathbf{B}$; 
it was also him who introduced both the vector calculus and the standard alphabetical symbols into electromagnetism.
The duality should have been evident to Heaviside in his notation; he even introduced magnetic currents in addition to electric currents.
We now note that when the quantization of electric and magnetic charges is ignored,
the duality group is $\U(1)_D$ under which $\mathbf{E}\pm \i \mathbf{B}$ has charge $\pm1$.
Equivalently, it assigns the charge $\pm1$ depending on the circular polarization of light,
and a positive/negative helicity photon has charge $\pm1$.
In other words, the total $\U(1)_D$ charge is the total helicity of the photon.
That this $\U(1)_D$ can be implemented at the Lagrangian level was noted in \cite{Deser:1976iy,Deser:1981fr} in the late 70s and the early 80s.
Then already in the late 80s, the mixed $\U(1)_D$-gravitational anomaly was derived perturbatively in \cite{Endo:1987sv,Reuter:1987ju,Vainshtein:1988ww,Dolgov:1988qx}.
This in particular means that there is an anomalous generation of the total helicity of light when the spacetime Pontryagin density $\propto \tr R\wedge R$ is nonzero, with a very specific coefficient.
This line of investigations was recently revisited in \cite{Agullo:2016lkj}.
The $\U(1)_D$ symmetry of the Maxwell equation is also being revisited in the field of atomic and molecular physics too, see e.g.~a paper from 2013 \cite{bliokh2013dual} 
where mostly classical aspects were discussed.
}

In terms of the description which uses either $\check b$ or $\check b'$, the electromagnetic duality is not manifest.
However, we can regard the Maxwell theory as a chiral (self-dual) theory as follows. We introduce both $\check b$ and $\check b'$,
and reduces the degrees of freedom by imposing the self-duality equations \eqref{eq:MSD}.
Then the electromagnetic duality, or more generally $\SL(2,\BZ)$ duality, can be made manifest. 

We cannot formulate the Maxwell theory with manifest $\SL(2,\BZ)$ duality group only within $d=4$ dimensions.
There is an anomaly of the $\SL(2,\BZ)$ group~\cite{Witten:1995gf,Seiberg:2018ntt,Hsieh:2019iba}.
We still expect that it can be formulated as the boundary mode of a $d+1=5$ theory,
which is exactly what we do in this section.
This allows us to determine the $\SL(2,\BZ)$ anomaly.
The results of this section have been reported in the letter \cite{Hsieh:2019iba}. 
We provide more details of that letter and justify the claims made there.

\subsection{From $d=6$ to $d=4$}
The most concrete way to realize the Maxwell theory with manifest $\SL(2,\BZ)$ is to start from the $d=6$
chiral $2$-form theory. In $d+1=7$, we consider a 3-form field $\check A \in \check H^4(Y_7)$.
Now let us assume that the 7-dimensional manifold $Y_7$ is a $T^2$ fiber bundle
\beq
T^2 \to Y_7 \to Y_5.
\eeq
We describe $T^2$ by using a coordinate
\beq
z = s^1 + \tau s^2 , \qquad s^1 \sim s^1+1, \qquad s^2 \sim s^2 + 1,
\eeq
where $\tau$ is the complex moduli of $T^2$. The metric on $T^2$ is taken to be 
\beq
\frac{1}{\Im \tau} |\d z|^2 = \CG_{ij}\d s^i \d s^j, \label{eq:T2metric}
\eeq
where
\beq
(\CG_{ij} ) =  (\Im \tau)^{-1} \begin{pmatrix} 1 & \Re \tau \\ \Re \tau  & |\tau|^2 \end{pmatrix}. 
\eeq
The overall scale of the metric is taken so that the volume of the $T^2$ is independent of $\tau$.
We also assume that the fiber has a section. This means that we can assume that $0 \in T^2$ is unchanged under transition functions of the fiber bundle.

Associated to the $T^2$ bundle, there is a principal $\SL(2,\BZ)$ bundle $P$ on the base $Y_5$ which acts on $T^2$ in the usual way.
More precisely, let $s = (s^1, s^2)^T \in T^2$. Then an element of the $T^2$ bundle is described by a pair $(p, s) \in P \times T^2 $
with the equivalence relation $(pg ,g^{-1}s) \sim (p,s)$ for $g \in \SL(2,\BZ)$. The $T^2$ bundle is thus $P \times_{\SL(2,\BZ)} T^2$.
In the same way, we can define a local coefficients system $\widetilde{\BZ}^2 = P \times_{\SL(2,\BZ)} \BZ^2$ which is twisted by the $\SL(2,\BZ)$ bundle.
For the treatment of bundles acted by $\SL(2,\BZ)$, it is convenient to use totally antisymmetric matrices $\epsilon^{ij}$ and $\epsilon_{ij}$ with $\epsilon^{12}=-\epsilon^{21}=+1$ and $\epsilon_{12}=-\epsilon_{21}=-1$.
We raise and lower indices by using them.

Under the assumption of the existence of a section $0 \in T^2$, there exists a differential cohomology element $\check s = (\check s^1, \check s^2)^T \in \check H^1(Y_7, \widetilde{\BZ}^2)$, where the coefficients system $ \widetilde{\BZ}^2$ is
pulled back from the base $Y_5$ to the total space $Y_7$.
The field strength is given by $\DF_s = \d s = (\d s^1, \d s^2)^T$, where $s=(s^1,s^2)^T$ are the coordinates of the fiber $T^2$ as discussed above.
Also, the values $\DA_s \in C^0(Y_7,\widetilde{\BR}^2)$ are defined as $\DA_s = s$  mod $ \widetilde{\BZ}^2$.

We perform dimensional reduction of the 3-form field $\check A \in \check H^4(Y_7)$ from $Y_7$ to $Y_5$ along the fibers, and neglect all Kaluza-Klein (KK) modes.
KK modes have momenta labelled by $\widetilde{\BZ}^2$,
and hence the fields with zero momentum along $T^2$ are well-defined. 
Among these zero modes, we get a 2-form field $\check a$ as well as 1-form and 3-form fields. We neglect the 1-form and 3-form fields
and only keep the 2-form field. It is described by a differential cohomology element $\check  a = (\check a^1, \check a^2)^T \in \check H^3(Y_5,\widetilde{\BZ}^2)$.
In other words, we take $\check A$ to be
\beq
\check A =  \check s^i \Dp \check a_j =\epsilon_{ij} \check s^i \Dp \check a^j:= \check s \Dp \check a,\label{eq:redAa}
\eeq
where $\check a_i =\epsilon_{ij} \check a^j$ and the summation over repeated indices is implicit. Notice that it is invariant under $\SL(2,\BZ)$.
The field strength is given by $\DF_A = \d s^i \wedge (\DF_{a})_i$.

In this section we are interested in $\check a$ and its boundary modes.
We can define its quadratic refinement $\CQ(\check a)$ on $Y_5$ simply as the quadratic refinement $\CQ(\check A)$ on $Y_7$, where $\check A$ is given by \eqref{eq:redAa}.
If $Y_5$ is a boundary of $Z_6$ on which the $\SL(2,\BZ)$ bundle is extended, the total space $Y_7$ can be extended to the $T^2$ fiber bundle $Z_8$
and we get
\beq
\CQ(\check a) &= \int_{Z_8} \left( \frac{1}{2} \DF_A \wedge \DF_A +  \Sw \wedge \DF_A  +28\hat{A}_2(R) \right) \nonumber \\
& =  \int_{Z_8} \left( \frac{1}{2} ( \d s^i \wedge (\DF_{a})_i) \wedge ( \d s^j \wedge (\DF_{a})_j)+  \Sw \wedge( \d s^j \wedge (\DF_{a})_j) +28\hat{A}_2(R)  \right) \nonumber \\
& = \int_{Z_6} \left( -\frac{1}{2}   \epsilon^{ij} (\DF_{a})_i \wedge  (\DF_{a})_j  \right) = \int_{Z_6} \left( \frac{1}{2}   \epsilon_{ij} (\DF_{a})^i \wedge  (\DF_{a})^i  \right),
\eeq
where we have used the fact that there is no invariant 3-forms and 6-forms constructed from the Riemann tensor and hence the terms involving $\Sw$ and $\hat{A}_2$ vanish after the dimensional reduction.
This in particular implies that there is no perturbative gravitational anomaly in $d=4$ dimensions.

We take the background field $\check C \in \check H^4(Y_7,\widetilde \BZ^2) $ as
$
\check C = \check s^i \Dp \check c_i
$.
The product $\check C \Dp \check A$ is computed as
\beq
\check C \Dp \check A = - (\check s^i \Dp \check s^j) \Dp ( \check c_i \Dp \check a_j).
\eeq
Since $\check c$ and $\check a$ have no dependence on the $T^2$ coordinates, 
its integral over fibers $T^2$ may be given by 
$ - \int_{T^2} \d s^i \wedge \d s^j  ( \check c_i \Dp \check a_j) = -\epsilon^{ij}  \check c_i \Dp \check a_j =  \epsilon_{ij}  \check c^i \Dp \check a^j := \check c \Dp \check a$.
Therefore, the differential cohomology pairing $(\check C, \check A)$ is given by
\beq
(\check C, \check A) =  \int_{Y_5} \DA_{c \Dp a} :=(\check c, \check a).
\eeq

We also define the kinetic term of $\check a$ as the integral of $\DF_A \wedge \hodge \DF_A $ on $Y_7$.
On $T^2$ with the metric \eqref{eq:T2metric}, the Hodge dual is given by
\beq
\hodge \d s^i =  \epsilon^{ij} \CG_{jk}  \d s^k, \label{eq:T2hodge}
\eeq
where $ \CG_{jk} $ is defined above.
By using this matrix, the kinetic term is
\beq
-\frac{2\pi }{2e^2}\int_{Y_7} \DF_A \wedge \hodge \DF_A 
= -\frac{2\pi }{2e^2}\int_{Y_5} \CG_{ij} (\DF_a)^i \wedge \hodge (\DF_a)^j.
\eeq
Combining them, the action for $\check a$ in the presence of the background field $\check c$ is given by
\beq
-S =  -\frac{2\pi }{2e^2}\int_{Y_5} \CG_{ij} (\DF_a)^i \wedge \hodge (\DF_a)^j + 2\pi \i \kappa \left( \widetilde \CQ(\check a) + (\check c, \check a) \right) . \label{eq:MBA}
\eeq
Now most of the discussions in Sec.~\ref{sec:theory} and Sec.~\ref{sec:anomaly} can be applied without much change.
Some finer points will be discussed it in the next subsection.

At this point, we remark that the partition function of this theory in the limit $e^2 \to \infty$ is completely the same as the full $d+1=7$ dimensional theory compactified on $T^2$
which includes all KK modes as well as 1-form and 3-form fields, if we restrict the background fields to the form $\check C = \check s^i \Dp \check c_i$.
The reason is as follows.  The 1-form and 3-form fields $\check d$ and $\check e$ with zero KK momentum appear as 
\beq
\check A \supset (\epsilon_{ij} \check s^i \Dp \check s^j )  \Dp  \check d  +  \check e.
\eeq
Therefore, they are invariant under $\SL(2,\BZ)$. 
There is no pure gravitational anomaly in $d=4$ dimensions. 
Thus $\check d$ and $\check e$ can only contribute to the anomaly of the type discussed in Sec.~\ref{sec:nonchiral}. 
By restricting the background fields to be of the form $\check C = \check s^i \Dp \check c_i$, there is no contribution to the anomaly at all.

The KK modes in the bulk contribute to the KK modes of the boundary, because their momenta along the fiber $T^2$ correspond to each other. 
The boundary KK modes are massive fields and they do not contribute to the anomaly. 
Thus, we expect that there is no contribution to the anomaly from the KK modes in the bulk.
This does not necessarily mean that the KK modes of the bulk field give no contribution to the partition function. 
In fact, there are examples that KK modes give nonzero contributions to the $\eta$-invariant, such as $S^1$ fiber bundles without a section. 
However, they should not contribute to the anomaly, so they must be given by an integral of a local density, $\int_{Y} I$, where $I$ is some gauge invariant polynomial of the curvature tensors.
Bulk partition functions of this type do not contribute to the anomaly of the boundary theory because $\int_{Y} I$ makes perfect sense on a manifold with boundary.
Now, in the case of $d+1=5$ dimensions, there is no candidate for such local density $I$. Thus the contribution of the KK modes in the dimensional reduction $Y_7 \to Y_5$ is completely zero.
Therefore, the partition function of the full theory on $Y_7$ and the reduced theory \eqref{eq:MBA} on $Y_5$ are completely the same.

Let us see the duality equation for the localized field $\check b$ which appears on the boundary of the theory described by $\check a$. 
We take the ansatz that the localized mode $\check b$ near the boundary is given (at the differential form level) by
\beq
\DA_a^{(\L)} =  \d \left( e^{m\tau}  \right)   \wedge \DA_b.
\eeq
We have seen in \eqref{eq:ansatz2} that $\check B = \check s \Dp \check b$ satisfies the self-duality equation
\beq
 \hodge_{X_6} \DF_B =  \i  \kappa \DF_B
\eeq
on a 6-dimensional manifold $X_6$ which is now a $T^2$ bundle over a base $X_4$.
The field strength is given by $\DF_B =  \d s^i \wedge (\DF_b)_j$. By using \eqref{eq:T2hodge}, 
we get
\beq
 \epsilon^{ij} \CG_{jk}  \hodge_{X_4}  (\DF_b)_i = \i \kappa (\DF_b)_k.
\eeq
Let us define $\DF_b^\pm = \frac{1}{2}(\DF_b \pm \hodge_{X_4}\DF_b)$.
Then we get 
$
\i \kappa  (\DF^\pm_b)^2 =  \pm (\Im \tau)^{-1} \left(  (\DF^\pm_b)^1 +   \Re(\tau) (\DF^\pm_b)^2 \right) 
$
which can be written as
\beq
( \DF^\pm_b)^1 +   \left( \Re(\tau) \mp \i \kappa  \Im(\tau) \right)   (\DF^\pm_b)^2 =0.
\eeq
These are the same form as the equations  \eqref{eq:MSD},
where $(\DF_b)^2=\DF_{b,\text{there}}$ and $(\DF_b)^1=\DF_{b',\text{there}}$.

\subsection{The anomaly of $\SL(2,\BZ)$ duality group} \label{sec:DRED}
Basically as in \eqref{eq:etaArf} of Sec.~\ref{sec:anomaly}, the phase of the partition function of the theory \eqref{eq:MBA} consists of the one-loop contribution described by the 
$\eta$-invariant and the sum over nontrivial topological sectors which gives the Arf invariant. 
Here we discuss some detail of the computation of the $\eta$-invariant in the present case of twisted coefficients.

The following computation can be done in any dimensions $d+1 = 2p+3$ as in Sec.~\ref{sec:Signature}.
For the computation of the $\eta$-invariant which gives the one-loop contribution, we can assume that the gauge fields are topologically trivial and can be treated by
a differential form. We denote them as $\DA_a^i$. In the case of the electromagnetic duality, the index runs over $i=1,2$ on which $\SL(2,\BZ)$ acts.
But it is not difficult to generalize it to arbitrary twisted coefficients systems $\widetilde \BZ^h, \ \widetilde \BR^h$ etc.~and $i=1,\ldots, h$. 
We assume that there is a non-degenerate invariant tensor $\epsilon_{ij}$ which is antisymmetric if $p$ is odd and symmetric if $p$ is even,
$\epsilon_{ji} = (-1)^{p} \epsilon_{ij}$. We also assume that there is a positive definite metric $\CG_{ij}$ on the bundle $\widetilde \BR^h$
which is compatible with $\epsilon_{ij}$ in the sense that $\CG^{ij} = \epsilon^{ik} \epsilon^{jl} \CG_{kl}$ is the inverse matrix of 
$\CG_{kl}$. One can check that these assumptions are satisfied in the case of the $\SL(2,\BZ)$ bundle with $\CG$ given by \eqref{eq:T2metric}.

These assumptions in particular imply the following. Let us define a matrix $J^i_{~j} = \CG^{ik}\epsilon_{kj}$.
Then we get $(J^2)^i_{~j} = (-1)^{p}$ and $J^i_{~k} J^j_{~l} \CG^{kl} = \CG^{kl} $. For odd $p$, $J$ defines the complex structure of $\widetilde \BR^h$ which is compatible with the metric.
The eigenvalues of $J$ are $\pm \i^{p}  $. We define $V_{\pm } \subset \widetilde \BC^h $ (where $ \widetilde \BC^h = \BC \otimes \widetilde \BR^h$)
as the eigenspace of $J$ with eigenvalue $\pm \i^{p(p+2)}$. Here the exponent $p(p+2)$ (rather than $p$) is taken just for later convenience.  
Thus we can split $\widetilde\BC^h = V_+ \oplus V_-$.

The part of the action which is relevant for the one-loop determinant is
\beq
-S =  \frac{2\pi \i \kappa}{2} \int_Y \epsilon_{ij} \DA_a^i \wedge \d \DA_a^j =   \frac{2\pi \i \kappa}{2} \int_Y \CG_{i j}   \DA_a^i \wedge \hodge (  J^j_{~k} \hodge \d \DA_a^k ).
\eeq
The one loop factor is thus given by the determinant of the operator $ -\i \kappa  J  \hodge \d$. 
This operator acts on $\d^\dagger \Omega^{m}(Y, \widetilde \BC^h) \subset \Omega^{m-1}(Y, \widetilde \BC^h)$, which is the space of $(m-1)$-forms
with twisted local coefficients $\widetilde \BC^h$ and which is orthogonal to the space of closed forms. 

As in \eqref{eq:middleD}, we define an operator $\widetilde{\CD}^{\rm sig}_{\pm}$ which is now coupled to the bundle $V_\pm$ as
\beq
\widetilde{\CD}^{\rm sig}_{\pm }  =  \i^{p(p+2)+2 }  \hodge \d =  \mp J \hodge \d.
\eeq
It acts on $\d^\dagger \Omega^{m}(Y,  V_\pm)  \subset \Omega^{m-1}(Y,  V_\pm) $ on which $J = \pm  \i^{p(p+2)}$.

By the same computation which leads to \eqref{eq:oneloopeta}, we get the one-loop factor
\beq
{\det}'( -\i \kappa  J  \hodge \d)^{-\frac 12} = \CN_1 \exp \left( -\frac{2\pi \i \kappa}{8} \cdot   2\left(\eta(\widetilde{\CD}^{\rm sig}_{+}) -  \eta(\widetilde{\CD}^{\rm sig}_{-}) \right) \right).
\eeq
The $\eta$-invariant appearing here can be computed as follows. 
Let $\CS$ be the spin bundle of $Y$. Then we consider the bundle $\CS \otimes ( (\CS^* \oplus \CS^*) \otimes  V_\pm)$.
Let $\CD^{\rm sig}_\pm$ be the Dirac operator acting on this bundle. 
By the results of Sec.~\ref{sec:Signature}, the $\eta$-invariants of the operators $\CD^{\rm sig}_Y$ and $\widetilde \CD^{\rm sig}_Y$
are related as 
\beq
\eta(\CD^{\rm sig}_\pm ) = 2 \eta( \widetilde \CD^{\rm sig}_\pm) + \frac{1}{2} \sum_{i=0}^{\dim Y} \dim H^i(Y, V_\pm) .\label{eq:sigeta}
\eeq
In other words, $\eta(\widetilde \CD^{\rm sig}_\pm)$ is the same as the $\eta$-invariant of the Dirac operator acting on $\CS \otimes ( \CS^*  \otimes  V_\pm ) $
excluding the contribution of the zero modes.

As in \eqref{eq:etaArf}, the anomaly is given by
\beq
\CA(Y) = \kappa\left( -   \widetilde{\CQ}(\check c) - \frac{1}{8} \cdot 2 \left( \eta(\widetilde{\CD}^{\rm sig}_+) - \eta(\widetilde{\CD}^{\rm sig}_-)\right)  +  \Arf(Y) \right).\label{eq:EManomaly1}
\eeq
In the case of $p=1$ and $d=5$, the bundles $V_+$ and $V_-$ are complex conjugates of each other, and 
the bundle $\CS \otimes \CS^*$ is real. Then the $\eta$-invariants are related by $\eta(\widetilde{\CD}^{\rm sig}_-) = - \eta(\widetilde{\CD}^{\rm sig}_+)$.

Let us return to the description of the $\SL(2,\BZ)$ case realized as the $T^2$ bundle $T^2 \to Y_7 \to Y_5$.
We need to be more specific about the structure group which is used to define the quadratic refinement $\widetilde{\CQ}(\check c)$. 
To define the quadratic refinement, we need a spin structure on the total space $Y_7$ of the $T^2$ fiber bundle.
$\SL(2,\BZ)$ may be regarded as a Lorentz symmetry on $T^2$, and its spin cover is denoted as $\Mp(2,\BZ)$.
Then the spin structure on $Y_7$ requires that the Lorentz symmetry $\SO(D)$ of the base space, where $D=5$ is the dimension of the base manifold $Y_5$, is extended to
\beq
\frac{\Spin(D) \times \Mp(2,\BZ)}{\BZ_2}.
\eeq
We call such a lift of the Lorentz group as the $\text{spin-$\Mp(2 ,\bZ)$}$ structure. The existence of such structure is a sufficient condition for the definition of the quadratic refinement.

There is no perturbative anomaly of the $\text{spin-$\Mp(2 ,\bZ)$}$ structure, so all the anomalies are global anomalies.
It is detected by the bordism group $\Omega^{\text{spin-$\Mp(2 ,\bZ)$}}_5$.
This will be determined in Sec.~\ref{sec:bordism}.

By using the description of the theory as a dimensional reduction from $Y_7$ to $Y_5$, we can get another representation of the anomaly as follows.
When the background $\check C$ is zero, we have seen in $d+1=7$ that the anomaly of the chiral 2-form theory is
equal to that of $-28 \kappa$ copies of chiral fermions, where the sign is determined by whether the fermions have positive or negative chirality, see \eqref{eq:28}. 
Therefore, the part $ - \frac{1}{8} \cdot 2 \left( \eta(\widetilde{\CD}^{\rm sig}_+) - \eta(\widetilde{\CD}^{\rm sig}_-) \right) +  \Arf(Y) $
should coincide with the anomaly of $-28 \kappa$ copies of fermions reduced on $T^2$. 

The reduction of fermions on $T^2$ is described as follows.
First, notice that the bundle $\widetilde \BR^2$ is exactly the tangent bundle of $T^2$.
By a straightforward computation, one can check that the complex structure $J^i_{~j} = \CG^{ik}\epsilon_{kj} = -\epsilon^{ik} \CG_{kj}$
acts on the complex coordinate $z = s^1 + \tau s^2$ as
\beq
J \begin{pmatrix} z \\ \bar{z} \end{pmatrix} = \begin{pmatrix} \i z \\ -\i \bar{z} \end{pmatrix}.
\eeq
Recall that $V_-$ was defined as $J V_- = -\i^{p(p+2)} V_- = +\i V_-$, so it is the complex tangent bundle $T_\BC (T^2)$ of $T^2$ spanned by the basis vector $\frac{\partial}{\partial z} $.
$V_+$ is its complex conjugate. Let $\Gamma^1$ and $\Gamma^2$ be gamma matrices on $T^2$. 
By noticing that $\Gamma^1 + \i \Gamma^2 = \Gamma^1(1 - \i^{-1} \Gamma^1\Gamma^2)$,
the spin bundle on $T^2$ with negative chirality $\i^{-1} \Gamma^1\Gamma^2 = -1$ is $V_-^{1/2}$ while the bundle with positive chirality $\i^{-1} \Gamma^1\Gamma^2 = +1$ is $V_+^{1/2}$.
Thus, the spin bundle $\CS_{Y_7}$ on $Y_7$ is reduced to
\beq
\CS_{Y_7} \to (\CS \otimes V_+^{1/2}) \oplus ( \CS' \otimes  V_-^{1/2}).
\eeq
Here, $\CS$ and $\CS'$ are the spin bundles on $Y_5$, or more precisely they are representations of Clifford algebra as follows.
On $\CS_{Y_7}$, the representation of the gamma matrices is taken as 
\beq
\i^{-3} \Gamma^1 \cdots \Gamma^7=1.
\eeq
We refer the reader to Appendix~\ref{app:NC} for the conventions for more general dimensions.
Then $\CS$ and $\CS'$ are representations of gamma matrices given by
\beq
\CS: \i^{-2} \Gamma^1 \cdots \Gamma^5=+1, \qquad \CS': \i^{-2} \Gamma^1 \cdots \Gamma^5=-1.
\eeq
The fermions which take values in $\CS$ in 5-manifold $Y_5$ give positive chirality fermions on the boundary $X_4 = \partial Y_5$, while fermions with $\CS'$ give negative chirality fermions. 
This claim can be checked by explicitly finding a localized chiral fermion as \eqref{eq:chiralfermion}.

We do not necessarily have bundles $\CS$ or $V_+^{1/2}$ separately. However, the bundle $(\CS \otimes V_+^{1/2}) $ must exist,
and it gives a $\text{spin-$\Mp(2 ,\bZ)$}$ structure of $Y_5$. 

The $\eta$-invariant of $(\CS' \otimes V_-^{1/2})$ is the same as that of $(\CS \otimes V_+^{1/2})$.\footnote{
For a general bundle $E$ in  five dimensions, the $\eta$-invariants of $\CS \otimes E$ and $\CS' \otimes E$ are negative of each other.
Also, the $\eta$-invariants of $\CS \otimes E$ and $\CS \otimes E^*$ are negative of each other. Hence the $\eta$-invariants of
$\CS \otimes E$ and $\CS' \otimes E^*$ are the same. }
Let $\CD_+$ be the Dirac operator of the fermions which take values in $\CS \otimes V_+^{1/2}$. 
A single fermion on $Y_7$ gives two fermions on $Y_5$ which take values in  $\CS \otimes V_+^{1/2} $ and $\CS' \otimes  V_-^{1/2}$ respectively, and hence
the contribution to the boundary anomaly is given by
$ -2 \eta(\CD_Y)$. The contribution to the anomaly of the theory \eqref{eq:MBA} is $-28\kappa$ times this value, so we get
\beq
\kappa\left(- \frac{1}{8} \cdot 2 \left( \eta(\widetilde{\CD}^{\rm sig}_+) - \eta(\widetilde{\CD}^{\rm sig}_-)\right)  +  \Arf(Y) \right) = 56\kappa \eta(\CD_+).\label{eq:AEc}
\eeq
and \eqref{eq:EManomaly1} becomes
\beq
\CA =  \kappa\left( -   \widetilde{\CQ}(\check c) + 56 \eta(\CD_+) \right). \label{eq:EManomaly2}
\eeq
The equations \eqref{eq:EManomaly1} and \eqref{eq:EManomaly2} are the main results of this subsection.
They were announced in \cite{Hsieh:2019iba}.

\subsection{D3-brane in S-fold backgrounds}\label{sec:Sfold}

Now we apply the formulas obtained in the previous subsection to D3-branes in the background of S-folds.
They are codimension-6 planes in Type IIB or F-theory whose special cases are O3-planes. 
The F-theory geometry is given by
\beq
\BR^4 \times (\BR^6 \times T^2)/\BZ_k,
\eeq
where $T^2$ is the F-theory elliptic fiber. 

We focus on the case that the S-fold preserves $\CN \geq 3$ supersymmetry~\cite{Garcia-Etxebarria:2015wns,Aharony:2016kai}.\footnote{%
Less supersymmetric cases are also discussed in \cite{Borsten:2018jjm,Apruzzi:2020pmv}.
}
The $\BZ_k$ action is given as follows. Let $\vec{z} = (z^1,z^2,z^3)$ be complex coordinates of $\BR^6 \cong \BC^3$. Also let $w $ be the coordinate of $T^2$.
We define the $\BZ_k$ action as
\beq
(\vec{z}, w) \mapsto e^{2\pi \i j/k} (\vec{z}, w) \qquad (j=0,1,\cdots,k-1).
\eeq
The complex moduli parameter $\tau$ is assumed to be invariant under the $\BZ_k$ action, which is possible for $k=2,3,4,6$.

Let $v $ be the coordinate of the bundle $V_+$. We have seen that it is the anti-homomorphic tangent bundle of $T^2$, and hence it transforms under the $\BZ_k$ action as
\beq
(\vec{z}, v) \mapsto e^{2\pi \i j/k} (e^{2\pi \i j/k} \vec{z}, \, e^{-2\pi \i j/k}  v) \qquad (j=0,1,\cdots,k-1).
\eeq
This action needs to be lifted to the spin group, which corresponds to specifying the $\text{spin-$\Mp(2 ,\bZ)$}$ structure.
We use the uplift specified as follows. On the supercharge $Q$ of Type IIB string, the action is uplifted to
\beq
Q \to  e^{-\pi \i j/k} R(j/k)Q,
\eeq
where
\beq
R(t) =\exp\left(- \pi t \left(  \Gamma^1\Gamma^2 + \Gamma^3\Gamma^4 + \Gamma^5\Gamma^6 \right) \right), \qquad (0 \leq t \leq 1). \label{eq:Srotation}
\eeq
Then one can check that it preserves $\CN \geq 3$ supersymmetry. 
Another possible choice for even $k$ is to take $Q \to (-1)^j  e^{-\pi \i j/k} R(j/k)Q$
which gives a different spin structure and breaks more supersymmetry when $k>2$ \cite{Borsten:2018jjm};  we do not consider this case here.
Equivalently, we have defined the spin lift so that the S-fold in Type IIB string can be lifted to M-theory orbifolds which we studied in Sec.~\ref{sec:orb}.

Let us restrict our attention to the manifold $Y_7=(S^5 \times T^2)/\BZ_k$ and $Y_5 = S^5/\BZ_k$.
We want to compute the anomaly evaluated on $Y_5$. 
The matter content of a D3-brane follows from the $T^2$ compactification of an M5-brane.
By the analysis of dimensional reduction in Sec.~\ref{sec:DRED}, and also from the discussion in M-theory given in Sec.~\ref{sec:M}, we conclude that the total anomaly is given by
\beq
\CA(Y_5) =  -   \widetilde{\CQ}(\check c) + (56+4) \eta(\CD_+).
\eeq
where $56$ are from the Maxwell field and $4$ are from fermions.
Let $\Sq$ be the S-fold charge. By the result of \cite{Tachikawa:2018njr}, the value of $\Sq$ mod 1 is related to the anomaly as
\beq
\Sq  = - \CA(S^5/\BZ_k) \mod 1. \label{eq:Sqrel}
\eeq
The charge $\Sq$ was computed in \cite{Aharony:2016kai} using string dualities and we can compare the anomaly with the charge.
The result there is summarized as follows:
\beq
\begin{array}{c|cccc}
k & 2 & 3 & 4 & 6  \\
\hline 
\Sq&  \pm \frac{1}{4} & \pm \frac{1}{3} & \pm  \frac{3}{8} & -\frac{5}{12}  
\end{array} \label{eq:Sfoldcharge}
\eeq
where the signs $\pm$ correspond to different types of S-folds. For example, we have O3$^\pm$ and $\widetilde{\text{O3}}^\pm$ for $k=2$.

The relevant $\eta$-invariant $\eta(\CD_+)$ is computed in Appendix~\ref{sec:etacomp},
where the  spin structure above corresponds to $\CD_{s=1/2}$ in the notation of the appendix.
Looking up the results there, the anomaly for the case $\check c=0$ is given as follows:
\beq
\begin{array}{c|cccc}
k & 2 & 3 & 4 & 6  \\
\hline 
60 \eta(\CD_+) \hspace{-2mm} \mod 1&  \frac{1}{4} & \frac{1}{3} &  -\frac{3}{8} & \frac{5}{12}  
\end{array} \label{eq:list60}
\eeq
These values are consistent with the relation \eqref{eq:Sqrel}.

We can perform additional consistency checks by comparing \eqref{eq:EManomaly1} and \eqref{eq:EManomaly2}.
These equations require \eqref{eq:AEc}, 
\beq
 \Arf(Y_5) =  \frac{1}{2} \eta(\widetilde{\CD}^{\rm sig}_+) + 56  \eta(\CD_+ ).\label{eq:Arfetarel}
\eeq
where we have used $ \eta(\widetilde{\CD}^{\rm sig}_-)   =  -\eta(\widetilde{\CD}^{\rm sig}_+)  $.
This equality was a consequence of comparing the partition function of the 5-dimensional theory \eqref{eq:MBA} and the dimensional reduction of the
anomaly of the $d=6$ theory. 
This  was argued by a rather indirect argument, so let us test it directly in the current setting.
The following discussions may also be regarded as a consistency check of charges in \eqref{eq:Sfoldcharge} which are not realized by \eqref{eq:list60}.

The right hand side of \eqref{eq:Arfetarel} is computed by using the result of Appendix~\ref{sec:etacomp}.
Notice that the cohomologies with twisted real coefficients $H^i(S^5/\BZ_k, \widetilde \BR^2)$ are all zero,\footnote{
Elements of $H^i(S^5/\BZ_k, \widetilde \BR^2)$ would be represented by harmonic differential forms annihilated by $\d + \d^\dagger$.
By pulling back them to $S^5$ under $S^5 \to S^5/\BZ_k$, we would get harmonic forms on $S^5$, which is possible only if $i=0,5$.
The cases $i=0,5$ are eliminated by the nontrivial twisting $\widetilde \BR^2$.
}
and hence \eqref{eq:sigeta} gives $  \eta( \widetilde \CD^{\rm sig}_+) = \frac{1}{2} \eta(\CD^{\rm sig}_+ )$.
The values of $ \eta(\CD^{\rm sig}_+ )$ are computed in Appendix~\ref{sec:etacomp},
where the transformation of $V_+$ corresponds to $\CD^{\rm sig}_{t=1}$  in the appendix. We get the results:
\beq
\begin{array}{c|cccc}
k & 2 & 3 & 4 & 6  \\
\hline 
 \frac{1}{2} \eta(\widetilde{\CD}^{\rm sig}_+) + 56  \eta(\CD_+ ) \hspace{-2mm} \mod 1&  \frac{1}{2} & -\frac{1}{4} &  \frac{1}{8} & 0
\end{array} \label{eq:ToBeArf}
\eeq

Next let us determine the Arf invariant. Let us first recall the definition of the Arf invariant.
We consider the cohomology with twisted integer coefficient $H^3(Y_5, \widetilde \BZ^2)$. 
If we require $\check a \in \check H^3(Y_5, \widetilde \BZ^2)$ to be flat, it is completely determined by $[\CN_a] \in H^3(Y_5, \widetilde \BZ^2)$ up to gauge transformations.
Therefore, we regard $\check a$ just as an element of $H^3(Y_5, \widetilde \BZ^2)$.
Then $\widetilde \CQ(\check a)$ is the quadratic refinement of the torsion pairing on $H^3(Y_5, \widetilde \BZ^2)$ which is reduced to have $\widetilde \CQ(0) =0$.
The Arf invariant is defined as
\beq
\Arf(Y_5) =  \frac{1}{2\pi} {\rm arg} \left( \sum_{\check a \in H^3(Y_5, \widetilde \BZ^2)  }  \exp( 2\pi \i  \widetilde \CQ(\check a) )\right).
\eeq
Thus it depends on the quadratic refinement. 

The relevant cohomology group $H^3(S^5/\BZ_k, \widetilde \BZ^2) $ 
was determined in \cite{Aharony:2016kai}.
We do not directly compute the quadratic refinement $\CQ(\check a)$.
Rather, we determine it from the relation \eqref{eq:Sqrel} and the known results \eqref{eq:Sfoldcharge}.
We study each case $k=2,3,4,6$ separately. 

\paragraph{The case $k=2$.} 
This case is the standard O3-planes which are discussed in Sec.~\ref{sec:O3}. The cohomology is 
$H^3(S^5/\BZ_2, \widetilde \BZ^2) = \BZ_2 \times \BZ_2$. We denote the corresponding elements of $H^3(S^5/\BZ_2, \widetilde \BZ^2) $ simply as $(n_1,n_2)$ where $n_1, n_2 = 0,1 \mod 2$.
The result \eqref{eq:list60} and the relation \eqref{eq:Sqrel} imply that the S-fold with the trivial background $\check c=0$ corresponds to the O3$^-$ plane.
Then other O-planes $\check c = (1,1), (1,0)$ and $(0,1)$ corresponds to $\widetilde {\text O3}^\pm$ and ${\text O3}^+$ planes.
They have charge $\Sq = +1/4$. By requiring \eqref{eq:Sqrel}, we conclude that $\CQ(\check c)$ is given as $\CQ(n_1, n_2) = 1/2$ for $(n_1, n_2) \neq (0,0)$.
Namely, the quadratic refinement is given by
\beq
\widetilde \CQ(n_1, n_2) = \frac{1}{2}(n_1 n_2 + n_1 + n_2) \mod \BZ.
\eeq
One can see that $\widetilde \CQ(\check c_1 + \check c_2) -  \widetilde\CQ(\check c_1) - \widetilde \CQ(\check c_2)$ gives a non-degenerate pairing on $\BZ_2 \times \BZ_2$.
We can compute 
\beq
 \sum_{\check a \in H^3(S^5/\BZ_2, \widetilde \BZ^2)  }  \exp( 2\pi \i  \widetilde \CQ(\check a) ) = (+1)+(-1) +(-1) +(-1) = 2 \exp(\pi \i ).
\eeq
Thus we get $\Arf = 1/2$ mod 1. This agrees with \eqref{eq:ToBeArf}.

\paragraph{The case $k=3$.} 
In this case we have $H^3(S^5/\BZ_3, \widetilde \BZ^2) = \BZ_3$. We denote its elements as $n = 0, \pm 1 \mod 3$.
The S-fold with $\Sq = - 1/3$ corresponds to the trivial background $\check c=0$.
The S-fold with $\Sq = +1/3$ is reproduced only if $\widetilde \CQ(\check c) = 2/3 $ mod 1.
We may also expect a symmetry $n \to -n$ which comes from the center $-1 \in \SL(2,\BZ)$.
Therefore, we conclude that $\widetilde \CQ(\pm 1) = 2/3$. The quadratic refinement is hence given by
\beq
\CQ(n) = \frac{2}{3}n^2 \mod 1 .
\eeq
One can see that $\widetilde \CQ(\check c_1 + \check c_2) -  \widetilde\CQ(\check c_1) - \widetilde \CQ(\check c_2)$ gives a non-degenerate pairing on $\BZ_3 $.
By using it, we compute
\beq
 \sum_{\check a \in H^3(S^5/\BZ_3, \widetilde \BZ^2)  }  \exp( 2\pi \i  \widetilde \CQ(\check a) ) = (+1)+ e^{4\pi \i /3} + e^{4\pi \i /3} = \sqrt{3} \exp( - \pi \i/2).
\eeq
Thus we get $\Arf = -1/4$ mod 1. This agrees with \eqref{eq:ToBeArf}.

\paragraph{The case $k=4$.}
In this case we have $H^3(S^5/\BZ_4, \widetilde \BZ^2) = \BZ_2$. We denote its elements as $n = 0,  1 \mod 2$.
The S-fold with $\Sq = +3/8$ corresponds to the trivial background $\check c=0$.
The S-fold with $\Sq = -3/8$ is reproduced only if $\widetilde \CQ(\check c) = -3/4 = 1/4 $ mod 1.
Therefore, we conclude that $\widetilde \CQ( 1) = 1/4$. The quadratic refinement is hence given by
\beq
\CQ(n) = \frac{1}{4}n^2 \mod 1 .
\eeq
One can see that $\widetilde \CQ(\check c_1 + \check c_2) -  \widetilde\CQ(\check c_1) - \widetilde \CQ(\check c_2)$ gives a non-degenerate pairing on $\BZ_2 $.
By using it, we compute
\beq
 \sum_{\check a \in H^3(S^5/\BZ_4, \widetilde \BZ^2)  }  \exp( 2\pi \i  \widetilde \CQ(\check a) ) = (+1)+ e^{\pi \i /2}  = \sqrt{2} \exp(  \pi \i/4).
\eeq
Thus we get $\Arf = 1/8$ mod 1. This agrees with \eqref{eq:ToBeArf}.

\paragraph{The case $k=6$.}
In this case we have $H^3(S^5/\BZ_6, \widetilde \BZ^2) = 0$. Thus the Arf invariant is trivial, $\Arf =0$ mod 1.
This agrees with \eqref{eq:ToBeArf}.

We have confirmed the relation \eqref{eq:Arfetarel} for all $k=2,3,4,6$, as promised in \cite{Hsieh:2019iba}.
It would be interesting to give an independent computation of the quadratic refinement without relying on the values of $\Sq$ given in \eqref{eq:Sfoldcharge}.

\subsection{Classification of global anomalies}\label{sec:bordism}
The $d=4$ dimensional theories with spin-$\Mp(2 ,\BZ)$ structure do not have perturbative anomalies, and all the anomalies are global anomalies. 
The classification of global anomalies is done by the (torsion part of the) bordism group $\Omega^\text{spin-$\Mp(2 ,\bZ)$}_5$. 
As a final computation in this paper, let us determine this group, and compute the anomaly of the Maxwell theory for all the generators.

First we note that $\Omega^\text{spin-$\Mp(2 ,\bZ)$}_5$ can be identified as the twisted spin bordism group 
\beq
\Omega^\text{spin}_5(B\SL(2, \BZ), \xi)
\eeq
which is described as follows. 
$\xi = \widetilde \BR^2$ is a real vector bundle  over $B\SL(2, \BZ)$ associated to the $\SL(2,\BZ)$ bundle on $B\SL(2, \BZ)$.
The $\SL(2,\BZ)$ action preserves the orientation, and hence
$\xi$ has a vanishing first Stiefel-Whitney class. 
Then we consider maps $f: Y \rightarrow B\SL(2, \BZ)$ such that the spin-$\Mp(2 ,\BZ)$ structure on $Y$ corresponds to a spin structure on $TY \oplus f^*(\xi)$. 
Such $Y$ is an element of $\Omega^\text{spin}_5(B\SL(2, \BZ), \xi)$.

Then we can apply the Atiyah-Hirzebruch spectral sequence (AHSS) for this twisted spin bordism group.
In particular, the $E^2$ page of $\Omega^\text{spin}_5(B\SL(2, \BZ), \xi)$ is the same as that of the untwisted group 
$\Omega^\text{spin}_5(B\SL(2, \BZ))$ and is given by\footnote{
Elements of $\Omega^\text{spin}_k(B\SL(2, \BZ), \xi) $ may be constructed by Pontryagin-Thom construction as follows.
The space $B\SL(2, \BZ)$ has the bundle $\xi$, and we consider the Thom space $T(\xi)$ associated to $\xi$ which is obtained by collapsing all points at infinity of $\xi$ to a single point.
Then, we consider a spin manifold $Y_{k+2}$ with dimension $k+2$, and a map $F: Y_{k+2} \to T(\xi)$.
By taking $F$ sufficiently generic, we assume that the image of $Y_{k+2}$ intersects transversally to the zero section of $\xi$ in $T(\xi)$.
We take $Y_k$ to be the inverse image of the zero section, which is a $k$-manifold. Its normal bundle inside $Y_{k+2}$
is isomorphic to the pullback $f^*(\xi)$, where $f$ is the restriction of $F$ to $Y_{k} \subset Y_{k+2}$. 
Since $Y_{k+2}$ is spin, the bundle $TY_{k} \oplus f^*(\xi)$ has a spin structure. This construction, and its inverse, implies
that the group $\Omega^\text{spin}_k(B\SL(2, \BZ), \xi) $ is equivalent to $\widetilde \Omega^\text{spin}_{k+2}(T(\xi)) $,
where $\widetilde \Omega$ is the reduced group which, roughly speaking, does not care what happens away from the zero section of $\xi$. 
The AHSS for generalized cohomology is applied to it with the $E^2$ page given 
by $\widetilde H_{p+2} (T(\xi), \Omega^\text{spin}_q({\rm pt})) = H_{p} ( B\SL(2, \BZ), \Omega^\text{spin}_q(\rm{pt})) $ where we have used the Thom isomorphism theorem.
}
\beq
E_{p,q}^2 = H_p(B\SL(2, \BZ), \Omega^\text{spin}_q(\rm{pt}) ) .
\eeq
The cohomology groups of $B\SL(2, \BZ)$ with integer coefficients are known to be given as follows: \begin{align}
H_{2m-1}(B\SL(2, \BZ),\BZ) &= \BZ_{12}, & m &\geq 1  \nonumber \\
 H_{2m}(B\SL(2, \BZ),\BZ) &= 0 . & m&\geq 1, \label{eq:BSLhom}
\end{align}
see \cite{Seiberg:2018ntt} for references.
We note that these are the same as those of the lens space $S^5/\BZ_{12}$ in the range of our interest.
The spin bordism group of a point is given by
\beq
\begin{array}{c|ccccccc}
q&0&1&2&3&4&5&6 \\
\hline
\Omega^\text{spin}_q ( \rm{pt}) & \BZ & \BZ_2 & \BZ_2 & 0 & \BZ & 0 & 0
\end{array}
\eeq
By the  universal coefficients theorem, we get
\beq
\begin{array}{c|ccccccc}
p&0&1&2&3&4&5 \\
\hline
E_{p,5-p}^2 & 0 &\BZ_{12} & 0 & \BZ_2 & \BZ_{2} & \BZ_{12} 
\end{array}
\eeq
It tells us that the order of the group is bounded as follows:
\beq
|\Omega^\text{spin-$\Mp(2 ,\bZ)$}_5|=|\Omega^\text{spin}_5(B\SL(2, \BZ), \xi)|\leq
\prod_p |E^2_{p,5-p}|= 576. \label{eq:567}
\eeq

The group $\Mp(2 ,\BZ)$ is described as follows. 
The group $\SL(2,\BZ)$ is generated by generators $S$ and $T$ which satisfy $S^2=(T^{-1}S)^3$ and $S^4=1$. 
The group $\Mp(2,\BZ)$ is the spin cover of $\SL(2,\BZ)$ and it is generated by $S$ and $T$ with the relations $S^2=(T^{-1}S)^3$ and $S^8=1$.
This group admits a homomorphism
\beq
\Mp(2,\BZ) \to \BZ_{24}
\eeq
given by the abelianization. 
This means that we map $S, T$ to $s,t$ with an additional commutativity relation $st=ts$.
Then $S^2=(T^{-1}S)^3$ and $S^8=1$ give $s=t^3$ and $t^{24}=1$. Therefore the abelianization gives $\BZ_{24}$.
We can also define two homomorphisms
\beq
\BZ_{8} \to \Mp(2,\BZ), \qquad  \BZ_{3} \to \Mp(2,\BZ).
\eeq
The first homomorphism maps the generator of $\BZ_{8}$ to $S$.
The second one maps the generator of $\BZ_{3}$ to $(T^{-1}S)^4$.

Notice that $\text{spin-$\BZ_{24}$} = ( \text{spin-$\BZ_{8}$}) \times \BZ_3$
because $\BZ_{24} \cong \BZ_8 \times \BZ_3$ and the subgroup $\BZ_2$ is included in $\BZ_8$.
From the above homomorphisms, we get homomorphisms of bordism groups
\beq
\Omega^\text{spin-$\BZ_{8}$}_5\oplus\Omega^\text{spin}_5(B\BZ_3) \to \Omega^\text{spin-$\Mp(2 ,\bZ)$}_5 \to \Omega^\text{spin-$\BZ_{24}$}_5  . \label{eq:38Mp24}
\eeq
Their composition,
\beq
\Omega^\text{spin-$\BZ_{8}$}_5\oplus \Omega^\text{spin}_5(B\BZ_3)  \to \Omega^\text{spin-$\BZ_{24}$}_5  ,\label{eq:3824}
\eeq
is based on the isomorphism $\BZ_8 \times \BZ_3 \xrightarrow{\sim} \BZ_{24}$. 

The map \eqref{eq:3824} is known to be isomorphism \cite{Hsieh:2018ifc}.
Therefore, \eqref{eq:38Mp24} implies that the homomorphism $\Omega^\text{spin-$\BZ_{8}$}_5\oplus\Omega^\text{spin}_5(B\BZ_3) \to \Omega^\text{spin-$\Mp(2 ,\bZ)$}_5$ must be injective.
This implies that the orders of the groups are $|\Omega^\text{spin-$\Mp(2 ,\bZ)$}_5| \geq | \Omega^\text{spin-$\BZ_{8}$}_5| \times |\Omega^\text{spin}_5(B\BZ_3) |$. 
We have~\cite{Hsieh:2018ifc},\footnote{As in the case of $\SL(2,\BZ)$,
AHSS shows that $|\Omega^\text{spin-$\BZ_{8}$}_5| \leq 64$ and $|\Omega^\text{spin}_5(B\BZ_3)| \leq 9$.
We will later present explicit generators which can be detected by $\eta$-invariants computed in Appendix~\ref{sec:etacomp}.
Those generators saturate the above bounds. 
}
\beq
\Omega^\text{spin-$\BZ_{8}$}_5\oplus\Omega^\text{spin}_5(B\BZ_3)=(\bZ_{32}\oplus \bZ_{2})\oplus\bZ_{9}
\eeq
 and in particular $ | \Omega^\text{spin-$\BZ_{8}$}_5| \times |\Omega^\text{spin}_5(B\BZ_3) |=576$. 
Combining these facts with \eqref{eq:567}, we conclude that the homomorphism 
$\Omega^\text{spin-$\BZ_{8}$}_5\oplus\Omega^\text{spin}_5(B\BZ_3) \to \Omega^\text{spin-$\Mp(2 ,\bZ)$}_5$ is in fact an isomorphism
and hence
\beq
\Omega^\text{spin-$\Mp(2 ,\bZ)$}_5 = \bZ_{32}\oplus \bZ_{2} \oplus \bZ_{9} .
\eeq

More explicitly, the generators of each factor $\bZ_{32} $, $ \bZ_{2} $ and $ \bZ_{9} $ are given as follows~\cite{GilkeyBook,Hsieh:2018ifc}.
The lens space $S^5/\BZ_k$ can be embedded in $\BC^3/\BZ_k$. We can specify the spin or spin-$\BZ_{2k}$ structure
on the lens space by specifying how the $\BZ_{k}$ acts on the spinor on $\BC^3$. Let us consider a structure defined by the $\BZ_k$ action
which is specified by a parameter $s \in 1/2 + \BZ$,
\beq
\Psi  \to  e^{-2\pi \i j s/k} R(j/k) \Psi \qquad (j=0,1,\cdots,k-1).
\eeq
where $R(t)$ was defined in \eqref{eq:Srotation}. In Sec.~\ref{sec:Sfold}, we have considered the case $s=1/2$ to preserve $\CN=3$ supersymmetry.
However, we need more general cases for the generators of the bordism groups.
Let $(S^5/\BZ_k)_s$ be the lens space with the structure specified by the parameter $s$.
Then the generators are as follows:
\begin{equation}
\begin{array}{cll}
\BZ_{32} &: (S^5/\BZ_4)_{s=1/2}& \in \Omega^\text{spin-$\BZ_{8}$}_5,  \\
\BZ_2 &: (S^5/\BZ_4)_{s=3/2} +  9(S^5/\BZ_4)_{s=1/2}& \in \Omega^\text{spin-$\BZ_{8}$}_5,   \\
\BZ_{9} &: (S^5/\BZ_3)_{s=1/2}& \in \Omega^\text{spin}_5(B\BZ_3).
\end{array}\label{eq:BordismG}
\end{equation}
Here we have used the fact that $\Omega^\text{spin-$\BZ_{8}$}_5\oplus\Omega^\text{spin}_5(B\BZ_3) \to \Omega^\text{spin-$\Mp(2 ,\bZ)$}_5$ is  
an isomorphism due to the above discussion.
The last one $ (S^5/\BZ_3)_{s=1/2}$ is an element of $\Omega^\text{spin}_5(B\BZ_3)$ 
because $(S^5/\BZ_k)_{s =  k/2}$ is a spin manifold for odd $k$, and $1/2 \equiv k/2$ mod 1. 
The manifold $(S^5/\BZ_3)_{s=1/2}$ can be detected by the Dirac operator with $\BZ_3$ charge $1$,
and the manifolds $ (S^5/\BZ_4)_{s=1/2}$ and $ (S^5/\BZ_4)_{s=3/2}$ are detected by the Dirac operators with $\BZ_8$ charge $1$ and $3$.\footnote{
For the $\text{spin-$\BZ_{8}$}$ structure to be well-defined, fermion charges under $\BZ_8$ must be odd. 
For a charge $q$ fermion, the value of $s$ is effectively changed to $s \to qs$. Let $\eta(\CD^{(q)})$ be the $\eta$-invariant of the Dirac operator of a fermion with $\BZ_8$ charge $q$.
By using the values of the $\eta$-invariants in Appendix~\ref{sec:etacomp}, one can check that $\eta(\CD^{(3)})$ and $\eta(\CD^{(1)}) + 9\eta(\CD^{(3)})$
generate the dual of the bordism groups, $\Hom(\BZ_{32} ,\U(1))$ and $\Hom(\BZ_2,\U(1))$, where $\BZ_{32}$ and $\BZ_2$ are the ones appearing in
$\Omega^\text{spin-$\BZ_{8}$}_5 = \BZ_{32} \oplus \BZ_2$. 
A generator of $\Hom(\BZ_{9} ,\U(1))$ (where $\BZ_9=\Omega^\text{spin}_5(B\BZ_3)$) is $\eta(\CD^{(1)})$ in a similar notation.
They are precisely the dual basis of \eqref{eq:BordismG}.}

We have already tested \eqref{eq:Arfetarel} for $(S^5/\BZ_4)_{s=1/2}$ and $(S^5/\BZ_3)_{s=1/2}$,
so let us test it for the remaining generator $(S^5/\BZ_4)_{s=3/2}$. This means that we take the relevant bundle $\CS \otimes \sqrt{V_+}$
to be the bundle specified by $(S^5/\BZ_4)_{s=3/2}$. For the signature we consider the bundle $\CS \otimes \CS^* \otimes V_+$.
By the result of Appendix~\ref{sec:etacomp}, we get
\beq
 \frac{1}{2} \eta(\widetilde{\CD}^{\rm sig}_+) + 56  \eta(\CD_+ ) = - \frac{1}{8} \mod \BZ.
\eeq
Thus the Arf invariant must be $\Arf = - 1/8$ for \eqref{eq:Arfetarel} to be valid.
This value is reproduced if the quadratic refinement of the torsion pairing in $H^3(S^5/\BZ_4, \widetilde \BZ^2) =\BZ_2$ for the current spin structure is given by
\beq
\CQ(n) = - \frac{n^2}{4} \qquad (n \in \BZ_2).
\eeq
It would be interesting to confirm it by a direct computation of $\CQ$.
However, we remark that the only possible values of the Arf invariant for $\BZ_2$ is $\pm 1/8$,
so it is already a nontrivial check that the above value of $ \frac{1}{2} \eta(\widetilde{\CD}^{\rm sig}_+) + 56  \eta(\CD_+ ) $
coincides with either of $\pm 1/8$.

\section*{Acknowledgements}
The authors thank P. Boyle Smith and J. Davighi for reporting an error in footnote 33 in a previous version of the paper.
CTH and YT are in part supported  by WPI Initiative, MEXT, Japan at IPMU, the University of Tokyo.
CTH is also supported in part by JSPS KAKENHI Grant-in-Aid (Early-Career Scientists), No.19K14608.
YT is also supported in part by JSPS KAKENHI Grant-in-Aid (Wakate-A), No.17H04837 
and JSPS KAKENHI Grant-in-Aid (Kiban-S), No.16H06335.
KY is supported by JSPS KAKENHI Grant-in-Aid (Wakate-B), No.17K14265.


\appendix

\section{Notations and conventions}\label{app:NC}
\begin{itemize}
\item $\i = \sqrt{-1}$ : the imaginary unit.
\item  $\d$: the exterior differential.
\item $X,Y,Z, M, N, ...$ : generic symbols for manifolds. (For general discussions, the dimensions are $\dim X =d$, $\dim Y = d+1$, $\dim Z = d+2$.) 
\item $\overline{X}$ : the orientation reversal of a manifold $X$.
\item $ H^p(X,\BA)$ : the cohomology on $X$ with coefficients $\BA (=\BZ, \BR, \BR/\BZ)$.  Coefficients with a tilde, such as $\widetilde{\BA}$, stand for twisted coefficient systems.
\item $\Omega^p(X)$ : differential forms of degree $p$, i.e.~a $p$-form.
\item $\Omega_{\rm closed}^p(X)$ : closed differential $p$-forms. 
\item Square bracket $[x ]_\BA$ : the cohomology element corresponding to a cocycle $x $ with coefficients $\BA$. (The subscript $\BA$ may be omitted if it is clear from context.)
\item $ \check H^p(X)$ : the differential cohomology group on $X$.
\item $A,B,C, a,b,c, ...$ : generic symbols for gauge fields as used usually by physicists.
\item $\check A, \check B,\check C, \check a, \check b, \check c ...$ : generic symbols for gauge fields as differential cohomology elements.
\item $\check A = (\DN_A, \DA_A, \DF_A)$ : the triplet representation of a differential cohomology element $\check A$.
\item $\CZ $ : the partition function. 
\item $\CA \in \BR/\BZ$ : the phase  $=\frac{1}{2\pi \i } \log \CZ$ of the partition function $\CZ$ of the bulk anomaly theory on closed manifolds. We simply call this $\CA$ as \emph{the anomaly}. 
\item$\CQ \in \BR/\BZ$ : a quadratic refinement, possibly with $\CQ(0)\neq 0$. 
We also use $\widetilde \CQ(\check A) = \CQ(\check A) - \CQ(0)$.
\item $\eta(\CD)$ : the APS $\eta$-invariant for a Dirac operator $\CD$.
\item $\index (\CD)$ : the index of a Dirac operator $\CD$.
\item $\text{spin-}G$: a tangential structure on a manifold $X$ such that the tangent frame bundle $\SO(\dim X)$ is uplifted to $[\Spin(\dim X) \times G]/\BZ_2 $, for a group $G$. A choice of an injection $\BZ_2 \to G$ is assumed.
\item $\Gamma^I$: gamma matrices. $\Gamma^{I_1 I_2 \cdots I_m} = \frac{1}{m!} \sum_{\sigma} \sign(\sigma)\Gamma^{I_{\sigma(1)}} \cdots \Gamma^{I_{\sigma(m)}} $, where the sum is over all permutations $\sigma$.
\item $\Ch$: the chirality operator or the $\BZ_2$ grading on spin bundles (or Clifford modules).
\end{itemize}
We also have some remarks related to index theorems:
\begin{itemize}
\item In even spacetime dimensions $2m$, the gamma matrices $\Gamma^1, \cdots, \Gamma^{2m}$ and the chirality operator $\Ch$ are related as $\Ch = \i^{-m} \Ch^1 \cdots \Ch^{2m}$.
In odd dimensions $2m+1$, the gamma matrices are usually taken as $\i^{-m} \Gamma^1 \cdots \Gamma^{2m+1} = +1$. 
When the bundle with opposite representation $\i^{-m} \Gamma^1 \cdots \Gamma^{2m+1} = -1$ appears,
we explicitly mention it.
\item $\partial Y$, the boundary of a manifold $Y$, is taken with the orientation given by the following convention. If the neighborhood of the boundary has the form $(-\epsilon, 0] \times X \subset Y$ ,
then the oriented volume forms $\omega_Y$ and $\omega_X$ of $Y$ and $X = \partial Y$ are related as $\omega_Y = \d \tau \wedge \omega_X$, where $\tau \in (-\epsilon, 0]$ and
$\tau=0$ is the boundary. This is the standard orientation for Stokes' theorem, but it is different from the usual convention for the APS index theorem in the literature. This leads to a sign change in the APS index theorem in front of the $\eta$-invariant. 
Namely, the APS index theorem on a manifold $Z$ with boundary $Y$ is of the form
\beq
\index (\CD_Z) = \int_Z \text{(local density)} + \eta(\CD_Y).
\eeq
\item In the conventions of the gamma matrices and the $\eta$-invariant used above,
the anomaly of a chiral fermion with positive chirality $\Ch=+1$ is given by $\CA = - \eta$ for a Dirac fermion (i.e.~the bulk partition function is $\CZ = \exp(-2\pi \i \eta)$), 
and $\CA = - \frac{1}{2}\eta$ for a Majorana fermion (i.e.~the bulk partition function is $\CZ = \exp(-\pi \i \eta)$).
For negative chirality fermions, the sign is reversed. 
\end{itemize}

\section{Some sign factors in M-theory}\label{sec:Mthconv}
The purpose of this appendix is to fix various sign factors which appear in M-theory. 
In particular, let $F^{\rm M}_4$ and $F^{\rm M}_7$ be M-theory 4-form and 7-form field strength. 
We will determine the sign $s$ in the duality equation
\beq
\hodge F^{\rm M}_4 = s \i F^{\rm M}_7.
\eeq
We want to determine whether it is $s=1$ or $-1$.
After fixing some conventions of the M$p$-branes and the fields $F^{\rm M}_{p+2}$,
the value of $s$ is not a convention, but is fixed.

In this appendix, we are only concerned with sign factors, and hence we neglect topology of $p+1$-form fields $C^{\rm M}_{p+1}$
and treat them as differential forms. In particular, $F^{\rm M}_{p+2} = \d C^{\rm M}_{p+1}$.

\subsection{Convention of M$p$-branes and $(p+1)$-form fields }
We always use Euclidean signature for the metric unless otherwise stated.
For gamma matrices, we take $\Gamma^0$ to be imaginary antisymmetric and other $\Gamma^I~(I=1,\cdots,10)$ to be real symmetric.
(In Lorentzian signature, all gamma matrices are real.) Then we see that the matrix
\beq
C=\i \Gamma^0
\eeq
has the properties that
\beq
C \Gamma^I C^{-1} = -(\Gamma^I)^T = - (\Gamma^I)^*~~~(I=0,1,\cdots, 10).
\eeq
Moreover, we take them to satisfy
\beq
\i^{-5} \Gamma^0 \Gamma^1 \cdots \Gamma^{10}=1. \label{eq:gammaprod}
\eeq
We use these conventions for the gamma matrices.

For M5 and M2 branes, we use the following conventions. Let $Q^{\rm M}$ be the supercharge in 11-dimensions. 
An M$p$-brane preserves the subgroups $\SO(p+1) \times \SO(10-p) \subset \SO(11)$ of the Lorentz symmetry and half of the supersymmetry.
Then, if we put it on $x^{p+1}= \cdots x^{10}=0$ with the orientation determined by the volume form $\omega_{p+1}=\d x^ 0 \wedge \cdots \wedge \d x^p$, 
it is clear that the supercharges preserved by the M$p$-brane should be given by
\beq
Q^{ {\rm M}p} = (1 \pm \Gamma^{10} \Gamma^{9} \cdots \Gamma^{p+1} ) Q^{\rm M}.
\eeq
Here the ambiguity is only sign factors $\pm$ and not a general complex phase, because $Q^{\rm M}$ in Lorentz signature metric is real.
The sign just specifies which we call as M$p$-branes and  which as anti-M$p$-branes.
We use the convention that M$p$-branes (as opposed to anti-M$p$-branes) with the worldvolume orientation 
\beq
\omega_{p+1}=\d x^ 0 \wedge \cdots \wedge \d x^p \text{ : positive volume form} \label{eq:orient}
\eeq
are specified by the unbroken supercharges
\beq
Q^{ {\rm M}p} = (1 + \Gamma^{10} \Gamma^{9} \cdots \Gamma^{p+1} ) Q^{\rm M},
\eeq
or more explicitly
\beq
{\rm M}5 &: Q^{\rm M5}=(1 + \Gamma^{10}\Gamma^{9}\Gamma^{8}\Gamma^{7}\Gamma^{6})Q^{\rm M},    \label{eq:M5Q}\\
{\rm M}2 &: Q^{\rm M2}= (1 + \Gamma^{10} \Gamma^{9} \Gamma^{8}\Gamma^{7}\Gamma^{6} \Gamma^{5} \Gamma^{4} \Gamma^{3}) Q^{\rm M}. \label{eq:M2Q}
\eeq
These are just conventions. 

Let us slightly rephrase the above conditions.
The supersymmetry transformations are written as
\beq
\epsilon^T C Q^{\rm M}
\eeq
where $\epsilon$ is the supersymmetry parameter, and $C (=\i \Gamma_0)$ is the matrix defined above.
The supersymmetry parameter for M$p$-branes must satisfy,
\beq
\frac{1}{2}\epsilon^T_p C (1+ \Gamma^{10} \cdots \Gamma^{p+1}) Q^{\rm M}= \epsilon^T_p C Q^{\rm M}
\eeq
or 
\beq
\Gamma^{p+1} \cdots \Gamma^{10} \epsilon_p = (-1)^{p}\epsilon_p. \label{eq:unbrokensusy}
\eeq
This is the supersymmetry parameter relevant for M$p$-branes.

The sign convention of $(p+1)$-form fields $C_{p+1}^{\rm M}$ coupled to M$p$-branes is determined by the following requirement.
Let us consider the above M$p$-brane with the orientation of the worldvolume $\omega_{p+1}$ given by \eqref{eq:orient}. Let us also define
\beq
\delta_{10-p} (\vec{z}) := \delta(x^{p+1} ) \cdots \delta(x^{10} ) \d x^{p+1} \wedge \cdots \wedge \d x^{10},
\eeq
where
\beq
\vec{z}= (x^{p+1}, \cdots, x^{10}).
\eeq
Then the coupling of $C_{p+1}^{\rm M}$ to the M$p$-brane is given by
\beq
-S \supset 2\pi \i \int  C_{p+1}^{\rm M} = 2\pi \i \int C_{p+1}^{\rm M} \wedge \delta_{10-p} (\vec{z}). \label{eq:MCcoupling}
\eeq
The sign of $C_{p+1}^{\rm M}$ is defined by this coupling.

Including the kinetic term, the action contains
\beq
-S \supset - \frac{2\pi}{2}  \int \d C_{p+1}^{\rm M}  \wedge \hodge \d C_{p+1}^{\rm M}  +   2\pi \i \int C_{p+1}^{\rm M} \wedge \delta_{10-p} (\vec{z}),
\eeq
where we are using the Planck unit $2\pi \ell_{\rm M}=1$.
The equation of motion is
\beq
(-1)^{p+1} \d  \hodge F_{p+2}^{\rm M} + \i \delta_{10-p} (\vec{z})=0.
\eeq
Now suppose that we have the duality equation of field strength
\beq
 \hodge F_{p+2}^{\rm M}  = \i s_{p+2}  F_{9-p}^{\rm M},
\eeq
where $s_{p+2} = \pm 1$ are sign factors which we want to determine. 
Notice that we have already defined the sign convention for the fields $C_{p+1}$ and hence there is no freedom to modify this self-dual condition.

Then we get
\beq
\d F_{9-p}^{\rm M} = (-1)^p s_{p+2}  \delta_{10-p} (\vec{z}).
\eeq
Because $\hodge^2=1$ in odd dimensional Riemann manifold, we have $(\i s_{p+2}) ( \i s_{9-p})=1$ or $ s_{p+2}  s_{9-p} = -1$.
Let us set $s := s_4$. Then $s_7=-s$ and 
\beq
\int_{S^7} F_{7}^{\rm M} = s,  \qquad
\int_{S^4} F_{4}^{\rm M} = s , \label{eq:fluxess}
\eeq
where $S^{9-p}$ is the sphere surrounding the M$p$-brane.
We will see that the value of $s$ is given by $s=+1$.

Before computing $s$,
let us explain more about the structure of various signs and why they are important for the anomaly of M5-branes.
The signs of $C_{p+1}$ are defined by \eqref{eq:MCcoupling}, that is, $C_{p+1}$ and $ - C_{p+1}$ are distinguished by 
the coupling to M$p$-branes. The distinction between M$p$-branes and anti-M$p$-branes are defined by \eqref{eq:M5Q} and \eqref{eq:M2Q}.
They affect the computation of the anomaly in the following way. 

First, the supercharge \eqref{eq:M5Q} determines the chirality of the worldvolume fields of the M5-brane.
The chirality operator $\Ch^{\M5}$ on the M5-brane with the orientation $\omega_6= \d x^0 \wedge \d x^1 \wedge \d x^2 \wedge \d x^3 \wedge \d x^4 \wedge \d x^5$ is given as
\beq
\Ch^{\M5} = \i^{-3} \Gamma^0 \Gamma^1 \Gamma^2 \Gamma^3 \Gamma^4 \Gamma^5.
\eeq
By using \eqref{eq:gammaprod}, it can be written also as $\Ch^{\M5} = - \Gamma^6 \Gamma^7 \Gamma^8 \Gamma^9 \Gamma^{10}$.
Under this chirality operator, $Q^{\rm M5}$ has a definite chirality as
\beq
\Ch^{\M5}  Q^{\rm M5} = - Q^{\rm M5} .
\eeq
From this, the worldvolumes fermions $\chi \sim [Q^{\rm M5} , \phi]$ (where $\phi$ represent worldvolume scalars) has negative chirality $\Ch^{\M5} \chi = -\chi$ and so on.
The chirality of the worldvolume fields affects the sign of the anomaly.

Next let us explain \eqref{eq:M2Q}. 
We are interested in the M2-charge of the M-theory orbifold 
\beq
\BR^3 \times (\BR^8/\BZ_k).
\eeq
Here the orbifold action on the coordinate
\beq
\vec z = (z^1, z^2, z^3, z^4) = (x^3 + \i x^4, x^5 + \i x^6, x^7 + \i x^8, x^9 + \i x^{10}) \label{eq:orb2}
\eeq
is given by
\beq
\vec{z} \to e^{2\pi \i j /k} \vec{z} , \qquad (j =0,1,\cdots, k-1). \label{eq:orb1}
\eeq
The question is how to define the uplift of this action to spinors. 
We define the action in such a way that the supercharges preserved by this orbifold action is a subset of the supercharges \eqref{eq:M2Q}.
In other words, adding M2-branes to the orbifold does not break supersymmetry, while adding anti-M2-branes breaks it. 
The uplift of \eqref{eq:orb1} on spinors $\Psi$ is either $\Psi \to + \SR(j/k) \Psi$
or $\Psi  \to -\SR(j/k) \Psi$, where
\beq
\SR(t) = \exp\left(- \pi t \left(\Gamma^3\Gamma^4 + \Gamma^5\Gamma^6 + \Gamma^7\Gamma^8 + \Gamma^9\Gamma^{10} \right) \right).
\eeq
In fact, one can check
\beq
\SR(t)^{-1} \left( \Gamma^{1+2q} + \i \Gamma^{2+2q}\right) \SR(t) = e^{ 2\pi \i t} \left( \Gamma^{1+2q} + \i \Gamma^{2+2q}\right) \qquad (q=1,2,3,4).
\eeq
which corresponds to \eqref{eq:orb2}.
The sign ambiguity in $\Psi \to \pm \SR(j/m) \Psi$ is the standard one in going from $\SO$ to $\Spin$, and it determines the spin structure of $\BR^8/\BZ_k$.
By requiring that $\SR(j/k)$ preserves some of the charge $Q^{\M2}$, we conclude that the sign must be such that
\beq
\Psi \to +\SR(j/k)\Psi. \label{eq:orb3}
\eeq
For example, $\SR(1/2) = \Gamma^3\Gamma^4  \Gamma^5\Gamma^6  \Gamma^7\Gamma^8  \Gamma^9\Gamma^{10} $ and $\SR(1/2)Q^{\M2} = Q^{\M2}$,
and hence $+\SR(1/2)$ preserves the same supercharges as M2-branes, while $-\SR(1/2)$ preserves the same supercharges as anti-M2-branes.
The choice \eqref{eq:orb3} determines the spin structure of $\BR^8/\BZ_k$, and the spin structure affects the value of the $\eta$-invariant on $S^7/\BZ_k$.
In this way the choice \eqref{eq:M2Q} affects the anomaly of the orbifold.

\subsection{Supergravity background and supersymmetry}
M$p$-branes are realized as extremal black $p$-brane solutions in supergravity.
As we will see, the remaining supersymmetries (i.e.~Killing spinors in the extremal black brane solutions) depend on the sign of the flux $F_{9-p}$. Our strategy is to relate the remaining supersymmetries \eqref{eq:unbrokensusy}
and the fluxes \eqref{eq:fluxess} and determine the sign factor $s$.

The Killing spinor equation has the following schematic form
\beq
(D_I  + c_1 \Gamma_I \slashed{F}^{\rm M}_4 + c_2 \slashed{F}^{\rm M}_4 \Gamma_I)\epsilon = 0,
\eeq
where $\epsilon$ is the Killing spinor, $D_I$ is the covariant derivative, and 
\beq
 \slashed{F}^{\rm M}_4 =\frac{1}{4!}(F^{\rm M}_4)_{JKLM} \Gamma^{JKLM}, \label{eq:Fslash}
\eeq
where $ \Gamma^{JKLM} =\Gamma^{[J}\Gamma^K \Gamma^L \Gamma^{M]}$ is the product of gamma matrices with the indices $J,K,L,M$ antisymmetrized. 
The above form of the Killing spinor equation may be inferred just by simple considerations of Lorentz structure and a counting of mass dimensions 
if we recover the Planck scale $2\pi \ell_{\rm M}$.
It is much more nontrivial to determine the coefficients $c_{1,2}$. 
According to the equation (13.13) of \cite{ImamuraSUGRA} , they are given by $|c_1| = 1/24$, $|c_2| = 1/8$ and $c_1c_2<0$.
The overall sign of $c_1,c_2$ depends on the convention of $F^{\M}_4$.\footnote{
By comparing (13.12) of \cite{ImamuraSUGRA} and \eqref{eq:SUGRA} of this paper, we see that our $C_3$ is the negative of $A_3$ of \cite{ImamuraSUGRA}.
Then the values of $c_1$ and $c_2$ in this paper are $c_1=-1/24$ and $c_2=1/8$. In particular, $c_2$ is positive.
This will be consistent with what we will find later. }

Let us consider the Killing equation when the Lorentz index $I$ is in the direction parallel to the M$p$-brane, which we denote by the Greek letter $\mu$.
Then by translational invariance, we have $\partial_\mu \epsilon = 0$. However, the covariant derivative $D_\mu $ is still nonzero.
The metric of the extremal black $p$-brane solution is of the form
\beq
\d s^2 = E(r)^2 ( \d x_0^2 + \cdots + \d x_p^2) + F(r)^2(\d x_{p+1}^2+\cdots + \d x_{10}^2),
\eeq
where $r=|\vec{z}|$. In this metric, we can take the orthonormal frame $e_I^a$ as $e_\mu^a = E(r) \delta^a_\mu$ as long as $\mu$ is in the tangent direction.
Then the spin connection $\omega_{\mu IJ}$ is
\beq
\frac{1}{4} \Gamma^{IJ} \omega_{\hat \mu IJ} = \frac{1}{2} \Gamma_{\hat \mu} \Gamma^{\hat r}   F(r)^{-1}  \partial_r \log E(r),
\eeq
where $\Gamma^{\hat \mu} $ and $ \Gamma^{\hat r} $ are gamma matrices in the directions $x^\mu$ and $r$ normalized in such a way that $(\Gamma^{\hat \mu} )^2 = (\Gamma^{\hat r} )^2=1$,
and $ \omega_{\hat \mu IJ} = E(r)^{-1}  \omega_{ \mu IJ}$.
Therefore, the Killing spinor equation is simplified to
\beq
\left( \frac{1}{2} \Gamma_{\hat \mu} \Gamma^{\hat r}   F(r)^{-1}  \partial_r \log E(r) + c_1  \Gamma_{\hat \mu} \slashed{F}^{\rm M}_4 + c_2 \slashed{F}^{\rm M}_4  \Gamma_{\hat \mu}   \right)\epsilon =0.
\eeq
In the M5 case ($p=5$), the ${F}^{\rm M}_4$ does not contain $\mu$ components and hence $\slashed{F}^{\rm M}_4 \Gamma^\mu = + \Gamma^\mu \slashed{F}^{\rm M}_4$.
On the other hand, in the M2 case ($p=2$), the term $\hodge {F}^{\rm M}_4  \propto {F}^{\rm M}_7$ does not contain $\mu$ and $r$,
and hence schematically $ {F}^{\rm M}_4 \sim \d x^{\mu_1} \wedge \d x^{\mu_2} \wedge \d x^{\mu_3} \wedge \d r$. Thus we get 
$\slashed{F}^{\rm M}_4 \Gamma^\mu = - \Gamma^\mu \slashed{F}^{\rm M}_4$.
Therefore, we get
\beq
\epsilon  = -Z(r)^{-1} (c_1 \pm c_2) \Gamma^{\hat r}  \slashed{F}^{\rm M}_4 \epsilon, \label{eq:kill1}
\eeq
where $Z(r) =  \frac{1}{2}  F(r)^{-1}  \partial_r \log E(r) $,
and the sign in $\pm c_2$ depends on the sign in $\slashed{F}^{\rm M}_4 \Gamma^\mu = \pm \Gamma^\mu \slashed{F}^{\rm M}_4$.

It is possible to rewrite $\slashed{F}^{\rm M}_4$ in terms of $\slashed{F}^{\rm M}_7$ which is defined in the similar way as in \eqref{eq:Fslash}.
Let us first notice that
\beq
\frac{1}{4!}\epsilon^{I_1 \cdots I_{7} J_1 \cdots J_4}\Gamma_{J_1 \cdots J_4} = -\i \Gamma^{I_1 \cdots I_7},
\eeq
which follows from \eqref{eq:gammaprod}. Also, we have $F_4^{\rm M} = \i s \hodge F_7^{\rm M}$ (because $\hodge F_4^{\rm M} = \i s F_7^{\rm M}$ and $\hodge^2=1$)
which is explicitly written as
\beq
(F_4^{\rm M} )_{J_1 \cdots J_4} = (\i s) \cdot \frac{1}{7!}(F_7^{\rm M})^{I_1 \cdots I_7} \epsilon_{I_1 \cdots I_{7} J_1 \cdots J_4} .
\eeq
Therefore,
\beq
\slashed{ F }^{\rm M}_4 &= \frac{1}{4!} (F^{\rm M}_4 )_{J_1 \cdots J_4} \Gamma^{J_1 \cdots J_4} 
= \frac{1}{4! 7!} (\i s)(F^{\rm M}_7 )^{I_1 \cdots I_7}\epsilon_{I_1 \cdots I_{7} J_1 \cdots J_4} \Gamma^{J_1 \cdots J_4}  = \frac{1}{ 7!}  s (F^{\rm M}_7 )^{I_1 \cdots I_7} \Gamma_{I_1 \cdots I_7} \nonumber \\
& = s \slashed{F}^{\rm M}_7.
\eeq

For the M$p$-brane solution, the flux $F_{9-p}$ is such that
\beq
\int_{S^{9-p}}{F}^{\rm M}_{9-p}=s
\eeq
as we have seen in the previous subsection.
Then we have
\beq
\Gamma^{\hat r} \slashed{F}^{\rm M}_{9-p} = s | {F}^{\rm M}_{9-p}| \Gamma^{p+1} \cdots \Gamma^{10},
\eeq
where the factor $ | {F}^{\rm M}_{9-p}| \sim 1/r^{9-p}$ is a positive function of $r$. 
Therefore \eqref{eq:kill1} is written for the M$p$-brane solution as
\beq
\epsilon = \left( K_p  Z(r)^{-1} | {F}^{\rm M}_{9-p}| \right) \Gamma^{p+1} \cdots \Gamma^{10} \epsilon, \label{eq:kill2} 
\eeq
where $K_p$ is defined by
\beq
K_p  = \left\{ \begin{array}{ll}
- s (c_1+c_2),  & (p=5), \\
-(c_1-c_2), & (p=2).
\end{array}
\right.
\label{eq:kill3}
\eeq

Now let us notice that the sign of $Z(r) =  \frac{1}{2}  F(r)^{-1}  \partial_r \log E(r) $ is independent of whether we consider branes or anti-branes, or whether we consider
M2 or M5, because gravity is always an attractive force. (Its absolute value depends on whether we consider M2 or M5.)
More explicitly, $\log E(r) \sim - 1 /r^{8-p}$ and hence $Z(r)>0$.
Therefore, the sign of the right-hand-side of \eqref{eq:kill2} is determined simply by the sign of $ K_p$.
By requiring that \eqref{eq:kill2} coincides with \eqref{eq:unbrokensusy}, we get $Z(r) = |K_p |  | {F}^{\rm M}_{9-p}|$ and 
\beq
\sign(K_p) = (-1)^{p}. \label{eq:kill4}
\eeq
The actual values of $c_1$ and $c_2$ are such that $|c_2| > |c_1|$ and hence $\sign(c_2 \pm c_1) = \sign(c_2)$.
Then \eqref{eq:kill3} and \eqref{eq:kill4} give $c_2>0$ and
\beq
s = +1,
\eeq
which is what we wanted to show.

\section{Cohomology pairing and $\eta$-invariant modulo 1 on lens spaces}\label{sec:lens}

In this appendix we compute the differential cohomology pairing on $S^7/\BZ_k$ and the $\eta$-invariants mod 1 which are required in Sec.~\ref{sec:orb},
and provide the values already quoted in \eqref{eq:summaryvalue}.
The strategy for the computation is to find a manifold $Z$ whose boundary is $Y = S^7/\BZ_k$.
For our purposes, $Z$ does not have to be  spin; a $\spin^c$ structure suffices.
We also extend $\CL_1$ and $\check C_1$ to $Z$, and use this $Z$ to compute the 
quantities appearing in \eqref{eq:summaryvalue}.
Most of the discussions here can be generalized for $S^{2m-1}/\BZ_k$ without much difficulty, so we take $m$ to be general.
Then we obtain a formula for the $\eta$-invariant mod 1 for lens spaces of general dimensions when the Dirac operator is coupled to general flat line bundles.
We note that we basically follow the discussion in \cite{Sethi:1998zk}.
We also emphasize that the computation in this section only gives the $\eta$-invariants mod 1,
and not the $\eta$-invariants themselves. This is enough for the purposes of Sec.~\ref{sec:orb}, but may not be enough for some other purposes,
such as using the $\eta$-invariant of the signature operator $\CD^{\rm sig}$ which is multiplied by $1/8$.
A different computation of the $\eta$-invariants of lens spaces which gives their values as real numbers will be presented in Appendix~\ref{sec:etacomp}.

\subsection{The geometry and the differential cohomology pairing}
First, we consider $\mathbb{CP}^{m-1}$ and define a line bundle $\CO(r)$ for an arbitrary integer $r \in \BZ$ as
\beq
[ S^{2m-1} \times \BC]/\U(1),
\eeq
where the $\U(1)$ acts as\footnote{
This line bundle $\CO(r)$ can also be represented as $[ (\BC^{m} \setminus \{ 0 \}) \times \BC]/\BC^*$ which makes manifest the fact that
it is a holomorphic line bundle over $\mathbb{CP}^{m-1}$. The holomorphic sections of $\CO(r)$ are degree $r$ polynomials of $\vec{z}$ 
as can be seen from this definition of $\CO(r)$. This is a well-known fact in algebraic geometry.
}
\beq
S^{2m-1} \times \BC \ni (\vec{z}, u) \mapsto ( e^{\i \alpha } \vec{z}, e^{ \i r \alpha } u) \qquad (\alpha \in \BR).
\eeq
We denote the equivalence class of $(\vec{z},u)$ under the equivalence relation $(\vec{z}, u) \sim ( e^{\i \alpha } \vec{z}, e^{ \i r \alpha } u)$
as $[\vec{z}, u]$. Then $\CO(r) =\{ [\vec{z}, u] \}$.

Now we take the total space of $\CO( - k)  $, and consider its subspace given by 
\beq
Z & = \{ [\vec{z}, u ]   \mid  |u| \leq 1\},  \\
Y &=  \{ [\vec{z}, u ]   \mid  |u| = 1\} . \label{eq:8dextension}
\eeq
On $Y$, we can fix ``the gauge symmetry'' $(\vec{z}, u) \sim ( e^{\i \alpha } \vec{z}, e^{ - \i k \alpha } u)$ by taking $u=1$.
Then the remaining gauge transformation is generated by $[\vec{z}, 1] = [e^{ 2\pi  \i /k } \vec{z}, 1] $. 
Therefore, we conclude that $Y = S^{2m-1}/\BZ_k$. 
The manifold $Z$ has this lens space as the boundary, $Y =\partial Z$. One can also check that the orientation of $Z$ as a complex manifold
is compatible with the orientation of $S^{2m-1}/\BZ_k$ induced from the standard orientation of $S^{2m-1}$.

On $Z$, we define a line bundle $\CL_{s}$ ($s \in \BZ$) by
\beq
\CL_s = \{ [\vec{z}, u, v ] \} /  (\vec{z}, u, v ) \sim  (e^{\i \alpha}\vec{z},  e^{- \i k \alpha} u, e^{- \i s \alpha}v ) . 
\eeq
Notice that $\CL_s = \CL_1^{\otimes s}$.
The line bundle $\CL_1$ extends the one defined in \eqref{eq:L0onlens} from $Y =S^{2m-1}/\BZ_k$ to $Z$. This is a pullback of $\CO(-1)$ from $\mathbb{CP}^{m-1}$ to the total space of $\CO(-k)$.

Let us consider a connection on $\CL_1$ which becomes the flat connection in $Y = S^{2m-1}/\BZ_k$.
We represent the connection by using a differential cohomology element $\check A \in \check H^2(Z)$.
Consider the holonomy $\exp(2\pi \i \int \DA_A)$ of this connection around a loop 
\beq
[e^{ 2\pi \i  t/k }\vec{z}_0, u_0]  \qquad ( 0 \leq t \leq 1)  \label{eq:loopinY}
\eeq
in $Y=S^{2m-1}/\BZ_k$ for a fixed $(\vec{z}_0,u_0)$.
The parallel transport of an element $[\vec{z}_0, u_0,v] $ of $\CL_1$ is given by $[e^{ 2\pi \i  t/k }\vec{z}_0, u_0,v] $.
From the fact that
\beq
( e^{2\pi \i /k}z_0,u_0, v) \sim (z_0, u_0, e^{2\pi \i /k}v),
\eeq
we can see that the holonomy of the flat connection around 
the loop is $e^{2\pi \i /k}$. 

Next consider a two dimensional disk
\beq
D=\{ [\vec{z}_0, u] ; |u| \leq 1 \} \subset Z \qquad (\vec{z}_0 \text{: fixed} ).
\eeq
Notice that the loop \eqref{eq:loopinY} is equal to $\partial D$ because $[e^{ \i  t/k }\vec{z}_0, u_0] =[\vec{z}_0, e^{ \i  t }u_0 ]$.
From the above holonomy, we see that its curvature integral on the disk is given by\footnote{The holonomy
constrains $\int_D \DF_A \in \frac{1}{k} + \BZ$. We can modify the connection if necessary by using the connection whose curvature is localized on $\mathbb{CP}^{m-1} =\{[\vec{z},u=0] \}\subset Z$ 
to get \eqref{eq:cnn}. }
\beq
\int_D \DF_A  = \frac{1}{k} .\label{eq:cnn}
\eeq
In particular, $\CL_1^{\otimes k}$ has a connection $k \check A$ which is trivial on $Y$ and has the curvature integral given by $\int_D k \DF_{A} =1$.
This implies that the connection $k\check A$ can be continuously deformed (without changing the boundary values) to a connection which is trivial on $Y$ and whose curvature is localized on 
\beq
M= \{ [\vec{z}, u=0] \} \subset Z. \label{eq:subW}
\eeq
This $M$ is isomorphic to $\mathbb{CP}^{m-1}$.
The localization of the curvature means $ k \DF_{A} \sim \delta(M)$ where $\delta(M)$ is the delta function localized on $M$.\footnote{
In the terminology of algebraic geometry, $M$ is a divisor class associated to the bundle $\CL_1^{\otimes k}$.}

We compute
\beq
&\int_Z (\DF_{A})^m =\frac{1}{k} \int_Z (\DF_{A})^{m-1} \delta(M) = \frac{1}{k} \int_M (\DF_{A})^{m-1} \nonumber \\
&=\frac{1}{k} \int_M (c_1(\CL_1))^{m-1}  = \frac{1}{k} \int_{\mathbb{CP}^{m-1} } c_1(\CO(-1) )^{m-1}  = \frac{(-1)^{m-1}}{k} . \label{eq:Hintersection}
\eeq
where we have used the fact that $\CL_1$ restricted to $M \cong \mathbb{CP}^{m-1}$ is $\CO(-1)$,
and also used the standard fact that $\int_{\mathbb{CP}^{m-1} } c_1(\CO(t) )^{m-1} = t^{m-1}$ for any $t \in \BZ$.

Now we have done all the preparations to compute the pairing of $\check C_1$ on $Y = S^7/\BZ_k$.
In this paragraph we restrict our attention to the original case $m=4$. 
We define $\check C_1$ as $\check C_1 = \check A \Dp \check A$ which indeed gives $[\DN_{C_1}] = c_1(\CL_1)^2$ on $Y$
as we have defined in \eqref{eq:NC1onlens}.
Then the pairing on $Y$ is given by
\beq
(\check C_1, \check C_1) = \int_Z (\DF_{C_1})^2 = \int_Z (\DF_{A})^4 = - \frac{1}{k}.\label{eq:pair1}
\eeq
Thus we have obtained the first equation in \eqref{eq:summaryvalue}.
The pairing takes values in $\BR/\BZ$, so this equation is meaningful only mod 1.

\subsection{The $\eta$-invariant}
Next we want to go to the computation of the $\eta$-invariant.
In this appendix we are interested not in $\eta$ itself, but only $\eta \mod 1$ 
for the purposes of computing anomalies. Thus we can use the APS index theorem to compute it by integrating the corresponding characteristic class on $Z$.

One point which we need to be careful is the following. 
We want to use the manifold \eqref{eq:8dextension}. It is not always a spin manifold depending on the values of $k$ (and $m$).
However, the manifold $Z$ (or any complex manifold) is a $\spin^c$ manifold.
For the purpose of computing the $\eta$-invariant, it is enough to have a $\spin^c$ structure.

First let us discuss why $Z$ is not necessarily $\spin$. 
If we restrict to the submanifold $M \cong \mathbb{CP}^{m-1}$ defined in \eqref{eq:subW},
the tangent bundle of $Z$ splits as $TZ = T \mathbb{CP}^{m-1} \oplus \CO(-k)$. The reason is that the normal bundle to $M$ in $Z$ is $\CO(-k)$ which follows from the definition of $Z$
as the total space of the $\CO(-k)$ bundle over $\mathbb{CP}^{m-1}$.
It is also well-known that if we add a trivial bundle $\underline{\BC} $ to $T \mathbb{CP}^{m-1}$, we get a sum of $m$ copies of $\CO(1)$,
\beq
T \mathbb{CP}^{m-1} \oplus \underline{\BC} \cong m \CO(1). \label{eq:TCPsplit}
\eeq
The bundle $m \CO(1) \oplus \CO(-k)$ has a spin lift only if $m+k$ is even. This is the obstruction to the existence of a spin structure on $Z$.

Instead of spin structure, we consider the following bundle on a complex manifold $Z$. The following discussion is generally true for any complex manifold and is well-known in algebraic geometry. 
Let ${T}^*Z$ be the complex cotangent bundle of the complex manifold $Z$,
and let $\overline{T}^* Z$ be its complex conjugate bundle. (This $\overline{T}^* Z$ is isomorphic to the complex tangent bundle $T Z$ by introducing an explicit hermitian metric on $T Z$.
Using the bundle $\overline{T}^* Z$ may be more natural in the context of Dolbeault complex without an explicit hermitian metric.) We define
\beq
\CS_c = \sum_{\ell  =0}^{\dim_\BC Z } \wedge^\ell (\overline{T}^* Z),
\eeq
where $\wedge^\ell E$ of a vector space $E$ means the $\ell$-th antisymmetric product of $E$.
This is $\spin^c$ by the following reason. Let $\Gamma_I ~(I=1,\ldots, 2\dim_\BC W)$ be the gamma matrices of $\dim_\BR W = 2 \dim_\BC W$ dimensional Clifford algebra.
We take $a_i = (\Gamma_{2i -1} + \i \Gamma_{2i})/2$ and $a_i^\dagger = (\Gamma_{2i -1} - \i \Gamma_{2i})/2$.
They may be regarded as creation and annihilation operators with $\{a_i, a_j^\dagger \} = \delta_{ij}$.
Then we can find a representation of the Clifford algebra by first taking $\ket{0}$ such that $a_i \ket{0}=0$, and then consider
$a_{i_1}^\dagger \ldots a_{i_\ell}^\dagger \ket{0}$. By this construction, we can see that $\CS_c$ defined above is a $\spin^c$ bundle (or more precisely an irreducible Clifford module) on which the Clifford algebra acts. 
More explicitly, on the bundle $\CS_c$, the actions of $a_{i}$ and $a_i^\dagger$ are given by $a_{i}^\dagger = \d \bar{z}^i \wedge$ and $a_i = \iota_{\bar{\partial}_i}$, respectively,
where $\d \bar{z}^i \wedge $ is an operator which acts on a differential form $\omega$ as $\omega \to \d \bar{z}^i \wedge \omega$, and $ \iota_{\bar{\partial}_i}$ is its adjoint.
The `chirality', or equivalently the $\BZ_2$ grading of the bundle, is determined by the degree $\ell$ of $ \wedge^\ell (\overline{T}^* W)$ mod 2.

The canonical line bundle of a complex manifold $W$ is defined by 
\beq
\CK = \wedge^{\dim_\BC W} (T^*W).
\eeq
Then, if there exists a square root $\CK^{1/2}$ of the canonical line bundle, we can define a spin bundle as
\beq
\CS = \CK^{1/2} \otimes \CS_c.
\eeq
Therefore, $\CK^{1/2}$ measures the difference between $\CS$ and $\CS_c$.

Let us return to the case of our manifold \eqref{eq:8dextension}.
Near the boundary, the manifold $Z = \{ [\vec{z},u] \}$ is embedded as a subspace of $\BC^m/\BZ_k $ by taking $\vec{w}=u^{1/k}\vec{z}$ as the complex coordinates of 
 $\BC^4/\BZ_k $ as far as $\vec{w} \neq 0$. 
 The equivalence relation is $\vec{w} \sim e^{2\pi \i/k}\vec{w}$.\footnote{In fact,
our manifold $Z$ is a blowup of $\BC^m/\BZ_k$ at the singular point.}
The tangent bundle $TZ$ is described by tangent vectors $\Delta \vec{w}$ with the equivalence relation $(\vec{w}, \Delta \vec{w}) \sim e^{2\pi i/k}(\vec{w}, \Delta \vec{w})$.
From this, we see that $TZ$ near the boundary is $m \CL_{-1} = (\CL_{1})^{-1} \oplus \cdots \oplus (\CL_{1})^{-1}$, i.e.~the sum of $m$ copies of $\CL_1^{-1}$.
The canonical bundle near the boundary is $\CK = \CL_1^{\otimes m} = \CL_m$. A square root of $\CK$ exists near the boundary if $m$ is even,
and we take it to be $\CK^{1/2} =\CL_1^{\otimes (m/2)} $.
Now we define 
\beq
\CS' = \CL_{1}^{\otimes (m/2)} \otimes \CS_c
\eeq
on $Z$.  We soon explain the case of odd $m$.

When restricted to the boundary $Y = \partial Z$, the bundle $\CS'$ gives a spin structure of $Y = S^{2m-1}/\BZ_k$. 
The spin structure is not unique when $k$ is even, because we can take a line bundle associated to a homomorphism $\BZ_k \to \BZ_2$ and modify the spin bundle by this line bundle.
However, the spin structure of $\CS'$ coincides with the one which is realized in the M-theory orbifold (which is given in \eqref{eq:orb3} of Appendix~\ref{sec:Mthconv})
as one can check by representing all bundles as a sum of powers of $\CL_1$.

Inside $Z$, $\CS'$ is not spin, but $\spin^c$. However, that is not a problem in computing the $\eta$-invariant on $Y = S^{2m-1}/\BZ_k$.
We multiply the bundle by $\CL_1^s$ to get
\beq
\CS_s : = \CL_1^{s} \otimes \CS' = \CL_1^{\otimes (s+m/2) } \otimes \CS_c .
\eeq
This is a $\spin^c$ bundle on $Z$. For the purpose of the computation of \eqref{eq:summaryvalue},
we need to consider the $\eta$-invariant of the bundles with $s=0$ and $s=1$.

Although it is not directly relevant to \eqref{eq:summaryvalue}, let us also comment on the case of odd $m$.
In this case, $\CS'$ is not well-defined. In fact, for odd $m$ and even $k$, $S^{2m-1}/\BZ_k$ is not a spin manifold as we explained before.
However, the bundle $\CS_s$ is well-defined if we take $s +m/2 $ to be integer.
The following computation is valid also for these cases. 

Recall that $Z$ is described as the total space of the $\CO(-k)$ bundle on $\mathbb{CP}^{m-1}$,
and $\CL_1$ is the pullback of $\CO(-1)$ to this total space. In particular, 
the canonical bundle $\CK$ of $Z$ is topologically given by $\CK = \CK_{\mathbb{CP}^{m-1}} \otimes \CO(k) = \CO(k-m)$.
Here we have used the fact that the canonical bundle of $\mathbb{CP}^{m-1}$ is given by $ \CK_{\mathbb{CP}^{m-1}} = \CO(-m)$, which follows from \eqref{eq:TCPsplit}.
The bundles $\CO(s)$ here are understood as a pullback of $\CO(s)$ from $\mathbb{CP}^{m-1}$ to $Z$.
In fact, $\CO(k) =\CL_1^{\otimes (-k)}$ is trivial near the boundary $Y = S^{2m-1}/\BZ_k$, so this canonical bundle reduces to what we have discussed above,
that is, $\CK = \CL_1^{\otimes m}$ near the boundary.
The expression $\CK = \CL_1^{\otimes (m -k)}$ is valid even inside $Z$.

Therefore, at the level of differential forms of the curvature (which is what is necessary for the computation of $\eta$ by using the higher dimensional manifold),
we can split the bundle as
\beq
\CS_s= \CL_1^{\otimes (s+m/2) } \otimes \CS_c = \CL_1^{ \otimes (s + k/2)} \otimes \CS, \label{eq:formalsplitting}
\eeq
where we have formally set $\CS = K^{1/2} \otimes \CS_c$. Such a formal expression is valid when we consider the characteristic polynomial of the curvature in the index theorem.
Let $\hat{A}(R)$ be the A-roof genus of $Z$ defined explicitly in terms of the Riemann curvature tensor $R$.
Let $\DF_A$ be the curvature  of the connection $\check A$ on $\CL_1$ which was considered in the computation of \eqref{eq:Hintersection}.
We can now compute the $\eta$-invariant on $Y = S^{2m-1}/\BZ_k$ by using the APS index theorem.
Let $\CD_s$ be the Dirac operator acting on to the (positive chirality part of) $\CS_s$ restricted to $Y$. From \eqref{eq:formalsplitting} we get
\beq
-\eta(\CD_s) \equiv  \int_Z \exp \left( (s+\frac{1}{2}k )\DF_A \right) \hat{A}(R) \quad \mod 1.
\eeq

We can further simplify this expression as follows. The tangent bundle $TZ$ is topologically the same as $T\mathbb{CP}^{m-1} \oplus \CO(-k)$.
The characteristic class does not change even if we add a trivial bundle $\underline{\BC}$, and we have the splitting \eqref{eq:TCPsplit}
Therefore, we get
\beq
TW \oplus \underline{\BC} = \CL_1^{k} \oplus  \CL_1^{-1} \oplus \cdots \oplus \CL_1^{-1} : = \CL_1^{\oplus (k, -1,\cdots,-1)} ,
\eeq
where there are $m$ copies of $\CL_1^{-1}$.
Let $\hat{A}( \CL_1^{\oplus (k, -1,\cdots,-1)} ) $ be the corresponding A-roof genus which is equal to $\hat{A}(R)$ (up to the continuous deformation of the connection
from the Levi-Civita connection on $TZ$ to the connection on $ \CL_1^{\oplus (k, -1,\cdots,-1)}$ determined by the connection $\check A$ on $\CL_1$). 
This can be represented as a polynomial of $\DF_A$. Explicitly, it is given by
\beq
\hat{A}(\CL_1^{\oplus (k, -1,\cdots,-1)}) = \left( \frac{k\DF_A/2}{\sinh(k\DF_A/2)} \right)  \left( \frac{\DF_A/2}{\sinh(\DF_A/2)} \right)^m.
\eeq 
The $\eta$ is now given by
\beq
-\eta(\CD_s) \equiv  \int_Z \exp \left( (s+ \frac{k}{2} )\DF_A \right) \hat{A}(\CL_1^{\oplus (k, -1,\cdots,-1)}).\label{eq:rslt1}
\eeq
The integrand is just a polynomial of $\DF_A $. Moreover, we know from \eqref{eq:Hintersection} that
\beq
\int_Z (\DF_A)^m = \frac{(-1)^{m-1}}{k}. \label{eq:rslt2}
\eeq
By using these results, we can compute the desired $\eta$-invariant.

The above result can be summarized as follows. Let us define power series of variables $y,x_1,\cdots,x_{m+1}$ 
which we denote (by abuse of notation) as $\ch(y)$ and $\hat{A}(x_1, \cdots, x_{m+1})$, and $p(x_1,\cdots,x_{m+1})$, by the following formulas:
\beq
&\ch(y) = \sum_{i} \ch_i(y) = e^y, \nonumber \\
&\hat{A}(x_1,\cdots,x_{m+1}) =\sum_i \hat{A}_i(x_1,\cdots,x_{m+1})= \prod_{i=1}^{m+1}  \left( \frac{ x_i /2}{\sinh(x_i /2)} \right) , \nonumber \\
& p (x_1,\cdots,x_{m+1}) = \sum_i p_i (x_1,\cdots,x_{m+1}) = \prod_{i=1}^{m+1} (1+x_i^2),
\eeq
where $\ch_i$ is the degree $i$ part of $\ch$, and $\hat{A}_i$ and $p_i$ are the degree $2i$ parts of $\hat{A}$ and $p$, respectively.
$\hat{A}$ can be expanded by $p$ as
\beq
\hat{A}_0=1, \qquad \hat{A}_1 = - \frac{1}{24}p_1, \qquad \hat{A}_2 = \frac{7p_1^2 - 4p_2}{5760}, \quad \ldots .
\eeq
By using these notations, the $\eta(\CD_s)$ is obtained from \eqref{eq:rslt1} and \eqref{eq:rslt2} as
\beq
(-1)^m \eta(\CD_s) \equiv \frac{1}{k} \sum_{i+2j = m} \ch_i \left(s+ k/2 \right) \hat{A}_j(k,1,\cdots,1) \mod 1.
\eeq
This result was obtained by Gilkey \cite{GilkeyOdd} by a different method.

Now let us restrict to the case $m=4$. We get
\beq
p_1(k,1,1,1,1) = k^2+4, \qquad p_2(k,1,1,1,1) = 4k^2 + 6.
\eeq
Then we can compute the following, where all equalities are valid mod 1:
\beq
12( \eta(\CD_s) - \eta(\CD_0) )&\equiv \frac{12}{k} \left( \ch_4( s+k/2) - \ch_4(k/2) \right) - \frac{1}{2k}p_1 \left(  \ch_2( s+k/2) - \ch_2(k/2)  \right) \nonumber \\
 & \equiv \frac{s^2( k^2- 2 + s^2  )}{2k} .\label{eq:s0}
\eeq
By putting $s=1$, we get the second equation in \eqref{eq:summaryvalue}. 
We can perform a consistency check of this result. For the background labeled by $s$,
the gauge field $s \check A$ gives a 3-form background $\check C = (s \check A) \Dp (s \check A)  = s^2 \check C_1$.
Therefore, we have
\beq
12( \eta(\CD_s) - \eta(\CD_0) ) = - \widetilde{\CQ}( s^2 \check C_1) = - s^2  \widetilde{\CQ}( \check C_1) - \frac{s^2(s^2-1)}{2} (\check C_1, \check C_1) .
\eeq
By using $ -   \widetilde{\CQ}( \check C_1) = (k^2-1)/2k$ and $(\check C_1, \check C_1) = -1/k$,
this equation gives the same result as \eqref{eq:s0} for general $s$.

We also compute
\beq
30\eta(\CD_0) - \frac{K_k(K_k-k)}{2k} &\equiv \frac{30}{k} \ch_4(k/2) - \frac{5}{4k}p_1 \ch_2(k/2) + \frac{7p_1^2 - 4p_2}{192k} - \frac{K_k(K_k-k)}{2k} \nonumber \\
& = \frac{k^2-1}{24k} -\frac{k(k+1)(k-1)}{6} \nonumber \\
& \equiv \frac{k^2-1}{24k}.
\eeq
This is the last equation in \eqref{eq:summaryvalue}.

\section{Computation of the $\eta$-invariant on lens spaces}\label{sec:etacomp}

In this appendix we present methods for computing the values of the $\eta$-invariant on lens spaces $S^{2m-1}/\BZ_k$
which are different from the method in Sec.~\ref{sec:lens}. We compute them as elements of $\BR$ rather than $\BR/\BZ$.

The basic setup is as follows. Let
\beq
\vec{z} = (z^1, \cdots, z^m) \in \BC^m.
\eeq
We consider the lens space defined by dividing the sphere $S^{2m-1} = \{ | \vec{z} | =1 \} $ by the $\BZ_k$ action
\beq
\vec{z} \to e^{2\pi \i j /k} \vec{z} , \qquad (j =0,1,\cdots, k-1). \label{eq:zkz}
\eeq
We denote this space as $S^{2m-1}/\BZ_k$.

On this space, we consider a $\spin^c$ connection as follows. 
Let $\CS_{\BC^m}  $ be the trivial $\spin^c$ bundle on $\BC^m$. 
We denote the coordinate of the fiber $\CS_{\BC^m}  $ as $\Psi$. 
Then we define the action of $\BZ_k$ by
\beq
\Psi \to e^{-2\pi \i j s /k} \SR(j/k) \Psi \qquad (j =0,1,\cdots, k-1). \label{eq:Psitr}
\eeq
where $s \in \frac{1}{2} \BZ$ is a parameter which specifies the $\spin^c$ connection, and 
\beq
\SR(\alpha) = \exp\left(- \pi \alpha \left(  \Gamma^1\Gamma^2 + \cdots+ \Gamma^{2m-1}\Gamma^{2m} \right) \right) \qquad (0 \leq \alpha \leq 1).
\eeq
Here $\Gamma^I ~(I=1,\cdots, 2m)$ are gamma matrices on $\BC^m \cong \BR^{2m}$.
For this action to be a $\BZ_k$ action, we must have
\beq
 e^{-2\pi \i s } \exp\left(- \pi \left(  \Gamma^1\Gamma^2 + \cdots+ \Gamma^{2m-1}\Gamma^{2m} \right) \right)=1,
\eeq
or equivalently
\beq
s  \in \frac{m}{2} + \BZ.
\eeq
Then we define a bundle on $S^{2m-1}/\BZ_k$ as,
\beq
\CS_s = \{ (\vec{z}, \Psi) \mid ~  |\vec{z} | =1, ~~ \Ch \Psi = + \Psi \} /\BZ_k,
\eeq
where $\Ch$ is the chirality operator on $\CS_{\BC^m}  $.
This is a $\spin^c$ bundle on $S^{2m-1}/\BZ_k$.

We may use the notation that $\CS_s  = \CS \otimes \CL_s$, where $\CS = \CS_{s=0}$ is the spin bundle defined by the above construction with $s=0$,
and $\CL_s$ is a line bundle. This splitting is possible only for even $m$, but we may also formally use such splitting for odd $m$.
Such a formal splitting is possible if we are only concerned with curvatures of the bundles. 

We also consider bundles associated to bi-spinor fields $\Phi$. The relevant bundle on $\BC^m$ for the bi-spinor is $\CS_{\BC^m} \otimes \overline{\CS_{\BC^m}} \otimes \CL_t$,
which means that $\Phi$ transforms as
\beq
\Phi \to e^{-2\pi \i j t /k} \SR(j/k) \Phi \SR(j/k)^{-1} \qquad (j =0,1,\cdots, k-1). \label{eq:Phitr}
\eeq
Here $t \in \BZ $ is an integer parameter.
We construct the corresponding bundle on $S^{2m-1}/\BZ_k$ by restricting the bi-spinor to $\Ch \Phi = \Phi$.
We denote the bundle as $\CS^{\rm sig}_t$; see Sec.~\ref{sec:Signature} for the details of the signature index theorem.

Let $\CD_s$ and $\CD^{\rm sig}_t$ be the Dirac operator acting on sections of $\CS_s$ and $\CS^{\rm sig}_t$, respectively. 
We want to compute their $\eta$-invariant. 
We sometimes denote the $\eta$-invariant on a manifold $Y$ as $\eta(Y)$
if the operator $\CD$ is clear from the context, or if $\CD$ is a general operator.

We present two methods: 
In Sec.~\ref{sec:torus}, we use  $\bZ_k$ orbifolds of the torus $T^{2m}$ to deduce the $\eta$-invariant.
This method only uses the standard APS index theorem, but is only applicable to $k=2,3,4,6$.
In Sec.~\ref{sec:equiveta}, we introduce and utilize the equivariant APS index theorem.
This requires the reader to learn an additional mathematical machinery, but it allows the computation for arbitrary $k$.
These two subsections can be read mostly independently.

\subsection{APS index theorem on $T^{2m}/\BZ_k$}\label{sec:torus}
Here we describe a method which only uses the APS index theorem, but is restricted to the cases $k=2,3,4,6$.
It is done by computing the index on $T^{2m}/\BZ_k$.
The method here was used e.g.~in \cite{Witten:2015aba,Tachikawa:2018njr}.

The (singular) manifold $T^{2m} /\BZ_k$ is defined as follows. 
We first consider a torus $T^{2m}$ whose complex coordinate is denoted as $\vec{z} = (z^1, \cdots, z^m)$.
It satisfies equivalence relations of the form $\vec{z} \sim \vec{z} + \vec{p}+\tau \vec{q}$,
where $\vec{p}$ and $\vec{q}$ are vectors whose components are integers, and $\tau$ is the standard period matrix of the torus.
For $k=2,3,4,6$ with specific $\tau$ for each $k$, we can act $\BZ_k$ on the torus as in \eqref{eq:zkz}.
Then we get $T^{2m}/\BZ_k$. We can define $\spin^c$ bundles in the same way as above.

The APS index theorem  on $T^{2m}/\BZ_k$ can be  applied as follows. 
First, notice that the manifold $T^{2m}/\BZ_k$ has orbifold singularities. 
Let $B_i$ be a small ball which is centered around a orbifold singularity labelled by $i$.
The boundary is $\partial B_i = S^{2n-1}/\BZ_{\ell_i}$ where $\ell_i$ is a divisor of $k$ and depends on $i$.
The index is defined as the APS index on a manifold which is obtained by subtracting $B_i$ from $T^{2m}/\BZ_k$,
\beq
Z = T^{2m}/\BZ_k - \bigsqcup_i B_i \label{eq:T2mZk}.
\eeq
This is nonsingular, and has as the boundary
\beq
\partial Z = \bigsqcup_i  \overline{B_i}= \bigsqcup_i  \overline{S^{2n-1}/\BZ_{\ell_i}}
\eeq
where the overline means orientation reversal.

By the APS index theorem, we get
\beq
\index (T^{2m}/\BZ_k ) = \eta (\partial Z ) = - \sum_i \eta( S^{2n-1}/\BZ_{\ell_i}),
\eeq
where we denote the $\eta$-invariant of the operator $\CD_s$ or $\CD^{\rm sig}_t$ on a manifold $Y$ as $\eta(Y)$, and also similarly for the index.
The minus sign is due to the fact that $\eta(\overline{Y}) = - \eta(Y)$.

More explicit form is determined by careful examination of the geometry. 
In the case $m=1$, one can see directly the following. $T^2/\BZ_2$ has four orbifold points with $\ell_i=2$.
$T^2/\BZ_3$ has three orbifold points with $\ell_i=3$. $T^2/\BZ_4$ has two orbifold points with $\ell_i=4$, and one orbifold point with $\ell_i=2$.
$T^2/\BZ_6$ has one orbifold point with $\ell_i=6$, one orbifold point with $\ell_i=3$, and one orbifold point with $\ell_i=2$.

Notice that if $i$ is an orbifold point of $T^2/\BZ_k$ with $\BZ_{\ell_i}$ orbifold singularity, its pre-image in $T^2$ consists of $k/\ell_i$ points.
Notice also that if $i$ and $j$ are two points in $T^2$ with orbifold singularity of type $\BZ_{\ell_i}$ and $\BZ_{\ell_j}$ respectively,
then $\{i\} \times \{j\} \in T^2 \times T^2 =T^4$ is a fixed point with $\BZ_\ell$ orbifold singularity, where $\ell$ is given by the greatest common divisor of $\ell_i$ and $\ell_j$.
From these facts, one can determine the orbifold points of $T^{2m}/\BZ_k$ from the knowledge of $T^2/\BZ_k$.

The answer is given by
\beq
-\index( T^{2m}/\BZ_2) & =  2^{2m}  \eta( S^{2m-1}/\BZ_{2} ), \nonumber  \\
-\index( T^{2m}/\BZ_3) & =  3^{m}  \eta( S^{2m-1}/\BZ_{3} ), \nonumber \\
-\index( T^{2m}/\BZ_4) & =  2^{m}  \eta( S^{2m-1}/\BZ_{4} ) + \frac{2^{2m} -2^m}{2} \eta( S^{2m-1}/\BZ_{2} ),  \nonumber \\
-\index( T^{2m}/\BZ_6) & =   \eta( S^{2m-1}/\BZ_{6} ) + \frac{3^m -1}{2}   \eta( S^{2m-1}/\BZ_{3} )  + \frac{2^{2m} -1}{3} \eta( S^{2m-1}/\BZ_{2} ) .
\eeq
By solving these equations, we can determine the $\eta$-invariants as
\beq
\eta( S^{2m-1}/\BZ_{2} ) & =    - 2^{-2m}\index( T^{2m}/\BZ_2), \nonumber \\
 \eta( S^{2m-1}/\BZ_{3} ) & =    - 3^{-m}\index( T^{2m}/\BZ_3),\nonumber \\
 \eta( S^{2m-1}/\BZ_{4} )  & = - 2^{-m} \index( T^{2m}/\BZ_4)  + \frac{2^{-m} -2^{-2m}}{2} \index( T^{2m}/\BZ_2 ), \nonumber \\
 \eta( S^{2m-1}/\BZ_{6} )  & =  -\index( T^{2m}/\BZ_6)  
 + \frac{1- 3^{-m} }{2}\index( T^{2m}/\BZ_3)    + \frac{1-2^{-2m} }{3}\index( T^{2m}/\BZ_2). \label{eq:etafromindex}
\eeq

The APS index theorem is valid under some boundary condition. In the present case of \eqref{eq:T2mZk}, the boundary condition 
is given by the following condition. We can shrink $B_i$ to zero size, and we extend the zero modes to the entire $T^{2m}/\BZ_k$.
The APS boundary condition is such that the extended zero modes remain finite at the orbifold points. 
This in turn implies that the zero modes come from the zero modes on $T^{2m}$ which are invariant under the $\BZ_k$ action.

Therefore the index on $T^{2m}/\BZ_k$ counts the net number of zero modes on $T^{2m}$ which are invariant under $\BZ_k$.
The zero modes on $T^{2m}$ are just constant modes $\partial_I \Psi=0$. The reason is that the Dirac operator is given by $\i \Gamma^I \partial_I$,
and $0=(\i \Gamma^I \partial_I)^2 \Psi= - \partial^2\Psi$ which implies that $0 = \int_{T^{2m}}\Psi^\dagger (- \partial^2\Psi) = \int_{T^{2m}}\partial_I \Psi^\dagger \partial_I\Psi $.
Therefore, we just need to count the number of components of the spinor field $\Psi$ which are
invariant under $\BZ_k$.

Let us consider the spinor field $\Psi$ which transforms like \eqref{eq:Psitr}. 
We consider operators $ \i^{-1} \Gamma^{2i-1}\Gamma^{2i}$ for $i=1,\cdots,m$. We denote the eigenvalues of $\i^{-1} \Gamma^{2i-1}\Gamma^{2i}$ as $\sigma_i =\pm 1$.
The chirality operator $\Ch$ is given by
\beq
\Ch = \prod_{i=1}^m \left( \i^{-1} \Gamma^{2i-1}\Gamma^{2i} \right), 
\eeq
and hence its eigenvalues are $\prod_{i=1}^m \sigma_i$.
The condition that $\Psi$ is invariant under \eqref{eq:Psitr} is given by
\beq
s + \frac{1}{2} \sum_{i=1}^m \sigma_i  \equiv 0 \mod k.
\eeq
We need to count the number of components of $\Psi$ satisfying this condition, and also determine the eigenvalues of $\Ch$ of these components. 
For this purpose, we rewrite the equation as
\beq
 \sum_{i=1}^m \frac{ 1-\sigma_i }{2} \equiv s+\frac{m}{2} \mod k.
\eeq
Thus, the number of components which has $\sigma_i = -1$ is of the form $kj+s+\frac{m}{2}$ for $j \in \BZ$.
The chirality of such components is $\Ch = (-1)^{kj + s + \frac{m}{2}}$.
Therefore, the index is given by using the binomial coefficient as
\beq
\index( T^{2m}/\BZ_k) =  \sum_{j} (-1)^{kj + s + \frac{m}{2}}  \begin{pmatrix}  m \\ kj + s + \frac{m}{2} \end{pmatrix}.
\eeq
By putting this formula into \eqref{eq:etafromindex}, we get the values of the $\eta$-invariant.

The index of the operator acting on the bi-spinor field $\Phi$ transforming as \eqref{eq:Phitr} is more complicated.
But the basic idea is the same. We just count the number of components which are invariant under \eqref{eq:Phitr}.

We list some examples. For the operator $\CD_s$, we get 
\beq
\begin{array}{c|cccc}
k & 2 & 3 & 4 & 6  \\
\hline\hline
\index (T^{4}/\BZ_k)_{s=0}   & -2 & -2 & -2 & -2 \\
\eta (S^3/\BZ_k)_{s=0} &  \frac{1}{8} & \frac{2}{9} &  \frac{5}{16} & \frac{35}{72} \vspace{0.1cm}   \\
\hline 
\index (T^{4}/\BZ_k)_{s= \pm 1}  & 2 & 1 & 1 & 1 \\
\eta (S^3/\BZ_k)_{s= \pm 1} & - \frac{1}{8} & -\frac{1}{9}  & - \frac{1}{16} &  \frac{5}{72} \vspace{0.1cm} \\
\hline\hline
\index (T^{6}/\BZ_k)_{s=1/2}   & 4 & 3 & 3 & 3 \\
\eta (S^5/\BZ_k)_{s=1/2} & - \frac{1}{16} & -\frac{1}{9} & - \frac{5}{32} & - \frac{35}{144}   \vspace{0.1cm} \\
\hline
\index (T^{6}/\BZ_k)_{s=3/2}   & -4 & 0 & -1 & -1 \\
\eta (S^5/\BZ_k)_{s=3/2} &  \frac{1}{16} &0 & - \frac{3}{32} & - \frac{5}{16}   \vspace{0.1cm} \\
\hline\hline
\index (T^{8}/\BZ_k)_{s=0}   & 8 & 6 & 6 & 6 \\
\eta (S^7/\BZ_k)_{s=0} & - \frac{1}{32} & -\frac{2}{27} & - \frac{9}{64} & - \frac{329}{864} \vspace{0.1cm} \\
\hline 
\index (T^{8}/\BZ_k)_{s= \pm 1}  & - 8 & - 3 & -4 & - 4 \\
\eta (S^7/\BZ_k)_{s= \pm 1} & \frac{1}{32} & \frac{1}{27}  & \frac{1}{64} & - \frac{119}{864} \\
\end{array}.
\eeq
For the operator $\CD^{\rm sig}_t$, we get
\beq
\begin{array}{c|cccc}
k & 2 & 3 & 4 & 6  \\
\hline\hline
\index (T^{4}/\BZ_k)_{t=0}   & 0 & -2 & -2 & -2 \\
\eta (S^3/\BZ_k)_{t=0} & 0 & \frac{2}{9} &  \frac{1}{2} & \frac{10}{9} \vspace{0.1cm} \\
\hline\hline
\index (T^{6}/\BZ_k)_{t=1}   & 0 & 3 & 4 & 3 \\
\eta (S^5/\BZ_k)_{t=1} & 0 & -\frac{1}{9}  & -\frac{1}{2} & - \frac{14}{9} \vspace{0.1cm}  \\
\hline
\index (T^{6}/\BZ_k)_{t=3}   & 0 & 0 & -4 & 0 \\
\eta (S^5/\BZ_k)_{t=3} & 0 & 0 & \frac{1}{2} & 0 \vspace{0.1cm}  \\
\hline\hline
\index (T^{8}/\BZ_k)_{t=0}   & 0 & 6 & 8 & 6 \\
\eta (S^7/\BZ_k)_{t=0} & 0 & -\frac{2}{27}  & -\frac{1}{2} & - \frac{82}{27} \\
\end{array}.
\eeq
Some of the results here were announced and used in \cite{Hsieh:2019iba}.

\subsection{Equivariant index theorem}\label{sec:equiveta}

The equivariant index theorem states the following. Let $Z$ be a manifold with boundary $\partial Z = Y$.
Suppose that a group $G$ acts on the space $Z$ (and any vector bundle on $Z$ in which we are interested). 
We consider an element $g \in G$ such that the fixed points of the $g$ action on $Z$ are isolated points $p \in \BZ$ which are not at the boundary, $p \notin Y$.
In particular, $g$ acts freely on $Y$. 

Let $\CD_Z$ be a Dirac operator which acts on sections of a bundle $\CS_Z$ on $Z$.
Let $\Ch$ be the chirality (or $\BZ_2$ grading) operator $\{ \CD_Z, \Ch \} =0$. 
Then the index can be defined as $\index (\CD_Z) = \tr ( \Ch e^{- \tau \CD_Z^2})$, where the trace is over the space spanned by the modes of $\CD_Z$,
and $\tau >0$ is an arbitrary positive constant. We can modify this definition by including $g \in G$ as
\beq
\index (\CD_Z,g) = \tr ( g \Ch e^{- \tau \CD_Z^2}), \label{eq:heatindex}
\eeq
where on the right hand side $g$ acts on the modes of $\CD_Z$.

We also define the $\eta$-invariant twisted by $g$ as follows. Consider the Dirac operator $\CD_Y$ on $Y$ which is constructed from $\CD_Z$ as described in Sec.~\ref{sec:APS}.
Let $\psi_j$ be eigenmodes of $\CD_Y$ which is in an irreducible representation $R_j(g)$ of $G$. Any mode in a single irreducible representation has the same eigenvalue $\lambda_j$.
Then we define
\beq
\eta(\CD_Y, g) = \frac{1}{2} \left(  \sum_j   \sign(\lambda_j) \tr R_j(g) \right)_{\rm reg},
\eeq
where the subscript ${\rm reg}$ means an appropriate regularization. 

If a point $p$ is fixed by $g$, this element $g$ acts on the fiber $(\CS_Z)_{p}$ of the bundle $\CS_Z$ by some matrix.
We denote this matrix as $\rho_p(g)$. Also, $g$ acts on the fiber $(TZ)_p$ of the tangent bundle $TZ$ and we denote this matrix as $\tau_p(g)$.

Now we can state the equivariant index theorem~\cite{Donnelly}. The index $\index (\CD_Z,g)$ for $g$ satisfying the above conditions is given by
\beq
\index (\CD_Z,g) = \eta(\CD_Y, g)  + \sum_{p \in \{ \text{fixed points}  \}  } \frac{ \tr  (\Ch\rho_p(g) )  }{\det( 1 -\tau_p(g) ) }. \label{eq:EIT}
\eeq
The sum is over the fixed points of $g$ on $Z$. The trace $ \tr  (\Ch\rho_p(g) )$ may also be called the supertrace of $\rho_p(g)$ under the $\BZ_2$ grading $\Ch$.

The above equivariant index theorem may be understood as follows. We may try to prove the index theorem by the heat kernel method,
which means that we use the expression \eqref{eq:heatindex} and take the limit $\tau \to +0$.
Very roughly speaking, the boundary condition used in the APS index theorem is such that the boundary modes with $\sign(\lambda_i) = +1$ contributes to 
\eqref{eq:heatindex} with $\Ch=+1$, and the boundary modes with $\sign(\lambda_i) = -1$ contributes to \eqref{eq:heatindex} with $\Ch=-1$.
This gives the boundary contribution $ \eta(\CD_Y, g)$ to the index. The bulk contribution is understood as follows. 
We can regard $H = \CD_Z^2$ as a Hamiltonian of a quantum mechanical particle living on $Z$~\cite{AlvarezGaume:1983at}. Then $e^{ - \tau \CD_Z^2}$ is the Euclidean time evolution operator. 
Within a very short time $\tau \to +0$, it is very hard for a particle to go from a point $p \in Z$ to another point $ g \cdot p \in Z$ unless these points are the same, $p = g \cdot p$.
This implies that only the fixed points $p = g \cdot p$ contribute in the heat kernel method.\footnote{This intuition is not valid on the boundary $Y$, because the APS boundary condition is non-local.
This is the reason that we have the contribution $ \eta(\CD_Y, g) $ even if $g$ acts freely on $Y$.}  
Near each fixed point $p$, we can approximate the manifold $Z$ by a flat space $\BR^D$ such that $p $ corresponds to $ 0 \in \BR^{D}$, where $D = \dim Z$. 
Then  the trace $ \tr ( g \Ch e^{- \tau \CD_Z^2})$ near the point $p $ is given by
\beq
& \tr  (\Ch\rho_p(g) )  \int \frac{\d^D x\d^{D} k}{(2\pi)^{D} } e^{-\i k \cdot (\tau_p(g)x) } e^{- \tau k^2} e^{\i k \cdot x}  \nonumber \\
&=\tr  (\Ch\rho_p(g) )  \int \d^{D}k\, e^{- \tau k^2} \delta\left( (1-\tau_p(g) )k \right)  \nonumber \\
& =  \frac{  \tr  (\Ch\rho_p(g) )  }{\det( 1 -\tau_p(g) ) }.
\eeq
This is what appears in \eqref{eq:EIT}.

Now suppose that $G$ is a finite group whose elements, except for the identity element $1 \in G$, satisfy the above conditions.
We can use the equivariant index theorem to compute the $\eta$-invariant on a manifold $Y/G$ which is smooth because of the assumption that $G$ acts freely on $Y$.
The $\eta$-invariant on this manifold is given by
\beq
\eta(\CD_{Y/G} ) = \frac{1}{ |G|} \sum_{ g \in G}  \eta(\CD_Y, g) ,
\eeq
where $|G|$ is the number of elements of the finite group $G$. The $ \eta(\CD_Y, g) $ for $g \neq 1$ is given by \eqref{eq:EIT}, while for $g=1$
we simply use the ordinary APS index theorem
\beq
\index ( \CD_Z,1) =  \eta(\CD_Y, 1)  + \int_Z \ch(F) \hat{A}(R) 
\eeq
where $F$ and $R$ are gauge and Riemann curvatures on $Z$ which are relevant to the index of the Dirac operator $\CD_Z$.
Therefore, we get
\beq
\eta(\CD_{Y/G} ) = -  \frac{1}{|G|} \left(  \sum_{g \neq 1} \sum_{p \in \{ \text{fixed points}  \}  } \frac{ \tr  (\Ch\rho_p(g) )  }{\det( 1 -\tau_p(g) ) } 
+  \int_Z \ch(F) \hat{A}(R) 
-  \sum_{g } \index(\CD_Y, g) \right) .
\eeq
This is the general expression.

In a special case that there are no zero modes of $\CD_Z$ and no curvature ($F=0 $ and $R=0$), the formula becomes simple.
This is the case for $Z = B^{2m} = \{ \vec{z} \in \BC^{m} ; |\vec{z}| \leq 1 \} $, $Y = S^{2m-1}$ and all the backgrounds are trivial.\footnote{
More precisely, we take $Z$ to be a hemisphere which is extended by a cylinder so that the neighborhood of the boundary
is of a product form $(-\epsilon, 0] \times S^{2m-1}$ with a product metric. The Riemann curvature is nonzero, but Pontryagin classes are zero.
There are no zero modes, that is, not just the APS index is zero,
but that each of the numbers of positive and negative chirality modes is zero. This fact follows from a vanishing theorem
on manifolds with positive Ricci scalar curvature~\cite{APS2}. The vanishing theorem can be shown from the equation 
$(\i \slashed{D})^2 = -\nabla_\mu \nabla^\mu + \frac{1}{4}R$, where $R$ is the Ricci scalar. This equation can be proved by a straightforward computation.
The APS boundary condition is such that zero modes can be extended to an infinite cylinder region with square normalizable eigenfunctions. 
Then by using $0 = \int \Psi^\dagger (-\nabla_\mu \nabla^\mu + \frac{1}{4}R)\Psi$ for a zero mode $\Psi$, we get $\Psi=0$, so there are no zero modes.
}
We take $G = \BZ_k$ which acts as \eqref{eq:zkz}. There is only a single fixed point $p =0 \in B^{2m}$.
On this point we get
\beq
\det( 1 -\tau(j) ) = |1 - e^{2\pi \i j /k}|^{2m},
\eeq
where $j \in \BZ_k$.
The matrix $\rho(g)$ is determined from \eqref{eq:Psitr} or \eqref{eq:Phitr}.
In each case the trace $ \tr  (\Ch\rho_p(g) )$ is given by
\beq
 \tr  (\Ch\rho(j) ) = \left\{ \begin{array}{ll}  
 e^{-2\pi \i j s/k} ( e^{- \pi \i j /k} -  e^{\pi \i j /k}   )^m, & \text{(Dirac)}, \\
  e^{-2\pi \i j s/k} ( e^{-\pi \i j /k} -  e^{\pi \i j /k}   )^m  ( e^{-\pi \i j /k} +  e^{\pi \i j /k}   )^m,& \text{(signature)}.
 \end{array} \right.
\eeq
where we have used the fact that the chirality is given by
$
\Ch = \prod_{i=1}^m \left( \i^{-1} \Gamma^{2i-1}\Gamma^{2i} \right).
$
Therefore, the equivariant index theorem gives
\beq
\eta(\CD_s) &= -\frac{\i^{-m}}{k} \sum_{j=1}^{k-1} \frac{e^{-2\pi \i j s/k} }{( 2 \sin( \pi  j/k) )^m } , \\
\eta(\CD^{\rm sig}_t) &= -\frac{\i^{-m}}{k} \sum_{j=1}^{k-1} \frac{e^{-2\pi \i j t/k} }{( \tan( \pi  j/k) )^m }.
\eeq

As examples, we list the values for the case of $m=4$, $s=0$ and $t=0$.
\begin{equation}
\begin{array}{c|ccccccccc}
k & 2 & 3 & 4 & 5 & 6 & 7 & 8  & 9 & 10 \\
\hline
\eta(\CD_{s=0}, S^7/\BZ_k) & -\frac{1}{32}& -  \frac{2}{27}& -  \frac{9}{64}&  - \frac{6}{25}& -  \frac{329}{864}&  -\frac{4}{7}& - \frac{105}{128}&  - \frac{92}{81}&  -\frac{1221}{800}\\
\eta(\CD^{\rm sig}_{t=0}, S^7/\BZ_k ) & 0& - \frac{2}{27}& -\frac{1}{2}&  -\frac{36}{25}&  -\frac{82}{27}&   -\frac{38}{7}&   -\frac{35}{4}&   -\frac{1064}{81}&   -\frac{468}{25}
\end{array}.
\end{equation}
Notice the agreement with the results in Sec.~\ref{sec:torus} for $k=2,3,4,6$.
The present method is valid for any $k$.

\section{Non-unitary counterexamples to cobordism classification}\label{sec:nonunitary}

The recent understanding of the anomaly and the corresponding invertible phases states that 
unitary topological invertible phases are in bijection to the Pontryagin dual of the bordism classes \cite{Kapustin:2014tfa,Kapustin:2014dxa,Freed:2016rqq,Yonekura:2018ufj}.
This statement is often called the cobordism classification of the invertible phases.
In particular, the partition function of a unitary topological invertible phase is a bordism invariant.

Here we present a simple class of non-unitary invertible topological phases whose partition function is not a bordism invariant;
in particular, its partition function on $S^D$ is $-1$.\footnote{%
Such non-unitary counterexamples in $D=4k+1$-dimensions were also treated briefly in Examples 6.11 and 6.15 of Freed's wonderful lecture notes \cite{FreedLectures}.
We also note that the subtlety of anomalies of non-unitary theories was discussed recently in \cite{Chang:2020aww}. 
}
These examples illustrate the necessity of the unitarity condition in the cobordism classification.
In this Appendix, $D$ is the spacetime dimensions of the bulk invertible phase, which was denoted by $d+1$ in the main part of the text.

\subsection{The simplest example}
The simplest example is given by a ``massive $bc$ ghost system'' in $D=1$ dimensions,
\beq
\CL = b \left( \i \frac{\d }{\d t} - m  \right) c  = - b \left(  \frac{\d }{\d \tau} +m  \right) c ,
\eeq
where $t$ is the time coordinate, and $\tau = \i t$ is the Euclidean time coordinate. 
$b$ and $c$ obey Fermi-Dirac statistics, but they are not spinors;
we will discuss their representations under Lorentz symmetry later for the case of general dimensions $D$, but for $D=1$ one can think of them just as scalars.
We regularize this theory by Pauli-Villars regularization with Pauli-Villars mass $M$,
and we take $m = - M $ and $M \to \infty$. Because $|m| \to \infty$, its Hilbert space is one-dimensional and spanned by the ground state.
We need only $\SO(D)$ symmetry to define this theory, and no spin structure is necessary.

When $|M| = | m| $, the partition function on $S^1$ is given by 
\beq
\CZ(S^1) = \frac{\det( \frac{\d }{\d \tau} +m) }{\det (\frac{\d }{\d \tau} +M)} = \frac{m}{M}.
\eeq
It is $\CZ(S^1) = -1$ for $m=-M$.
Obviously $S^1$ is trivial in the bordism group $\Omega^1_{\SO}(\text{pt})$. This gives a counterexample to the cobordism classification.

The point is that $b$ and $c$ obey the periodic boundary conditions and the zero mode contribute the above factor $m/M$. (Nonzero modes do not contribute to the phase of the partition function.)
If we instead consider a massive fermion, the circle $S^1_{\text{NS}}$ which is trivial in $\Omega^1_{\spin}(\text{pt})$ has the anti-periodic boundary condition, and in that case 
we get $\CZ(S^1_{\text{NS}})=+1$. This is consistent with the cobordism classification. For the periodic boundary condition,
we get $\CZ(S^1_{\text{R}})=-1$ which is of course consistent because $S^1_{\text{R}}$ is the nontrivial element of $\Omega^1_{\spin}(\text{pt}) = \BZ_2$.

\subsection{Abstract description}
The abstract description of the system which is closely related to the above one was discussed in \cite{Freed:2016rqq}.
Here we reproduce a sketch of the argument; we refer the reader to \cite{Freed:2016rqq,Yonekura:2018ufj} for more details on the axioms used below. 

First we give general discussions.
If we have two Hilbert spaces $\CH_A$ and $\CH_B$, then $\CH_A \otimes \CH_B$ and $\CH_B\otimes \CH_A$ are isomorphic.
We want a way to identify their elements, so we want to have a map
\beq
\tau : \CH_A \otimes \CH_B \to \CH_B \otimes \CH_A.
\eeq
such that
\beq
\tau :  \ket{a} \otimes \ket{b} \mapsto \epsilon_{a,b} \ket{b} \otimes \ket{a},
\eeq
where $\epsilon_{a,b}$ is a sign factor.
Mathematically, we need such $\tau$ to define the symmetric monoidal category of super vector spaces.

Physically, we might expect the states $\ket{a}$ and $\ket{b}$ to have statistics ${\rm deg} (a)$ and ${\rm deg} (b)$.\footnote{
On non-compact spaces, there are physical states which have more nontrivial statistics than signs ${\rm deg}(a) =\pm 1$, such as anyon particles in 3-dimensions.
However, on compact spaces, it seems that such generalized statistics do not arise and hence we assume that it is described just by signs.
}
Mathematically this is the $\BZ_2$ grading in super vector spaces.
${\rm deg}$ is even for Bose-Einstein statistics and odd for Fermi-Dirac statistics. Then we expect
\beq
\epsilon_{a,b} = (-1)^{{\rm deg}(a ){\rm deg}(b )  }.
\eeq

Now we want to compute the partition function on $S^1$. For this purpose, we first consider an interval $I=[0,\beta]$ and glue the two ends.
The amplitude on the interval $I$ is 
\beq
e^{-\beta H} = \sum e^{-\beta E_a} \ket{a} \otimes \bra{a} \in \CH_A \otimes \CH_A^*
\eeq
in the obvious notation. To glue the two ends we need to exchange $ \ket{a} \otimes \bra{a}$ to $\bra{a} \otimes  \ket{a}  $ and
then use a natural map (gluing) $\bra{a} \otimes \ket{b} \to \bra{a} b \rangle$. However, this exchange gives the sign
\beq
\tau :   \ket{a} \otimes \bra{a} \mapsto (-1)^{{\rm deg}(a) } \bra{a} \otimes  \ket{a}  .
\eeq
Let us also consider another quantity, which we denote as ${\rm f}(a)$ and is defined by
\beq
(-1)^F \ket{a} = (-1)^{{\rm f}(a)} \ket{a}.
\eeq
Here $(-1)^F \in \Spin(D)$ is in the center. This quantity determines whether the state $\ket{a}$ is a spinor or not.

Now the partition function on $S^1$ is given by\footnote{
We implicitly assume $\bra{a}a\rangle=1$ on the Hilbert space on the space which is a single point ${\rm pt}$. 
For the case of the massive $bc$ ghost system, this may be true since its quantization on ${\rm pt}$
can be done in the same way as the massive fermion. However, the nontrivial $S^1$ partition function implies
that the Hilbert space on ${\rm pt} \sqcup {\rm pt}$ may have a negative inner product.
}
\beq
\CZ(S_{\rm{R}}^1) &= \sum e^{-\beta E_a} (-1)^{{\rm deg}(a) } , \\
\CZ(S_{\rm{NS}}^1) &= \sum e^{-\beta E_a} (-1)^{{\rm deg}(a) + {\rm f}(a)},
\eeq
where we have used the fact that $S_{\rm{NS}}^1$ has an additional insertion of $(-1)^F \in \Spin(D)$.
What makes the usual thermal partition function positive definite is the spin-statistics theorem ${\rm deg}(a) + {\rm f}(a)=0 \mod 2$.

The massive $bc$ system considered above has ${\rm deg}(\Omega) = 1$ but ${\rm f}(\Omega)=0$,
where $\ket{\Omega}$ is the ground state, which is the only state in the limit of large mass gap.
Therefore, we get $\CZ(S^1)=-1$.

\subsection{Generalizations to odd dimensions}
We generalize the above system to theories in dimension $D=2n+1$.
The following discussion is generally valid for $D = 4\ell+1$, but for $D=4\ell+3$ we will need to restrict to some specific dimensions
as we discuss later.

Consider a manifold $Y$ with $\dim Y = D=2n+1$.
We take $c$ and $b$ to be sections of $\CS \otimes \CS^*$, where $\CS$ is the spin bundle on $Y$ and $\CS^*$ is the dual to $\CS$. 
For this theory itself, we do not need spin structure because of the relation
\beq
\CS \otimes \CS^* \cong \sum_{i=0}^n \wedge^{2i} T^* Y.
\eeq
Notice that only even degrees appear.
There is an isomorphism $\wedge^{2i} T^* Y \cong \wedge^{D-2i} T^* Y$, so we could have used odd degree instead.
Sec.~\ref{sec:Signature} for the details.

The Lagrangian of the theory is 
\beq
\CL = - b( \slashed{D}+m)c.
\eeq
Here gamma matrices $\Gamma^I$ of $\slashed{D} = \Gamma^I D_I$ only act on the first factor $\CS$ in $\CS \otimes \CS^*$,
while the covariant derivative $D_I$ acts on both factors. Thus $\CD := \i \slashed{D}$ is the Dirac operator
relevant for the signature index theorem, which we discussed in detail in Sec.~\ref{sec:Signature}.
However, we remark that $\CD^{\rm sig}_Y$ in Sec.~\ref{sec:Signature} acts on $\CS \otimes  (\CS^* \oplus \CS^*)$,
so $\eta(\CD^{\rm sig}_Y)$ of that section is twice the $\eta(\CD)$ of this appendix. 

As before, we take $m = -M$ and $M \to \infty$. Then the partition function is
\beq
\CZ(Y) = \exp( -2 \pi \i \eta(\CD)). \label{eq:nonunitaryPT}
\eeq

The zero modes of the operator $\CD$ is given by the zero modes on $\sum_{i=0}^n \wedge^{2i} T^* Y$.
Let us define
\beq
b = \sum_{i=0}^{n} \dim H^{2i}(Y,\BR),
\eeq
which is the sum of the Betti numbers of even dimensional cohomology. 
Then the number of zero modes is given by $b$, and 
\beq
\eta(\CD) = \frac{b}{2} + \frac{1}{2} \sum_{\lambda \neq 0} \frac{\lambda}{|\lambda|},
\eeq
where the sum is over nonzero eigenvalues of $\CD$.
By the discussion in Sec.~\ref{sec:Signature}, this sum over nonzero eigenvalues is the same as $\eta(\widetilde \CD^{\rm sig}_Y)$ of that section
and hence we get $\eta(\CD)  = b/2 + \eta(\widetilde \CD^{\rm sig}_Y)$. 

Suppose that $Y$ is a boundary of a $D+1$-manifold $Z$.
By the signature index theorem \eqref{eq:signind} and the fact that $\eta(\CD)  = b/2 + \eta(\widetilde \CD^{\rm sig}_Y)$,
we have
\beq
 \eta(\CD)  =  \frac{1}{2} \left( b+\sigma(Z) - \int_Z L \right). 
\eeq
Here the signature is defined by the pairing on the relative cohomology $H^\bullet ( Z,  \partial Z, \BR )$.

Let us consider the partition function on $Y = S^D$.
It has $b=1$ coming from $H^0(S^D,\BR)$.
We emphasize that  $H^D(S^D,\BR)$ does not contribute because we summed only over even-degree cohomology.
We can take the $(D+1)$-manifold as $Z = B^{D+1}$, which is the $(D+1)$-dimensional ball. 
The signature $\sigma(Z)$ and the Hirzebruch polynomial $L$ (for a round sphere metric) are zero on the ball $B^{D+1}$.
We conclude that
\beq
\eta(\CD_{S^D}) = \frac{1}{2}
\eeq
and hence
\beq
\CZ(S^D) = -1.
\eeq
Notice that $S^D = \partial B^{D+1}$ is trivial in the bordism groups of both $\SO$ and $\Spin$.

The above discussion was general for any $D=2n+1$.
For $D=4\ell+1$, there is no perturbative gravitational anomaly in $d=4\ell$ and hence the invertible theory on $D=4\ell+1$ is topological.
In fact, in these dimensions nonzero modes always cancel and we have
\begin{equation}
\CZ(Y) = (-1)^{b}=(-1)^{\sum_i \dim H^{2i}(Y)}. 
\end{equation}
It is interesting that the partition function is determined by Betti numbers.

In dimensions $D=7$ and $D =8\ell+3$, we can combine the above theory with the anomaly theories relevant to fermions and self-dual 2-form fields to cancel the perturbative anomaly.
Then we get a topological theory. The anomaly theories for fermions and 2-form fields are unitary, and hence they do not change the above conclusion $\CZ(S^D)=-1$.

\subsection{Remark on the anomaly polynomial and the Euler characteristic class} 
Some theories we have discussed above, such as the ones in $D=4\ell+1$-dimensions, are topological theories in the sense that partition functions are topological invariants. 
But they can still be interpreted to have anomaly polynomials which are given by the Euler characteristic class.
Let us explain this point. 

To see the point clearly, and also to generalize the situation slightly,
we consider a bundle $P$ with the structure group $\Spin(D+1)$ which is not necessarily associated to the tangent bundle.
More precisely, we consider a $(D+1)$-manifold $Z$ with a bundle whose structure group is given by $[\Spin(D+1)_1 \times \Spin(D+1)_2]/\bZ_2$, where the first
$\Spin(D+1)_1$ is the Lorentz group and the second $\Spin(D+1)_2$ is the one we have introduced above. 
Below we consider as if the group is $\Spin(D+1)_1 \times \Spin(D+1)_2$ just for notational simplicity, but the discussions below make sense even if we divide the group by $\bZ_2$. 

Let $\CT_{\pm}$ be the spin bundles with positive and negative chirality associated to $P$.
We consider Dirac operators $\cD_{Z,\CT^*_{\pm}}$ coupled to $\CT^*_{\pm}$.
Namely, it acts on sections of $\CS_{Z} \otimes \CT^*_{\pm}$, where $ \CS_{Z}  = \CS_+ \oplus \CS_-$ is the spin bundle on the manifold $Z$ associated to the tangent bundle $TZ$.
Suppose that the manifold $Z$ has a boundary $\partial Z =Y$. We denote the relevant APS $\eta$-invariants as $\eta(\cD_{Y,\CT^*_{\pm}} ) $.
Here the Dirac operator $\cD_{Y,\CT^*_{\pm}}$ acts on sections of $\CS_Y \otimes \CT^*_{\pm}$ where $\CS_Y := \CS_+|_{Y} = \CS_-|_{Y}$.
By using the APS index theorem, one can see that\,\footnote{
Let $\pm y_i~ (i=1,\cdots,(D+1)/2)$ be the Chern roots of the bundle $P$ in the vector representation of $\Spin(D+1)_2$.
This means that we regard the curvature 2-form (multiplied by $\i/2\pi$), $\frac{\i}{2\pi}F$, in the vector representation
to have eigenvalues $\pm y_i$. Then, the $\frac{\i}{2\pi}F$ in the spinor representations $\CT^*_{\pm}$ have eigenvalues 
$
\frac12 (\pm y_1 \pm y_2 \cdots \pm y_{(D+1)/2} ),
$
such that the product of the signs $(\pm)$ of the coefficients is equal to $+1$ for $\CT^*_{+}$ and $-1$ for $\CT^*_{-}$.
From this fact, we see that the Chern characters of $\CT^*_{\pm}$ are given by
$(\tr_{\CT^*_{+}} + \tr_{\CT^*_{-}})  \exp\left( \frac{\i}{2\pi}F \right) = \prod_{i} (e^{y_i/2}+ e^{-y_i/2})$
and $(\tr_{\CT^*_{+}} - \tr_{\CT^*_{-}})  \exp\left( \frac{\i}{2\pi}F \right) = \prod_{i} (e^{y_i/2}- e^{-y_i/2})$.
In particular, taking the $(D+1)$-form part, we get $(\tr_{\CT^*_{+}} - \tr_{\CT^*_{-}})  \exp\left( \frac{\i}{2\pi}F \right)|_{(D+1)\text{-form}}  = \prod_i y_i :=E$ which is by definition
the Euler characteristic class. 
}
\beq
&{\rm index}(\cD_{Z,\CT^*_{+}}) - {\rm index}(\cD_{Z,\CT^*_{-}}) = \int_Z E + \left( \eta(\cD_{Y,\CT^*_{+}} ) -\eta(\cD_{Y,\CT^*_{-}} )  \right) \\
&{\rm index}(\cD_{Z,\CT^*_{+}}) + {\rm index}(\cD_{Z,\CT^*_{-}}) = \int_Z I(p, p') +  \left( \eta(\cD_{Y,\CT^*_{+}} ) + \eta(\cD_{Y,\CT^*_{-}} )  \right)
\eeq
where $E$ is the Euler characteristic class of $P$, and $I(p,p')$ is some polynomial of the Pontryagin classes of the tangent bundle (denoted as $p=\{p_1,p_2,\cdots\}$) and 
the Pontryagin classes of the bundle $P$ (denoted as $p'=\{p'_1,p'_2,\cdots\}$).
From the above formulas, we get
\beq
-\eta(\cD_{Y,\CT^*_{+}} ) =  \frac12 \int_Z \left(E +  I(p, p')\right)  - {\rm index}(\cD_{Z,\CT^*_{+}})  
\eeq
In particular, we see that $\int_Z \left(E +  I(p, p')\right)$ is divisible by 2 if the boundary is empty, $\partial Z = \varnothing$. 

Now let us consider the case that the bundle $P$ is associated to the tangent bundle so that $\CT_{\pm} = \CS_{\pm}$.
In this case, $\CS :=\CS_Y = \CT_{+}|_{Y} = \CT_{+}|_{Y}  $.
The above result implies that $\frac12  \left(E +  I(p, p)\right)$ (where we have set $p'=p$) is the anomaly polynomial for the theory \eqref{eq:nonunitaryPT}.
Notice that if $D+1 = 4\ell+2$, the term $\int I(p,p')$ is zero.
Also, for a hemisphere which has the standard metric, we have 
\beq
 \exp\left[2\pi \i \cdot  \frac12 \int_{\rm hemisphere} \left(E +  I(p, p') \right) \right]= -1,
 \eeq
 since the Euler number of the hemisphere is 1. 
 More generally, the Euler number is a topological invariant even for manifolds with boundaries. (We usually need a term on the boundary related to extrinsic curvature,
 but the APS index theorem assumes that the region near the boundary is of cylindrical form, so the extrinsic curvature is zero.) 
 Thus the theory \eqref{eq:nonunitaryPT} (for $D=4\ell+1$) is topological even though it has a nontrivial anomaly polynomial $\frac12 E$.
 
Let us remark that the theories considered here can still be related to bordism theory in a certain generalization. 
As we have seen above, the polynomial $  \frac12  \left(E +  I(p, p')\right) $ satisfies the integrality condition $  \frac12 \int \left(E +  I(p, p')\right) \in \bZ$
on closed manifolds. This is valid for any $P$ not necessarily associated to the tangent bundle. In particular, it is valid in the special case
that the bundle $P \times_{\rho} \bR^{D+1}$ (where $\rho$ is the vector representation of $\Spin(D+1)_2$) is only stably isomorphic to the tangent bundle $TZ$, that is,
$P \times_{\rho} \bR^{D+1} \oplus \bR^K \simeq TZ \oplus \bR^K$ for some $K$.
By the result of \cite{Yamashita:2021cao}, the theory in $D$-dimensions considered above give an element of the Anderson dual to a certain bordism theory, $(I\Omega^G)^{D+1}(\text{pt})$. 
Here, we need to take the sequence of groups $G=\{G_k\}$ in that paper to be $G_k=\SO(k)$ for $k \leq D+1$, 
and $G_k = \SO(D+1)$ for $k >D+1$ so that we can allow the Euler density $E$ as a possible characteristic class. 

Finally, we note that the choice of $G=\{G_k\}$ above does not satisfy the conditions\footnote{%
The relevant conditions are that 1) the image of $G_k \to \mathrm{O}(k)$ includes $\SO(k)$, and that 2) the commutative diagram $ \begin{array}{c@{}c@{}c}
G_k &\to &\mathrm{O}(k) \\
\downarrow & & \downarrow\\
G_{k+1} & \to & \mathrm{O}(k+1)
\end{array}$ is a pull-back diagram.
} discussed by Freed and Hopkins~\cite{Freed:2016rqq}, 
which are required to formulate  reflection positivity. 
Therefore, there is no contradiction with the results of \cite{Freed:2016rqq,Yonekura:2018ufj}, 
which relates unitary invertible theories with Anderson/Pontryagin duals of bordism groups for $G=\{G_k\}$ 
satisfying the conditions of Freed and Hopkins.

\bibliographystyle{ytphys}
\bibliography{ref}

\end{document}